%% file: these-arXiv.tex
\RequirePackage{snapshot}
\documentclass[12pt]{book}

\usepackage[T1]{fontenc}
\usepackage[utf8x]{inputenc}
\usepackage{these}

\usepackage[nottoc]{tocbibind}
\french
\usepackage{amsmath,amssymb,amsfonts}
\usepackage{dsfont,epigraph}
\allowdisplaybreaks[1]

\usepackage{stmaryrd,wasysym,upgreek,latexsym}
\usepackage{calc}
\usepackage{empheq}
\usepackage{slashed}
\usepackage{subfig}
\usepackage{multicol}
\usepackage{lscape}

\usepackage[dvips]{graphicx}
\usepackage[dvips,all]{xy}

\usepackage[hyperref,thmmarks,amsmath]{ntheorem}
\usepackage{fabmesthm}
\usepackage{fabmacromath}

\usepackage{shorttoc}

\usepackage[vflt]{floatflt}
\usepackage{afterpage}

\newcommand{\lrtimes}{\super{\ltimes}{\rtimes}}

\renewcommand{\tau}{\uptau}

\def\today{%
        \the\day\the\month\the\year}

\input{entete}

\begin{document}

\version{}
\pagedegarde

\include{resume}

\tableofcontents
\listoffigures


\include{remerciements}
\include{intro-arXiv}
\include{renormalisation-arXiv}
\include{matrixbase-arXiv}
\include{phi4-arXiv}
\include{GN-arXiv}
\include{conclusion-arXiv}

\appendix
\include{GNapp-arXiv}

\bibliographystyle{myalpha}
\bibliography{biblio-articles,biblio-books}
\end{document}

%% file: entete.tex
 \etablissement{
  \raisebox{-30pt}[0pt][0pt]{
     \hspace*{-15pt}\makebox[0pt][l]{
       \includegraphics{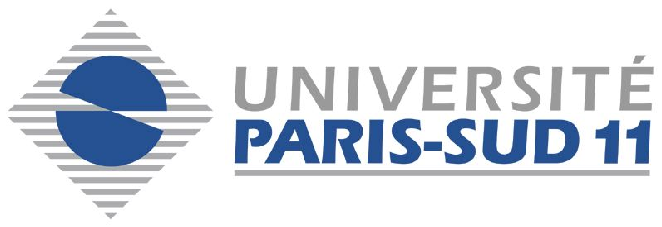}
     }}
 }

\universite{Université Paris 11} 
\discipline{physique théorique}
\auteur{Fabien Vignes-Tourneret} 
\adresseelectronique{fabien.vignes@th.u-psud.fr}

\titre{Renormalisation des th\'eories de champs non commutatives}  
\directeur{Vincent Rivasseau} 
\datesoutenance{14 septembre 2006}

\jury{%
  Costas& Bachas & président\\%
&&\\    
  Vincent& Rivasseau & directeur\\%
&&\\            
  Krzysztof& Gaw\c{e}dzki & rapporteur \\%
  Raimar& Wulkenhaar & rapporteur\\%
&&\\
  Alain& Connes & examinateur \\%
  Christoph& Kopper & examinateur%
}


%% file: resume.tex
\pagestyle{empty}
\vfill
\begin{center}
  \textbf{\large Résumé}
\end{center}
La physique des très hautes énergies nécessite une description cohérente des quatre forces fondamentales. La géométrie non commutative représente un cadre mathématique prometteur qui a déjà permis d'unifier la relativité générale et le modèle standard, au niveau classique, grâce au principe de l'action spectrale. L'étude des théories quantiques de champs sur des espaces non commutatifs est une première étape vers la quantification de ce modèle. Celles-ci ne sont pas simplement obtenues en récrivant les théories commutatives sur des espaces non commutatifs. En effet, ces tentatives ont révélé un nouveau type de divergences, appelé mélange ultraviolet/infrarouge, qui rend ces modèles non renormalisables. H.~Grosse et R.~Wulkenhaar ont montré, sur un exemple, qu'une modification du propagateur restaure la renormalisabilité. L'étude de la généralisation de cette méthode est le cadre de cette thèse. Nous avons ainsi étudié deux modèles sur espace de Moyal qui ont permis de préciser certains aspects des théories non commutatives. En espace $x$, la principale difficulté technique est due aux oscillations de l'interaction. Nous avons donc généralisé les résultats de T.~Filk afin d'exploiter au mieux ces oscillations. Nous avons pu ainsi distinguer deux types de mélange, renormalisable ou pas. Nous avons aussi mis en lumière la notion d'orientabilité : le modèle de Gross-Neveu non commutatif orientable est renormalisable sans modification du propagateur. L'adaptation de l'analyse multi-échelles à la base matricielle a souligné l'importance du graphe dual et représente un premier pas vers une formulation des théories de champs indépendante de l'espace sous-jacent.
\english

\vfill
\begin{center}
  \textbf{\large Abstract}
\end{center}
Very high energy physics needs a coherent description of the four fundamental forces. Non-commutative geometry is a promising mathematical framework which already allowed to unify the general relativity and the standard model, at the classical level, thanks to the spectral action principle. Quantum field theories on non-commutative spaces is a first step towards the quantification of such a model. These theories can't be obtained simply by writing usual field theory on non-commutative spaces. Such attempts exhibit indeed a new type of divergencies, called ultraviolet/infrared mixing, which prevents renormalisability. H.~Grosse and R.~Wulkenhaar showed, with an example, that a modification of the propagator may restore renormalisability. This thesis aims at studying the generalization of such a method. We studied two different models which allowed to specify certain aspects of non-commutative field theory. In $x$ space, the major technical difficulty is due to oscillations in the interaction part. We generalized the results of T.~Filk in order to exploit such oscillations at best. We were then able to distinguish between two mixings, renormalizable or not. We also bring the notion of orientability to light : the orientable non-commutative Gross-Neveu model is renormalizable without any modification of its propagator. The adaptation of multi-scale analysis to the matrix basis emphasized the importance of dual graphs and represents a first step towards a formulation of field theory independent of the underlying space.

\french
\vfill

%% file: remerciements.tex
\pagestyle{empty}
\chapter*{Remerciements}
\addcontentsline{toc}{chapter}{\protect\numberline{}Remerciements}
\chaptermark{Remerciements} 

Il y a déjà cinq jours que j'ai soutenu et c'est avec un grand soulagement que je rédige ces remerciements.

\medskip
Je voudrais tout d'abord remercier Vincent d'avoir accepté de diriger mon travail durant ces trois dernières années. Il a fait preuve d'une patience et d'une disponibilité à toute épreuve. Grâce à ses qualités d'enseignant et de pédagogue, il m'a appris la renormalisation, l'analyse multi-échelles, l'art de faire des bornes et surtout celui de la recherche. Ses qualités humaines en font un homme exceptionnel dont la compagnie est très stimulante. Le choix d'un sujet et (surtout) d'un directeur de thèse n'est pas toujours facile mais l'impression que j'ai eue la première fois que j'ai rencontré Vincent a été déterminante. Ce jour-là, j'ai fait le bon choix. \textit{A posteriori} je n'aurai pu rêver meilleur directeur. J'éspère que notre collaboration et notre amitié se poursuivront encore longtemps.

\medskip
Je remercie le Laboratoire de Physique Théorique d'Orsay, notamment Dominique Schiff et Hendrik-Jan Hilhorst, pour m'avoir offert des conditions de travail excellentes. Merci aussi à l'équipe administrative et technique sans qui le temps dédié à la recherche serait encore plus court.

\medskip
Mes plus sincères remerciements vont aux membres de mon jury de thèse. Costas Bachas l'a présidé alors qu'il n'est pas un spécialiste du développement en espace des phases. Je l'en remercie. Pour moi, il fut très agréable de soutenir devant l'un de mes anciens professeurs.

Ce sont les résultats de Raimar Wulkenhaar et Harald Grosse qui ont inspiré notre travail. La collaboration avec Raimar est très agréable et enrichissante. Je le remercie d'avoir accepté d'écrire un rapport sur ma thèse.

Merci également à Krzysztof Gaw\c{e}dzki qui s'est plongé dans les théories de champs non commutatives et a écrit un rapport très complet en un minimum de temps.

Je suis très heureux que Christoph Kopper ait été membre de mon jury. Il s'est rapidement intéressé à mon travail. Son expertise du modèle de Gross-Neveu est précieuse.

Enfin, toute ma gratitude va à Alain Connes qui s'est intéressé très tôt à notre travail. Il n'a malheureusement pas pu être présent le jour de ma soutenance mais a accepté de diriger une estimation de ma thèse lorsque nous étions à Cambridge. Il m'a par ailleurs beaucoup soutenu pour mes candidatures de postdoc. Pour tout cela, je le remercie et suis très fier de le compter parmi les membres du jury.

\medskip
Durant les derniers mois de ma thèse, j'ai eu le plaisir de collaborer étroitement avec Jacques Magnen. Ce fut un réel plaisir. Jacques impressionne par le nombre d'idées qu'il a à la seconde et par sa capacité à prendre de la hauteur. Il est l'un des pionniers de l'analyse multi-échelles, j'ai beaucoup appris à son contact. Je le remercie de m'avoir aidé et soutenu lors de ces derniers mois difficiles et souhaite vivement que notre collaboration future soit aussi fructueuse.

Si mes années de thèse se sont aussi bien déroulées, c'est aussi grâce aux personnes qui m'entourent au laboratoire. Je voudrais tout d'abord remercier Thierry pour sa disponibilité, sa rigueur et son esprit critique. Il est aussi un grand expert de \LaTeX{} ce qui ne fut pas dénué d'intérêt lors de la phase de rédaction.

Un grand merci à Emmanuel pour m'avoir tant aidé. Sa clarté d'esprit et son expertise en informatique me furent précieuses. Il m'a également donné l'envie de jouer de la guitare et est maintenant un véritable ami que  la recherche française n'a malheureusement pas su retenir.

Je remercie également Jean-Christophe Wallet pour sa bonne humeur et pour m'avoir rappelé que dans \og{}physique mathématique\fg{}, il y a aussi \og{}physique\fg{}.

Merci à Razvan Gurau pour son dynamisme et à Marco Maceda pour nos échanges intéressants.

Ces trois années au LPT n'auraient pas été si agréables sans la présence des jeunes mariés qui ont animé nos pauses-café de conversations de toutes sortes.

\medskip
Je souhaite aussi remercier tous les enseignants qui, au long de ma scolarité, m'ont tant appris. Ils m'ont soutenu et ont développé ma curiosité et mon intérêt pour la physique et les mathématiques. Grâce à eux, apprendre a toujours été un plaisir.

\medskip
Comment exprimer ma gratitude et ma reconnaissance à ma famille et mes amis ? Ils m'ont soutenu, écouté et se sont intéressés à mon travail. C'est grâce à eux que j'en suis arrivé là. Ce fut tellement agréable (peut-être pas réellement sur le moment) de conclure mes années d'étude par une soutenance publique où bon nombre d'entre eux furent présents. Je remercie tout particulièrement mes parents (tous) de m'avoir transmis leurs valeurs, ma grand-mère Éva qui a fait le voyage depuis Grenoble, mes grand-parents Elvyre et Yves qui ont grandement contribué à mon éducation, mes frères pour m'avoir appris le partage.

\medskip
Je terminerai en exprimant l'amour, la gratitude et le respect que j'ai pour Sandra, la femme qui partage ma vie depuis déjà sept ans. Elle a subi toutes mes études supérieures et notamment ces trois dernières années qui ne furent pas les plus faciles. Elle m'a soutenu et aidé durant les moments de doute et de stress qui accompagnent une thèse. Sans elle, rien de tout cela n'aurait été possible.

\cleardoublepage%
\pagestyle{myheadings}


%% file: intro-arXiv.tex
\renewcommand{\theequation}{\arabic{equation}}
\chapter*{Introduction}
\addcontentsline{toc}{chapter}{\protect\numberline{}Introduction}
\chaptermark{Introduction}
\epigraph{Une utopie est une réalité en puissance.}{Édouard Herriot}
\begin{rem}
  Avant de commencer, je voudrais expliquer une convention lexicale que j'ai utilisée tout au long de ce manuscrit. J'ai employé la première personne du pluriel (nous) quand j'ai voulu rendre compte de travaux que j'ai effectués avec d'autres ou de discussions. J'ai écrit à la première personne du singulier quand il s'agissait de ne donner que ma propre opinion, qui n'engage que moi. Enfin, l'emploi de la troisième personne du singulier (on) se réfère à un groupe de personnes ou une communauté plus ou moins bien définie.  
\end{rem}

La description du monde par la physique théorique repose sur deux théories extrêmement bien vérifiées expérimentalement. La relativité générale d'Einstein décrit l'interaction entre l'espace-temps (classique) et la matière et l'énergie de l'univers. Cette théorie dont la première pierre fut la relativité restreinte, unifie l'espace et le temps et en fait un objet dynamique. Le modèle standard unifie les interactions électro-faible et forte et rend compte de la physique à l'échelle des constituants élémentaires de la matière. Du point vue de la physique théorique, la nature est ainsi séparée en deux domaines distincts qui obéissent respéctivement aux lois des deux théories précédentes. Une conséquence est que notre conception de l'espace-temps est ambiguë. En théorie des champs, l'espace est une donnée \emph{à priori}. Au contraire, en relativité générale, la distribution de matière et d'énergie détermine l'espace-temps (au moins sa géométrie) qui, à son tour, modifie cette distribution. L'objet est donc statique d'un côté, dynamique de l'autre.

Il y a au moins deux bonnes raisons de ne pas se satisfaire de cette situation. La première provient de considérations esthétiques. Quiconque apprécie la beauté des mathématiques ne peut adhérer à l'idée que la nature ait apparemment décidé d'unifier seulement trois des forces qui la régissent. L'état de la physique théorique des hautes énergies ne peut donc qu'être dû à notre incapacité à faire mieux. Il y a de fortes chances que des voix s'élèvent pour protester contre un tel argument. De quel droit érige-t-on en principe physique fondateur l'esthétique mathématique, notion subjective au demeurant ? Personne ne peut ni ne doit s'affranchir du couperet de l'expérience. Mais c'est, entre autres choses, le manque d'expérience qui fait défaut à la physique théorique moderne. Et je ne pense pas seulement au LHC mais également à notre incapacité à reproduire les phénomènes violents de l'univers. Dans cette situation, il faut se baser sur des principes ; les symétries et l'esthétique mathématique ont d'ailleurs toujours guidé les physiciens. Tout en n'étant pas aussi extrémiste que Dirac, je ne peux m'empêcher de le citer :
\begin{quotation}
  Une théorie mathématiquement belle a plus de chance d'être correcte qu'une théorie  inélégante, même si cette dernière décrit correctement les résultats expérimentaux. (P.~A.~M.~Dirac)
\end{quotation}
La seconde raison repose sur une constatation simple : il existe des phénomènes pour lesquels la gravitation et, au moins, une des trois forces du modèle standard sont pertinentes. Il s'agit par exemple de l'effondrement d'une étoile à neutrons ou des premières années de l'univers. Alors si on essaie de réunir ces deux théories, on rencontre un problème de taille.\\

En effet, la conclusion habituelle est que la structure de l'espace-temps est modifiée à très courte échelle. L'argument \og{}physique\fg{} standard qui conduit à ce résultat est le suivant. Il faut d'abord remarquer que la perception que nous avons de l'espace-temps est uniquement due aux évènements qui y prennent place. Ainsi, pour sonder la structure fine de l'espace, il est nécessaire de considérer une distribution de matière localisée sur une région très restreinte de l'espace. Puis pour tester sa présence, nous devons utiliser des particules de longueur d'onde inférieure au diamètre de cette distribution. En-deça d'un certain diamètre, l'énergie nécessaire est si grande qu'elle crée un trou noir dont le rayon est supérieur à celui de la distribution de matière, empêchant alors toute observation. Pour un observateur, l'espace à petite échelle perd sa continuité. Ceci se traduit par des relations d'incertitude sur les coordonnées de l'espace-temps. Cet argument est souvent utilisé pour justifier d'une éventuelle non commutativité de l'espace à courte échelle. En effet, dans \cite{Doplicher1994tu}, il a été montré que les relations d'incertitude mentionnées ci-dessus peuvent se déduire d'une algèbre d'opérateurs non commutative qui remplacerait les opérateurs $\hat{x}^{\mu}$ habituels.

Personnellement je ne crois pas en cet argument pour la raison suivante. Le domaine d'applicabilité d'une théorie est fixé par l'expérience. Ainsi la mécanique quantique et la relativité générale n'ont pas réellement de recouvrement. L'argument précédent repose sur une extrapolation de la relativité générale à un domaine où elle n'a pas été testée. La seule conclusion que l'on peut en tirer est que la relativité générale ne s'applique pas à ces échelles. De plus, en supposant que l'espace-temps devienne flou à courte échelle, cela n'entraîne pas nécessairement qu'il soit \ncf{}. D'autres théories telles que la théorie des cordes, ont aussi pour conséquence des relations d'incertitude sur la position. La justification de la pertinence de la géométrie \ncv{} pour la physique se trouve ailleurs.

Étant admise la nécessité d'unifier le modèle standard et la gravitation (ou au moins de leur trouver un cadre commun), il me semble que le meilleur argument pour la géométrie \ncv{} (outre son extraordinaire richesse mathématique) est le suivant. La géométrie \ncv{} est une reformulation et une généralisation de la géométrie ordinaire en des termes algébriques et d'analyse fonctionnelle. Elle utilise les mêmes outils que la mécanique quantique (les opérateurs sur un espace de Hilbert) et contient toute la géométrie classique. Elle pourrait donc être le cadre approprié à une description unifiée des quatre forces fondamentales. Pour preuve du potentiel de la géométrie \ncv{}, mentionnons tout d'abord les travaux de M.~Dubois-Violette, R.~Kerner et J.~Madore qui ont unifié le champ de Higgs et les champs de jauge standards en un champ de jauge unique \cite{Dubois-Violette1988ir,Dubois-Violette1989vq}. L'unification du champ de Higgs et du champ de jauge du modèle standard sur un espace quadri-dimensionnel à deux feuillets  constitue le modèle de Connes-Lott \cite{Connes1990qp}. Encore plus fort, le modèle de Connes-Chamseddine \cite{Chamseddine1996zu} unifie les champs de Yang-Mills-Higgs et le graviton. Malheureusement une telle unification est aujourd'hui restreinte au niveau classique. Remarquons également que les espaces \ncf{}s ont, en un certain sens, un caractère universel. Par là, j'entends que la théorie des cordes qui a également pour ambition d'unifier toutes les forces connues, a pour limite des théories de champs sur espaces \ncf{}s \cite{a.connes98noncom,Schomerus1999ug,Seiberg1999vs}.\\

L'excellent accord entre les prédictions du modèle standard et l'expérience nous montre que d'une part, celui-ci est très bien choisi et d'autre part que la théorie des champs est un outil puissant et efficace. Au coeur de cet outil se trouve la renormalisation qui permet de donner un sens aux infinis de la théorie. Il est donc naturel d'essayer d'utiliser ces mêmes outils pour la quantification des modèles \ncf{}s. Plutôt que d'attaquer la gravitation quantique de front, il est plus prudent d'étudier d'abord la structure du groupe de renormalisation sur des espaces \ncf{}s simples et fixés. Les théories de champs sur espace \ncf{} sont alors une étape intermédiaire entre l'échelle de la QCD et celle de la gravité quantique. L'idée de considérer des théories de champs sur des espaces \ncf{}s n'est pas nouvelle. Elle remonte à Schrödinger, Heisenberg et Peierls mais le premier article concernant une algèbre \ncv{} représentant l'espace-temps est dû à Snyder \cite{snyder47quant-space-time}. La motivation principale concernait les divergences ultraviolettes de la théorie des champs. L'espoir était qu'une théorie écrite sur un espace flou (sans points) ne présenterait plus ces divergences. À l'époque, l'idée n'a rien donné et la réussite de la renormalisation a fait oublier cette approche. Depuis les travaux d'Alain Connes sur la géométrie \ncv{} et l'apparition des théories de champs \ncv{}s en théorie des cordes, l'intérêt de la communauté des physiciens théoriciens pour les théories de champs \ncv{}s est ravivé. Encore une fois, l'espoir est né d'écrire une théorie sans divergence ultraviolette.

L'espace \ncf{} le plus simple et le plus étudié (du point de vue de la théorie des champs) est le plan de Moyal. Malgré son nom, cet espace peut être défini en n'importe quelle dimension. Il s'agit d'une déformation de l'espace plat $\R^{n}$ où les coordonnées satisfont les relations de commutations
\begin{align}
  \lsb x^{\mu},x^{\nu}\rsb=&\imath\Theta^{\mu\nu}
\end{align}
avec $\Theta$ une matrice anti-symétrique $n\times n$. Ses entrées ont la dimension d'une aire et leurs racines représentent une longueur minimale. Pour écrire une théorie de champs sur le plan de Moyal, on remplace habituellement l'algèbre des fonctions sur $\R^{n}$ par l'algèbre engendrée par la relation de commutation précédente. Il est alors possible de définir un isomorphisme entre cette dernière algèbre et les fonctions sur $\R^{n}$ munies d'un produit \ncf{}. C'est le produit de Moyal dont nous verrons une définition précise au chapitre \ref{cha:base-matr}. Le lagrangien d'une théorie des champs sur espace de Moyal consisterait donc en le lagrangien ordinaire où le produit point par point est remplacé par le $\star$-produit de Moyal. Par exemple, le modèle $\phi^{4}_{n}$ \ncf{} est donné par
\begin{align}
  S[\phi] =&\int d^nx \Big( -\frac{1}{2} \partial_\mu \phi
\star \partial^\mu \phi  + \frac{1}{2} m^2
\,\phi \star \phi
+ \frac{\lambda}{4} \phi \star \phi \star \phi \star
\phi\Big)(x).\label{eq:phi4naif}
\end{align}
Le produit de Moyal a pour principale caractéristique d'être non local. Néanmoins sa non commutativité a pour conséquence que le vertex de la théorie \eqref{eq:phi4naif} n'est invariant que sous les permutations cycliques des champs. Cette invariance restreinte nous incite à représenter les graphes de Feynman associés par des graphes à rubans. On peut alors facilement faire la distinction entre graphes planaires et non planaires.

La non localité du $\star$-produit permet de comprendre la découverte de Minwalla, Van Raamsdonk et Seiberg \cite{MiRaSe}. En effet, ils ont montré que non seulement le modèle \eqref{eq:phi4naif} n'est pas fini dans l'ultraviolet mais encore qu'il présente un nouveau type de divergences qui le rendent non renormalisable. Dans l'article \cite{Filk1996dm}, Filk a calculé les règles de Feynman relatives au modèle \eqref{eq:phi4naif}. Il a montré que les amplitudes planaires sont égales à celles de la théorie commutative. Par contre, les graphes non planaires donnent lieu à des oscillations qui couplent les pattes internes et externes. L'exemple typique est celui du tadpole non planaire :
\begin{align} 
  \raisebox{-0.4\totalheight}{\includegraphics{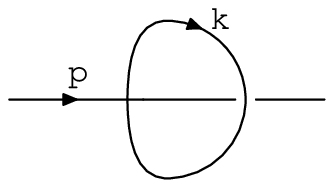}}&=\frac{\lambda}{12} \int \frac{d^4k}{(2\pi)^4} \frac{e^{ip_{\mu} k_{\nu}
      \Theta^{\mu\nu} }}{k^2 + m^2}\notag\\ 
  &=  \frac{\lambda}{48\pi^2}  \sqrt{\frac{m^2}{(\Theta p)^2}}  K_1(\sqrt{m^2
    (\Theta p)^2})\seq_{p\to 0}p^{-2}.
\end{align}
Si $p\neq 0$, le tadpole non planaire est fini mais, à $p$ petit, il diverge comme $p^{-2}$. Autrement dit, si nous mettons une coupure ultraviolette $\Lambda$ à l'intégrale sur $k$, les limites $\Lambda\to\infty$ et $p\to 0$ ne commutent pas. C'est le phénomène de mélange ultraviolet/infrarouge. Une chaîne de tadpoles non planaires, insérée dans de plus grands graphes, peut faire diverger n'importe quelle fonction (à six points ou plus). Or cette divergence n'est pas locale et ne peut donc pas être absorbée dans une redéfinition de la masse. C'est ce qui rend le modèle non renormalisable. Nous verrons au chapitre \ref{cha:GN} que le mélange UV/IR se traduit par un couplage des différentes échelles de la théorie. Nous verrons également qu'il convient de distinguer plusieurs types de mélanges.

Le mélange UV/IR a été étudié en détails par plusieurs groupes. Tout d'abord, Chepelev et Roiban \cite{Chepelev2000hm} ont démontré un comptage de puissance pour plusieurs modèles scalaires. Ils ont ainsi identifié précisemment les graphes divergents et ont pu classé les divergences de la théorie par les données topologiques de ses graphes. Puis V.~Gayral \cite{Gayral2004cs} a montré la présence de mélange UV/IR sur toutes les déformations isospectrales (il s'agit de généralisations courbes des espaces de Moyal et du tore \ncf{}). Pour cela, il a considéré un modèle scalaire du type \eqref{eq:phi4naif} et a découvert des contributions à l'action effective qui divergent lorsque les moments externes tendent vers zéro. Le mélange UV/IR est donc une caractéristique générale des théories \ncv{}s, au moins sur les déformations.\\

La situation est restée en l'état jusqu'à ce que H.~Grosse et R.~Wulkenhaar découvre le moyen de définir une théorie \ncv{} renormalisable. Nous détaillerons leurs résultats au chapitre \ref{cha:base-matr} mais le message principal est le suivant. En ajoutant un terme harmonique au lagrangien \eqref{eq:phi4naif},
\begin{align}
    S[\phi] =&\int d^4x \Big( -\frac{1}{2} \partial_\mu \phi
\star \partial^\mu \phi +\frac{\Omega^2}{2} (\tilde{x}_\mu \phi )\star (\tilde{x}^\mu \phi ) + \frac{1}{2} m^2
\,\phi \star \phi
+ \frac{\lambda}{4} \phi \star \phi \star \phi \star
\phi\Big)(x)\label{eq:phi4Omega}
\end{align}
où $\xt=2\Theta^{-1} x$ , le modèle, en dimension quatre, est renormalisable à tous les ordres de perturbation \cite{GrWu04-3}. Nous verrons au chapitre \ref{cha:le-modele-phi4_4} que ce terme supplémentaire fournit une coupure infrarouge et permet de découpler les différentes échelles du problème. La théorie, que nous noterons $\Phi^{4}_{4}$, ne présente alors plus de mélange UV/IR. Ce résultat est capital car il ouvre la voie à d'autres théories des champs \ncv{}s. Dans la suite, nous appellerons \emph{vulcanisation}\footnote{TECHNOL. Opération consistant à traiter le caoutchouc naturel ou synthétique par addition de soufre, pour en améliorer les propriétés mécaniques et la résistance aux variations de température, Trésor de la Langue Française informatisé, \href{http://www.lexilogos.com/}{http://www.lexilogos.com/}.} la procédure consistant à rajouter un terme au lagrangien d'une théorie \ncv{} pour la rendre renormalisable.\\

Langmann et Szabo ont remarqué que l'interaction quartique avec produit de Moyal est invariante sous une transformation de dualité. Il s'agit d'une symétrie entre l'espace des moments et l'espace direct. La partie interaction du modèle \eqref{eq:phi4Omega} s'écrit (voir équation \eqref{eq:int-Moyal-even})
\begin{align}
  S_{\text{int}}[\phi]=&\int d^{4}x\,\frac{\lambda}{4}(\phi\star\phi\star\phi\star\phi)(x)\\
  =&\int\prod_{a=1}^{4}d^{4}x_{a}\,\phi(x_{a})\,V(x_{1},x_{2},x_{3},x_{4})\label{eq:Vx}\\
  =&\int\prod_{a=1}^{4}\frac{d^{4}p_{a}}{(2\pi)^{4}}\,\hat{\phi}(p_{a})\,\hat{V}(p_{1},p_{2},p_{3},p_{4})\label{eq:Vp}
  \intertext{avec}
  V(x_{1},x_{2},x_{3},x_{4})=&\frac{\lambda}{4}\frac{1}{\pi^{4}\det\Theta}\delta(x_{1}-x_{2}+x_{3}-x_{4})\cos(2(\Theta^{-1})_{\mu\nu}(x_{1}^{\mu}x_{2}^{\nu}+x_{3}^{\mu}x_{4}^{\nu}))\notag\\
  \hat{V}(p_{1},p_{2},p_{3},p_{4})=&\frac{\lambda}{4}(2\pi)^{4}\delta(p_{1}-p_{2}+p_{3}-p_{4})\cos(\frac 12\Theta^{\mu\nu}(p_{1,\mu}p_{2,\nu}+p_{3,\mu}p_{4,\nu}))\notag
\end{align}
où on a utilisé une transformée de Fourier \emph{cyclique} : $\hat{\phi}(p_{a})=\int dx\,e^{(-1)^{a}\imath p_{a}x_{a}}\phi(x_{a})$. La transformation
\begin{align}
  \hat{\phi}(p)\leftrightarrow\pi^{2}\sqrt{|\det\Theta|}\,\phi(x),&\qquad p_{\mu}\leftrightarrow\xt_{\mu}  
\end{align}
échange \eqref{eq:Vx} et \eqref{eq:Vp}. Par ailleurs, la partie libre du modèle \eqref{eq:phi4naif} n'est pas covariante sous cette dualité. La vulcanisation ajoute un terme au lagrangien qui rétablit cette symétrie. Le modèle \eqref{eq:phi4Omega} est ainsi covariant sous la dualité de Langmann-Szabo :
\begin{align}
  S[\phi;m,\lambda,\Omega]\mapsto&\Omega^{2}\,S[\phi;\frac{m}{\Omega},\frac{\lambda}{\Omega^{2}},\frac{1}{\Omega}].  
\end{align}
Par symétrie, le paramètre $\Omega$ est donc confiné dans l'intervalle $\lsb 0,1\rsb$. Notons qu'à $\Omega=1$, le modèle est invariant.

L'interprétation du terme harmonique supplémentaire n'est pas encore trouvée. Néanmoins, la procédure de vulcanisation a déjà permis de prouver la renormalisabilité de plusieurs autres modèles sur espace de Moyal tels que $\phi^{4}_{2}$ \cite{GrWu03-2}, $\phi^{3}_{2,4}$ \cite{Grosse2005ig,Grosse2006qv} et les modèles LSZ \cite{Langmann2003if,Langmann2003cg,Langmann2002ai}. Ces derniers sont du type
\begin{align}
    S[\phi] =&\int d^nx \Big( \frac{1}{2} \bar{\phi}\star(-\partial_\mu+\xt_{\mu}+m)^{2}\phi
+ \frac{\lambda}{4} \bar{\phi} \star \phi \star \bar{\phi} \star\phi\Big)(x).\label{eq:LSZintro}
\end{align}
En comparant ce modèle à \eqref{eq:phi4Omega}, on s'aperçoit qu'ici le terme supplémentaire est formellement équivalent à un champ magnétique de fond uniforme. La tentation est grande de l'interpréter comme tel. Ce modèle est également invariant sous la dualité sus-mentionnée et est soluble exactement. Remarquons que l'emploi d'une interaction complexe comme celle du modèle \eqref{eq:LSZintro} rend la dualité de Langmann-Szabo plus naturelle. Elle ne nécessite plus l'introduction d'une transformée de Fourier cyclique. Les modèles $\phi^{3}$ ont aussi été considérés à $\Omega=1$ où ils présentent une structure soluble.\\

Outre son intérêt pour la quantification de la gravitation, la théorie des champs \ncv{} pourrait bien nous renseigner et peut-être même résoudre un des problèmes majeurs de la théorie des champs commutative. Celle-ci souffre de divergences ultraviolettes et infrarouges mais aussi de la divergence de la série perturbative. Par exemple, le modèle $\phi^{4}_{4}$ est asymptotiquement libre dans l'infrarouge. En présence d'une coupure ultraviolette, on peut construire sa limite infrarouge. La construction de la limite ultraviolette impose une théorie libre dans l'infrarouge. De la même façon, les théories asymptotiquement libres dans l'ultraviolet, comme Yang-Mills ou Gross-Neveu, possèdent une zone de couplage fort dans l'infraouge responsable du confinement des quarks pour Yang-Mills. Le fait que le flot de la constante de couplage soit non borné est une des raisons qui empêchent la construction de ces modèles et la resommation de leurs séries des perturbations.

Par ailleurs, Grosse et Wulkenhaar \cite{GrWu04-2} ont calculé, à une boucle, le flot de la constante de couplage du modèle $\Phi^{4}_{4}$ \eqref{eq:phi4Omega}. Le flot du paramètre supplémentaire $\Omega$ régule celui de la constante de couplage. Dans ce modèle, $\Omega$ va de $\Omega_{\text{bare}}<1$ (fixé) à $\Omega_{\text{ren}}=0$ et le rapport $\frac{\lambda_{\text{bare}}}{\Omega^{2}_{\text{ren}}}$ est constant. Ce modèle $\Phi^{4}$ flotte de $\lambda_{\text{bare}}<\infty,\Omega_{\text{bare}}<1$ à $\lambda_{\text{ren}}>0,\Omega_{\text{ren}}=0$. Le flot de la constante de couplage est borné ce qui devrait permettre de définir ce modèle non perturbativement. Remarquons également qu'à $\Omega=1$, les fonctions $\beta_{\lambda}$ et $\beta_{\Omega}$ s'annulent. Ceci est certainement relié aux structures intégrables rencontrées à $\Omega=1$.\\

Les questions auxquelles nous aimerions répondre sont : dans quelle mesure la procédure de vulcanisation est générale ? S'applique-t-elle à d'autres espaces que le plan de Moyal, à d'autres modèles que $\phi^4$ ? Et surtout comment interpréter le terme supplémentaire ? Cette thèse a pour but de donner quelques éléments de réponse.

Pour étudier finement le comportement d'un modèle sous l'action du groupe de renormalisation, l'analyse multi-échelles est très utile. De plus, elle permet une étude constructive ultérieure. C'est donc la méthode que nous emploierons tout au long de cette thèse. Le chapitre \ref{cha:un-theoreme-bphz} rappelle en quoi consiste la renormalisation perturbative à la BPHZ. Il présente l'analyse multi-échelles et ce qu'elle nous apprend sur la série perturbative. Entre autres, nous verrons qu'il est naturel (du point de vue du groupe de renormalisation) d'organiser les termes de la perturbation en une infinité de constantes de couplage effectives. Cette multi-série est également un objet mathématique bien mieux adapté à la renormalisation que la série renormalisée habituelle.

Dans le chapitre \ref{cha:base-matr}, nous donnerons une définition précise de l'algèbre de Moyal. Nous verrons que l'espace de Schwartz est stable par le produit de Moyal et que nous pouvons étendre cette algèbre par dualité en une sous-algèbre de $\cS'$. Puis nous introduisons la base matricielle, une base de fonctions où le produit de Moyal devient un produit matriciel ordinaire. C'est dans cette base que Grosse et Wulkenhaar ont établit la preuve de la renormalisabilité du modèle $\Phi^{4}$ à tous les ordres. Nous ferons un résumé de leur travaux et les mettrons en perspéctive avec le reste de cette thèse. Nous présenterons également l'étude multi-échelles que nous avons effectuée dans \cite{Rivasseau2005bh} au sujet du modèle $\Phi^{4}$ dans la base matricielle. Celle-là a permis d'adapter l'analyse multi-échelles à la base matricielle et de préciser le rôle central des graphes duaux. Étant donné l'importance du propagateur dans les théories de champs \ncv{}s, nous exposerons les résultats obtenus dans l'article \cite{toolbox05} au sujet de généralisations du propagateur du modèle \eqref{eq:phi4Omega}. Celles-ci seront notamment utiles pour étudier le modèle de Gross-Neveu \ncf{}. Nous insisterons sur la pertinence de ces études de propagateurs en regard du travail de Grosse et Wulkenhaar, particulièrement celui de \cite{GrWu03-1}. Enfin j'expliquerai pourquoi je pense que l'étude des théories de champs dans la base matricielle est importante et a un fort potentiel.

Le chapitre \ref{cha:le-modele-phi4_4} est principalement dédié à la preuve de la renormalisabilité perturbative du modèle $\Phi^{4}_{4}$ en espace $x$. La preuve ayant déjà été établie par Grosse et Wulkenhaar dans la base matricielle, ce chapitre a pour principal but de nous familiariser avec les techniques de renormalisation \ncv{} en espace $x$. Ceci nous permettra de comparer avec le modèle de Gross-Neveu, techniquement plus ardu. Bien qu'à long terme la base matricielle me semble plus adaptée aux théories \ncv{}s que l'espace direct, celui-ci permet de comparer les comportements des modèles commutatifs et \ncf{}s. De plus, les méthodes de théories constructives \cite{Riv1} ne sont pas encore développées dans la base matricielle. L'essentiel de ce chapitre est adapté de l'article \cite{xphi4-05}. Nous verrons comment nous avons généralisé les résultats de Filk \cite{Filk1996dm} au cas où le propagateur ne conserve pas l'impulsion. Nous finissons ce chapitre en expliquant la renormalisabilité d'un modèle LSZ modifié qui est une légère généralisation de \eqref{eq:LSZintro}.

Le chapitre \ref{cha:GN} concerne le modèle de Gross-Neveu \ncf{}. Nous démontrons qu'il est renormalisable à tous les ordres de perturbation \cite{RenNCGN05}. Les principales caractéristiques de ce modèle sont les suivantes. La vulcanisation est complètement équivalente à écrire le modèle dans un champ magnétique de fond uniforme. C'est une théorie fermionique donc plus \og{}physique\fg{} que $\Phi^4$ du point de vue des champs de matière. Nous verrons que ce modèle présente du mélange UV/IR même après vulcanisation. Néanmoins les graphes qui possèdent ce mélange sont renormalisables. Mais ceci implique l'ajout d'un contreterme du type $\psib\imath\gamma^0\gamma^1\psi$ au lagrangien du modèle massif. Enfin cette étude permet de mettre en lumière le rôle de la notion d'orientabilité (voir section \ref{sec:graph-orientNCphi4}).



%% file: renormalisation-arXiv.tex
\numberwithin{equation}{section}
\chapter[Un théorème BPHZ]{Un théorème BPHZ de B à Z}
\label{cha:un-theoreme-bphz}
\epigraph{Ce qui est affirmé sans preuve peut être nié sans preuve.}{Euclide de Mégare}

Dans ce chapitre, nous allons principalement considérer le modèle $\phi^{4}_{4}$. Dans la première section, nous évoquons quelques généralités sur la théorie des champs et plus particulièrement sur la renormalisation perturbative. Nous insistons notamment sur les différents types de divergences ultraviolettes. Puis, en deuxième section, nous introduisons l'analyse multi-échelles qui est l'outil principal utilisé pendant cette thèse. Nous exposons ce que cette méthode peut nous apprendre sur la série des perturbations. Nous montrons comment on peut obtenir des bornes uniformes pour les graphes complètement convergents et, en section $3$, comment prouver que l'amplitude renormalisée, donnée par la formule de forêts de Zimmermann, de n'importe quel graphe de la théorie est finie et quelle borne nous pouvons en donner. Enfin nous définissons la multi-série effective et plaidons en sa faveur.

\section{Renormalisation perturbative}
\label{sec:renphi4}

Dans tout ce chapitre, nous prendrons comme exemple le modèle $\phi^{4}_{D}$ dans le cadre de l'intégrale fonctionnelle. Nous nous placerons toujours sur l'espace euclidien $\R^{D}$. La fonction de partition de ce modèle est 
\begin{align}
  Z[J]=&\int d\mu_{C}(\phi)\,e^{-\frac{\lambda}{4!}\int d^Dx\,\phi^{4}(x)+\imath\int J(x)\phi(x)d^Dx}\label{eq:Zphi4}  
\end{align}
où $C=(-a\Delta+m^{2})^{-1}$ est la covariance associée à la mesure gaussienne $d\mu$ et $a$ est un paramètre réel qui vaut $1$ pour l'instant. Les quantités que nous souhaitons définir sont les fonctions de corrélations (fonctions de Schwinger). Pour tout $N\in\N$ pair,
\begin{align}
  S_{N}(y_{1},\dotsc,y_{N})=&\prod_{i=1}^N-\imath\frac{\delta}{\delta J(y_{i})}Z[J]\Big|_{J\equiv 0}\label{eq:fctcorreldef}\\
  =&\int d\mu_{C}(\phi)\,\prod_{i=1}^{N}\phi(y_{i})e^{-\frac{\lambda}{4!}\int d^Dx\,\phi^{4}(x)}.\notag
\end{align}

\subsection{De la nécessité d'une coupure}
\label{sec:de-la-necessite}

Les fonctions de Schwinger autant que la fonction de partition sont des quantités particulièrement mal définies. Par le théorème de Bochner-Minlos, il existe une mesure fonctionnelle gaussienne supportée par l'espace $\cS'(\R^{D})$ des distributions tempérées. Dans \cite{reed73}, il a été montré que les distributions typiques du support sont en fait relativement régulières (par exemple, en dimension $2$, l'ensemble des $T\in\cS'(\R^{2})$ telles que $\forall\veps>0,\,(-\Delta+m^{2})^{-\veps}T$ soit localement de carré intégrable est de mesure $1$). Le support reste cependant un espace de distributions et l'expression $\phi^{4}(x)$ qui fait intervenir le produit de distributions n'a pas de sens. Une façon de régulariser cette situation consiste à introduire une coupure $\kappa$ dans la covariance \cite{colella73}.

La covariance $C$ est donnée par
\begin{align}
  \hat{C}(p,q)=&\frac{1}{p^{2}+m^{2}}\delta_{p+q,0}=\int_{0}^{\infty} dt\,e^{-t(p^{2}+m^{2})}\delta_{p+q,0},\label{eq:covp}\\
  C(x,y)=&\int_{0}^\infty\frac{dt}{(4\pi t)^{D/2}}\,e^{-\frac{(x-y)^2}{4t}-tm^2}.\label{eq:covx}
\end{align}
L'intégrale \eqref{eq:covx} est bien définie sauf quand $x=y$ en dimension $D\ges 2$. C'est le problème des divergences ultraviolettes. Le fait que l'expression \eqref{eq:covx} soit mal définie si $x=y$ est équivalent au fait que la covariance $C(p)$ ne soit pas intégrable en dimensions supérieures à $2$ pour $p$ proche de l'infini autrement dit dans la région ultraviolette. Quels que soient $x$ et $y$, l'intégrale \eqref{eq:covx} est bien définie pour $t$ proche de l'infini. C'est la zone infrarouge protégée par la masse $m$. Si $x=y$, l'intégrale diverge dans la zone $t$ proche de $0$, c'est la région ultraviolette. Ainsi la région infrarouge correspond à $p$ petit, $\labs x-y\rabs$ grand ou $t$ grand. La région ultraviolette correspond, au contraire, à $p$ grand, $\labs x-y\rabs$ petit ou $t$ petit. Dans tout ce chapitre, nous considérerons un modèle massif $m>0$. La zone infrarouge ne pose donc pas de difficultés. La seule coupure nécessaire est ultraviolette : pour $\kappa\in\R_{+}^{*}$, nous définissons
\begin{align}
  \hat{C}_{\kappa}(p)=&\int_{\kappa}^\infty dt\,e^{-t(p^2+m^2)}=\frac{1}{p^{2}+m^{2}}e^{-\kappa(p^2+m^2)}.\label{eq:covpreg}
\end{align}
Quel que soit $\kappa>0$, l'expression \eqref{eq:covpreg} est intégrable ce qui signifie que sa transformée de Fourier $C_{\kappa}(x-y)$ est bien définie en $x=y$. La limite $\kappa\to 0$ redonne la covariance initiale.

Il a été montré dans \cite{colella73,Riv1} que la mesure gaussienne associée à la covariance $C_{\kappa}$ est supportée par les fonctions $\cC^\infty $ à croissance au plus logarithmique. Tout polynôme dans les champs, comme $\phi^{4}(x)$, est ainsi bien défini. Néanmoins les fonctions $\phi$ n'appartiennent typiquement pas à $L^{4}(\R^{D})$ si bien que $\int_{\R^D}\phi^{4}(x)dx$ n'existe pas. On peut remédier à ce problème en restreignant l'intégrale à un domaine compact $\Lambda\subset\R^D$. Le point de départ du modèle $\phi^{4}$ est donc la fonction de partition régularisée
\begin{align}
  Z_{\Lambda,\kappa}[J]=&\int d\mu_{C,\kappa}(\phi)\,e^{-\frac{\lambda}{4!}\int_{\Lambda} d^Dx\,\phi^{4}(x)+\imath\int_{\Lambda} J(x)\phi(x)d^Dx}.\label{eq:Zreg}
\end{align}

\subsection{Séries perturbatives}
\label{subsec:series-perturbatives}

L'équation \eqref{eq:Zreg} est un point de départ. Nous souhaitons trouver un moyen de définir les limites $\Lambda\to\R^D$ et $\kappa\to 0$. Quel que soit le moyen utilisé pour définir ces limites, il consiste en une perturbation autour de $\lambda=0$. Dans la suite, les coupures sont sous-entendues mais omises pour alléger les notations. Pour calculer des grandeurs physiques ou, historiquement, pour définir le modèle \eqref{eq:Zphi4}, on a écrit
\begin{align}
    S_{N}(y_{1},\dotsc,y_{N})=&\int d\mu_{C}(\phi)\,\prod_{i=1}^{N}\phi(y_{i})\sum_{n=0}^{\infty}\frac{1}{n!}\Big(\frac{-\lambda}{4!}\int\phi^4(x)dx\Big)^n.\label{eq:SNpert}
\end{align}
Jusque là, tout va bien. Puis on a interverti les intégrations sur $x$ et la somme sur $n$ avec l'intégrale fonctionnelle. Grâce aux coupures $\Lambda$ et $\kappa$, les intégrations sur $x$ et $\phi$ sont absoluement convergentes et peuvent être échangées. Nous verrons par la suite que la seule série qui ait un sens aux limites $\Lambda\to\R^D$ et $\kappa\to 0$ (la série \emph{renormalisée}\footnote{La \emph{multi-série} effective a également un sens (voir section \ref{sec:la-serie-effective}).}) n'est généralement pas convergente et donc certainement pas absoluement convergente. Ainsi l'échange de la somme sur $n$ et de l'intégrale fonctionnelle n'est pas autorisée.

\emph{À coeur vaillant, rien d'impossible}, nous échangeons la somme et l'intégrale fonctionnelle et prenons l'expression suivante comme définition du modèle :
\begin{align}
    S_{N}(y_{1},\dotsc,y_{N})=&\sum_{n=0}^{\infty}\frac{1}{n!}\lbt\frac{-\lambda}{4!}\rbt^{n}\int\prod_{j=1}^{n}dx_{j}\int d\mu_{C}(\phi)\,\prod_{i=1}^{N}\phi(y_{i})\prod_{j=1}^{n}\phi^{4}(x_{j}).\label{eq:SNpertinverse}
\end{align}
À partir de là, le théorème de Wick nous permet de calculer les fonctions de Schwinger ordre par ordre. Pour le modèle $\phi^{4}$, les règles de Feynman associées sont :\\

\begin{itemize}
\item $\frac{-\lambda}{4!}\int dx$ par vertex :\hspace{.5cm}\raisebox{-2ex}{\includegraphics{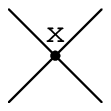}}\;,
\item $C(x,y)$ par ligne :\hspace{.5cm}\includegraphics{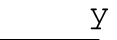}.
\end{itemize}
\medskip

Les théories de champs quantiques souffrent de deux types de divergences ultraviolettes. La première apparait au niveau des graphes individuels. Considérons l'exemple de la fonction à quatre points connexe jusqu'à l'ordre $2$ en espace des moments :
\begin{align}
  S^{\text{c}}_{4}(p_{1},\dotsc,p_{4})=&-\lambda\hspace{1em}\raisebox{-4ex}{\includegraphics{xphi4-fig.10}}+\frac{(-\lambda)^{2}}{2}\hspace{1em}\raisebox{-4ex}{\includegraphics{xphi4-fig.11}}\notag\\
  &\notag\\
  &+\lb1\leftrightarrow 3\rb+\lb 1\leftrightarrow 4\rb+\cO(\lambda^{3})\label{eq:S4c}
\end{align}
où $p=p_{1}+p_{2}=-p_{3}-p_{4}$. L'amplitude associée au graphe $G$ d'ordre $2$ (la bulle) vaut
\begin{align}
  A_{G}(p_{1},\dotsc,p_{4})=&\int d^Dq\,\frac{1}{q^2+m^2}\frac{1}{(p+q)^2+m^2}  
\end{align}
En dimension $D\ges 4$, $A_{G}$ est divergente dans la région $|q|\gg 1$ à $p$ fixé. Nous verrons dans la section suivante comment traiter les divergences des graphes individuels.

\begin{figure}[htbp!]
  \centering
  \includegraphics{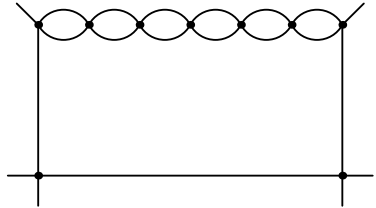}
  \caption{Renormalon}
  \label{fig:renormalon6pts}
\end{figure}

Le second type de divergences concerne la série perturbative dans son ensemble. Considérons l'ordre $n$ de la fonction à $4$ points renormalisée (encore une fois nous verrons ce que cela signifie dans la section suivante). À grands moments externes, en dimension $4$ :
\begin{align}
  S_{4}^{R,n}(p)\seq_{p\to\infty}&\lambda_{R}^{n}\Big(\beta_{2}\log\frac{p}{m}\Big)^{n-1}.
\end{align}
Ce comportement, inséré dans une boucle convergente comme celle du graphe à $6$ points de la figure \ref{fig:renormalon6pts}, contribue à l'ordre $n$ de la fonction à $6$ points par :
\begin{align}
  \lambda_{R}^{n}\int\frac{d^4p}{(p^2+m^2)^{3}}\Big(\beta_{2}\log\frac{p}{m}\Big)^{n-3}\simeq&(n-3)!\lambda_{R}^{n}\beta_{2}^{n-3}.  
\end{align}
Ces contributions ne sont pas sommables (ni même Borel sommables car elles s'ajoutent avec le même signe) et conduisent donc à des difficultés pour resommer la série des perturbations. Nous reviendrons sur les renormalons plus tard et expliquerons leur origine.
\medskip

\noindent
Une autre façon de voir le même problème consiste à étudier le comportement de l'ordre $n$ de la fonction à $4$ point nue (bare en anglais) à moments externes nuls et en présence d'une coupure $p$ (les moments internes sont inférieurs à $p$) :
\begin{align}
  S_{4,p}^{B,n}(0)\seq_{p\to\infty}&(-\lambda_{B})^{n}\Big(\beta_{2}\log\frac{p}{m}\Big)^{n-1}\label{eq:S4B}
\end{align}
où $\lambda_{B}=\lambda$. Nous verrons dans la section suivante que la fonction à $4$ points à moments $0$ définit la constante de couplage renormalisée $\lambda_{R}$. Ainsi en sommant les contributions du type \eqref{eq:S4B}, on obtient
\begin{align}
  -\lambda_{R}=&-\lambda_{B}+\sum_{n=2}^\infty(-\lambda_{B})^{n}\Big(\beta_{2}\log\frac{p}{m}\Big)^{n-1}=\frac{-\lambda_{B}}{1+\lambda_{B}\beta_{2}\log\frac{p}{m}}.\label{eq:ren=bare}
\end{align}
À $\lambda_{B}>0$ fixée, la constante de couplage renormalisée $\lambda_{R}$ tend vers $0$ quand la coupure $p$ tend vers l'infini. C'est la trivialité de $\phi^{4}$. La théorie renormalisée semble être libre.\\
Enfin, nous pouvons aussi inverser la relation \eqref{eq:ren=bare} et obtenir
\begin{align}
    -\lambda_{B}=&\frac{-\lambda_{R}}{1-\lambda_{R}\beta_{2}\log\frac{p}{m}}.\label{eq:bare=ren}
\end{align}
À $\lambda_{R}>0$ fixée, pour $p$ suffisamment grand, $\lambda_{B}$ devient de l'ordre de $1$ et la série perturbative n'a plus de sens. C'est le fameux fantôme de Landau. Ces phénomènes sont les trois facettes d'une même médaille et sont regroupés sous le terme de problème des renormalons. Nous verrons dans la suite que les renormalons sont intimement liés à la façon dont on renormalise la série nue \ie{} à la façon dont on traite le problème de la divergence des graphes individuels.

\subsection{Le théorème BPHZ}
\label{sec:le-theoreme-bphz}

La renormalisation perturbative fut inventée vers 1950 par Feynman, Schwinger, Dyson, Tomonaga et d'autres (voir \cite{Dys1} par exemple). Il s'agit de redéfinir les paramètres observables de la théorie en incluant les divergences dans des paramètres nus inobservables. Il fallut environ dix ans pour faire de cette idée un théorème solide : le théorème BPHZ. Pour le modèle $\phi^4_{4}$, il peut être formulé de la façon suivante :
\begin{thm}[BPHZ pour $\phi^{4}_{4}$]\label{thm:bphzphi4comm}
  Il existe trois séries formelles en un paramètre $\lambda_{R}\in\R_{+}$ telles que si nous remplaçons, dans \eqref{eq:Zreg}, $-\lambda$ par $-\lambda_{R}+\sum_{n=2}^{\infty}c_{n}(\Lambda,\kappa)(-\lambda_{R})^{n}$, $m^{2}$ par $m_{R}^{2}+\sum_{n=1}^{\infty}d_{n}(\Lambda,\kappa)(-\lambda_{R})^{n}$ et $a$ par $a_{R}+\sum_{n=2}^{\infty}e_{n}(\Lambda,\kappa)(-\lambda_{R})^{n}$, toute fonction de Schwinger est finie ordre par ordre en $\lambda_{R}$ aux limites $\Lambda\to\R^{4}$ et $\kappa\to 0$.
\end{thm}
Nous allons donner les grandes lignes de la preuve de ce théorème. On dit qu'une théorie est renormalisable s'il existe un nombre fini de paramètres à redéfinir pour la rendre finie ordre par ordre. Dans le modèle $\phi^4_{4}$, ces paramètres sont la constante de couplage, la masse et le coefficient de la fonction d'onde $a$. Les constantes $\lambda_{R}$, $m_{R}$ et $a_{R}$ sont fixées par trois expériences. S'il fallait fixer un nombre infini de paramètres, la théorie perdrait sa prédictibilité. Ainsi, pour démontrer le théorème \ref{thm:bphzphi4comm}, il faut d'abord prouver qu'il n'y a qu'un nombre fini de fonctions divergentes dans le modèle. Pour cela, on démontre une borne supérieure sur les amplitudes des graphes : c'est le comptage de puissance.\\

Considérons un graphe $G$ d'ordre $n$ avec $N$ pattes externes. En moments, les fonctions delta aux vertex nous apprennent qu'il y a un moment indépendant par boucle. Dans la région où tous les moments internes tendent vers l'infini de la même façon, l'amplitude se comporte, en dimension $D$, comme
\begin{align}
  A_{G}\simeq&\int_{0}^{K}\frac{d|p|\,p^{DL-1}}{p^{2I}}\simeq K^{DL-2I}
\end{align}
où $L$ est le nombre de boucles du graphe, $I$ le nombre de lignes internes et $K$ une coupure ultraviolette. En utilisant $4n=2I+N$ et $L=I-n+1$ (pour $G$ connexe), on obtient
\begin{align}
  A_{G}\simeq K^{-\omega},\;\omega=(4-D)n+\frac{D-2}{2}N-D.\label{eq:degsuperfconv}
\end{align}
$\omega$ est appelé degré superficiel de convergence. Si $D>4$, quel que soit $N$, il existera toujours un $n$ à partir duquel $\omega$ sera négatif et $A_{G}$ divergera. Toute fonction diverge, la théorie est non renormalisable. Si $D<4$, au contraire, il existe un ordre minimal à partir duquel toutes les fonctions convergent. Un nombre fini de graphes divergent, le modèle est super-renormalisable. Enfin si $D=4$, un nombre infini de graphes divergent mais ils concernent seulement les fonctions à $2$ et $4$ points. Le modèle est dit juste renormalisable. Bien sûr, il faudrait obtenir une borne plus précise pour prendre en compte toutes les sous-divergences des sous-graphes de $G$. Nous verrons que l'analyse multi-échelles nous permet de l'obtenir très naturellement.\\

\begin{figure}[htbp!]
  \centering
  \includegraphics{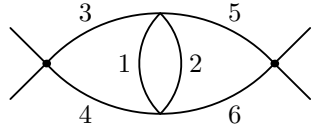}
  \caption{Le graphe \emph{oeil}}
  \label{fig:oeil}
\end{figure}

Restreignons-nous maintenant à la dimension $4$. Seules les fonctions à deux et quatre points divergent (nous ne nous intéresserons pas aux graphes du vide). Nous devons donc prouver que leurs divergences peuvent être absorbées dans une redéfinition des constantes du modèle. Considérons par exemple la fonction à quatre points dont le développement à l'ordre $\lambda_{B}^2$ est donné en \eqref{eq:S4c}. Notons $\tau$ l'opération de Taylor consistant à évaluer l'amplitude d'un graphe à moments externes nuls. Pour la bulle, nous écrivons
\begin{align}
  A_{G}=&\tau A_{G}+(1-\tau)A_{G}.  
\end{align}
Montrons que $\tau A_{G}$ peut être absorbé dans une redéfinition de la constante de couplage.
\begin{align}
    \tau A_{G}=&\prod_{i=1}^{4}\hat{C}(p_{i})\,\delta(p_{1}+\dotsb+p_{4})\int\frac{d^{4}q}{(q^{2}+m^{2})^{2}},\\
    S_{4}^{\text{c}}(p_{1},\dotsc,p_{4})=&\Big(-\lambda_{B}+\frac{3}{2}(-\lambda_{B})^{2}\int\frac{d^{4}q}{(q^{2}+m^{2})^{2}}\Big)\hspace{1em}\raisebox{-4ex}{\includegraphics{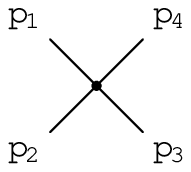}}\notag\\
    &+\frac{(-\lambda)^{2}}{2}(1-\tau)\hspace{1em}\raisebox{-4ex}{\includegraphics{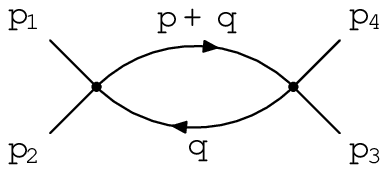}}\notag\\
    &+(1-\tau)\Big(\lb1\leftrightarrow 3\rb+\lb 1\leftrightarrow 4\rb\Big)+\cO(\lambda^{3}).
\end{align}
Nous pouvons alors définir $-\lambda_{R}\defi-\lambda_{B}+\frac{3}{2}(-\lambda_{B})^{2}\int\frac{d^{4}q}{(q^{2}+m^{2})^{2}}+\cO(\lambda_{B}^{3})$. L'intégrale sur $q$ doit être considérée comme régularisée par une coupure quelconque. Il faut encore montrer que les restes de Taylor $(1-\tau)$ sont finis :
\begin{align}
  (1-\tau)A_{G}=&\prod_{i=1}^{4}\hat{C}(p_{i})\,\delta(p_{1}+\dotsb+p_{4})\int\frac{1}{(q^{2}+m^{2})((p+q)^{2}+m^{2})}-\frac{1}{(q^{2}+m^{2})^{2}} d^{4}q\notag\\
  =&-\prod_{i=1}^{4}\hat{C}(p_{i})\,\delta(p_{1}+\dotsb+p_{4})\int\frac{p^{2}+2p\cdot q}{((p+q)^{2}+m^{2})(q^{2}+m^{2})^{2}}d^{4}q.
\end{align}
La limite $\kappa\to 0$ de l'intégrale sur $q$ est alors finie. Le développemement de la fonction connexe à quatre points s'écrit donc
\begin{align}
  S_{4}^{\text{c}}(p_{1},\dotsc,p_{4})=&-\lambda_{R}\hspace{1em}\raisebox{-4ex}{\includegraphics{xphi4-fig10.eps}}+\text{contributions finies}+\cO(\lambda_{R}^{3}).
\end{align}
Bien sûr, il faut être capable de redéfinir les constantes du modèle à chaque ordre de perturbation. Soit un graphe $G$ avec $N$ pattes externes. Son amplitude amputée en espace
des moments est notée $\hat{A}_{G}(k_{1},\dotsc,k_{N})$. La conservation des
moments aux vertex implique la conservation des moments entrants. Ainsi
$\hat{A}_{G}$ se décompose de la façon suivante :
\begin{equation}
  \label{eq:amplamput-TF}
  \hat{A}_{G}(k_{1},\dotsc,k_{N})=\delta(k_{1}+\dotsb+k_{N})\hat{g}(k_{1},\dotsc,k_{N-1}).
\end{equation}
On définit le contreterme $-\tau_{G}\hat{A}$ associé au graphe $G$ par
\begin{equation}
  \label{eq:contreterme}
  -\tau_{G}\hat{A}=-\delta\big(\sum_{i=1}^{N}k_{i}\big)\sum_{j=1}^{D(G)}\frac{1}{j!}
  \frac{d^{j}}{dt^{j}}\ \hat{g}(tk_{1},\dots,tk_{N-1})\Big|_{t=0}
\end{equation}
où $D(G)$ est le plus grand nombre entier inférieur ou égal à $-\omega(G)$. Pour le modèle $\phi^{4}_{4}$, cela consiste à effectuer un développement de Taylor à l'ordre $0$ pour la fonction à quatre points, aux ordres $0$ et $2$ (l'ordre $1$ est nul par parité) pour la fonction à deux points. Il faut évidemment montrer que les premiers termes de ces développements peuvent être absorbés dans une redéfinition des constantes du modèle pour tous les graphes à tous les ordres. Nous verrons aux chapitres \ref{cha:le-modele-phi4_4} et \ref{cha:GN} comment nous le démontrons dans les cas des modèles $\Phi^{4}$ et Gross-Neveu \ncf{}s. Ici, supposons que nous l'ayons fait. Le schéma de soustraction \eqref{eq:contreterme} nous permet alors de définir
\begin{subequations}
  \begin{align}
    S_{4}^{\text{c}}(0,0,0,0)\defi&-\lambda_{R},\\
    S_{2}^{\text{c}}(0,0)\defi&\frac{1}{m_{R}^{2}},\\
    \frac{d^{2}}{dp^{2}}S_{2}^{\text{c}}(p,-p)\Big|_{p=0}\defi&-\frac{2a_{R}}{m^{4}}.
  \end{align}\label{eq:schemasoustract}
\end{subequations}
Ces définitions donnent $\lambda_{R}$, $m_{R}$ et $a_{R}$ en termes de séries formelles en $\lambda_{B}$. La combinatoire à vérifier est alors la suivante. Il faut inverser les trois séries définies par les équations \eqref{eq:schemasoustract} :
\begin{subequations}
  \begin{align}
    -\lambda_{R}=-\lambda_{B}+\sum_{n=2}^{\infty}c_{n}^{B}(-\lambda_{B})^{n}\Longleftrightarrow& -\lambda_{B}=-\lambda_{R}+\sum_{n=2}^{\infty}c_{n}(-\lambda_{R})^{n}\\
     m^{2}_{R}=m^{2}_{B}+\sum_{n=1}^{\infty}d_{n}^{B}(-\lambda_{B})^{n}\Longleftrightarrow& m^{2}_{B}=m^{2}_{R}+\sum_{n=1}^{\infty}d_{n}(-\lambda_{R})^{n}\\
     a_{R}=a_{B}+\sum_{n=2}^{\infty}e_{n}^{B}(-\lambda_{B})^{n}\Longleftrightarrow& a_{B}=a_{R}+\sum_{n=2}^{\infty}e_{n}(-\lambda_{R})^{n}
  \end{align}\label{eq:serieform}
\end{subequations}
puis vérifier que la théorie exprimée en fonction des constantes renormalisées est finie ordre par ordre en $\lambda_{R}$ (les coefficients $c_{n},d_{n}$ et $e_{n}$ sont ceux du théorème \ref{thm:bphzphi4comm}). Dans ce but, la formule de Zimmermann est très utile. Elle permet d'exprimer les termes de la série renormalisée (\ie{} écrite en termes des constantes renormalisées) par une formule compacte. Soit le modèle \eqref{eq:Zreg} où nous avons remplacé $\lambda_{B}$, $m_{B}$ et $a_{B}$ par les trois séries formelles \eqref{eq:serieform} :
\begin{align}
  Z_{\Lambda,\kappa}[J]=&\int d\mu_{C(a_{R},m_{R},\lambda_{R}),\Lambda,\kappa}(\phi)\,e^{-\frac{1}{4!}(\lambda_{R}-\sum_{n=2}^{\infty}c_{n}(-\lambda_{R})^{n})\int_{\Lambda} d^Dx\,\phi^{4}(x)+\imath\int_{\Lambda} J(x)\phi(x)d^Dx}.
\end{align}
En développant les fonctions de Schwinger en puissances de $\lambda_{R}$, on obtient une série identique à la série nue mais où l'amplitude de chaque graphe $G$ devient
\begin{align}
  A_{G}^{R}=&\sum_{\cF}\prod_{g\in\cF}(-\tau_{g})A_{G}\label{eq:Zimmforest}
\end{align}
où $A_{G}$ est l'amplitude nue. La somme contient toutes les \textbf{forêts} de sous-graphes divergents (à deux et quatre pattes externes). Rappelons qu'une forêt est un ensemble de sous-graphes tel que $\forall g,g'\in\cF$, soit $g\cap g'=\emptyset$ soit $g\subset g'$ ou $g'\subset g$. Par exemple, le graphe $G$ de la figure \ref{fig:oeil} contient les sous-graphes divergents suivants : $G_{1}=\lb 1,2\rb,G_{2}=\lb 1,2,3,4\rb,G_{3}=\lb1,2,5,6\rb$ et $G=\lb 1,2,3,4,5,6\rb$. Ses forêts divergentes sont donc : $\emptyset,\lb G\rb,\lb G_{1}\rb,\lb G_{2}\rb,\lb G_{3}\rb,\lb G_{1},G\rb,\lb G_{2},G\rb,\lb G_{3},G\rb,\lb G_{1},G_{2}\rb,\lb G_{1},G_{3}\rb$,\\$\lb G_{1},G_{2},G\rb,\lb G_{1},G_{3},G\rb$. Les sous-graphes $G_{2}$ et $G_{3}$ ne peuvent pas appartenir à la même forêt.
Pour vérifier que la formule de Zimmermann \eqref{eq:Zimmforest} donne une ampitude renormalisée finie, nous renvoyons à la section \ref{sec:bornes-et-finitude} où nous avons utilisé la classification des forêts dans le cadre de l'analyse multi-échelles.



\section{Analyse multi-échelles}
\label{sec:multiscalephi4}

Dans cette section, nous présentons l'analyse multi-échelles et donnons une borne supérieure sur les amplitudes montrant que seuls les graphes à $0$, $2$ et $4$ points divergent. De plus, nous démontrons un théorème de Weinberg uniforme sur les graphes complètement convergents.

L'analyse multi-échelles est un outil  puissant qui permet d'étudier
précisemment la renormalisabilité d'une théorie. De plus elle fournit des
bornes précises sur les amplitudes. Cette technique suit les idées de
K.~Wilson \cite{Wilson1973jj} sur le groupe de renormalisation. 

\subsection{Espace des phases}
\label{sec:espace-des-phases}

Au lieu d'effectuer l'intégrale fonctionnelle d'un seul coup, nous commençons par séparer les degrés de liberté de fréquences différentes. La renormalisation va alors se faire des échelles les plus hautes (ultraviolettes) vers les échelles les plus basses (infrarouges). Pour cela, nous découpons le propagateur en tranches de moments. Il est pratique d'utiliser une progression géométrique de raison $M>1$ :
\begin{align}
  C_{\rho}(x,y)=&\int_{\kappa=M^{-2\rho}}^\infty\frac{dt}{(4\pi t)^{D/2}}\,e^{-\frac{(x-y)^2}{4t}-tm^2}=\sum_{i=0}^{\rho}C^{i},\label{eq:propslice}\\
  C^{0}(x,y)=&\int_{1}^\infty\frac{dt}{(4\pi t)^{D/2}}\,e^{-\frac{(x-y)^2}{4t}-tm^2},\notag\\
  C^{i}(x,y)=&\int_{M^{-2i}}^{M^{-2(i-1)}}\frac{dt}{(4\pi t)^{D/2}}\,e^{-\frac{(x-y)^2}{4t}-tm^2}.\notag
\end{align}
La mesure étant gaussienne, il existe une décomposition associée \cite{Salmhofer} :
\begin{align}
  \phi_{\rho}=&\sum_{i=0}^{\rho}\phi^{i},\;d\mu_{\rho}(\phi_{\rho})=\bigotimes_{i=0}^{\rho}d\mu_{i}(\phi^{i}).\label{eq:factorgauss}
\end{align}
C'est principalement la décomposition du propagateur qui nous sera utile dans la suite. Néanmoins il est intéressant de garder à l'esprit que chaque $\phi^{i}$ correspond à une théorie avec des coupures infrarouge et ultraviolette, de plus en plus proche de l'ultraviolet quand $i$ augmente.

\begin{figure}[htbp]
  \centering
     \raisebox{2cm}{\subfloat[Un graphe de $\phi^{4}$]{\label{fig:exgraph}\includegraphics[scale=0.9]{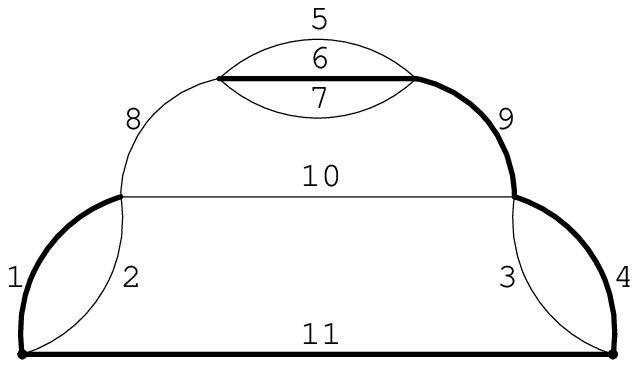}}}
     \subfloat[L'arbre de \og Gallavotti-Nicol\`o\fg]{\label{fig:GNarbre}\includegraphics[scale=1]{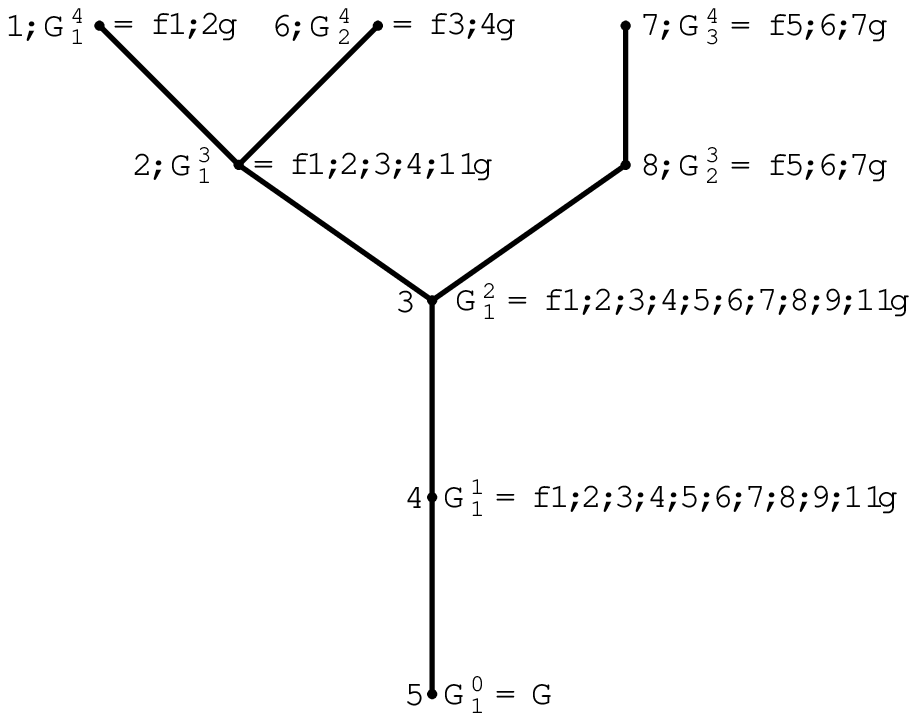}}\\
     \subfloat[Exemple d'attribution des échelles]{\label{fig:exscale}\includegraphics[scale=0.7]{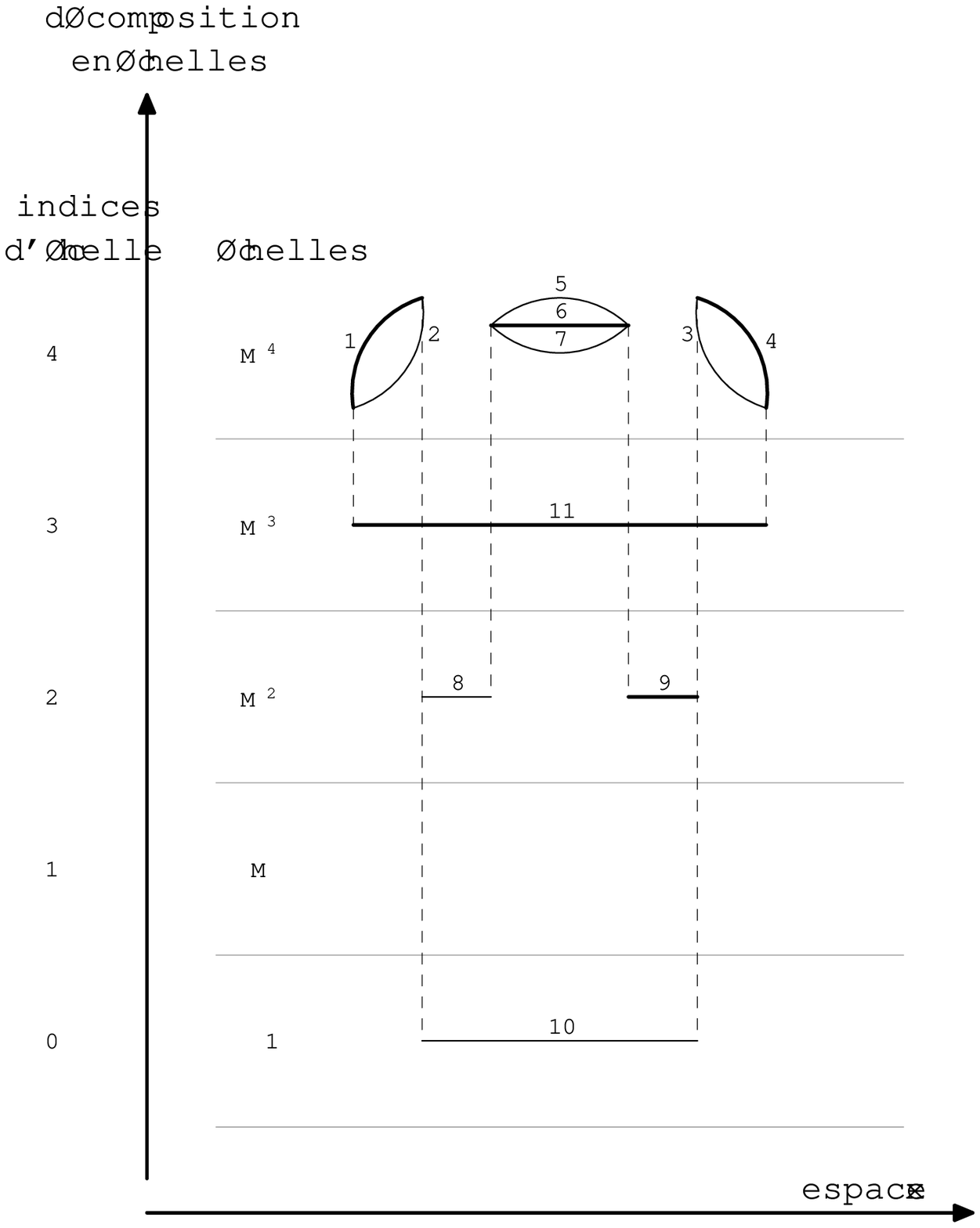}} 
     \caption{Les outils de l'analyse multi-échelles}
  \label{multiscale-tools}
\end{figure}

La première borne importante concerne le propagateur régularisé dans une tranche $i$ :
\begin{lemma}\label{lem:bornepropaphi4comm}
  Pour tout $0\les i\les\rho$, il existe $K,k\in\R_{+}$ tels que
  \begin{align}
    C^{i}(x,y)\les&KM^{(D-2)(i+1)}\,e^{-kM^{i+1}|x-y|}.\label{eq:propibound}
  \end{align}
\end{lemma}
\begin{proof}
  Pour $i\in\lnat 1,\rho\rnat$, on utilise simplement
  \begin{align}
    C^{i}(x,y)=&\int_{M^{-2i}}^{M^{-2(i-1)}}\frac{dt}{(4\pi t)^{D/2}}\,e^{-\frac{(x-y)^2}{4t}-tm^2}\\
    \les&K_{1}(M^{-2(i-1)}-M^{-2i})\sup_{M^{-2i}\les t\les M^{-2(i-1)}}\frac{e^{-\frac{(x-y)^2}{4t}}}{t^{D/2}}\notag\\
    \les&K_{2}M^{(D-2)i}\,e^{-k_{1}M^{2i}(x-y)^{2}}\notag\\
    \les&KM^{(D-2)(i+1)}\,e^{-kM^{i+1}|x-y|}.\notag
  \end{align}
  Pour la première tranche $i=0$ (la tranche infrarouge), il faut utiliser la masse pour intégrer sur $t$ :
  \begin{align}
    C^{0}(x,y)=&\int_{1}^{\infty}\frac{dt}{(4\pi t)^{D/2}}\,e^{-\frac{(x-y)^2}{4t}-tm^2}\\
    \les&\sup_{t}e^{-\frac{(x-y)^2}{4t}-tm^2/2}\times\int_{1}^{\infty}\frac{dt}{(4\pi t)^{D/2}}\,e^{-tm^2/2}\notag\\
    \les&K\,e^{-k|x-y|}.\notag
  \end{align}
\end{proof}
\begin{rem}
  Dans le cas d'une théorie sans masse, le propagateur dans la tranche infrarouge ($i=0$) n'obéit pas à la borne \eqref{eq:propibound}. Il est alors nécessaire de découper également la zone $1\les t<\infty$. Le propagateur, dans ces nouvelles tranches, obéit à la borne \eqref{eq:propibound} avec $i\to -i$. On retrouve le fait que dans une théorie massive, la masse \og{}arrête\fg{} le flot alors qu'en l'abence de masse, le flot est non trivial dans l'infrarouge.
\end{rem}
Considérons un graphe $G$ amputé avec $I(G)$ lignes internes. La décomposition \eqref{eq:propslice} des propagateurs conduit à 
\begin{align}
  A_{G}=&\sum_{\mu\in\N^{I(G)}}A_{G}^\mu,\\
  A_{G}^\mu=&\int\prod_{\nu\in G}dx_{\nu}\,\prod_{l\in G}C^{i_{l}}(x_{l},y_{l})
\end{align}
où $\mu$, une attribution d'échelles (ou d'indices), est une liste d'entiers naturels $i_{l}$ correspondant, pour chaque ligne interne $l$, à la tranche $i_{l}$ du propagateur de cette ligne. Une paire $(G,\mu)$ peut alors être représentée comme l'exemple de la figure \ref{fig:exscale}. Sur cette figure, l'espace $x$ est représenté par la direction horizontale tandis que les échelles du graphe suivent la direction verticale. De manière générale, lignes et vertex jouent des rôles duaux dans cet \og{}espace de phase\fg{}. Chaque propagateur se situe dans une tranche de \og{}moments\fg{} et joint deux vertex. Au contraire, chaque vertex se situe en un point $x$ fixé et joint quatre demi-lignes situées \emph{à priori} dans quatre tranches différentes. Les lignes verticales correspondant aux vertex du graphe sont dessinées en pointillés pour les distinguer des propagateurs.

Dans l'espace des phases $(x,t)$ ($t$ est le paramètre de Schwinger et non le temps!), les divergences ultraviolettes à la limite $\kappa\to 0$ (ou $\rho\to\infty$) se traduisent par la divergence de la somme sur les attributions $\mu$. En anticipant sur la suite et pour illustrer la dualité entre l'espace $x$ et l'espace (unidimensionnel) des échelles, nous verrons que tout comme à chaque propagateur $C_{l}$ est associée une décroissance entre $x_{l}$ et $y_{l}$, à chaque vertex, pour les sous-graphes convergents, est associée une décroissance entre les échelles des lignes qui lui sont accrochées. Ces décroissances duales nous permettront, pour un graphe complètement convergent, d'effectuer la somme sur les attributions d'échelles et d'obtenir une amplitude finie pour ces graphes.

Pour les sous-graphes divergents, ceux pour lesquels la somme sur $\mu$ diverge, l'opération de renormalisation $1-\tau$ (voir section \ref{sec:le-theoreme-bphz}) permettra d'obtenir des facteurs supplémentaires de décroissance exponentielle dans les échelles de ces sous-graphes. La somme sur les attributions d'indices des amplitudes renormalisées sera alors convergente.

\subsection{Graphes complètement convergents}
\label{sec:chepa}

Nous allons démontrer un théorème de Weinberg uniforme. Cela nous donnera l'occasion d'introduire (naturellement j'éspère) différents outils et structures de l'analyse multi-échelles. Commençons par définir quelques notations graphiques.

Soit $G$ un graphe connexe. $G$ est en fait une paire $(\cV,\cI)$ composée des ensembles $\mathbf{\cV(G)}=\cV$ de ses vertex et $\mathbf{\cI(G)}=\cI$ de ses lignes. Si $G'=(\cV',\cI')$, nous noterons $\cV(G')=\cV'$ et $\cI(G')=\cI'$. Le cardinal de $\cV(G)$ sera $\mathbf{n(G)}$, celui de $\cI(G)$ sera $\mathbf{I(G)}$.

Nous noterons également $\mathbf{\cV_{\cI}(G)}$ l'ensemble des vertex accrochés aux lignes de $\cI$. Soit $\nu\in\cV(G)$, définissons $\mathbf{\cI_{\nu}(G)}$ comme l'ensemble des lignes internes de $G$ acrochées au vertex $\nu$.

Un \textbf{sous-graphe} $g$ de $G$ ($g\subset G$) est une paire $(\cV_{\ki}(G),\ki)$ avec $\ki\subset\cI(G)$. Un sous-graphe est donc composé d'un sous-ensemble de lignes de $G$ et des vertex aux extrémités de ces lignes. Avec cette définition, un sous-graphe n'est pas nécessairement connexe.

Pour tout graphe $G$, l'ensemble $\cV(G)$ est l'union disjointe de deux sous-ensembles. Soit $\mathbf{c_{G}}$ la fonction de $\cV(G)$ dans $\N$ telle que $\forall\nu\in\cV(G)$, $c_{G}(\nu)$ est la \textbf{coordination} du vertex $\nu$ dans $G$ (c'est le nombre de lignes de $G$ accrochées au vertex $\nu$). Un vertex $\nu$ d'un graphe amputé (ou d'un sous-graphe) $g$ de la théorie $\phi^{4}$ sera dit \textbf{interne} si $c_{g}(\nu)=4$. Il sera \textbf{externe} si $1\les c_{g}(\nu)\les 3$. Nous noterons $\mathbf{\cV_{i}(G)}$ (resp. $\mathbf{\cV_{e}(G)}$) l'ensemble des vertex internes (resp. externes) de $G$ : $\cV(G)=\cV_{i}(G)\cup\cV_{e}(G)$ avec $\cV_{i}(G)\cap\cV_{e}(G)=\emptyset$. Nous noterons $\mathbf{N(G)}$ le nombre de pattes externes (amputées) de $G$ : $N(G)=\sum_{\nu\in\cV_{e}(G)}4-c_{G}(\nu)$. Remarquons que tous les vertex externes d'un graphe amputé sont intégrés. Par défaut, nous noterons $z_{\nu},\,\nu\in\cV_{e}(G)$ les positions de ces vertex.\\

Nous définissons $A_{G}\defi\int\prod_{\nu\in\cV_{e}(G)}dz_{\nu}\,A_{G}(\{z_{\nu}\})$. Quel que soit $\nu_{0}\in\cV_{e}(G)$, l'invariance par translation du modèle implique que $A_{G}(\{z_{\nu}\})$ est indépendant de $z_{\nu_{0}}$.

Nous pouvons maintenant énoncer le théorème :
\begin{thm}[Weinberg uniforme]\label{thm:Weinberg}
  Soit $G$ un graphe de $\phi^{4}_{4}$, connexe, amputé et complètement conver\-gent (\ie{} $\forall g\subset G$, $\omega(g)>0$). Soit $f\in L^{1}(\R^{4})$. Quel que soit $\nu_{0}\in\cV_{e}(G)$, il existe $K\in\R_{+}$ tel que
  \begin{align}
    \int\prod_{\nu\in\cV_{e}(G)}dz_{\nu}\,f(z_{\nu_{0}})A_{G}(\{z_{\nu}\}_{\nu\neq\nu_{0}})\les&K^{n(G)}\|f\|_{1}.\label{eq:Weinbbound}
  \end{align}
\end{thm}
Le théorème de Weinberg original \cite{weinberg60} affirme que l'amplitude d'un graphe complètement convergent est finie. Ici nous démontrons en plus que l'amplitude est bornée par $K^{n}$. L'adjectif \emph{uniforme} signifie que la constante $K$ ne dépend pas de $n(G)$. La démonstration du théorème \ref{thm:Weinberg} sera l'occasion de continuer à introduire les outils de l'analyse mutli-échelles. 
\begin{proof}
  Soit $G$ un graphe connexe amputé et $\mu$ une attribution d'indices. Soit $\nu_{0}\in\cV_{e}(G)$. On a :
  \begin{align}
    A^{\mu}_{G}(z_{\nu_{0}})\defi&\int\prod_{\substack{\nu\in\cV(G)\tq\\\nu\neq\nu_{0}}}dx_{\nu}\prod_{l\in\cI(G)}C_{l}^{i_{l}}.  
  \end{align}
  L'invariance par translation du modèle montre que $A_{G}(z_{\nu_{0}})$ est indépendante de $z_{\nu_{0}}$. C'est pourquoi nous avons besoin d'une fonction test pour intégrer sur $z_{\nu_{0}}$.

Le lemme \ref{lem:bornepropaphi4comm} implique que l'amplitude est bornée par
\begin{align}
  |A^{\mu}_{G}|\les&\int dz_{\nu_{0}}|f(z_{\nu_{0}})|\int\prod_{\substack{\nu\in\cV(G)\tq\\\nu\neq\nu_{0}}}dx_{\nu}\prod_{l\in\cI(G)}M^{(D-2)i_{l}}e^{-kM^{i_{l}+1}|x_{l}-y_{l}|}.
\end{align}
La structure minimale nécessaire pour intégrer sur les positions des vertex est un arbre générateur $\cT(G)$. Il s'agit d'un ensemble connexe de $n(G)-1$ lignes de $\cI(G)$ sans boucle. Par exemple, sur la figure \ref{fig:exgraph}, un arbre est représenté en gras. Il est pratique de choisir un arbre générateur enraciné \ie{} un arbre avec un vertex marqué appelé racine. Nous choisissons $\nu_{0}$ comme racine. Cet arbre représente simplement un ensemble minimale de fonctions $\prod_{l\in\cT}e^{-kM^{i_{l}+1}|x_{l}-y_{l}|}$ permettant d'intégrer sur tous les vertex de $G$ sauf un. On a alors
\begin{align}
  |A^{\mu}_{G}|\les&\prod_{l\in\cI(G)}M^{(D-2)(i_{l}+1)}\int dz_{\nu_{0}}|f(z_{\nu_{0}})|\int\prod_{\substack{\nu\in\cV(G)\tq\\\nu\neq\nu_{0}}}dx_{\nu}\prod_{l\in\cT(G)}e^{-kM^{i_{l}+1}|x_{l}-y_{l}|}\\
  \les&K_{1}^{n(G)}\|f\|_{1}\prod_{l\in\cI(G)}M^{(D-2)(i_{l}+1)}\prod_{l\in\cT(G)}M^{-D(i_{l}+1)}.\label{eq:borneinterm1}
\end{align}
En dimension $D\ges 2$, chaque ligne $l$ apporte un facteur de divergence $M^{(D-2)i_{l}}$. Cependant chaque ligne d'arbre fournit également un bon facteur $M^{-Di_{l}}$. Il est donc nécessaire de choisir un arbre avec les lignes les plus hautes possible pour optimiser la borne \eqref{eq:borneinterm1}.

En section \ref{sec:optim-de-larbre}, se trouve la procédure précise d'optimisation de l'arbre. Remarquons toutefois que cette optimisation est faite de telle sorte que l'arbre contienne les lignes les plus basses. Néanmoins la procédure est identique dans notre cas. Une simple induction de l'échelle de la coupure $\rho$ vers l'échelle $0$ permet de choisir un arbre optimisé $\cT_{\mu}$. Nous commençons par l'échelle $\rho$. Nous choisissons le plus grand nombre de lignes qui ne forment pas de boucles. Si l'ensemble de ces lignes forme un arbre, $\cT_{\mu}$ est complet et uniquement constitué de lignes d'échelle $\rho$. Sinon nous ajoutons à l'ensemble précédent le maximum de lignes d'échelle $\rho-1$ qui ne forment pas de boucle et ainsi de suite. L'arbre (en gras) de la figure \ref{fig:exgraph} est un exemple d'arbre optimisé. Remarquons qu'il existe généralement plusieurs arbres optimisés.\\

La borne \eqref{eq:borneinterm1} devient
\begin{align}
    |A^{\mu}_{G}|\les&K_{1}^{n(G)}\|f\|_{1}\prod_{l\in\cI(G)}\prod_{i=0}^{i_{l}}M^{D-2}\prod_{l\in\cT_{\mu}(G)}\prod_{j=0}^{i_{l}}M^{-D}.\label{eq:borneapparitionfacteurs}
\end{align}
Dans l'équation précédente, tout se passe comme si, chaque ligne $l$ du graphe fournissait un facteur $M^{D-2}$ par étage entre les étages $i_{l}$ et $0$. De même, toute ligne d'arbre donne $M^{-D}$ par étage entre $i_{l}$ et $0$. Ainsi, dans le comptage de puissance, la contribution de chaque ligne $l$ n'apparait qu'à partir de l'échelle $i_{l}$. Il est alors naturel de considérer :
\begin{defn}[Composantes connexes]\label{def:CC}
  Soit $G$ un graphe connexe amputé et $\mu$ une attribution d'échelles. Nous définissons $G^{i}$ comme l'ensemble des lignes de $G$ d'indices supérieurs ou égaux à $i$. Pour tout $i\in\lnat 0,\rho\rnat$,
  \begin{align}
    \cI^{i}(G)\defi&\lb l\in\cI(G)\tqs i_{l}\ges i\rb,\\
    G^{i}\defi&(\cV_{\cI^{i}},\cI^{i}(G))\subset G.
  \end{align}
Le sous-graphe $G^{i}$ n'est, en général, pas connexe. Soit $N_{c}(G^{i})$ le nombre de composantes connexes de $G^{i}$,
\begin{align}
  G^{i}\defi&\bigcup_{k=1}^{N_{c}(G^{i})}G^{i}_{k}.  
\end{align}
Les $G^{i}_{k},\,i\in\lnat 0,\rho\rnat,k\in\lnat 1,N_{c}(G^{i})\rnat$ sont les \textbf{composantes connexes} de $G$ ou ses sous-graphes \textbf{quasi-locaux}.
\end{defn}
Les composantes connexes sont les sous-graphes de $G$ qui \og{}apparaissent\fg{} quand on parcourt les échelles de $\rho$ à $0$. La figure \ref{fig:CC} montre les composantes connexes du graphe de la figure \ref{fig:exgraph}.

Soit $g\subset G$. Il existe un moyen simple de déterminer si $g$ est une composante connexe pour la paire $(G,\mu)$. Soit $\mathbf{\cI_{e}(g)}$ l'ensemble des lignes de $\cI(G)\setminus\cI(g)$ dont au moins une extrémité appartient à $\cV_{e}(g)$. Ce sont les pattes externes du sous-graphe $g$. Soient
\begin{subequations}
  \begin{align}
    i_{g}(\mu)\defi&\min_{l\in\cI(g)}i_{l},\\
    e_{g}(\mu)\defi&\max_{l\in\cI_{e}(g)}i_{l}
  \end{align}
\end{subequations}
alors $g$ est une composante connexe de $(G,\mu)$ si et seulement si $i_{g}(\mu)>e_{g}(\mu)$. Un sous-graphe est une composante connexe si toutes ses lignes internes sont au-dessus de toutes ses pattes externes. Par convention, les pattes externes de $G$ sont d'échelle $-1$ si bien que $G$ est toujours une composante connexe. La borne \eqref{eq:propibound} indique qu'un propagateur à l'échelle $i$ a une extension spatiale bornée par $M^{-i}$. Pour une composante connexe $g$, l'extension spatiale de ses propagateurs internes est au plus $M^{i_{g}}$. Celle de ses pattes externes est au moins $M^{-e_{g}}>M^{-i_{g}}$. Ainsi la taille typique d'une composante connexe d'échelle $i$ est d'ordre $M^{-i}$. Celle-ci est reliée à son environnement extérieur par des propagateurs plus longs et apparaît donc comme quasi-ponctuelle du point de vue de l'extérieur d'où le qualificatif de quasi-local.\\
\begin{figure}[htbp!]
  \centering
  \includegraphics[scale=.8]{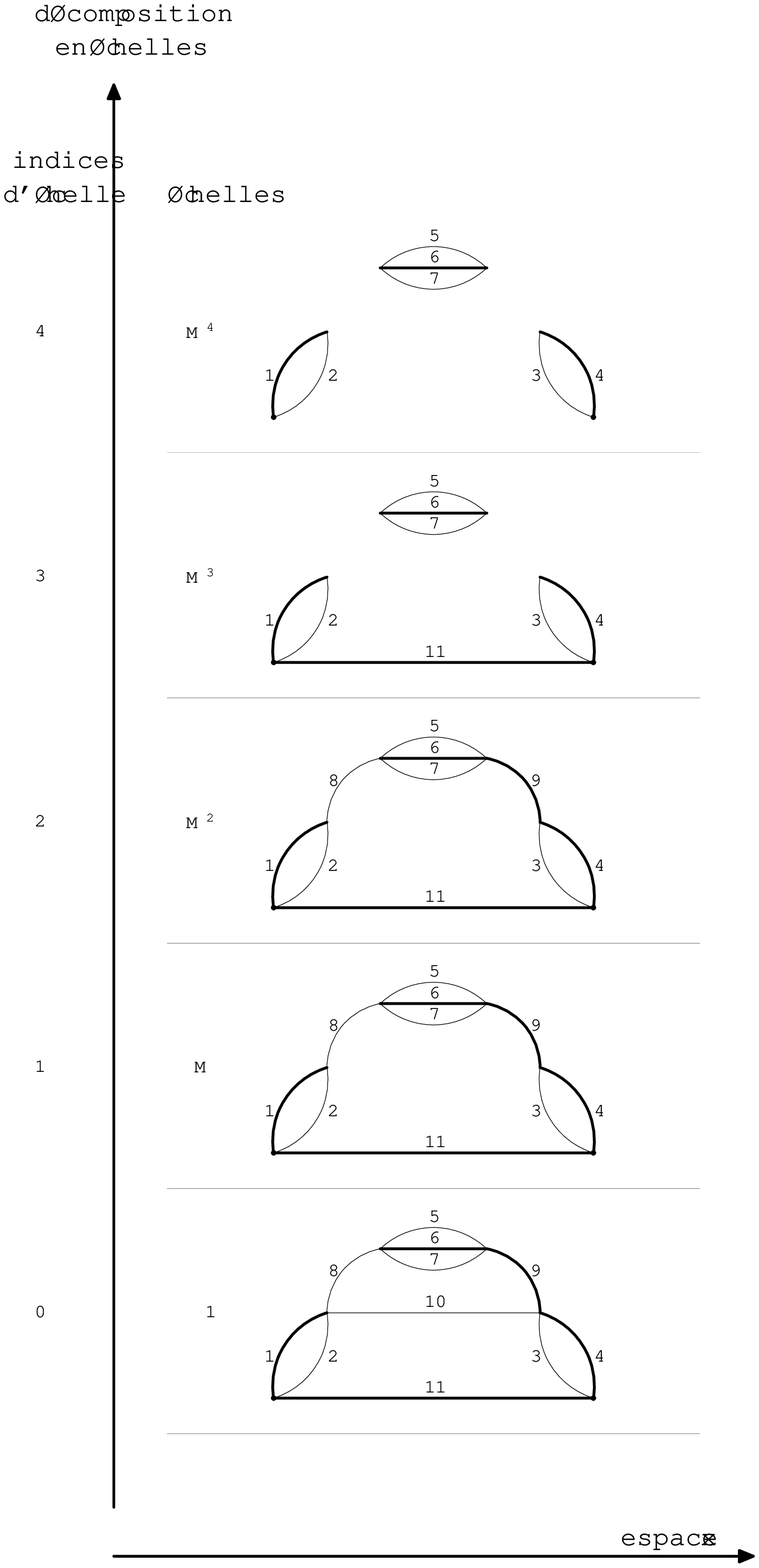}
  \caption{Composantes connexes}
  \label{fig:CC}
\end{figure}

Par construction, deux sous-graphes quasi-locaux de $G$ sont soit disjoints soit inclus l'un dans l'autre. Ainsi l'ensemble des composantes connexes d'un graphe $G$ donné forme une forêt (au sens de Zimmermann). $G$ lui-même étant toujours quasi-local, cette forêt n'est en fait constituée que d'un seul arbre dont la racine est le graphe $G$ lui-même. Nous l'appellerons par la suite arbre de Gallavotti-Nicol\`o bien qu'il soit légèrement différent de celui qu'utilisent Gallavotti et Nicol\`o \cite{GaN,Gallavotti1985qc,Gallavotti1985ib}. La figure \ref{fig:GNarbre} représente l'arbre des composantes connexes du graphe de la figure \ref{fig:CC}. Chaque noeud de cet arbre est une composante connexe, chaque lien est une relation d'inclusion.\\

Revenons maintenant à la borne \eqref{eq:borneapparitionfacteurs}. Nous allons montrer qu'elle s'écrit naturellement en termes des composantes connexes.
\begin{align}
    |A^{\mu}_{G}|\les&K_{1}^{n(G)}\|f\|_{1}\prod_{l\in\cI(G)}\,\prod_{i=0}^{i_{l}}M^{D-2}\;\prod_{l\in\cT_{\mu}(G)}\,\prod_{j=0}^{i_{l}}M^{-D}\tag{\ref{eq:borneapparitionfacteurs}}\\
    =&K_{1}^{n(G)}\|f\|_{1}\prod_{l\in\cI(G)}\,\prod_{\substack{(i,k)\tq\\l\in\cI(G^{i}_{k})}}M^{D-2}\;\prod_{l\in\cT_{\mu}(G)}\,\prod_{\substack{(i,k)\tq\\l\in\cI(G^{i}_{k})}}M^{-D}\notag\\
    =&K_{1}^{n(G)}\|f\|_{1}\prod_{(i,k)}\,\prod_{l\in\cI(G^{i}_{k})}M^{D-2}\;\prod_{(i,k)}\,\prod_{l\in\cI(G^{i}_{k})\cap\cT_{\mu}(G)}M^{-D}
\end{align}
L'arbre $\cT_{\mu}$ a été optimisé de telle sorte qu'il contienne les lignes les plus hautes possibles. Il n'est pas difficile de se rendre compte, à partir de la définition \ref{def:CC} des composantes connexes, que l'arbre $\cT_{\mu}$ est sous-arbre dans chaque composante connexe. Nous noterons $\mathbf{\cT^{i}_{k}}$ la restriction de $\cT_{\mu}$ à la composante connexe $G^{i}_{k}$ : $\cT^{i}_{k}=\cT_{\mu}\cap\cI(G^{i}_{k})$. Le nombre de lignes dans une composante $G^{i}_{k}$ est $I(G^{i}_{k})$. Le cardinal de $\cT^{i}_{k}$ est $n(G^{i}_{k})-1$. On a donc
\begin{align}
    |A^{\mu}_{G}|\les&K_{1}^{n(G)}\|f\|_{1}\prod_{\substack{(i,k)\in\\\lnat 0,\rho\rnat\times\lnat 1,N_{c}(G^{i})\rnat}}M^{-\omega(G^{i}_{k})},\label{eq:comptpuissfactor}\\
    \omega(G^{i}_{k})=&Dn(G^{i}_{k})+(2-D)I(G^{i}_{k})-D=(4-D)n(G^{i}_{k})+\frac{D-2}{2}N(G^{i}_{k})-D.\label{eq:degsuperfconvCC}
\end{align}
Nous retrouvons le degré superficiel de convergence \eqref{eq:degsuperfconv} mais cette fois-ci pour toutes les composantes connexes de $G$. Remarquons également que le comptage de puissance \eqref{eq:comptpuissfactor} fait uniquement intervenir les sous-graphes \emph{quasi-locaux}. Les sous-graphes de $G$ qui ont au moins une ligne interne plus basse que leurs pattes externes ne participent pas à la borne \eqref{eq:comptpuissfactor}.

Encore une fois, si $D>4$, la théorie n'est pas renormalisable : $\forall N\in\N,\exists n_{0}\in\N\tq\forall n>n_{0},\,\omega<0$. Si $D<4$, le modèle est super-renormalisable : les graphes divergents, qui contribuent à un nombre fini de fonctions, sont en nombre fini. Enfin, en dimension $4$, la théorie est juste renormalisable. Le degré superficiel de convergence ne dépend pas du nombre de vertex et le nombre de fonctions divergentes est fini.\\

Pour démontrer le théorème \ref{thm:Weinberg}, il reste à prouver que nous pouvons effectuer la somme sur les attributions d'échelles au prix d'une constante à la puissance $n(G)$. Nous nous restreindrons à la dimension $4$ à partir de maintenant. Par hypothèse, tous les sous-graphes de $G$ sont complètement convergents. Ses composantes connexes le sont donc aussi quelle que soit l'attribution $\mu$. Cela signifie que, quels que soient $i,k$, $N(G^{i}_{k})\ges 6$ ce qui entraîne
\begin{align}
  \omega(G^{i}_{k})=&N(G^{i}_{k})-4\ges\frac{N(G^{i}_{k})}{3}.
\end{align}
Soit $\nu\in\cV(G)$. Nous définissons
\begin{subequations}
  \begin{align}
    e_{\nu}(\mu)=&\max_{l\in\cI_{\nu}(G)}i_{l},\\
    i_{\nu}(\mu)=&\min_{l\in\cI_{\nu}(G)}i_{l}.
  \end{align}\label{eq:indicesvertex}
\end{subequations}
Quels que soient $i\in\N$ et $\nu\in\cV(G)$, si $i\les i_{\nu}(\mu)$ alors $\nu$ appartient à une unique composante connexe sinon $\nu$ n'appartient à aucune composante connexe. De plus, $\nu$ est un vertex externe si $i_{\nu}<i\les e_{\nu}$. On a donc
\begin{align}
    |A^{\mu}_{G}|\les&K_{1}^{n(G)}\|f\|_{1}\prod_{(i,k)}M^{-\omega(G^{i}_{k})}\les K_{1}^{n(G)}\prod_{(i,k)}M^{-N(G^{i}_{k})/3}\notag\\
    \les&K_{1}^{n(G)}\|f\|_{1}\prod_{(i,k)}\prod_{\substack{\nu\in\cV(G^{i}_{k})\tq\\e_{\nu}(\mu)< i\les i_{\nu}(\mu)}}M^{-1/3}.\label{eq:1}
    \intertext{La dernière équation est une inégalité car le nombre de pattes externes d'un sous-graphe est supérieur ou égal à son nombre de vertex externes.}
    |A^{\mu}_{G}|\les&K_{1}^{n(G)}\|f\|_{1}\prod_{\nu\in\cV(G)}\prod_{\substack{(i,k)\tq\\e_{\nu}(\mu)< i\les i_{\nu}(\mu)}}M^{-1/3}=K_{1}^{n(G)}\prod_{\nu\in\cV(G)}M^{-(e_{\nu}-i_{\nu})/3}.\label{eq:2}
\end{align}
Pour les graphes complètement convergents, l'intégration sur les positions a donné une décroissance exponentielle dans la longueur des lignes verticales (les vertex). Cette décroissance est duale de celle des propagateurs. Elle permet de sommer sur les échelles, l'invariance par translation des décroissances obtenues étant brisée par l'échelle $-1$ des pattes externes. En effet,
\begin{align}
  \prod_{\nu\in\cV(G)}M^{-(e_{\nu}-i_{\nu})/3}\les&\prod_{\nu\in\cV(G)}\ \prod_{(l,l')\in\cI_{\nu}\times\cI_{\nu}}M^{-|i_{l}-i_{l'}|/18}\label{eq:3}
\end{align}
De la même façon que l'intégration sur les positions des vertex du graphe est faite en choisissant un arbre parmi les lignes horizontales, la somme sur les attributions est faite en choisissant un arbre parmi les lignes verticales qui connecte les lignes (donc les échelles) du graphe. On obtient alors
\begin{align}
  \prod_{l\in\cI(G)}\sum_{i_{l}=0}^{\infty}|A^{\mu}_{G}|\les&K_{1}^{n(G)}K_{2}^{I(G)}\|f\|_{1}\les K^{n(G)}\|f\|_{1}
\end{align}
ce qui achève la preuve du théorème.
\end{proof}
\begin{rem}
  Dans \cite{Feldman1985xv}, le théorème \ref{thm:Weinberg} a été démontré pour une très large classe de modèles. En effet, pour toute théorie vérifiant les hypothèses :
  \begin{enumerate}
  \item $\forall l\in\cI(G)$, $|C_{l}(x,y)|\les K'\int\frac{e^{\imath k\cdot(x-y)}}{(k^{2}+m^{2})^{p(l)}}d^{d}k$ avec $p(l)>0$,
  \item la coordination des vertex est arbitraire,
  \item $\forall g\subset G$, il existe $\veps>0$ tel que le degré superficiel de divergence
    \begin{align}
      -\omega(g)=&(I(g)-n(g)+1)d-2\sum_{l\in\cI(g)}p(l)
      \intertext{obéisse à la borne}
      \omega(g)\ges&2\veps N(g),\label{eq:bornedegconv}
    \end{align}
  \end{enumerate}
  le théorème \ref{thm:Weinberg} est vrai. La condition \eqref{eq:bornedegconv} est très naturelle. Elle est notamment vérifiée par toutes les théories renormalisables connues. Néanmoins, qualifions de renormalisable un comptage de puissance qui est convergent à partir d'un nombre minimum de pattes externes. Ainsi un modèle renormalisable en ce sens a un degré superficiel de convergence qui s'écrit 
  \begin{align}
    \omega=&f(N),\,\exists N_{0}\in\N\tq\forall N>N_{0},\,f(N)>0.\label{eq:conditionrenorm}
  \end{align}
Ceci n'implique pas \eqref{eq:bornedegconv}. Rappelons que toute les théories renormalisables connues obéissent bien sûr à \eqref{eq:conditionrenorm} mais aussi à \eqref{eq:bornedegconv}. Si toutefois nous avions affaire à un modèle renormalisable ne vérifiant que \eqref{eq:conditionrenorm}, la borne que nous obtiendrions dépendrait du comportement précis de $f(N)$. Par exemple, si $f(N)$ est seulement bornée (inférieurement) par une constante, le facteur $K^{n(G)}$ dans \eqref{eq:Weinbbound} deviendrait $I(G)!$.
\end{rem}
\begin{rem}
  Considérons le cas d'une théorie super-renormalisable comme $\phi^{4}_{2}$. À partir de \eqref{eq:degsuperfconvCC}, on a $\omega=2n-2$ qui, pour $n\ges 2$, vérifie $\omega\ges n$. On a alors
  \begin{align}
    |A^{\mu}_{G}|\les&\prod_{(i,k)}M^{-n(G^i_{k})}=\prod_{(i,k)}\prod_{\nu\in\cV(G^i_{k})}M^{-1}\\
    =&\prod_{\nu\in\cV(G)}\prod_{\substack{(i,k)\tq\\\nu\in\cV(G^i_{k})}}M^{-1}=\prod_{\nu\in\cV(G)}M^{-i_{\nu}(\mu)}\notag\\
    \les&\prod_{\nu\in\cV(G)}\prod_{l\in\cI_{\nu}(G)}M^{-i_{l}/4}.\label{eq:4}
  \end{align}
  Si nous considérons les décroissances exponentielles verticales comme des \emph{ressorts} verticaux, nous remarquons, en comparant \eqref{eq:4} à \eqref{eq:3}, qu'un sous-graphe d'un modèle super-renormalisable est \og{}attaché\fg{} au sol (l'échelle $0$) alors qu'un sous-graphe d'une théorie juste renormalisable n'est attaché qu'à ses vertex externes. Cette image intuitive sera utile pour la renormalisation (voir section \ref{sec:ampl-renorm}).
\end{rem}

Concluons cette section en notant que le théorème \ref{thm:Weinberg} est en fait valable pour de nombreuses attributions correspondant à des graphes qui ne sont pas complètement convergents. En effet, le comptage de puissance \eqref{eq:comptpuissfactor} ne fait intervenir que les sous-graphes quasi-locaux. Soit un graphe $G$ connexe, pas nécessairement complètement convergent. Qualifions de \emph{convergente} une attribution telle que tous les sous-graphes à deux et quatre points ne soient pas quasi-locaux. Alors
\begin{align}
  \Big|\sum_{\mu\text{ convergentes}}A_{G}^{\mu}\Big|\les&K^{n(G)}.\label{eq:muconv}
\end{align}
Ce résultat est non trivial (la somme est non vide) uniquement si $G$ est superficiellement convergent car, par définition, $G$ est toujours quasi-local. Évidemment, pour un graphe $G$ complètement convergent, toutes les attributions sont convergentes et le théorème \ref{thm:Weinberg} est alors un cas particulier de \eqref{eq:muconv}.

\section{Amplitudes renormalisées}
\label{sec:ampl-renorm}

L'analyse mutli-échelles nous a permis de montrer qu'un graphe sans sous-graphe divergent est fini mais aussi que son amplitude est uniformément bornée par $K^{n}$. Plus important, le comptage de puissance \eqref{eq:comptpuissfactor} et l'équation \eqref{eq:muconv} nous apprennent que les divergences ne peuvent provenir que des composantes connexes. Et c'est heureux. En effet, les composantes connexes sont quasi-locales. Leur extension spatiale est d'ordre $M^{i_{g}}$ alors que celle de leurs propagateurs externes est plutôt $M^{e_{g}}$. Du point de vue de l'extérieur, les composantes connexes apparaissent comme \emph{quasi-locales}. Cette qualité est d'autant plus prononcée que la différence $i_{g}-e_{g}$ est grande ce qui est d'ailleurs l'origine des divergences ultraviolettes. Ainsi il n'est pas étonnant que ces divergences puissent être absorbées par un contreterme (exactement) local : la \og{}différence\fg{} entre l'amplitude d'un graphe quasi-local et un contreterme local est certainement convergente. Pour illustrer ce fait, nous revisiterons plus loin l'exemple du graphe bulle dans le cadre de l'analyse multi-échelles.

L'autre point important est que les sous-graphes quasi-locaux ont une structure en arbre pour la relation d'inclusion : c'est l'arbre de Gallavotti-Nicol\`o. Ainsi les composantes connexes divergentes (pour $\phi^{4}_{4}$, ce sont les sous-graphes quasi-locaux à deux et quatre points) forment une forêt. Un sous-ensemble d'un arbre est une forêt qui peut être aussi un arbre si le sous-ensemble contient la racine du premier arbre. Cette structure résoud un des problèmes les plus épineux de la renormalisation perturbative : les sous-graphes divergents à intersection non triviale (\og{}overlapping divergences\fg{} ou divergences enchevêtrées). Très tôt, il a été remarqué qu'une renormalisation qui utilise des contretermes dans le lagrangien donne lieu à des contributions indicées par des forêts \cite{Bogoliubov1957gp,bogoliubov59,Hepp1966eg,hepp69,Zimmermann1969jj}. Ces forêts de contretermes rendent les amplitudes renormalisées finies. Mais, par définition, les sous-graphes enchevêtrés ne peuvent pas appartenir à une même forêt. Les sous-graphes quasi-locaux sont les seuls à devoir être renormalisés et sont naturellement organisés en forêts. Le problème des divergences  \og{}overlappantes\fg{} n'apparaît donc tout simplement pas en analyse multi-échelles car il n'y a pas besoin d'associer un contreterme à de tels graphes.

\subsection{Contretermes}
\label{sec:contretermes}

Nous allons maintenant illustrer sur l'exemple de la bulle la procédure d'extraction de la partie divergente d'un graphe, autrement dit la définition des contretermes. 

L'analyse multi-échelles a été principalement développée en espace $x$. Cela permet de l'utiliser pour la théorie constructive. La procédure de prise de contreterme à moments zéro (le schéma de soustraction BPHZ) se traduit en espace
$x$ par un déplacement des pattes externes en un même point. Précisemment, soient $x_{1},\dotsc,x_{N}$ les points externes d'un graphe $G$. Si nous notons $a(x_{1},\dots,x_{N})$ la fonction test contre laquelle
$A_{G}(x_{1},\dots,x_{N})$ est intégrée et $\tau$ l'opérateur qui met à zéro les moments externes dans l'amplitude amputée $A_{G}$, nous avons la propriété suivante :
\begin{prpt}[Contreterme en espace $x$]\label{prpt:contreterme}
  \begin{align}
    &\ \int dx_{1}\dots dx_{N}\ (\tau
    A_{G})(x_{1},\dots,x_{N})\,a(x_{1},\dots,x_{N})\label{eq:taustar}\\
    =&\ \int dx_{1}\dots dx_{N}\ A_{G}(x_{1},\dots,x_{N})\,
    (\tau^{\ast}a)(x_{1},\dots,x_{N})\nonumber
  \end{align}
  avec $(\tau^{\ast}a)(x_{1},\dots,x_{N})=\sum_{j=1}^{D(G)}\frac{1}{j!}
  \frac{d^{j}}{dt^{j}}\ a(x_{1}(t),\dots,x_{N}(t))\big|_{t=0}$ et
  $x_{i}(t)=x_{N}+t(x_{i}-x_{N})$.
\end{prpt}
\begin{rem}
  L'équation (\ref{eq:taustar}) ne définit pas l'opérateur $\tau^{*}$ de
  manière unique. En effet on aurait pu remplacer $x_{N}$ par n'importe quel
  point parmi les $x_{i}$. C'est l'invariance par
  translation de $A_{G}$ qui est responsable de cette ambiguïté.
\end{rem}
\begin{proof}
  On montre ici comment le second membre de (\ref{eq:taustar}) est égal au
  premier.
  \begin{align}
    &\ \int dx_{1}\dots dx_{N}\ A_{G}(x_{1},\dots,x_{N})\,
    (\tau^{\ast}a)(x_{1},\dots,x_{N})\nonumber\\
    =&\sum_{j=1}^{D(G)}\frac{1}{j!}\frac{d^{j}}{dt^{j}}\ \int dx_{1}\dots
    dx_{N}\ A_{G}(x_{1},\dots,x_{N})\,
    a(x_{1}(t),\dots,x_{N}(t))\big|_{t=0}\label{eq:interm1}
  \end{align}
  où $x_{i}(t)=x_{N}+t(x_{i}-x_{N})$. $A_{G}$ étant invariante par
  translation, (\ref{eq:interm1}) devient
  \begin{align}
    &\sum_{j=1}^{D(G)}\frac{1}{j!}\frac{d^{j}}{dt^{j}}\ \int dx_{1}\dots
    dx_{N}\ A_{G}(x_{1}-x_{N},\dots,0)\,
    a(x_{1}(t),\dots,x_{N}(t))\big|_{t=0}\nonumber\\
    =&\sum_{j=1}^{D(G)}\frac{1}{j!}\frac{d^{j}}{dt^{j}}\ \int dx_{1}\dots
    dx_{N}\, \frac{d^{N-1}p}{(2\pi)^{d(N-1)}}\, \frac{d^{N}k}{(2\pi)^{dN}}\ 
    \hat{A}(p_{i})e^{-\imath p\cdot(x-x_{N})}\hat{a}(k_{i})e^{\imath k\cdot
      x(t)}\big|_{t=0}\label{eq:interm2}
  \end{align}
  avec $d^{N}k=\prod_{i=1}^{N}dk_{i}$ et $k\cdot
  x(t)=\sum_{i=1}^{N}k_{i}x_{i}(t)$. L'équation (\ref{eq:interm2}) contient
  l'intégrale
  \begin{align}
    \int dx_{1}\dots dx_{N-1}\, \exp\imath\sum_{i=1}^{N-1}x_{i}\lbt
    tk_{i}-p_{i}\rbt=(2\pi)^{d(N-1)}\prod_{i=1}^{N-1}\delta\lbt
    tk_{i}-p_{i}\rbt.\nonumber
  \end{align}
  Après intégration sur les $p_{i},\, i\in\lnat 1,N-1\rnat$, on a
  \begin{align}
    &\ \int dx_{1}\dots dx_{N}\ A_{G}(x_{1},\dots,x_{N})\,
    (\tau^{\ast}a)(x_{1},\dots,x_{N})\nonumber\\
    =&\sum_{j=1}^{D(G)}\frac{1}{j!}\frac{d^{j}}{dt^{j}}\ \int dx_{N}\,
    \frac{d^{N}k}{(2\pi)^{dN}}\ 
    \hat{A}(tk_{i})\hat{a}(k_{i})\exp\imath\sum_{i=1}^{N}k_{i}x_{N}\big|_{t=0}\nonumber\\
    =&\sum_{j=1}^{D(G)}\frac{1}{j!}\frac{d^{j}}{dt^{j}}\ \int
    \frac{d^{N}k}{(2\pi)^{d(N-1)}}\ 
    \delta\big(\sum_{i=1}^{N}k_{i}\big)\hat{A}(tk_{i})\hat{a}(k_{i})\big|_{t=0}\nonumber\\
    =&\int \frac{d^{N}k}{(2\pi)^{d(N-1)}}\ 
    (\tau\hat{A}_{G})(k_{1},\dots,k_{N})\,
    \hat{a}(k_{1},\dots,k_{N})\nonumber\\
    =&\int d^{N}x\ (\tau
    A_{G})(x_{1},\dots,x_{N})\,a(x_{1},\dots,x_{N}).\nonumber
  \end{align}
\end{proof}

Considérons donc l'exemple du graphe bulle. Notons $i_{1},i_{2}$ les indices des propagateurs internes et $e_{1},\dotsc,e_{4}$ les indices des propagateurs externes. Nous écrirons qu'un sous-graphe $g$ est \textbf{dangereux} pour l'attribution $\mu$ si, d'une part, il est divergent au sens habituel (par exemple, s'il a deux ou quatre pattes externes dans $\phi^{4}_{4}$) et si, d'autre part, $i_{g}(\mu)>e_{g}(\mu)$. Un sous-graphe de $\phi^{4}_{4}$ est donc dangereux s'il a deux ou quatre pattes externes et s'il est quasi-local. Dans toute la suite, nous omettrons l'étoile de l'opérateur $\tau^{*}$. L'amplitude du graphe est
\begin{align}
  A_{G}=&\int dx_{1}dx_{2}\,C^{i_{1}}(x_{1},x_{2})C^{i_{2}}(x_{1},x_{2})C^{e_{1}}(x_{1},y_{1})C^{e_{2}}(x_{1},y_{2})C^{e_{3}}(x_{2},y_{3})C^{e_{4}}(x_{2},y_{4})\label{eq:taylsubstract}\\
  =&\int dx_{1}dx_{2}\,C^{i_{1}}(x_{1},x_{2})C^{i_{2}}(x_{1},x_{2})C^{e_{1}}(x_{1},y_{1})C^{e_{2}}(x_{1},y_{2})\notag\\
  &\times\Big\{C^{e_{3}}(x_{1},y_{3})C^{e_{4}}(x_{1},y_{4})\notag\\
  &+\int_{0}^{1}ds\,(x_{2}-x_{1})\cdot\nabla\big(C^{e_{3}}(x_{1}+s(x_{2}-x_{1}),y_{3})C^{e_{4}}(x_{1}s(x_{2}-x_{1}),y_{4})\big)\Big\}\notag\\
  =&\tau A_{G}+\int dx_{1}dx_{2}\,C^{i_{1}}(x_{1},x_{2})C^{i_{2}}(x_{1},x_{2})C^{e_{1}}(x_{1},y_{1})C^{e_{2}}(x_{1},y_{2})\notag\\
  &\times\int_{0}^{1}ds\,(x_{2}-x_{1})\cdot\nabla\big(C^{e_{3}}(x_{1}+s(x_{2}-x_{1}),y_{3})C^{e_{4}}(x_{1}s(x_{2}-x_{1}),y_{4})\big).\notag
\end{align}
À partir de la définition \eqref{eq:propslice} du propagateur dans une tranche, il est facile de montrer la borne suivante :
\begin{align}
   \big| (x_{2}-x_{1})\cdot\nabla C^{e_{3}}(x_{1}+s(x_{2}-x_{1}),y_{3})\big|\les&|(x_{2}-x_{1})|M^{3e_{3}}e^{-kM^{e_{3}}|x_{1}-y_{3}+s(x_{2}-x_{1})|}\\
   \les&|(x_{2}-x_{1})|M^{3e_{3}}e^{-kM^{e_{3}}|x_{1}-y_{3}|}\notag.
\end{align}
Pour la deuxième inégalité, nous avons utilisé la forte décroissance des propagateurs internes. En utilisant la même borne pour le deuxième propagateur externe, il n'y a plus de dépendance en $s$ dans l'intégrale et celle-ci peut être bornée par $1$. Finalement, l'amplitude renormalisée est bornée par
\begin{align}
  (1-\tau)A_{G}\les&KM^{-2|i_{1}-i_{2}|}\cO(1)\,M^{-(\max(i_{1},i_{2})-\max(e_{3},e_{4}))}.
\end{align}
Supposons que la bulle soit dangereuse \ie{} $\min(i_{1},i_{2})>\max(e_{1},e_{2},e_{3},e_{4})$. Considérons $i_{1}>i_{2}$, le facteur $M^{-2(i_{1}-i_{2})}$ vient du fait qu'à l'échelle $i_{1}$, la ligne d'échelle $i_{2}$ n'apparaît pas encore si bien que nous avons affaire à une fonction à six points qui converge quadratiquement. À l'échelle $i_{2}$, la bulle est complète et sa divergence est logarithmique. C'est le facteur $\cO(1)$. Nous constatons ainsi que l'opération $1-\tau$ a fourni un facteur supplémentaire $M^{-(\max(i_{1},i_{2})-\max(e_{3},e_{4}))}$. Le point important ici est qu'il permet d'effectuer la somme sur $i_{2}$ qui est logarithmiquement divergente dans $\tau A$. La renormalisation fournit assez de décroissance \og{}verticale\fg{} pour effectuer la somme sur les attributions d'échelles.

Revenons un instant sur le facteur $M^{-(\max(i_{1},i_{2})-\max(e_{3},e_{4}))}$. Lorsque le graphe à renormaliser est plus grand que la bulle, la distance $|x_{2}-x_{1}|$ (la distance parcourue par les propagateurs déplacés) ne peut généralement être bornée que par $M^{-i_{G}}$. De plus, le (mauvais) facteur $M^{\max(e_{3},e_{4}))}$ n'est pas non plus représentatif du cas général. En effet, nous ne pouvons pas choisir de déplacer les propagateurs externes les plus bas car le choix du vertex auquel nous les déplaçons est dicté par la forêt complète des contretermes. Le meilleur facteur qu'on puisse obtenir est donc $M^{-(i_{G}-e_{G})}$ qui est inférieur à $1$ si le graphe $G$ est quasi-local.\\

Le contreterme associé à la bulle est de la forme
\begin{align}
  \tau A_{G}=&\int dx_{1}\,\phi_{e_{1}}(x_{1})\phi_{e_{1}}(x_{1})\phi_{e_{3}}(x_{1})\phi_{e_{4}}(x_{1})\int dx_{2}\sum_{\substack{i_{1},i_{2}\tq\\i_{G}>e_{G}}}C^{i_{1}}(x_{1},x_{2})C^{i_{2}}(x_{1},x_{2}).\label{eq:taubulle}
\end{align}
L'intégrale sur $x_{2}$ est en fait indépendante de $x_{1}$ grâce à l'invariance par translation. Ainsi le contreterme $\tau A_{G}$ est de la forme d'un vertex local (en $x$) multiplié par un réel positif. Mais ce nombre dépend de l'échelle des champs au vertex à cause de la restriction sur la somme sur les attributions. La partie divergente de la bulle ne peut donc pas être absorbée dans une unique constante renormalisée mais peut l'être dans $\rho+1$ constantes \emph{effectives} qui dépendent d'un indice d'échelle. Pour pouvoir absorber la divergence de la bulle dans une unique constante renormalisée, il faudrait relâcher la contrainte $i_{G}>e_{G}$ et donc rajouter à \eqref{eq:taubulle} la somme sur les attributions telles que $i_{G}\les e_{G}$.

Nous qualifierons d'\textbf{utiles} les contretermes correspondant à des sous-graphes dangereux. Ils sont utiles dans le sens où ils annulent réellement une divergence. Par contre, nous avons vu dans la section précédente qu'aucune divergence n'est associée aux graphes dont au moins une patte interne est plus basse qu'une patte externe. Ainsi les contretermes correspondant à de tels graphes seront qualifiés d'\textbf{inutiles}. Pour exprimer la série perturbative en terme d'une unique constante de couplage renormalisée, nous sommes obligés d'extraire la partie locale des sous-graphes qui ne sont pas quasi-locaux et donc d'inclure dans la constante nue les contretermes inutiles.

Les contretermes inutiles sont finis car la somme sur leurs indices internes est bornée par l'indice d'au moins une de leurs pattes externes. Mais, dans le cas d'un graphe $g$ à quatre points, cette somme logarithmique se comporte comme $e_{g}$. Si ce sous-graphe est inséré un grand nombre de fois dans une boucle convergente, on aura
\begin{align}
  \sum_{e_{g}}M^{-ke_{g}}e_{g}^{n}\simeq&K^{n}n!.
\end{align}
Or nous avons vu en section \ref{subsec:series-perturbatives} que la fonction à quatre points renormalisée souffre du problème des renormalons \ie{} des contributions en $n!$. Le terme de renormalon prend maintenant tout son sens car nous venons de voir que ces $n!$ proviennent des contretermes inutiles. En toute rigueur, nous avons seulement constaté que les contretermes inutiles fournissent des contributions du type renormalons mais non que tous les renormalons leur sont dûs. En fait, nous donnerons dans la section \ref{sec:bornes-et-finitude} un théorème qui prouve qu'en l'absence de contretermes inutiles, les amplitudes \og{}renormalisées utilement\fg{} n'ont pas de renormalon.\\

Dans les deux sections suivantes, nous allons donner les grandes lignes de la preuve de la finitude des amplitudes renormalisées. Pour cela nous devons tout d'abord écrire la formule de forêts de Zimmermann dans le cadre de l'analyse multi-échelles. Ceci constituera le lemme de classification des forêts. Puis nous expliquerons comment ce lemme permet d'une part de montrer la finitude ordre par ordre de la série
renormalisée et d'autre part de donner une borne sur l'amplitude des graphes renormalisés. Le théorème que nous montrerons sera
\begin{thm}[BPHZ uniforme]\label{thm:BPHunif}
  Soit $G$ un graphe de $\phi^{4}_{4}$, connexe, amputé. Soit $h\in L^{1}(\R^{4})$. Soit $f(G)=\max_{\text{forêts divergentes $\cF$ de $G$}}\card\cF$. Quel que soit $\nu_{0}\in\cV_{e}(G)$, il existe $K\in\R_{+}$ telle que l'amplitude renormalisée $A_{G}^{R}$ (donnée par la formule (\ref{eq:Zimmforest})) soit bornée par
  \begin{align}
        \int\prod_{\nu\in\cV_{e}(G)}dz_{\nu}\,h(z_{\nu_{0}})A^{R}_{G}(\{z_{\nu}\}_{\nu\neq\nu_{0}})\les&K^{n(G)}f(G)!\|h\|_{1}.\label{eq:boundBPHunif}
  \end{align}
\end{thm}
Nous pouvons facilement nous convaincre que le nombre d'éléments d'une forêt de sous-graphes divergents d'un graphe donné est borné par $f(G)\les n(G)$.

\subsection{La forêt qui cache l'arbre}
\label{sec:la-foret-qui-cache}

Nous allons montrer comment organiser la somme sur les forêts divergentes en fonction de la structure de l'arbre de Gallavotti-Nicol\`o. Pour simplifier les arguments, nous nous restreindrons aux graphes ne contenant que des sous-divergences à quatre points. Ce qui suit est extrait de \cite{Riv1}.

Soit un graphe $G$ à quatre points ou plus ne contenant que des sous-divergences à quatre points. Soit une attribution d'échelle $\mu$. L'ensemble des sous-graphes quasi-locaux divergents forme une forêt notée $D_{\mu}$. Rappelons que pour rendre l'amplitude de $G$ finie, il est seulement nécessaire d'extraire les parties divergentes de ces graphes-là. Par ailleurs, nous avons remarqué grâce à l'exemple de la bulle que la prise de partie locale d'un graphe découple les lignes internes des pattes externes. Soit $g\subset G$, nous avons $\tau_{g}A_{G}=A_{G/g}\tau A_{g}$ avec $G/g$ le graphe $G$ où on a réduit $g$ à un point. Le fait que chaque $\tau_{g}$ découple $g$ de ses pattes externes nous incite à redéfinir une nouvelle forêt $D_{\mu}$ qui prenne en compte cette factorisation.

\begin{figure}[htbp!]
  \centering
  \subfloat[Un oeil multi-étages]{\label{fig:oeilmultiscale}\includegraphics{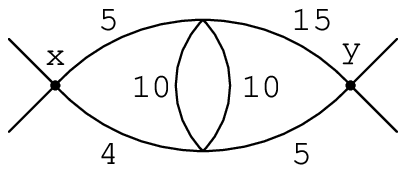}}\hspace{2cm}
  \subfloat[Factorisation d'un contreterme]{\label{fig:CTfactor}\includegraphics[scale=1]{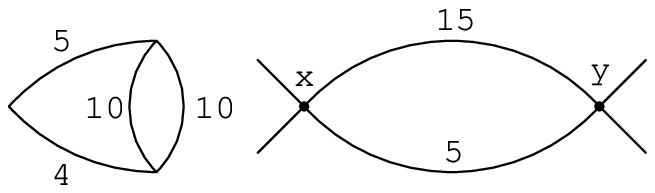}}
  \caption{De forêt inoffensive à forêt dangereuse}
  \label{fig:inofftodanger}
\end{figure}

Prenons l'exemple de la figure \ref{fig:oeil}. Sur la figure \ref{fig:oeilmultiscale}, nous avons attribué une échelle à chaque ligne de ce graphe. Considérons la forêt $\cF=\{G_{1},G_{2},G\}$ (voir la figure \ref{fig:oeil} et le paragraphe au-dessus). Dans $\cF$ et avec une telle attribution, ni $G_{1}$ ni $G_{2}$ ne sont dangereux. Cependant, l'application de l'opérateur $\tau_{G_{2}}$ déplace les lignes d'échelles $15$ et $5$ au point $x$ (voir figure \ref{fig:CTfactor}). Le graphe $G_{1}$ ne possède alors plus que deux pattes externes, plus basses que ses lignes internes. Remarquons que si le sous-graphe $G_{1}\subset G_{2}$ n'avait eu aucune patte externe en commun avec les pattes externes de $G_{2}$, l'application de $\tau_{G_{2}}$ n'aurait pas changé la nature dangereuse ou pas de $G_{1}$. Les seuls changements proviennent des sous-graphes $G_{1}\subset G_{2}$ qui ont des pattes externes en commun. De plus, il est facile de se rendre compte que si $G_{2}$ est dangereux alors $\tau_{G_{2}}$ ne change pas non plus la nature de $G_{1}$.

Voici comment définir une forêt dangereuse qui dépend d'une forêt de graphes. Nous écrirons qu'un sous-graphe $g$ est \textbf{compatible} avec une forêt $\cF$ si $\cF\cup\{g\}$ est une forêt. Nous définissons également $B_{\cF}(g)$ comme l'\textbf{ancêtre} de $g$ dans $\cF\cup\{g\}$ et $A_{\cF}(g)$ comme les \textbf{enfants} de $g$ dans $\cF$ : $A_{\cF}(g)=\cup_{h\subset g,\,h\in\cF}h$.

Nous pouvons alors définir deux sous-forêts de $\cF$ correpondant aux parties dangereuses et inoffensives de $\cF$. La forêt inoffensive $T_{\mu}(\cF)$ est le complémentaire dans $\cF$ de la forêt dangereuse $D_{\mu}(\cF)$ :
\begin{align}
    D_{\mu}(\cF)=&\lb g\in\cF\tqs i_{g}(\cF)>e_{g}(\cF)\rb\\
    \notag\\
    \text{avec }i_{g}(\cF)=&\min\lb i_{l}(\mu)\tqs l\in g\setminus A_{\cF}(g)\rb,\\
    e_{g}(\cF)=&\max\lb i_{l}(\mu)\tqs l\in\cI_{e}(g)\cap B_{\cF}(g)\rb\notag
\end{align}
$\cI_{e}(g)$ étant l'ensemble des lignes externes de $g$. Cette définition généralise les indices $i_{g}$ et $e_{g}$ car $i_{g}=i_{g}(\emptyset)$ et $e_{g}=e_{g}(\emptyset)$. De plus, elle généralise la notion de quasi-localité au graphe réduit $g/A_{\cF}(g)$.\\

Nous vérifions facilement que si $\cF_{1}\subset\cF_{2}$ alors $\forall g\in\cF_{1},\,i_{g}(\cF_{1})\les i_{g}(\cF_{2})$
et $e_{g}(\cF_{1})\ges e_{g}(\cF_{2})$. Nous avons le lemme suivant :
\begin{lemma}
  \begin{align}
    i_{g}(\cF)=&i_{g}(T_{\mu}(\cF)\cup\{g\}),\\
  e_{g}(\cF)=&e_{g}(T_{\mu}(\cF)\cup\{g\}).
  \end{align}
\end{lemma}
La preuve n'est pas très compliquée à partir des définitions précédentes. Néanmoins ce lemme est capital car il  implique $T_{\mu}(T_{\mu}(\cF))=T_{\mu}(\cF)$ si bien que l'ensemble $\cF^{D}(G)$ des forêts divergentes de quadrupèdes se décompose en classe d'équivalences sous l'action du projecteur $T_{\mu}$ :
\begin{align}
    \cF^{D}(G)=\bigcup_{\cF\tq T_{\mu}(\cF)=\cF}\lb\cF'\tqs T_{\mu}(\cF')=\cF\rb.
\end{align}
Les forêts satisfaisant $T_{\mu}(\cF)=\cF$ forment l'ensemble $I(\mu)$ des forêts inoffensives. À partir d'une forêt $\cF\in I(\mu)$, on peut construire la forêt maximale $\cF\cup H_{\mu}(\cF)$ qui satisfait $T_{\mu}(\cF\cup H_{\mu}(\cF))=\cF$ :
\begin{align}
    H_{\mu}(\cF)=&\lb g\subset G\tqs g\text{ compatible avec }\cF\text{ et }g\in D_{\mu}(\cF\cup\{g\})\rb.
\end{align}
Le lemme suivant caractérise alors les classes d'équivalences de $\cF^{D}(G)$ :
\begin{lemma}[Classification des forêts]
  Pour toute forêt $\cF\in I(\mu)$, on a 
  \begin{align}
    \cF\cup H_{\mu}(\cF)\in&\cF^{D}(G),\\
    \forall\cF'\in\cF^{D}(g), T_{\mu}(\cF')=\cF\Longleftrightarrow&\cF\subset\cF'\subset\cF\cup H_{\mu}(\cF).
  \end{align}
\end{lemma}
Nous renvoyons à \cite{Riv1} pour la preuve. Le fait important est que ce lemme permet de récrire l'opérateur de Zimmermann en factorisant un produit d'opérateurs $1-\tau$ correspondant aux graphes dangereux dans chaque forêt de sous-graphes inoffensifs incluant la forêt vide :
\begin{align}
  R=&\sum_{\cF\in\cF^{D}(G)}\ \prod_{g\in\cF}(-\tau_{g})=\sum_{\cF\in I(\mu)}\ \prod_{g\in\cF}(-\tau_{g})\prod_{h\in H_{\mu}(\cF)}(1-\tau_{h}).
\end{align}

\subsection{Bornes et finitude}
\label{sec:bornes-et-finitude}

Pour montrer le théoème \ref{thm:BPHunif}, nous écrivons d'abord l'amplitude renormalisée sous la forme
\begin{align}
  A_{G}^{R}=&\sum_{\cF\in\cF^{D}(G)}A_{G,\cF}^{R}\\
  A_{G,\cF}^{R}=&\sum_{\mu\tq\cF\in I(\mu)}\ \prod_{g\in\cF}(-\tau_{g})\prod_{h\in H_{\mu}(\cF)}(1-\tau_{h})A_{G}^{\mu}.
\end{align}
Pour chaque forêt $\cF$, nous pouvons montrer une borne du type \eqref{eq:boundBPHunif} puis conclure car le nombre de forêts divergentes de quadrupèdes est borné par $8^{n(G)}$ (voir \cite{deCalan1981ps,DeCalan1983dk}).\\

Nous appliquons tout d'abord les opérateurs $\tau_{g}$ pour tout $g\in\cF$. Ceux-ci déplacent les pattes externes de $g$ et internes à $B_{\cF}(g)$ en un point noté $\nu_{e}(g)$. Puis nous appliquons les opérateurs $1-\tau_{h}$ pour $h\in H_{\mu}(\cF)$. Cela donne lieu à des différences de propagateurs que nous évaluons par un développement de Taylor semblable à l'équation \eqref{eq:taylsubstract}. Remarquons que la ligne qu'on déplace doit se trouver dans $\cI_{e}(h)\cap B_{\cF}(h)$ car les lignes externes communes à $h$ et $B_{\cF}(h)$ ont été déplacées en un même point après l'opérateur $\tau_{B_{\cF}(h)}$. Le produit des $1-\tau_{h}$ se factorise en
\begin{align}
  \prod_{g\in\cF\cup\{G\}}\ \prod_{h\in H_{\mu}(\cF)\tq B_{\cF}(h)=g}(1-\tau_{h})  
\end{align}
et chaque produit d'opérateurs agit uniquement sur $g/A_{\cF}(g)$. Ainsi les différences de propagateurs rapportent au moins $M^{i_{h}(\cF)-e_{h}(\cF)}$. Puis, comme dans le cas de la bulle, nous bornons les propagateurs interpolés par une borne du type \eqref{eq:propibound}. Les intégrales sur les paramètres d'interpolation sont alors bornées par $1$.

L'intégrande est alors prêt pour l'intégration sur les vertex internes. Il faut garder à l'esprit le fait que l'amplitude de $G$ est factorisée dans les $g/A_{\cF}(g),\,g\in\cF$, après application des opérateurs $\tau$. Pour intégrer les vertex internes de G, nous choisissions un arbre $\cT$ dont la restriction à $g/A_{\cF}(g)$ est un arbre générateur $\cT_{g}$ de $g/A_{\cF}(g)$. Ce choix est possible, comme dans le cas de l'arbre de Gallavotti-Nicol\`o, car $\cF\cup\{G\}$ est un arbre. Puis nous demandons que pour tout $g\in\cF$, $\cT_{g}$ soit sous-arbre dans toutes les composantes connexes $(g/A_{\cF}(g))^{i}_{k}$.

L'intégration sur les vertex internes produit
\begin{align}
  \prod_{g\in\cF\cup\{G\}}\prod_{(i,k)}M^{-\omega((g/A_{\cF}(g))^{i}_{k})}
\end{align}
où $\omega$ est donné par l'équation \eqref{eq:degsuperfconvCC} à $D=4$. En ajoutant les facteurs de décroissance verticale obtenus grâce aux opérateurs $1-\tau_{h}$, on obtient
\begin{align}
  |A_{G,\cF}^{R}|\les&\sum_{\mu\tq\cF\in I(\mu)}\ \prod_{g\in\cF\cup\{G\}}\prod_{(i,k)}M^{-\omega'((g/A_{\cF}(g))^{i}_{k})}\label{eq:omegaZimm}
  \intertext{où }
  \omega'((g/A_{\cF}(g))^{i}_{k})=&\max\lb 1,\omega((g/A_{\cF}(g))^{i}_{k})\rb
  \intertext{sauf si $g\in\cF$ et $(g/A_{\cF}(g))^{i}_{k}=g/A_{\cF}(g)$ auquel cas}
  \omega'((g/A_{\cF}(g))^{i}_{k})=&\omega((g/A_{\cF}(g))^{i}_{k})=0.
\end{align}

Soit $i_{\text{max}}(\mu)$ l'indice le plus haut de l'attribution $\mu$. À partir d'une fraction de la décroissance verticale \eqref{eq:omegaZimm}, nous pouvons extraire $M^{-\delta\,i_{\text{max}}}$. Puis en fixant un indice de $g/A_{\cF}(g)$, on somme sur les échelles des $h\in H_{\mu}(\cF)$ :
\begin{align}
    \prod_{g\in\cF\cup\{G\}} K^{n(g/A_{\cF}(g))}\les&K^{n(G)}.
\end{align}
Puis la somme logarithmique sur les échelles des sous-graphes inoffensifs est certainement bornée par $i_{\text{max}}(\mu)$. Enfin la somme sur $i_{\text{max}}(\mu)$ donne
\begin{align}
  \sum_{i_{\text{max}}} (i_{\text{max}})^{\card\cF}M^{-\delta\,i_{\text{max}}}\les&(\card\cF)!\,K^{\card\cF}
\end{align}
ce qui prouve le théorème \ref{thm:BPHunif}. Encore une fois, la preuve montre un peu plus. Si on considère l'opérateur de Taylor restreint aux graphes dangereux \ie{} $\cF=\emptyset$ et sachant que $H_{\mu}(\emptyset)=D_{\mu}$, on a
\begin{thm}[Renormalisation utile]
  \begin{align}
    |A_{G}^{\text{UR}}|=&\sum_{\mu}\prod_{h\in D_{\mu}}(1-\tau_{h})A_{G}^{\mu}\les K^{n(G)}.
  \end{align}
\end{thm}
Ce dernier théorème montre que les renormalons sont bien dus uniquement aux contretermes inutiles.

Pour démontrer le théorème \ref{thm:BPHunif} en toute généralité \ie{} avec les sous-graphes à deux points, il faudrait utiliser le formalisme des graphes $1$-particule irréductible. Sinon, on aurait affaire à des graphes dont le cardinal des forêts divergentes atteindrait $3n/2$ ce qui empêcherait d'obtenir un rayon de convergence fini dans le plan de Borel.

\section{La série effective}
\label{sec:la-serie-effective}

Pour définir une théorie des champs, au sens mathématique du terme, par exemple par l'intermédiaire de sa fonction de partition, il est nécessaire d'utiliser des méthodes perturbatives. Ces techniques sont également un bon moyen de calculer des grandeurs physiques. Le point de départ est la série \emph{nue} \eqref{eq:SNpertinverse}. Sachant que l'on souhaite prendre la limite $\kappa\to 0$, cette série n'est clairement pas le bon objet à considérer. Les méthodes de renormalisation perturbative du type BPHZ permettent de réorganiser la série nue en une série entière en une constante \emph{renormalisée} $\lambda_{R}$. La série renormalisée est finie ordre par ordre en $\lambda_{R}$. Néanmoins, elle n'est pas sommable à cause du problème des renormalons. C'est la définition même de la série renormalisée qui crée ces contributions d'ordre $n!$ rendant le rayon de convergence de la série nul. Pour les théories asymptotiquement libres, la série peut être Borel sommable mais pour les autres, telles que $\phi^{4}_{4}$, ce n'est à priori pas le cas. Ces problèmes de resommation sont importants pour définir le modèle au-delà de la perturbation, notamment en régime de couplage fort.\\

Nous avons constaté, dans la section précédente, que les renormalons sont uniquement dus aux contretermes inutiles. En effet, l'analyse multi-échelles nous apprend que seuls les graphes quasi-locaux, \ie{} ceux dont toutes les lignes internes sont \og{}au-dessus\fg{} de toutes leurs pattes externes, conduisent à des divergences. Les contretermes associés à ces graphes sont dits utiles. Les autres, non seulement ne servent pas à annuler une divergence mais encore créent les renormalons. Pour obtenir une série finie ordre par ordre sans renormalon, nous sommes amenés à ne pas introduire de contretermes inutiles.

Nous avons vu, avec l'exemple du graphe bulle, que les contretermes utiles dépendent de l'échelle de la plus haute patte externe (voir équation  \eqref{eq:taubulle}). En effet, un contreterme est utile s'il correspond à la partie locale d'un graphe $g$ quasi-local \ie{} dont l'attribution $\mu(g)$ est telle que $i_{g}>e_{g}$. La somme sur $\mu(g)$ est alors contrainte par cette condition de quasi-localité. Ainsi, il n'est pas possible d'absorber les contretermes utiles dans la redéfinition d'une unique constante de couplage renormalisée. Il est nécessaire d'introduire une infinité de constantes de couplage, indicées par l'échelle de la plus haute patte du vertex correspondant. Ces constantes sont intimement reliées à la philosophie du groupe de renormalisation à la Wilson. La constante effective $\lambda_{i}$ à l'échelle $i$ correspond à la partie locale des graphes à quatre points quasi-locaux dont toutes les lignes internes ont un indice supérieur ou égal à $i+1$. Or le \og{}découpage\fg{} de la théorie en tranches de moments induit une décomposition de l'intégrale fonctionnelle par la factorisation de la mesure gaussienne (voir équation \eqref{eq:factorgauss}) : pour tout $0\les i\les\rho$, l'intégration sur le champ $\phi^{i}$ crée les propagateurs $C^{i}$ d'échelle $i$. Si nous intégrons tous les champs de $\rho$ à $i+1$, nous obtenons une théorie \og{}effective\fg{} dont la constante de couplage est $\lambda_{i}$.\\

Pour résumer et préparer le lecteur au formalisme qui suit, nous pouvons écrire que la série effective est obtenue à partir de la série nue en développant certaines constantes nues en termes des constantes effectives et des contretermes utiles. De même, en extrayant les contretermes inutiles des constantes effectives, on a la série renormalisée. Ainsi nous avons :
\begin{subequations}
  \begin{align}
    \text{constante nue}=&\text{constantes effectives $+$ contretermes utiles}\\
    \text{constante renormalisée}=&\text{constantes effectives $-$ contretermes inutiles}
    \intertext{ce qui est compatible avec la définition habituelle \eqref{eq:schemasoustract} :}
    \text{constante nue}=&\text{constante renormalisée $+$ tous les contretermes.}
  \end{align}
\end{subequations}
Ainsi de la même façon qu'on définit la constante renormalisée comme une série formelle en la constante nue, nous allons donner les constantes effectives en termes de la constante nue. Encore une fois, nous nous restreindrons aux graphes sans sous-graphe à deux points. Toute fonction de Schwinger connexe à $N$ points peut s'exprimer comme une série formelle en la constante nue :
\begin{align}
    C^{\rho}_{N}=&\sum_{G}\sum_{\mu\in\lnat 0,\rho\rnat^{I(G)}}(-\lambda_{\rho})^{n(G)}A_{G,\mu}\label{eq:bareseries}
\end{align}
où la somme sur $G$ est restreinte aux graphes connexes à $N$ points sans sous-graphes à deux points. Nous allons démontrer le théorème suivant concernant l'existence de la série effective :
\begin{thm}[Série effective]\label{sec:la-serie-effective-1}
  Il existe $\rho+1$ séries formelles en $\lambda_{\rho}=\lambda^{\rho}_{\rho}$, appelées $\lambda^{\rho}_{\rho-1},\dotsc,\lambda^{\rho}_{-1}$, telles que la série formelle \eqref{eq:bareseries} s'exprime comme
  \begin{align}
    C_{N}^{\rho}=&\sum_{G\tq N(G)=4}\ \sum_{\mu\in\lnat 0,\rho\rnat^{I(G)}}\prod_{\nu\in\cV(G)}(-\lambda^{\rho}_{e_{\nu}(\mu)})A_{G,\mu}^{UR},\label{eq:CnEffSeries}\\
    A_{G,\mu}^{UR}=&\prod_{h\in D_{\mu}(G)}(1-\tau_{h})A_{G,\mu}
  \end{align}
où l'exposant $\rho$ dans les constantes de couplage rappelle que la théorie est définie avec une coupure $\rho$, où $e_{\nu}(\mu)$ est défini en \eqref{eq:indicesvertex} et $D_{\mu}(G)$ est la forêt des sous-graphes quasi-locaux à quatre pattes externes de G. Les constantes effectives obéissent à la définition récursive :
\begin{align}
  -\lambda^{\rho}_{i}=&-\lambda^{\rho}_{i+1}+\sum_{H\tq N(H)=4}\ \sum_{\substack{\mu(H)\tq\\ i_{H}=i+1>e_{H}}}\prod_{\nu\in\cV(H)}(-\lambda^{\rho}_{e_{\nu}(\mu)})\prod_{h\in D_{\mu}(H)}(1-\tau_{h})\,\tau_{H}A_{H,\mu}\label{eq:ConstEffInd}
\end{align}
\end{thm}
L'équation \eqref{eq:ConstEffInd} définit (par substitutions répétées) chaque constante effective comme une série formelle en $\lambda_{\rho}$. Cette récurrence s'arrête à $\lambda_{-1}$ car c'est la dernière constante pour laquelle la somme sur les graphes $H$ est non vide. Si nous appliquons la formule \eqref{eq:CnEffSeries} à la fonction connexe à quatre points et prenons sa partie locale, tous les termes sont nuls sauf le graphe trivial formé d'un seul vertex. Ses quatre pattes externes sont, par définition, d'échelle $-1$ si bien que la constante de couplage associée est $\lambda_{-1}$. Celle-ci est donc la partie locale, ou à moments externes nuls, de la fonction à quatre points connexe et peut ainsi être identifiée à la constante de couplage renormalisée.
\begin{proof}
  La preuve du théorème \ref{sec:la-serie-effective-1} s'effectue par récurrence. Nous allons montrer que si nous ne renormalisons \emph{utilement} une fonction connexe que jusqu'à l'échelle $i$, nous obtenons une série intermédiaire qui est
  \begin{align}
    C_{N}^{\rho}=&\sum_{G\tq N(G)=4}\ \sum_{\mu\in\lnat 0,\rho\rnat^{I(G)}}\prod_{\nu\in\cV(G)}(-\lambda^{\rho}_{\max(i,e_{\nu}(\mu))})A_{G,\mu}^{UR,i},\label{eq:EffInterm}\\
    A_{G,\mu}^{UR,i}=&\prod_{h\in D^{i}_{\mu}(G)}(1-\tau_{h})A_{G,\mu},\\
    D^{i}_{\mu}(G)=&\lb h\in D_{\mu}(G)\tqs i_{h}>i\rb.
  \end{align}
L'hypothèse de récurrence est clairement vérifiée à $i=\rho$ où \eqref{eq:EffInterm} se réduit à \eqref{eq:bareseries}. Si nous prouvons le passage de $i+1$ à $i$, le théorème sera démontré dans la mesure où, pour $i=-1$, \eqref{eq:EffInterm} devient \eqref{eq:CnEffSeries}. Supposons donc vraie l'hypothèse de récurrence à l'échelle $i+1$. Nous allons la démontrer à l'échelle $i$ en ajoutant et soustrayant les contretermes qui permettent de passer de $A_{G,\mu}^{UR,i+1}$ à $A_{G,\mu}^{UR,i}$. Ceux-ci correspondent aux graphes quasi-locaux $H$ à quatre pattes vérifiant $i_{H}=i+1$.
\begin{align}
  C_{N}^{\rho}=&\sum_{G\tq N(G)=4}\ \sum_{\mu\in\lnat 0,\rho\rnat^{I(G)}}\prod_{\nu\in\cV(G)}(-\lambda^{\rho}_{\max(i+1,e_{\nu}(\mu))})\prod_{h\in D^{i+1}_{\mu}(G)}(1-\tau_{h})A_{G,\mu}\\
  =&\sum_{G,\mu(G)}\prod_{\nu\in\cV(G)}(-\lambda^{\rho}_{\max(i+1,e_{\nu}(\mu))})\prod_{h\in D^{i+1}_{\mu}(G)}(1-\tau_{h})\prod_{H\in D^{i}_{\mu}(G)\setminus D^{i+1}_{\mu}(h)}(1-\tau_{H}+\tau_{H})A_{G,\mu}\notag\\
  =&\sum_{G,\mu(G)}\prod_{\nu\in\cV(G)}(-\lambda^{\rho}_{\max(i+1,e_{\nu}(\mu))})\prod_{h\in D^{i+1}_{\mu}(G)}(1-\tau_{h})\notag\\
  &\times\sum_{p\in\Part\lbt D_{\mu}^{i}(G)\setminus D_{\mu}^{i+1}(G)\rbt}\prod_{H\in p}\tau_{H}\prod_{H'\in\comp p}(1-\tau_{H'})A_{G,\mu}.\notag
\end{align}
Pour un ensemble dénombrable $A$, nous avons noté $\Part(A)$ l'ensemble des parties de $A$. Les sous-graphes $H'\in\comp p$ sont régularisés. Par contre, les $H\in p$ ne le sont pas. Néanmoins, les opérateurs $\tau_{H}$ \og{}détachent\fg{} le sous-graphe $H$ de $G$ et, en notant $G/p$ le graphe $G$ réduit par tous les graphes de $p$\footnote{Cette projection est possible car les éléments de $p$ sont disjoints.}, nous avons
\begin{align}
  \prod_{H\in p}\tau_{H}A_{G,\mu}=\Big(\prod_{H\in p}\tau_{H}A_{H}\Big)A_{G/p,\mu'}
\end{align}
où $\mu'$ est simplement la restriction de l'attribution $\mu$ aux lignes de $G/p$. Les parties locales des sous-graphes $H$ vont être absorbées dans une redéfinition des constantes effectives. Pour cela, nous regroupons les graphes $G$ tels que $G/p$ soit un graphe fixé :
\begin{align}
    C_{N}^{\rho}=&\sum_{G',\mu'}\Big(\sum_{G,\mu}\ \sum_{\substack{p\in\Part\lbt D_{\mu}^{i}(G)\setminus D_{\mu}^{i+1}(G)\rbt\tq\\G/p=G'}}\ \prod_{\nu\in\cV(G)}(-\lambda^{\rho}_{\max(i+1,e_{\nu}(\mu))})\prod_{h\in D^{i+1}_{\mu}(G)}(1-\tau_{h})\notag\\
  &\times\prod_{H\in p}\tau_{H}\prod_{H'\in\comp p}(1-\tau_{H'})A_{G,\mu}\Big).
\end{align}
Les graphes $G$ vérifiant $G/p=G'$, sont obtenus à partir de $G'$ en remplaçant chaque vertex $\nu\in\cV(G')\tq e_{\nu}(\mu')\les i$ par un graphe $H\tq i_{H}=i+1$ (voir figure \ref{fig:projection}). Aucun des ces graphes ne peut contenir d'autres graphes quasi-locaux dont l'indice minimum est $i+1$. Ainsi chaque $H\in p$ correspond à un vertex, et un seul, de G'.
\begin{figure}[htbp!]
  \centering
  \includegraphics[scale=.9]{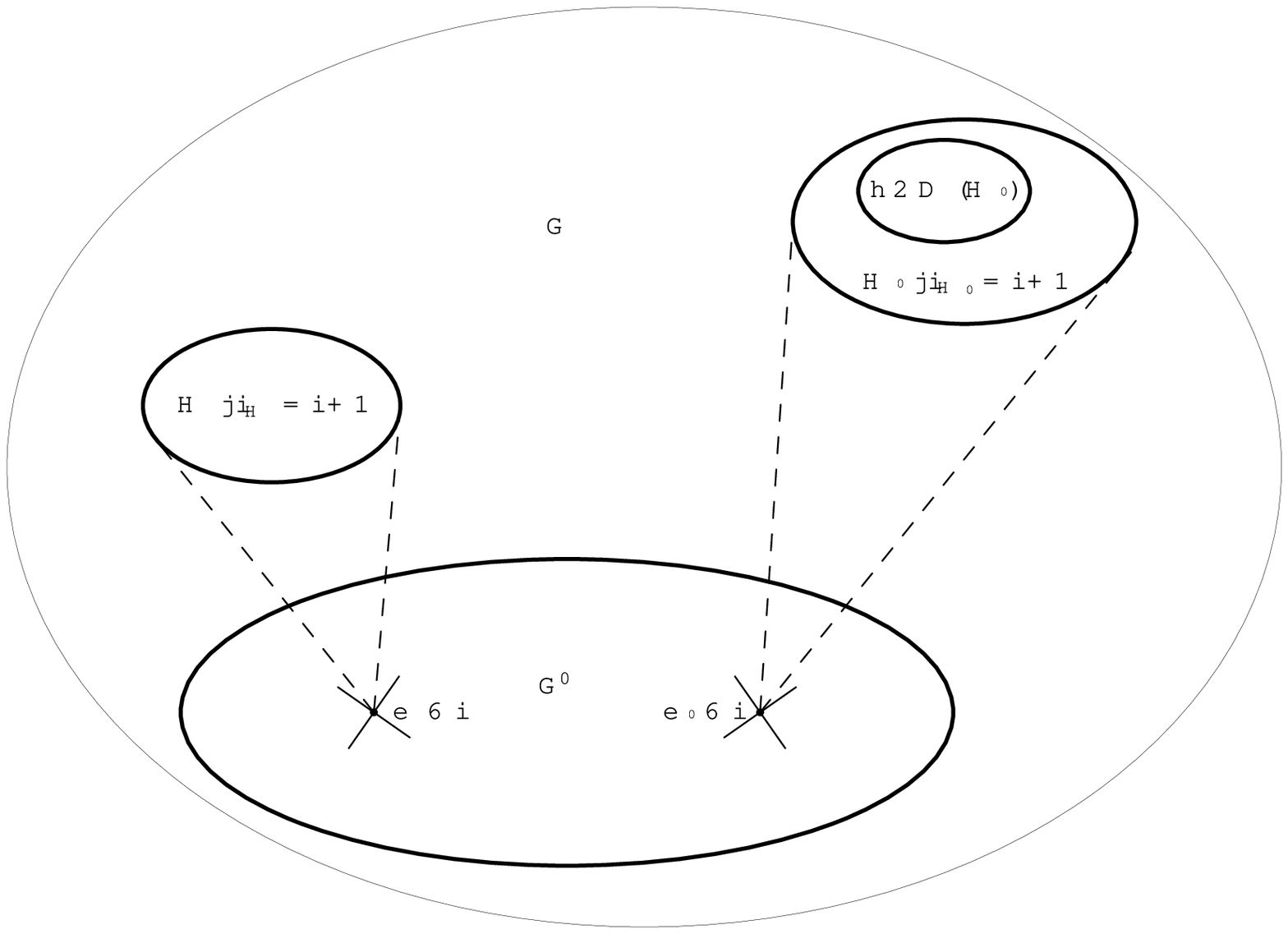}
  \caption{Projection des sous-graphes dangereux}
  \label{fig:projection}
\end{figure}
 En remarquant que
\begin{align}
  \prod_{H\in p}\tau_{H}=&\prod_{\substack{\nu\in\cV(G')\tq\\e_{\nu}(\mu')\les i}}\tau_{H_{\nu}},\\
  \prod_{\nu\in\cV(G)}(-\lambda^{\rho}_{\max(i+1,e_{\nu}(\mu))})=&\prod_{\substack{\nu\in\cV(G')\tq\\e_{\nu}(\mu')> i}}(-\lambda^{\rho}_{\max(i+1,e_{\nu}(\mu))})\prod_{\substack{\nu\in\cV(G')\tq\\e_{\nu}(\mu')\les i}}\ \prod_{\nu'\in\cV(H_{\nu})}(-\lambda^{\rho}_{\max(i+1,e_{\nu'}(\mu))}),\\
  \prod_{h\in D^{i+1}_{\mu}(G)}(1-\tau_{h})=&\prod_{\substack{\nu\in\cV(G')\tq\\e_{\nu}(\mu')\les i}}\ \prod_{h\in D^{i+1}_{\mu}(H_{\nu})}(1-\tau_{h})\ \prod_{h\in D^{i+1}_{\mu'}(G')}(1-\tau_{h})\\
\text{et }\prod_{H'\in\comp p}(1-\tau_{H'})=&\prod_{H'\in D_{\mu'}^{i}(G')\setminus D_{\mu'}^{i+1}(G')}(1-\tau_{H'}),
\end{align}
la série effective se récrit comme
\begin{align}
    C_{N}^{\rho}=&\sum_{G',\mu'}\prod_{\substack{\nu\in\cV(G')\tq\\e_{\nu}(\mu')> i}}(-\lambda^{\rho}_{\max(i+1,e_{\nu}(\mu))})\lb\prod_{\substack{\nu\in G'\tq\\e_{\nu}(\mu')\les i}}\Big(\sum_{\substack{H_{\nu},\mu(H_{\nu})\tq\\ i_{H_{\nu}}=i+1}}\ \prod_{\nu'\in\cV(H_{\nu})}(-\lambda^{\rho}_{\max(i+1,e_{\nu'}(\mu))})\right.\notag\\
  &\left.\times\prod_{h\in D^{i+1}_{\mu}(H_{\nu})}(1-\tau_{h})\Big)\,\tau_{H_{\nu}}A_{H_{\nu}}\rb\prod_{h\in D^{i}_{\mu'}(G')}(1-\tau_{h})A_{G',\mu'}.\label{eq:CNi+1}
\end{align}
La somme sur les graphes $H_{\nu}$ commence par le graphe trivial formé d'un unique vertex dont toutes les pattes externes sont plus basses que $i$. La constante de couplage associée est donc $-\lambda^{\rho}_{i+1}$. Ainsi la somme sur $H_{\nu}$ est égale au membre de droite de l'équation \eqref{eq:ConstEffInd} et l'équation \eqref{eq:CNi+1} est égale à \eqref{eq:EffInterm}.
\end{proof}

Pour que la série effective soit utile, il faut pouvoir prendre la limite $\rho\to\infty$. Nous allons donner un exemple illustrant le fait que la limite $\rho\to\infty$ de n'importe quelle constante $\lambda_{i}^{\rho}$ existe en tant que série en $\lambda_{-1}$. Tout d'abord, remarquons qu'il existe une définition non récursive des constantes effectives :
\begin{align}
  -\lambda_{i}=&-\lambda_{\rho}+\sum_{\substack{H,\mu(H)\tq\\i_{H}\ges i+1>e_{H}}}(-\lambda_{\rho})^{n(H)}\tau_{H}A_{H},\label{eq:lambdaidirect}\\
  -\lambda_{-1}=&-\lambda_{\rho}+\sum_{H}(-\lambda_{\rho})^{n(H)}\tau_{H}A_{H}\label{eq:lambdaRen}
\end{align}
où, à chaque fois, la somme sur $H$ est restreinte aux graphes connexes à quatre points. Pour exprimer $\lambda_{i}$ en fonction de $\lambda_{-1}$, il faut inverser \eqref{eq:lambdaRen} et substituer $\lambda_{\rho}$ en tant que série formelle en $\lambda_{-1}$ dans \eqref{eq:lambdaidirect}. À l'ordre $2$ en $\lambda_{-1}$, on a 
\begin{align}
    -\lambda_{i}=&-\lambda_{-1}-\lambda_{-1}^{2}\sum_{j_{1},j_{2}}\tau\raisebox{-4ex}{\includegraphics{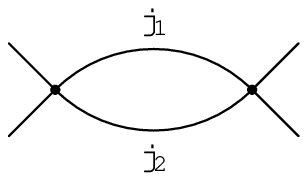}}+\lambda_{-1}^{2}\sum_{\min(j_{1},j_{2})\ges i+1}\tau\raisebox{-4ex}{\includegraphics{xphi4-fig.17}}+\cO(\lambda_{-1}^{3})\notag\\
    =&-\lambda_{-1}-\lambda_{-1}^{2}\sum_{\min(j_{1},j_{2})< i+1}\tau\raisebox{-4ex}{\includegraphics{xphi4-fig17.eps}}+\cO(\lambda_{-1}^{3}).\label{eq:flot1}
\end{align}
Si $j_{1}=j_{2}$, la somme sur $j_{1}$ est logarithmiquement divergente mais bornée par $i$. Si $j_{1}\neq j_{2}$, la somme sur $\max(j_{1},j_{2})$ est convergente car, jusqu'à $\min(j_{1},j_{2})$, on a affaire à une fonction à six points. Puis la somme sur $\min(j_{1},j_{2})$ est logarithmiquement divergente mais bornée par $i$. Ainsi, à l'ordre 2, la limite $\rho\to\infty$, à $i$ fixé, existe. En fait, l'argument se généralise à tous les ordres. On obtient, pour $\lambda_{i}$, la somme de tous les graphes dont l'indice minimum est inférieur ou égal à $i$ plus la somme de toutes les forêts inutiles. Ainsi la multi-série effective, entre les séries nue et renormalisée, pas tout à fait à mi-chemin cependant, est finie ordre par ordre, ne présente aucun renormalon et ne fait pas intervenir les forêts de Zimmermann. C'est donc certainement la bonne série à considérer. Par exemple, le modèle de Gross-Neveu commutatif, asymptotiquement libre, est Borel sommable en tant que série renormalisée mais, en tant que multi-série effective, est sommable (le rayon de convergence de chaque série en $\lambda_{i}$ est fini) \cite{Riv1}. De plus, les idées qui sous-tendent la définition de cette série sont les mêmes que celles du groupe de renormalisation de Wilson. La série effective semble donc en être la formulation la plus adaptée.\\

Remarquons enfin que la définition récursive \eqref{eq:ConstEffInd} des constantes effectives engendre un flot discret de la constante de couplage. Par exemple, l'équation \eqref{eq:flot1} donne, à des constantes positives près,
\begin{align}
  -\lambda_{i}=&-\lambda_{-1}-i\lambda_{-1}^{2}\simeq\frac{-\lambda_{-1}}{1-i\lambda_{-1}}.
\end{align}
À la limite $i\to\rho$, $i$ se comporte comme le logarithme de la coupure $M^{\rho}$ et nous retrouvons l'équation \eqref{eq:bare=ren}.

\begin{rem}
  Dans les chapitres \ref{cha:le-modele-phi4_4} et \ref{cha:GN}, nous ne prouverons pas explicitement les théorèmes BPHZ pour les modèles $\Phi^{4}_{4}$ et Gross-Neveu \ncf{}s. Nous démontrerons l'existence et la finitude ordre par ordre de la série effective. Pour retrouver la série renormalisée, il faut exprimer les constantes effectives en terme de la constante renormalisée. Cette opération, standard, fait apparaître les contretermes inutiles. La finitude, ordre par ordre, de la série renormalisée est alors vérifiée grâce à la classification des forêts.
\end{rem}


%% file: matrixbase-arXiv.tex
\chapter{Dans la base matricielle}
\label{cha:base-matr}
\epigraph{La difficulté n'est pas de comprendre les idées nouvelles, elle est d'échapper aux idées anciennes qui ont poussé leurs ramifications dans tous les recoins de l'esprit.}{Keynes}

Dans ce chapitre, nous allons énoncer et commenter les résultats que nous avons obtenus dans la base matricielle. Il existe une base de fonctions sur $\R^{D}_{\theta}$ où le produit de Moyal devient un produit matriciel ordinaire (voir section \ref{sec:defin-et-propr-Moyal}). Cette base permet donc d'exprimer une théorie des champs sur un espace de Moyal comme un modèle de matrices \emph{dynamiques} (voir section \ref{sec:phi4-matrixbase}). La preuve originale de la renormalisabilité de $\Phi^{4}$ a été obtenue dans cette base par H.~Grosse et R.~Wulkenhaar  \cite{GrWu04-3,GrWu03-2,GrWu03-1}. Un des avantages de cette base est qu'elle rend l'interaction très simple et élimine les difficultés liées aux oscillations. Malheureusement le propagateur devient très compliqué. C'est justement une étude approfondie de différents propagateurs dans la base matricielle que nous avons effectuée dans \cite{toolbox05}. Cette étude fournit toutes les bornes nécessaires pour calculer le comptage de puissance de diverses théories non commutatives : $\Phi^{4}$, LSZ (modifié ou non), Gross-Neveu. Par ailleurs, dans \cite{Rivasseau2005bh}, nous avons adapté les méthodes d'analyse multi-échelles à la base matricielle et redémontré ainsi le comptage de puissance de $\Phi^{4}$ de manière plus simple et plus efficace que dans la preuve originale.
\newpage
\section{L'algèbre de Moyal}
\label{sec:Moyal}

Nous commençons par définir l'algèbre de Moyal. Ce qui suit est principalement basé sur \cite{Gracia-Bondia1987kw}.

\subsection{Définitions et propriétés}
\label{sec:defin-et-propr-Moyal}

\paragraph{L'algèbre $\cA_{\Theta}$}
\label{sec:lalgebre-ca_theta}

L'algèbre de Moyal $\cA_{\Theta}$ est l'espace vectoriel des fonctions lisses à décroissances rapides $\cS(\R^{D})$ muni du produit \ncf{} défini par : $\forall f,g\in\cS_{D}\defi\cS(\R^{D})$,
\begin{align}
  (f\star_{\Theta} g)(x)=&\int_{\R^D} \frac{d^{D}k}{(2\pi)^{D}}d^{D}y\, f(x+{\textstyle\frac 12}\Theta\cdot
  k)g(x+y)e^{\imath k\cdot y}\\
  =&\frac{1}{\pi^{D}\labs\det\Theta\rabs}\int_{\R^D} d^{D}yd^{D}z\,f(x+y)g(x+z)e^{-2\imath y\Theta^{-1}z}\label{eq:moyal-def}
\end{align}
où $\Theta$ est une matrice anti-symétrique non dégénérée de dimension $D$ (ce qui implique $D=2N$). Cette algèbre joue le rôle des \og{}fonctions sur l'espace de Moyal $\R^{D}_{\theta}$\fg{}. Dans la suite nous écrirons souvent $f\star g$ au lieu de $f\star_{\Theta}g$ et utiliserons les notations et définitions suivantes : $\forall f,g\in\cS_{D}$, $\forall j\in\lnat 1,2N\rnat$,
\begin{align}
  (\mu^{j}f)(x)=&x^{j}f(x),\qquad \partial_{j}f=\frac{\partial}{\partial x^{j}},\quad\dt^{j}f=\frac{\imath}{2}\Theta^{jk}\partial_{k},\label{eq:def1}\\
    \langle f,g\rangle=&\int f(x)g(x)dx,\quad dx\defi (2\pi)^{-N}d^{2N}x.\label{eq:def2}
\end{align}
Nous écrirons
\begin{align}
  (\scF f)(x)=&\int f(t)e^{-\imath tx}dt  
\end{align}
pour la transformation de Fourier et
\begin{align}
  (f\diamond g)(x)=&\int f(x-t)g(t)e^{2\imath x\Theta^{-1}t}dt  
\end{align}
pour la convolution twistée.
\begin{prop}
  Si $f,g\in\cS_{D}$ alors $f\star g\in\cS_{D}$, $(f,g)\mapsto f\star g$ est bilinéaire continue et
  \begin{align}
    \partial_{j}(f\star g)=&\partial_{j}f\star g+f\star\partial_{j}g,\label{eq:Leibniz}\\
    \mu^{j}(f\star g)=&f\star\mu^{j}g+\imath\dt^{j}f\star g=\mu^{j}f\star g-\imath f\star\dt^{j}g.\label{eq:prod-x}
  \end{align}
\end{prop}
\begin{proof}
  La règle de Leibniz \eqref{eq:Leibniz} s'obtient en dérivant \eqref{eq:moyal-def} sous l'intégrale et \eqref{eq:prod-x} est un calcul direct également à partir de \eqref{eq:moyal-def}. En appliquant \eqref{eq:Leibniz} et \eqref{eq:prod-x} à \eqref{eq:moyal-def}, on prouve que $f\star g\in\cS_{D}$.

Soit $\alpha=(\alpha_{1},\dotsc,\alpha_{2N})\in\N^{2N}$, nous écrirons $\partial^{\alpha}=\partial_{1}^{\alpha_{1}}\dotsb\partial_{2N}^{\alpha_{2N}}$ et définissons de la même façon $\mu^{\alpha}$ et $\dt^{\alpha}$. Alors 
  \begin{align}
    \mu^{\alpha}\partial^{\gamma}(f\star g)=&\sum_{\beta\les\alpha}\sum_{\epsilon\les\gamma}(-\imath)^{|\beta|}\binom{\alpha}{\beta}\binom{\gamma}{\epsilon}\mu^{\alpha-\beta}\partial^{\gamma-\epsilon}f\star\dt^\beta\partial^{\epsilon}g.
  \end{align}
De \eqref{eq:moyal-def}, nous obtenons $\|f\star g\|_{\infty}\les K\|f\|_{1}\|g\|_{1}$ où $K=(\pi^{D}\labs\det\Theta\rabs)^{-1}$. Or la topologie de $\cS_{D}$ étant donnée par les seminormes $p_{\alpha\gamma}(f)=\|\mu^{\alpha}\partial^{\gamma}\|_{\infty}$ ou $q_{\alpha\gamma}(f)=\|\mu^{\alpha}\partial^{\gamma}\|_{1}$, la borne
\begin{align}
  p_{\alpha\gamma}(f\star g)\les&K'\sum_{\beta\les\alpha}\sum_{\epsilon\les\gamma}\binom{\alpha}{\beta}\binom{\gamma}{\epsilon}q_{\alpha-\beta,\gamma-\epsilon}(f)q_{0,\eta+\epsilon}(g)
\end{align}
où $\eta$ tient compte de l'inversion de la troisième relation de \eqref{eq:def1}, montre que $(f,g)\mapsto f\star g$ est jointement continue pour la topologie de $\cS_{D}$.
\end{proof}

Tout comme sur $\R^{D}$, la transformation de Fourier échange le produit et la convolution : 
\begin{align}
    \scF(f\star g)=&\scF(f)\diamond\scF(g)\label{eq:prodtoconv}\\
    \scF(f\diamond g)=&\scF(f)\star\scF(g)\label{eq:convtoprod}.
\end{align}
On montre aussi que le produit de Moyal et la convolution twistée sont \textbf{associatifs} :
\begin{align}
  ((f\diamond g)\diamond h)(x)=&\int f(x-t-s)g(s)h(t)e^{2\imath(x\Theta^{-1}t+(x-t)\Theta^{-1}s)}ds\,dt\\
  =&\int f(u-v)g(v-t)h(t)e^{2\imath(x\Theta^{-1}v-t\Theta^{-1}v)}dt\,dv\notag\\
  =&(f\diamond(g\diamond h))(x).
\end{align}
En appliquant \eqref{eq:convtoprod}, on a l'associativité du $\star$-produit. La conjugaison complexe est une \textbf{involution} dans $\cA_{\Theta}$
\begin{align}
  \overline{f\star_{\Theta}g}=&\bar{g}\star_{\Theta}\bar{f}.\label{eq:Moyal-involution}  
\end{align}
On a également 
\begin{align}
  f\star_{\Theta}g=&g\star_{-\Theta}f.\label{eq:Moyal-commutation}  
\end{align}
\begin{prop}[Trace]\label{prop:trace}
  Pour tous $f,g\in\cS_{D}$,
  \begin{align}
    \int dx\,(f\star g)(x)=&\int dx\,f(x)g(x)=\int dx\,(g\star f)(x)\label{eq:Moyal-trace}
  \end{align}
\end{prop}
\begin{proof}
  \begin{align}
    \int dx\,(f\star g)(x)=&\scF(f\star g)(0)=(\scF f\diamond\scF g)(0)\\
    =&\int\scF f(-t)\scF g(t)dt=(\scF f\ast\scF g)(0)=\scF(fg)(0)\notag\\
    =&\int f(x)g(x)dx\notag
  \end{align}
où $\ast$ est la convolution ordinaire.
\end{proof}

Dans les chapitres suivants, nous aurons besoin du lemme \ref{lem:Moyal-prods} pour calculer les termes d'interaction des modèles $\Phi^{4}_{4}$ et Gross-Neveu. Nous écrirons $x\wed y\defi 2x\Theta^{-1}y$.
\begin{lemma}\label{lem:Moyal-prods}
  Pour tout $j\in\lnat 1,2n+1\rnat$, soit $f_{j}\in\cA_{\Theta}$. Alors
  \begin{align}
    \lbt f_{1}\star_{\Theta}\dotsb\star_{\Theta}f_{2n}\rbt(x)=&\frac{1}{\pi^{2D}\det^{2}\Theta}\int\prod_{j=1}^{2n}
    dx_{j}f_{j}(x_{j})\,e^{-\imath
      x\wed\sum_{i=1}^{2n}(-1)^{i+1}x_{i}}\,e^{-\imath\varphi_{2n}},\\
    \lbt f_{1}\star_{\Theta}\dotsb\star_{\Theta}f_{2n+1}\rbt(x)=&\frac{1}{\pi^{D}\det\Theta}\int\prod_{j=1}^{2n+1}
    dx_{j}f_{j}(x_{j})\,\delta\Big(x-\sum_{i=1}^{2n+1}(-1)^{i+1}x_{i}\Big)\,e^{-\imath\varphi_{2n+1}},\\
  \forall p\in\N,\,\varphi_{p}=&\sum_{i<j=1}^{p}(-1)^{i+j+1}x_{i}\wed x_{j}.
  \end{align}
\end{lemma}
La démonstration est une simple récurrence.
\begin{cor}\label{cor:int-Moyal}
  Pour tout $j\in\lnat 1,2n+1\rnat$, soit $f_{j}\in\cA_{\Theta}$. Alors
  \begin{align}
    \int dx\,\lbt
    f_{1}\star_{\Theta}\dotsb\star_{\Theta}f_{2n}\rbt(x)=&\frac{1}{\pi^{D}\det\Theta}
    \int\prod_{j=1}^{2n}
    dx_{j}f_{j}(x_{j})\,\,\delta\Big(\sum_{i=1}^{2n}(-1)^{i+1}x_{i}\Big)e^{-\imath\varphi_{2n}},\label{eq:int-Moyal-even}\\
    \int dx\,\lbt f_{1}\star_{\Theta}\dotsb\star_{\Theta}f_{2n+1}\rbt(x)=&\frac{1}{\pi^{D}\det\Theta}\int\prod_{j=1}^{2n+1}
    dx_{j}f_{j}(x_{j})\,e^{-\imath\varphi_{2n+1}},\\
    \forall p\in\N,\,\varphi_{p}=&\sum_{i<j=1}^{p}(-1)^{i+j+1}x_{i}\wed x_{j}.
  \end{align}
\end{cor}

La cyclicité du produit, héritée de la proposition \ref{prop:trace} implique : $\forall f,g,h\in\cS_{D}$,
\begin{align}
  \langle f\star g,h\rangle=&\langle f,g\star h\rangle=\langle g,h\star f\rangle
\end{align}
et nous permet d'étendre par dualité l'algèbre de Moyal en une algèbre de distributions tempérées. 

\paragraph{Extension par dualité}
\label{sec:extens-par-dual}

Considérons d'abord le produit d'une distribution tempérée par une fonction de Schwartz. Pour $T\in\cS'_{D}$ et $h\in\cS_{D}$, nous définissons $\langle T,h\rangle\defi T(h)$ et $\langle T^\ast,h\rangle =\overline{\langle T,\overline{h}\rangle}$.
\begin{defn}\label{defn:Tf}
  Pour $T\in\cS'_{D}$, $f,h\in\cS_{D}$, nous définissons $T\star f$ et $f\star T$ par
  \begin{align}
    \langle T\star f,h\rangle=&\langle T,f\star h\rangle,\\
    \langle f\star T,h\rangle=&\langle T,h\star f\rangle.
  \end{align}
\end{defn}
La continuité du $\star$-produit implique que les membres de droite sont continus (et linéaires) en $h$. Ainsi $T\star f$ et $f\star T$ sont continues (comme composées d'applications continues) et linéaires en $h$ donc appartiennent à $\cS'_{D}$. Par exemple, l'identité $\bbbone$ en tant qu'élément de $\cS'_{D}$ est l'élément neutre pour le $\star$-produit : $\forall f,h\in\cS_{D}$,
\begin{align}
  \langle\bbbone\star f,h\rangle=&\langle\bbbone,f\star h\rangle\\
  =&\int(f\star h)(x)dx=\int f(x)h(x)dx\notag\\
  =&\langle f,h\rangle.\notag
\end{align}
Nous pouvons maintenant définir l'espace vectoriel $\cM$ comme intersection de deux sous-espaces $\cM_{L}$ et $\cM_{R}$ de $\cS'_{D}$.
\begin{defn}[Algèbre des multiplicateurs]\label{defn:M}
  \begin{align}
    \cM_{L}=&\lb S\in\cS'_{D}\tqs\forall f\in\cS_{D},\,S\star f\in\cS_{D}\rb,\\
    \cM_{R}=&\lb R\in\cS'_{D}\tqs\forall f\in\cS_{D},\,f\star R\in\cS_{D}\rb,\\
    \cM=&\cM_{L}\cap\cM_{R}.
  \end{align}
\end{defn}
Nous allons montrer que $\cM$ est une $\ast$-algèbre associative. Commençons par définir le produit dans $\cS'_{D}$ d'un élément de $\cS'_{D}$ par un élément de $\cM_{L}$ ou $\cM_{R}$ :
\begin{defn}\label{defn:TS}
  Pour $T\in\cS'_{D}$, $S\in\cM_{L}$ et $R\in\cM_{R}$,
  \begin{align}
    \langle T\star S,h\rangle=&\langle T,S\star h\rangle,\\
    \langle R\star T,h\rangle=&\langle T,h\star R\rangle.
  \end{align}
\end{defn}
Les membres de droite sont continus en $h$ (voir définition \ref{defn:Tf}). Comme $S\star h$ et $h\star R$ sont dans $\cS_{D}$, $T\star S$ et $R\star T$ sont des éléments de $\cS'_{D}$ ($\cS'_{D}$ est un $\cM$-bimodule). Puis nous montrons que $\cM$ est une algèbre :
\begin{prop}
  Pour $R,S\in\cM$, $R\star S\in\cM$.
\end{prop}
\begin{proof}
  Nous commençons par démontrer un lemme d'associativité :
  \begin{lemma}\label{lem:Sfg}
    Pour $S\in\cM_{L}$ et $f,g\in\cS_{D}$, on a $S\star (f\star g)=(S\star f)\star g$.
  \end{lemma}
  En effet, $\langle S\star (f\star g),h\rangle=\langle S,f\star(g\star h)\rangle=\langle (S\star f)\star g,h\rangle$ avec $h\in\cS_{D}$ et où on a utilisé l'associativité du $\star$-produit dans $\cS_{D}$.\\

  \noindent
  Soient $R,S\in\cM$ et $f,h\in\cS_{D}$,
  \begin{align}
    \langle (R\star S)\star f,h\rangle=&\langle R\star S,f\star h\rangle&&\text{(définition \ref{defn:Tf})}\\
    =&\langle R,S\star(f\star h)\rangle=\langle R,(S\star f)\star h\rangle&&\text{(lemme \ref{lem:Sfg})}\notag\\
    =&\langle R\star(S\star f),h\rangle.\notag
  \end{align}
  $S\star f\in\cS_{D}$ car $S\in\cM$ donc $R\star S\star f\in\cS_{D}$ et $R\star S\in\cM$ (la vérification de $f\star R\star S\in\cM$ est similaire). 
\end{proof}

Pour montrer que $\cM$ est associative, nous avons besoin du lemme intermédiaire suivant :
\begin{lemma}
  Pour $S,T\in\cM$ et $f,h\in\cS_{D}$, $(R\star S)\star f=R\star(S\star f)$ et $f\star(R\star S)=(f\star R)\star S$.
\end{lemma}
La démonstration est complètement similaire à celle du lemme \ref{lem:Sfg}. L'associativité du produit dans $\cM$ suit : $\forall R,S,T\in\cM$, $\forall h\in\cS_{D}$,
\begin{align}
    \langle (R\star S)\star T,h\rangle=&\langle T,h\star(R\star S)\rangle=\langle T,(h\star R)\star S\rangle\\
    =&\langle S\star T,h\star R\rangle=\langle R\star(S\star T),h\rangle.\notag
\end{align}
De plus, l'algèbre $\cM$ est munie d'une involution héritée de la conjugaison complexe dans $\cS_{D}$ : $\forall R,S\in\cM$ et $h\in\cS_{D}$,
\begin{align}
    \langle (R\star S)^\ast,h\rangle=&\overline{\langle R\star S,\overline{h}\rangle}=\overline{\langle R,S\star\overline{h}\rangle}\\
    =&\langle R^\ast,\overline{h}\star S^{\ast}\rangle=\langle S^{\ast}\star R^{\ast},h\rangle\notag\\
    \notag\\
    \text{car }\langle (S\star f)^{\ast},h\rangle=&\overline{\langle S\star f,\overline{h}\rangle}=\overline{\langle S,f\star\overline{h}\rangle}\\
    =&\langle S^{\ast},h\star\overline{f}\rangle=\langle\overline{f}\star S^{\ast},h\rangle.
\end{align}

En conclusion, $\cM$ est une $\ast$-algèbre associative. Elle contient, entre autres, l'identitié, les polynômes, la distribution $\delta$ et toutes ses dérivées. Ainsi la relation
\begin{align}
  \lsb x^{\mu},x^{\nu}\rsb=&\imath\Theta^{\mu\nu}, 
\end{align}
souvent donnée comme définition de l'espace de Moyal, est valable dans $\cM$ (mais pas dans $\cA_{\Theta}$).

\subsection{La base matricielle}
\label{sec:la-base-matricielle}

Ce qui suit est basé principalement sur \cite{Gracia-Bondia1987kw,gayral05,wulkenhaar04}. L'algèbre $\cA_\Theta$ possède une base naturelle constitu\'ee des fonctions propres $f_{mn},\,m,n\in\N^{D/2}$ de l'hamiltonien de Landau. Nous utiliserons la notation multi-indicielle 
\begin{align}
  m =& (m_1,\dotsc,m_{D/2})\in\N^{D/2},\\
  |m|\defi&\sum_{i=1}^{D/2}m_i,\\
  m!\defi&\prod_{i=1}^{D/2}m_i!.
\end{align}
Dans la suite de ce chapitre, nous utiliserons une matrice $\Theta$ donnée par
\begin{align}
  \Theta=\theta
  \begin{pmatrix}
    \mathbf{S}&0&\hdots&0\\
    0&\mathbf{S}&\hdots&0\\
    \vdots&\vdots&\ddots&\vdots\\
    0&0&\hdots&\mathbf{S}
  \end{pmatrix},\qquad \mathbf{S}=
  \begin{pmatrix}
    0&-1\\
    1&0
  \end{pmatrix}.\label{eq:Thetamatrixbase}
\end{align}
Soient $H_l=\frac 12(x_{2l-1}^2 + x_{2l}^2)$ pour $l=1,\dotsc,\frac{D}{2}$ et $H =\sum_{l=1}^{D/2} H_l$. La gaussienne
\begin{align}
  f_{00}(x)=&2^{D/2}\,e^{-2\theta^{-1}H}\label{eq:gaussfondam}
\end{align}
est idempotente\footnote{Quelle que soit $M\in\cM_{D}(\C)$ définie positive, $e^{-\frac 12 xMx}$ est idempotente dans $\cA_{\Theta}$.} dans $\cA_{\Theta}$ : $f_{00}\star f_{00}=f_{00}$. Nous définissons les fonctions de création et d'annihilation
\begin{align}
  a_{l}=&\frac{1}{\sqrt 2}(x_{2l-1}+\imath x_{2l}),\quad\bar{a}_{l}=\frac{1}{\sqrt 2}(x_{2l-1}-\imath x_{2l})
\end{align}
qui satisfont
\begin{align}
  \lsb a_{l},a_{l'}\rsb =& \lsb \bar{a}_{l},\bar{a}_{l'}\rsb =0,\quad\lsb\bar{a}_{l},a_{l'}\rsb=\theta\delta_{ll'}.
\end{align}
Les fonctions $f_{mn}$, définies par 
\begin{align}
  f_{mn}(x)=&\frac{1}{\sqrt{m!n!\theta^{m+n}}}\bar{a}^{\star m}\star f_{00}\star a^{\star n},\label{eq:fmndefn }
\end{align}
diagonalisent l'hamiltonien $H$ :
\begin{align}
H_l \star f_{mn}=&\theta(m_l+\frac 12) f_{mn},\quad f_{mn} \star H_l =\theta(n_l+\frac 12) f_{mn}.
\label{eq:moyal-haml}
\end{align}
Elles diagonalisent également l'hamiltonien de Landau $H_{L}^{\pm}=\frac 12(p\pm \xt)^{2}$ (où $\xt=2\Theta^{-1}x$) ce qui est à l'origine de plusieurs solutions exactes de théories des champs sur espace de phases non commutatif \cite{Langmann2002ai,Langmann2003cg,Langmann2003if}. De plus, elles constituent une base de $\cA_{\Theta}$. Plus précisement, la décomposition
\begin{align}
  \cA_{\Theta}\ni a(x)=&\sum_{m,n}a_{mn}f_{mn}(x)\label{eq:decomposition}
\end{align}
définit un isomorphisme d'algèbre de Fréchet entre $\cA_{\Theta}$ et l'algèbre des suites doublement indicées $\{a_{mn}\}$ à décroissances rapides :
\begin{align}
  \forall k\in\N,\,r_{k}(a)=&\lbt\sum_{m,n}(m+\frac 12)(n+\frac 12)\labs a_{mn}\rabs^{2}\rbt^{1/2}<\infty.
\end{align}\\
Dans la suite, nous aurons besoin de quelques propriétés des fonctions $f_{mn}$ ainsi que certaines identités. À partir de la définition \eqref{eq:fmndefn }, $\overline{f_{mn}}=f_{nm}$. Quelle que soit $f \in \cA_\Theta$, on a
\begin{align}
  (a_{l} \star f)(x)=& a_{l}(x) f(x) + \frac{\theta}{2} \frac{\partial
    f}{\partial \bar{a_{l}}}(x), & (f \star a_{l})(x) =& a_{l}(x) f(x) -
  \frac{\theta}{2} \frac{\partial f}{\partial \bar{a_{l}}}(x), \notag\\
  (\bar{a_{l}} \star f)(x) =&\bar{a_{l}}(x) f(x) - \frac{\theta}{2}
  \frac{\partial f}{\partial a_{l}}(x), & (f \star \bar{a_{l}})(x) =&
  \bar{a_{l}}(x) f(x) + \frac{\theta}{2} \frac{\partial f}{\partial
    a_{l}}(x)\label{af}
\end{align}
où $\frac{\partial}{\partial a_{l}} = \frac{1}{\sqrt{2}}(\partial_{2l-1} -
  \imath \partial_{2l} )$ et $\frac{\partial}{\partial \bar{a_{l}}}=
  \frac{1}{\sqrt{2}}(\partial_{2l-1} + \imath \partial_{2l} )$. \c{C}a implique $\bar{a_{l}}^{\star m_{l}} \star f_{00}=2^{m_{l}} \bar{a_{l}}^{m_{l}} f_{00}$, $f_{00} \star a_{l}^{\star n_{l}} =2^{n_{l}} a^{n_{l}} f_{00}$ et
\begin{align}
  a_{l} \star \bar{a_{l}}^{\star m_{l}} \star f_{00} &=
  \begin{cases}
    m_{l}\theta (\bar{a_{l}}^{\star (m_{l}-1)} \star f_{00}) &\text{pour } m_{l}\ges 1\\
    0 & \text{pour } m_{l}=0,
  \end{cases}\notag\\
  f_{00} \star a_{l}^{\star n_{l}} \star \bar{a_{l}} &=
  \begin{cases}
    n_{l}\theta (f_{00} \star a_{l}^{\star (n_{l}-1)})& \text{pour } n_{l}\ges 1 \\
    0 & \text{pour } n_{l} =0
  \end{cases}
\label{faa}
\end{align}
On en déduit
\begin{align}
    a_ {l}(x)f(x)&=\frac 12\lbt(a_ {l}
\star f)(x)+(f\star a_ {l})(x)\rbt,\label{eq:atoastar}\\
    \bar{a_ {l}}(x)f(x)&=\frac 12\lbt(\bar{a_ {l}}\star f)(x)+(f\star\bar{a_ {l}})(x)\rbt,\notag\\
     x_{2l-1}&=\frac{1}{\sqrt 2}\lbt a_ {l}+\bar{a_ {l}}\rbt,\label{eq:xtoa}\\
  x_{2l}&=\frac{-\imath}{\sqrt 2}\lbt a_ {l}-\bar{a_ {l}}\rbt,\notag\\
  \partial_{1}f&=\frac{1}{\theta\sqrt 2}\lb f\star(\bar{a_ {l}}-a_ {l})+(a_ {l}-\bar{a_ {l}})\star f\rb,\label{eq:dtoa}\\
  \partial_{2}f&=\frac{\imath}{\theta\sqrt 2}\lb f\star(\bar{a_ {l}}+a_ {l})-(\bar{a_ {l}}+a_ {l})\star f\rb.\notag
\end{align} 
Il vient de (\ref{faa}) et de l'idempotence de $f_{00}$
\begin{align}
  (f_{mn} \star f_{kl})(x) = \delta_{nk} f_{ml}(x).
\label{fprod}
\end{align}
Le point important est que la multiplication (\ref{fprod}) identifie le $\star$-produit
avec le produit de matrices ordinaire :
\begin{align}
  a(x) &= \sum_{m,n=0}^\infty a_{mn} f_{mn}(x),& b(x) &=
  \sum_{m,n=0}^\infty b_{mn} f_{mn}(x) \nonumber
  \\*
  \Rightarrow\quad (a\star b)(x) &= \sum_{m,n=0}^\infty (ab)_{mn}
  f_{mn}(x), & (ab)_{mn} &= \sum_{k=0}^\infty a_{mk} b_{kn}.
\label{fprodmat}
\end{align}
Finalement, en utilisant la propriété de trace de
l'intégrale et (\ref{faa}) on a
\begin{align}
  \int d^D x \, f_{mn}(x) &= \frac{1}{\sqrt{m! n!\, \theta^{m+n}}}
  \int d^Dx\, \big( \bar{a}^{\star m} \star f_{00} \star f_{00} \star
  a^{\star n} \big)(x) 
\nonumber
\\*
&= \delta_{mn} \int d^Dx f_{00}(x) = (2 \pi \theta)^{D/2}\delta_{mn}\;.
\label{intfmn}
\end{align}

\section{Un modèle de matrices dynamiques}
\label{sec:phi4-matrixbase}

Comme nous l'avons mentionné dans l'introduction de cette thèse, il ne suffit pas de remplacer le produit point à point par le produit de Moyal pour généraliser une théorie des champs commutative en une théorie non commutative. Le modèle que l'on obtiendrait serait non renormalisable à cause du phénomène de mélange UV/IR. Heureusement H.~Grosse et R.~Wulkenhaar ont découvert le moyen de modifier l'action de la théorie (\ncv{}) $\phi^{4}$ naïve pour la rendre renormalisable. Le bon modèle à considérer est donc
\begin{equation}\label{actionphi4-chapmatrix}
S[\phi] = \int d^4x \Big( -\frac{1}{2} \partial_\mu \phi
\star \partial^\mu \phi + \frac{\Omega^2}{2} (\tilde{x}_\mu \phi )
\star (\tilde{x}^\mu \phi ) + \frac{1}{2}\mu_{0}^2
\,\phi \star \phi
+ \frac{\lambda}{4} \phi \star \phi \star \phi \star
\phi\Big)(x)
\end{equation}
avec $\xt_\mu=2(\Theta^{-1}x)_{\mu}$. Nous nous placerons toujours dans le cas
euclidien. La métrique employée est donc
$g_{\mu\nu}=\delta_{\mu\nu}$. 

\subsection{De l'espace direct à la base matricielle}
\label{sec:de-lespace-direct}

En utilisant la décomposition \eqref{eq:decomposition} et les identités \eqref{eq:xtoa} à \eqref{intfmn}, nous pouvons exprimer l'action dans la base matricielle :
\begin{align}
  S[\phi]=&(2\pi)^{D/2}\sqrt{\det\Theta}\Big(\frac 12\phi\Delta\phi+\frac{\lambda}{4}\Tr\phi^{4}\Big)\label{eq:SPhi4matrix}
\end{align}
où $\phi=\phi_{mn},\,m,n\in\N^{D/2}$ et 
\begin{align}
    \Delta_{mn,kl}=&\sum_{i=1}^{D/2}\Big(\mu_{0}^{2}+\frac{2}{\theta}(m_{i}+n_{i}+1)\Big)\delta_{ml}\delta_{nk}\label{eq:formequadMatrixPhi4}\\
    &\hspace{-1.5cm}-\frac{2}{\theta}(1-\Omega^{2})\Big(\sqrt{(m_{i}+1)(n_{i}+1)}\,\delta_{m_{i}+1,l_{i}}\delta_{n_{i}+1,k_{i}}+\sqrt{m_{i}n_{i}}\,\delta_{m_{i}-1,l_{i}}\delta_{n_{i}-1,k_{i}}\Big)\prod_{j\neq i}\delta_{m_{j}l_{j}}\delta_{n_{j}k_{j}}.\notag
\end{align}
La matrice $\Delta$ (quadri-dimensionnelle) représente la partie quadratique du lagrangien. La première difficulté pour étudier le modèle de matrices \eqref{eq:SPhi4matrix} est le calcul de son propagateur $G$ défini comme l'inverse de la matrice $\Delta$ :
\begin{align}
  \sum_{r,s\in\N^{D/2}}\Delta_{mn;rs}G_{sr;kl}=\sum_{r,s\in\N^{D/2}}G_{mn;rs}\Delta_{sr;kl}=\delta_{ml}\delta_{nk}.
\end{align}
Inverser une matrice quadri-dimensionnelle n'est pas chose facile. Heureusement la forme \eqref{eq:Thetamatrixbase} de la matrice $\Theta$ implique l'invariance de l'action sous $SO(2)^{D/2}$. Il en résulte une loi de conservation se traduisant dans les indices de matrices par
\begin{align}
  \Delta_{mn,kl}=&0\iff m+k\neq n+l.\label{eq:conservationindices}
\end{align}
Nous avons évidemment la même contrainte sur les indices du propagateur. La matrice $\Delta$ est donc (seulement) tri-dimensionnelle et il faut donc inverser une infinité de matrices bi-dimensionnelles paramétrées par l'un des indices de $\Delta_{mn;kl}=\Delta_{m,m+h;l+h,l}=\Delta_{m;l}^{(h)}$. L'inversion fait intervenir les polynômes de Meixner. Le résultat est
\begin{align}
  \label{eq:propaPhimatrix}
  G_{m, m+h; l + h, l} 
&= \frac{\theta}{8\Omega} \int_0^1 d\alpha\,  
\dfrac{(1-\alpha)^{\frac{\mu_0^2 \theta}{8 \Omega}+(\frac{D}{4}-1)}}{  
(1 + C\alpha )^{\frac{D}{2}}} \prod_{s=1}^{\frac{D}{2}} 
G^{(\alpha)}_{m^s, m^s+h^s; l^s + h^s, l^s},
\\
 G^{(\alpha)}_{m, m+h; l + h, l}
&= \left(\frac{\sqrt{1-\alpha}}{1+C \alpha} 
\right)^{m+l+h} \sum_{u=\max(0,-h)}^{\min(m,l)}
   {\cal A}(m,l,h,u)\ 
\left( \frac{C \alpha (1+\Omega)}{\sqrt{1-\alpha}\,(1-\Omega)} 
\right)^{m+l-2u},\notag
\end{align}
où ${\cal A}(m,l,h,u)=\sqrt{\binom{m}{m-u}
\binom{m+h}{m-u}\binom{l}{l-u}\binom{l+h}{l-u}}$ et $C$ est une fonction de $\Omega$ : $C(\Omega)=
\frac{(1-\Omega)^2}{4\Omega}$. Nous avons déjà mentionné que le principal avantage de la base matricielle est qu'elle permet d'éviter l'exploitation, souvent difficile, des oscillations \eqref{eq:int-Moyal-even} dans l'interaction. En effet, dans cette base, $\phi^{\star 4}$ devient $\Tr\phi^{4}$. Cependant la contrepartie à cette simplification est la complexité du propagateur. Il suffit de comparer $(p^{2}+\mu_{0}^{2})^{-1}$ à \eqref{eq:propaPhimatrix}.\\

\begin{figure}[htb]
  \centering
  \includegraphics{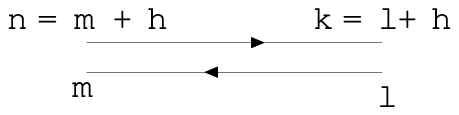}
  \caption{Propagateur matriciel}
  \label{fig:propamatrix}
\end{figure}

Soit $Z[J]$ la fonction de partition du modèle :
\begin{align}
  \label{eq:ZJ}
  Z[J]=&\int \prod_{a,b\in\N^{D/2}}d\phi_{ab}\,\exp-\Big(S[\phi]+(2\pi)^{D/2}\sqrt{\det\Theta}\sum_{m,n}\phi_{mn}J_{nm}\Big).
\end{align}
Pour démontrer la renormalisabilité perturbative de $\Phi^{4}$, Grosse et Wulkenhaar ont utilisé la méthode de Polchinski (voir \cite{Salmhofer}). Cela consiste à utiliser une équation différentielle reliant les amplitudes du modèle. Grâce à cette équation, on peut démontrer, de manière inductive, des bornes sur les amplitudes. Voici schématiquement comment on obtient cette équation. Tout d'abord, nous savons que les amplitudes nues divergent. Nous devons donc intoduire une régularisation\footnote{Grosse et  Wulkenhaar ont utilisé une fonction $K(m/\Lambda^{2})$ qui vaut $1$ si $m\les\Lambda^{2}$ et $0$ si $m\ges 2\Lambda^{2}$.} caractérisée par une coupure $\Lambda$. Puis nous autorisons les différentes constantes du modèle à dépendre de l'échelle arbitraire $\Lambda$ de telle sorte que $Z[J,\Lambda]$ soit en fait indépendant de $\Lambda$. L'équation de Polchinski est
\begin{align}
  \label{eq:Polchinski}
  \Lambda\partial_{\Lambda}Z[J,\Lambda]=&0.
\end{align}
Remarquons que le modèle de matrices \eqref{eq:SPhi4matrix} est \emph{dynamique} dans le sens où la partie cinétique du lagrangien n'est pas triviale. En effet, habituellement les modèles de matrices étudiés sont \emph{locaux}.
\begin{defn}
  Un modèle de matrice est dit \textbf{local} si $G_{mn;kl}=G(m,n)\delta_{ml}\delta_{nk}$ et \textbf{non local} sinon.
\end{defn}
Dans les théories matricielles, les graphes de Feynman sont représentés par des graphes à rubans. Ainsi le propagateur $G_{mn;kl}$ est représenté par la figure \ref{fig:propamatrix}.
Un modèle local correspond donc à un propagateur qui conserve les indices le long des trajectoires (simples lignes).

\subsection{Topologie des graphes à rubans}
\label{sec:topologie-des-graphes}

\begin{figure}[htbp]
  \centering 
  \subfloat[Planaire]{{\label{fig:ribongraph1}}\includegraphics[scale=1]{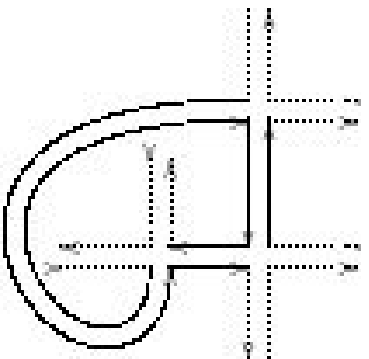}}\qquad
  \subfloat[Non planaire]{\label{fig:ribongraph2}\includegraphics[scale=1]{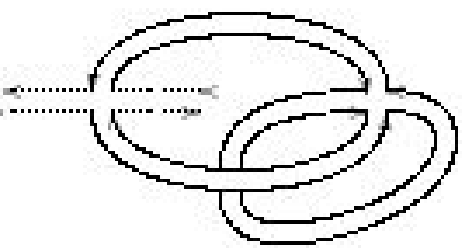}}
  \caption{Graphes à rubans}
  \label{fig:ribbon-examples}
\end{figure}

\begin{figure}[hbtp]
  \centering
\begin{minipage}[c]{3cm}
    \centering
    \includegraphics[width=3cm]{gt1.eps}
  \end{minipage}%
  \begin{minipage}[c]{2cm}
    \centering
  $ \Longrightarrow$
  \end{minipage}%
  \begin{minipage}[c]{2.6cm}
    \centering
    \includegraphics[width=2.6cm]{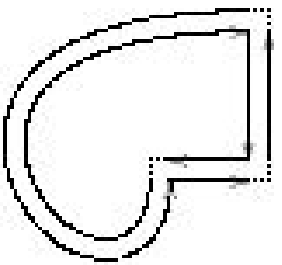}
  \end{minipage} \hspace{1cm}
   $\left .
     \begin{array}{c}
       $V=3$\\
       $I=3$\\
       $F=2$\\
       $B=2$
     \end{array}\rb
     \Longrightarrow\ g=0$\\
   \vspace{1cm}
   \begin{minipage}[c]{4cm}
     \centering
     \includegraphics[width=4cm]{gt3.eps}
   \end{minipage}%
   \begin{minipage}[c]{2cm}
     \centering
     $\Longrightarrow$
   \end{minipage}%
   \begin{minipage}[c]{3cm}
     \centering
     \includegraphics[width=3cm]{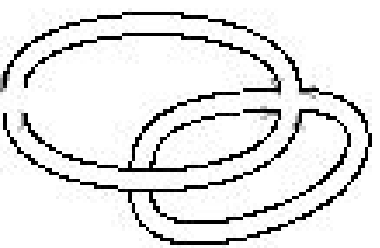}
   \end{minipage} \hspace{1cm}
    $\left .
      \begin{array}{c}
        $V=2$\\
        $I=3$\\
        $F=1$\\
        $B=1$
      \end{array}\rb
       \Longrightarrow\ g=1$
       \caption[Données topologiques]{Données topologiques de graphes à rubans}
       \label{fig:topo-ruban}
\end{figure}

Le comptage de puissance des modèles matriciels dépend des données topologiques des graphes. La figure \ref{fig:ribbon-examples} donne deux exemples de graphes à rubans.
Tout graphe à rubans peut être dessiné sur une variété bi-dimensionnelle. En fait, chaque graphe définit la surface sur laquelle on le dessine. En effet, soit un graphe $G$ avec $V$ vertex, $I$ propagateurs internes (doubles lignes) et $F$ faces (faites de lignes simples). La caractéristique d'Euler 
\begin{align}
  \chi=&2-2g=V-I+F\label{eq:Eulercar}
\end{align}
donne le genre $g$ de la variété. On peut s'en rendre compte en passant au \textbf{graphe dual}. Le dual d'un graphe $G$ est obtenu en échangeant faces et vertex. Les graphes duaux de la théorie $\Phi^{4}$ sont des quadrangulations des surfaces sur lesquelles ils sont dessinés (la valence d'un vertex devient la longueur de la face duale). De plus, chaque face du graphe direct brisée par des pattes externes devient dans le graphe dual un \textbf{point marqué}. Si parmi les $F$ faces d'un graphe, $B$ sont brisées, il peut être dessiné sur une variété de genre $g=1-\frac 12(V-I+F)$ avec $B$ points marqués. La figure \ref{fig:topo-ruban} donne les caractéristiques topologiques des graphes de la figure \ref{fig:ribbon-examples}.

 \paragraph{Graphe dual}\label{sec:graphe-dual}

Le graphe dual d'un graphe à rubans est obtenu en associant à chaque face un vertex et à chaque vertex une face. Toute ligne bordant deux faces voisines est remplacée par une ligne qui joint les deux vertex correspondants du graphe dual. Remarquons que le genre peut être calculé soit dans le graphe direct à partir de \eqref{eq:Eulercar} soit dans le graphe dual car le genre est invariant sous cette dualité. En effet, soient  $V'=F$, $F'=V$ les nombres de vertex et de faces du graphe dual (les quantités duales seront notées avec une apostrophe). Alors $\chi=V-I+F=V'-I+F$ ($I'=I$). Si le graphe direct est un graphe de $\phi^{4}$ \ie{} chaque vertex est de coordination $4$, nous avons $4=I_{f'}+N_{f'}$ pour toute face $f' \in F'$ où $I_{f'}$ et $N_{f'}$ sont les nombres d'arêtes et de valences externes appartenant à $f'$. La coordination des vertex du graphe dual est arbitraire.\\

\begin{figure}[hbt]
  \centering
  \includegraphics[scale=1]{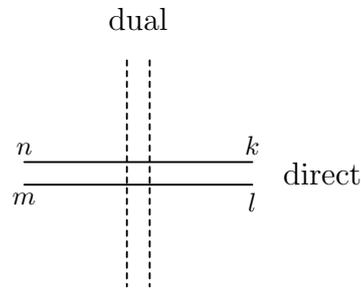}
  \caption{Propagateur dual}
  \label{fig:dualprop}
\end{figure}

Voici comment construire le dual d'un graphe. Tout d'abord, pour chaque face orientée du graphe direct, nous dessinons un vertex orienté en associant
\begin{itemize}
\item à chaque simple ligne d'un propagateur du graphe direct une valence interne du vertex dual,
\item à chaque valence externe du graphe direct une valence externe du vertex dual
\end{itemize}
en respectant l'ordre des flèches sur les trajectoires. Ensuite nous connectons les valences des vertex duaux par les propagateurs duaux, orthogonaux aux propagateurs directs (voir figure \ref{fig:dualprop}).

\begin{figure}[!htbp]
  \centering 
  \subfloat[Un graphe à ruban]{{\label{fig:ribbon-ex}}\includegraphics[scale=1]{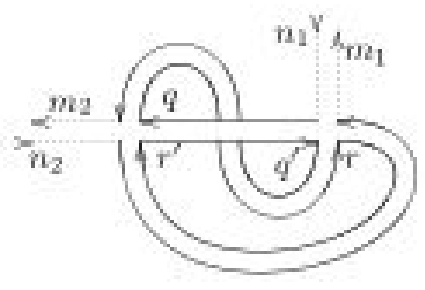}}\qquad
  \subfloat[Son dual]{\label{fig:ribbon-ex2}\includegraphics[scale=.8]{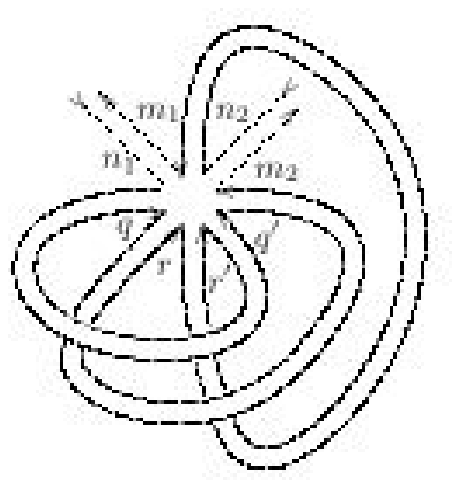}}\\
  \subfloat[Vertex dual]{\label{fig:dualvertex}\includegraphics[scale=.8]{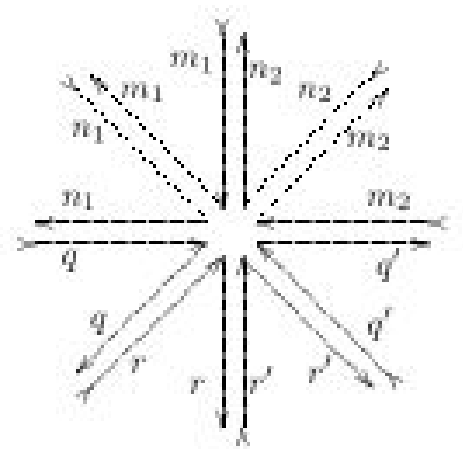}}
  \caption{Dualité}
  \label{fig:duality}
\end{figure}

Considérons l'exemple de la figure \ref{fig:ribbon-ex} qui n'a qu'une seule face. Les règles précédentes conduisent au vertex dual de la figure \ref{fig:dualvertex}. Puis nous connectons les valences par les propagateurs duaux \ie{} $n_1q$ avec $r'q'$, $qr$ avec $q'm_2$ et $n_2m_1$ avec $rr'$. Le résultat est la figure \ref{fig:ribbon-ex2}.

\subsection{Un comptage de puissance général}
\label{sec:comptage-matrix-general}

Dans \cite{GrWu03-1}, Grosse et Wulkenhaar ont utilisé la méthode de Polchinski pour démontrer un comptage de puissance très général. Ils ont considéré une théorie matricielle d'interaction $\Tr\phi^{4}$ avec un propagateur (quasiment) quelconque. La seule restriction imposée au propagateur est l'invariance par rotation \eqref{eq:conservationindices}. Les graphes qui sont solutions de l'équation de Polchinski sont reliés par des propagateurs modifiés\footnote{\label{fn:propatranche-polch}Dans la preuve originale, Grosse et Wulkenhaar ont utilisé une régularisation consistant en une fonction $K(m/\Lambda^{2})$ valant $1$ si $m\les\Lambda^{2}$ et $0$ si $m\ges 2\Lambda^{2}$. La coupure est donc directement appliquée aux indices matriciels. Ils auraient également pu utiliser une régularisation $\alpha$ consistant à introduire une fonction $K((\alpha\Lambda^{2})^{-1})$ qui restreint le paramètre de Schwinger : $\alpha\ges\Lambda^{-2}$. Dans ce schéma de régularisation, le propagateur $Q$ correspond, dans le cadre de l'analyse multi-échelles, au propagateur $G$ restreint dans une tranche $i$ telle que $M^{2i}\simeq\Lambda^{2}$.} $Q$ :
\begin{align}
  \label{eq:Qpropamatrix}
  Q_{mn;kl}=&\Lambda\partial_{\Lambda}G_{mn;kl}(\Lambda).
\end{align}
Ils ont introduit deux exposants $\delta_{0}$ et $\delta_{1}$ par
\begin{align}
  \max_{m,n,k,l}\labs Q_{mn;kl}(\Lambda)\rabs\les&C_{0}\lbt\frac{\mu}{\Lambda}\rbt^{\delta_{0}}\delta_{m+k,n+l},\label{eq:delta0}\\
  \max_{m}\sum_{l}\max_{n,k}\labs Q_{mn;kl}(\Lambda)\rabs\les&C_{1}\lbt\frac{\mu}{\Lambda}\rbt^{\delta_{1}}\label{eq:delta1}
\end{align}
où $\mu=\det^{-1/(2D)}\Theta$. Ces exposants encodent le comportement d'échelle (scaling) du propagateur régularisé. On peut alors séparer les modèles de matrices en deux classes :
\begin{defn}\label{defn:modele-regulier}
  Un modèle de matrices non local est dit \textbf{régulier} si $\delta_{0}=\delta_{1}=2$ et \textbf{anormal} sinon.
\end{defn}
On peut alors montrer le théorème suivant
\begin{thm}\label{thm:powcount-matrix}
  Soit $G$ un graphe de genre $g$ à $V$ vertex, $N$ pattes externes et $B$ faces brisées. Si $G$ est issu d'un modèle de matrice régulier alors son amplitude régulairsée $A(\Lambda)$ est bornée par
  \begin{align}
    \label{eq:borneampli-matrix}
    \labs A_{m_{1}n_{1};\dotsc;m_{N}n_{N}}(\Lambda)\rabs\les\lbt\frac{\mu}{\Lambda}\rbt^{\omega+D(2g+B-1)}P^{2V-N/2}\lsb\ln\frac{\Lambda}{\Lambda_{R}}\rsb,
  \end{align}
où $\omega=N\frac{D-2}{2}-D-V(D-4)$, $\Lambda_{R}$ l'échelle de la théorie renormalisée et $P^{2V-N/2}$ est un polynôme de degré $2V-N/2$.
\end{thm}
Ce comptage de puissance est, en dimension $4$, celui d'une théorie juste renormalisable. De plus, il montre que seuls les graphes planaires ($g=0$) avec une seule face brisée ($B=1$) sont potentiellement divergents. Cette restriction aux graphes planaires est très importante car elle met de côté les graphes non planaires qui sont responsables du mélange UV/IR.\\

En fait, Grosse et Wulkenhaar ont montré un théorème plus général pour des modèles réguliers ou non. Cependant, l'énoncé de ce théorème nécessite d'introduire de nouvelles notions graphiques ce qui nous obligerait à rentrer un peu trop dans les détails. De plus, les modèles que nous considererons seront réguliers. Toutefois notons que ce théorème général donne une condition nécessaire à la renormalisabilité des modèles de matrices. Il faut que $\delta_{0}$ et $\delta_{1}$ soient assez grands par rapport à la dimension de l'espace ou bien il ne faut pas que le modèle soit trop non local. Que signifie la localité dans un modèle de matrices ? Rappelons que la base permettant de passer des indices de matrices à l'espace $x$ est la base matricielle des $f_{mn}(x)$ :
\begin{align}
  Q(x,y;\Lambda)=&\sum_{m,n,k,l}f_{mn}(x)Q_{mn;kl}(\Lambda)f_{kl}(y).\label{eq:xtomatrix}
\end{align}
Les fonctions $f_{mn}$ sont des polynômes multipliés par la gaussienne fondamentale $f_{00}$ (voir \eqref{eq:gaussfondam} et \eqref{eq:fmndefn }). Celle-ci a une largeur $\cO(1)$ autour de l'origine. Les $f_{mn}$ ont plus généralement leur maximum en $|x|\simeq\Lambda$ \cite{GrWu03-2}. Ainsi si le propagateur n'est pas assez local, la corrélation entre des modes distants est trop forte. Autrement dit, si on considère $Q(\Lambda)$ comme équivalent à un propagateur dans une tranche $i$ (voir note de bas de page \ref{fn:propatranche-polch} page \pageref{fn:propatranche-polch}), la non localité du propagateur matriciel est équivalente à un couplage fort entre les échelles d'énergie. C'est le mélange UV/IR.\\

Le modèle $\phi^{4}$ sur $\R^{4}_{\theta}$ à $\Omega=0$ a $\delta_{0}=1$ et $\delta_{1}=0$. Les échelles se mélangent et le modèle n'est pas renormalisable. À $\Omega\neq 0$, $\delta_{0}=\delta_{1}=2$, le modèle est régulier (au sens de la définition \ref{defn:modele-regulier}) et renormalisable\footnote{Évidemment, le comptage de puissance \eqref{eq:borneampli-matrix} n'est qu'une condition nécessaire à la renormalisabilité du modèle. Dans \cite{GrWu04-3,GrWu03-2}, Grosse et Wulkenhaar ont également montré que les parties divergentes des amplitudes planaires régulières ($g=0$, $B=1$) sont de la forme du lagrangien initial.}. En fait, nous verrons au chapitre \ref{cha:le-modele-phi4_4} que le potentiel harmonique $\Omega\xt^{2}$ confine, à chaque échelle $i$, la théorie dans une boîte (lisse) de taille $M^{i}$. Ceci a pour effet de découpler les différentes échelles du problème.

Nous verrons au chapitre \ref{cha:GN} que le modèle de Gross-Neveu non commutatif présente un certain mélange entre les échelles d'énergie mais est cependant renormalisable. Il faudra donc distinguer entre mélange UV/IR renormalisable et non renormalisable (voir section \ref{sec:renorm-et-vulc}). Ce mélange entre les échelles qui persiste même après vulcanisation n'est pas surprenant dans un espace non commutatif \og{}défini\fg{} par $[x^{\mu},x^{\nu}]=\imath\Theta^{\mu\nu}$. Néanmoins il met en doute la sacro-sainte direction UV$\to$IR du groupe de renormalisation et le découplage entre les échelles que l'on constate dans la factorisation du comptage de puissance des composantes connexes en analyse multi-échelles. Le groupe de renormalisation dans la base matricielle va des indices infinis vers les indices nuls. Une seule direction est dangereuse. Il n'y a plus qu'un seul secteur à l'infini que l'on pourrait qualifier d'ultrarouge ou infraviolet. La base matricielle est-elle alors plus adaptée que les espaces $x$ ou $p$ à ces théories non commutatives ? Pour être totalement convaincu, il faudrait trouver comment la base matricielle distingue les mélanges UV/IR renormalisable et non renormalisable. Nous reviendrons sur l'influence des exposants $\delta_{0}$ et $\delta_{1}$ sur le comptage de puissance à la fin de la section \ref{sec:prop-et-renorm}.\\

Finalement, pour achever la preuve de la renormalisabilité perturbative de $\Phi^{4}_{4}$, Grosse et Wulkenhaar ont identifié les parties divergentes des amplitudes. La quasi-localité du propagateur améliore la situation prédite par le théorème \ref{thm:powcount-matrix} :
\begin{itemize}
\item les graphes planaires à quatre points avec un \emph{indice constant} le long des trajectoires sont marginaux,
\item les graphes planaires à deux points avec un \emph{indice constant} le long des trajectoires sont pertinents,
\item les graphes planaires à deux points avec un \emph{saut d'indice de $2$} le long des trajectoires sont marginaux.
\end{itemize}
Nous nous référons à \cite{GrWu04-3} pour les détails. Les trajectoires sont les lignes simples ouvertes du graphe. Il reste alors à identifier les parties divergentes de ces graphes. Ceci est fait par un développement de Taylor autour des indices externes nuls. Par exemple, la décomposition du cas marginal $m=l$ et $n=k$ du graphe suivant est
\begin{align}
&\includegraphics[scale=1]{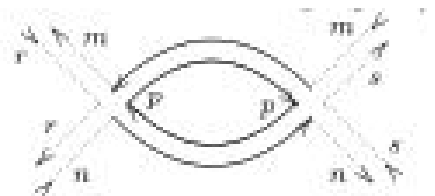}\label{4point-i}
\\*
&= \Bigg\{\int_{\Lambda}^{\Lambda_0} \frac{d\Lambda_2}{\Lambda_2} 
\int_{\Lambda_2}^{\Lambda_0} \frac{d\Lambda_1}{\Lambda_1} 
\sum_{p} \big(
Q_{m,p;p,m}(\Lambda_2)\,Q_{n,p;p,n}(\Lambda_1) 
-Q_{0,p;p,0}(\Lambda_2)\,Q_{0,p;p,0}(\Lambda_1) \big) 
\nonumber
\\*
&+ \int_{\Lambda_R}^{\Lambda} \frac{d\Lambda_2}{\Lambda_2} 
\int_{\Lambda_2}^{\Lambda_0} \frac{d\Lambda_1}{\Lambda_1} 
\sum_{p} 
Q_{0,p;p,0}(\Lambda_2)\,Q_{0,p;p,0}(\Lambda_1) \Bigg\}
+ \{
\Lambda_1 \leftrightarrow \Lambda_2\} +A_{00;00;00;00;00}(\Lambda_R)\;.\notag
\end{align}
Les graphes marginaux à deux points avec un saut d'indice de $2$ sont particulièrement importants. Par exemple,
\begin{align}
\sum_{p,p',q,q'}~\raisebox{-5ex}{\includegraphics[scale=.8]{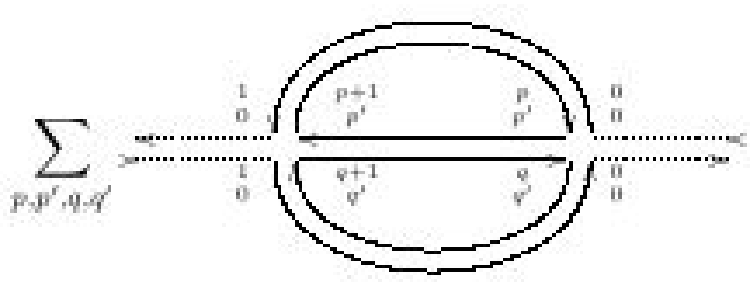}}
\end{align}
contribue à la renormalisation du paramètre $\Omega$.

\section{Analyse multi-échelles}
\label{sec:analyse-multi-echell-matrix}

Dans \cite{Rivasseau2005bh}, V.~Rivasseau, R.~Wulkenhaar et moi-même avons utilisé l'analyse multi-échelles pour redémontrer le comptage de puissance de la théorie $\Phi^{4}$ \ncv{}. Cette preuve a deux avantages principaux par rapport à la preuve originale \cite{GrWu04-3}. Elle est nettement plus courte et elle est exprimée dans le formalisme de l'analyse multi-échelles, première étape d'une étude constructive. De plus, notre preuve comble une très légère lacune de la preuve originale. En effet, dans \cite{GrWu04-3}, les exposants $\delta_{0}$ et $\delta_{1}$ (voir \eqref{eq:delta0} et \eqref{eq:delta1}) ont été estimés numériquement. Dans \cite{Rivasseau2005bh}, nous donnons des bornes analytiques sur le propagateur.

\subsection{Bornes sur le propagateur}
\label{sec:bornes-sur-le}

Soit $G$ un graphe à rubans de la théorie $\Phi^4_{4}$ avec $N$ pattes externes, $V$
vertex, $I$ lignes internes et $F$ faces, le genre est donc $g=1 - \frac 12
(V-I+F)$. Quatre indices $\{m,n;k,l\} \in\N^2$ sont associés à chaque ligne interne du graphe et deux indices à chaque ligne externe soit $4 I +2N =8V$ indices. Mais, à chaque vertex, l'indice de gauche d'un ruban est égal à l'indice de droite du ruban voisin (voir figure \ref{fig:dualvertex}). Ceci donne lieu à $4V$ identifications indépendantes qui permettent d'écrire les indices de tout propagateur en termes d'un ensemble $\mathcal{I}$ de $4V$ indices, quatre par vertex, par exemple l'indice de gauche de chaque demi-ruban.\\

L'amplitude du graphe est donc
\begin{align}  
  A_{G} = \sum_{\mathcal{I}} \prod_{\delta \in G}
  G_{m_{\delta}(\mathcal{I}),n_{\delta}(\mathcal{I});k_{\delta}(\mathcal{I}),l_{\delta}(\mathcal{I})}\;
    \delta_{m_{\delta}-l_{\delta},n_{\delta}-k_{\delta}} \;,
\label{IG}
\end{align}
où les quatre indices du propagateur $G$ de la ligne $\delta$ sont fonction de $\mathcal{I}$ et écrits\\
$\{m_{\delta}(\mathcal{I}),n_{\delta}(\mathcal{I}); k_{\delta}(\mathcal{I}),l_{\delta}(\mathcal{I})\} $. Nous décomposons chaque propagateur, donné par \eqref{eq:propaPhimatrix} :
\begin{align}  
G = \sum_{i=0}^{\infty}G^i\qquad \text{grâce à }\int_{0}^{1}d\alpha=\sum_{i=1}^{\infty}\int_{M^{-2i}}^{M^{-2(i-1)}}d\alpha.
\end{align}
Nous avons une décomposition associée de chaque amplitude
\begin{align}  
A_G &= \sum_{\mu} A_{G,\mu}\;,
\\
A_{G,\mu} &= \sum_{\mathcal{I}} \prod_{\delta \in G} G^{i_{\delta}}_{
m_{\delta}(\mathcal{I}),n_{\delta}(\mathcal{I});
k_{\delta}(\mathcal{I}),l_{\delta}(\mathcal{I})}  \;
\delta_{m_{\delta}(\mathcal{I})-l_{\delta}(\mathcal{I}),
n_{\delta}(\mathcal{I})-k_{\delta}(\mathcal{I})} \;,
\label{IGmu}
\end{align}
où $\mu=\{i_{\delta}\}$ parcourt toutes les attributions possible d'un entier positif $i_{\delta}$ pour toute ligne $\delta$. Nous avons prouvé les quatre propositions suivantes qui permettent de montrer que le modèle \eqref{eq:SPhi4matrix} est régulier au sens de la définition \ref{defn:modele-regulier} :
\begin{prop}
\label{thm-th1}
Pour $M$ suffisament grand, il existe une constante $K$ telle que, pour $\Omega\in [0.5,1]$, nous avons la borne uniforme
\begin{equation} 
    \label{th1}
    G^i_{m,m+h;l+h,l}\les 
    KM^{-2i} e^{-\frac{\Omega}{3}M^{-2i}\|m+l+h\|}.
  \end{equation}
\end{prop}
Cette borne montre que $\delta_{0}=2$.
\begin{prop}
\label{thm-th2}
Pour $M$ suffisament grand, il existe deux constantes $K$
et $K_{1}$ telles que, pour $\Omega \in [0.5,1]$, nous avons la borne uniforme
\begin{align} 
&   G^i_{m,m+h;l+h,l}
\nonumber
\\*
& \les K M^{-2i} e^{-\frac{\Omega}{4}M^{-2i}\|m+l+h\|} 
\prod_{s=1}^{\frac{D}{2}} \min\left( 1, 
\left(
    \frac{K_{1}\min(m^s,l^s,m^s+h^s,l^s+h^s)}{M^{2i}}
\right)^{\!\!\frac{|m^s-l^s|}{2}} \right).
\label{th2}
\end{align}
\end{prop}
Cette borne permet de montrer que seuls les graphes avec un indice constant le long des trajectoires ou avec un saut de $2$ sont divergents.
\begin{prop}\label{prop:bound3}
Pour $M$ suffisament grand, il existe une constante $K$ telle que, pour $\Omega \in [\frac 23,1]$, nous avons la borne uniforme
\begin{align}
\sum_{l =-m}^p G^i_{m,p-l,p,m+l} &\les 
K M^{-2i} \,e^{-\frac{\Omega}{4} M^{-2i} (\|p\|+\|m\|) }\;.
\label{thsum}
\end{align}
\end{prop}
Cette borne montre que le propagateur est quasi-local dans le sens où, à $m$ fixé, la somme sur $l$ ne coûte rien (voir figure \ref{fig:propamatrix}). Néanmoins les sommes que nous aurons à effectuer sont entrelacées (un même indice apparait généralement dans plusieurs propagateurs) si bien que nous aurons besoin de la proposition suivante.
\begin{prop}\label{prop:bound4}
Pour $M$ suffisament grand, il existe une constante $K$ telle que, pour $\Omega \in [\frac 23,1]$, nous avons la borne uniforme
  \begin{equation} \label{thsummax}
   \sum_{l=-m}^\infty\max_{p\ges\max(l,0)}G^i_{m,p-l;p,m+l}
\les  KM^{-2i}e^{-\frac{\Omega}{36}M^{-2i}\|m\|} \;.   
  \end{equation}
\end{prop}
Cette dernière borne montre que $\delta_{1}=2$. Ainsi, avec la proposition \ref{thm-th1}, cette dernière proposition prouve que le modèle \eqref{eq:SPhi4matrix} est régulier. Nous renvoyons à \cite{Rivasseau2005bh} pour les preuves de ces quatre propositions. Notons néanmoins que les contraintes sur le paramètre $\Omega$ dans les propositions \ref{thm-th1} à \ref{prop:bound4} sont d'origine purement technique. Les études numériques de Grosse et Wulkenhaar montrent qu'elles pourraient certainement être relâchées.\\
\begin{rem}
  Dans \cite{Rivasseau2005bh}, nous avons également calculé des bornes sur les propagateurs \emph{composites}. Ce sont des différences de propagateurs pris à indices externes différents. Ces différences apparaissent dans le développement de Taylor des amplitudes pour identifier les parties divergentes (les contretermes) \cite{GrWu04-3}. Ces bornes permettent de montrer que les amplitudes renormalisées sont finies.
\end{rem}

\subsection{Variables indépendantes}
\label{sec:vari-indep}

Une partie non négligeable des $4V$ indices initialement associés au graphe est déterminée par les indices externes et les fonctions delta dans (\ref{IG}). Les autres sont des indices de sommation. Le comptage de puissance consiste à trouver quels sont les indices pour lesquels la somme coûte $M^{2i}$ et ceux pour lesquels elle ne coûte que $\mathcal{O}(1)$ grâce à (\ref{thsum}). Le facteur $M^{2i}$ provient de (\ref{th1}) après avoir sommé sur un indice\footnote{Rappelons que chaque indice est en fait composé de deux indices, un pour chaque paire symplectique de $\R^{4}_{\theta}$.} $m\in\N^2$,
\begin{align}
\sum_{m^1,m^2=0}^\infty e^{- c M^{-2i}(m^1+m^2)} = \frac{1}{(1-e^{- c
    M^{-2i}})^2} = \frac{M^{4i}}{c^2} (1 + \mathcal{O}(M^{-2i}))\;.
\label{summ1m2}
\end{align}

Nous souhaitons d'abord utiliser les fonctions delta autant que possible pour réduire l'ensemble $\mathcal{I}$ à un ensemble minimal $\cI'$ d'indices indépendants. Pour cela, il est pratique d'utiliser les graphes duaux où la résolution des fonctions delta devient un problème classique d'attribution de moments. Pour la définition et la construction des graphes duaux, voir la section \ref{sec:topologie-des-graphes} page \pageref{sec:graphe-dual}.

Le graphe dual est composé des mêmes propagateurs que le graphe direct, seule la position des indices change. Alors que dans le graphe original, nous avons $G_{mn;kl}=\raisebox{-1ex}{\includegraphics[scale=.6]{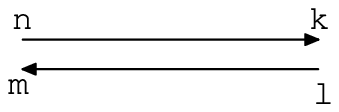}}$, la position des indices des propagateurs du graphe dual est
\begin{align}
G_{mn;kl} = \raisebox{-1ex}{\includegraphics[scale=.6]{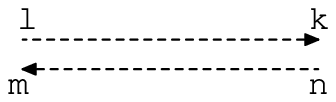}}\;.
\label{dual-assign}
\end{align}
La conservation $\delta_{l-m,-(n-k)}$ dans (\ref{IG}) implique que la différence entre les indices entrant et sortant d'un demi-propagateur attaché à un vertex dual \ie{} $l-m$, est conservée le long du propagateur. En fait, ces différences d'indices se comportent comme un \emph{moment angulaire} et la conservation des différences $\ell=l-m$ et $-\ell=n-k$ n'est rien d'autre que la conservation du moment angulaire grâce à la symétrie $SO(2)\times SO(2)$ de l'action \eqref{eq:SPhi4matrix}. Ainsi, en choisissant l'indice entrant comme indice de référence, le moment angulaire détermine l'indice sortant :
\begin{align}
\raisebox{-1.5ex}{\includegraphics[scale=1]{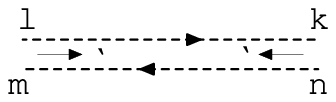}}&\qquad
l=m+\ell\;,~~ n=k+(-\ell).
\label{angmom}
\end{align}
De la même façon, des moments angulaires externes $\wp$ entrent dans le graphe dual par les valences externes. La cyclicité des vertex implique que la somme des moments angulaires entrant dans un vertex est nulle. Bien sûr la somme des moments externes est donc nulle également. Ainsi le moment angulaire dans le graphe dual se comporte exactement comme le moment habituel dans un graphe de Feynman ordinaire : les moments se conservent le long des lignes et à chaque vertex. Remarquons que cette conservation provient d'habitude de l'invariance par translation. Insistons également sur le fait que nous devons prendre en compte des contraintes de positivité pour les moments angulaires : $\ell\in\Z$ mais les indices $m,n,k,l$ sont dans $\N$.\\

Nous savons donc que le nombre de moments (différences d'indices) \emph{indépendants} est exactement le nombre de boucles $L'$ du graphe dual. Pour un graphe connexe, ce nombre vaut $L'= I-V'+1$. De plus, chaque indice à un vertex (dual) donné est seulement fonction des différences d'indices à ce vertex et d'un unique \emph{indice de référence}. Si le vertex dual est un vertex externe, nous choisissons un indice externe sortant comme indice de référence. Pour les vertex internes, nous expliquerons plus tard comment faire ce choix. Les indices de référence des vertex internes correspondent aux indices de boucles du graphe direct. Après utilisation des fonctions delta des proagateurs, le nombre d'indices indépendants à sommer est $V'-B + L' = I +
(1-B)$. Ici $B\ges 0$ est le nombre de faces brisées du graphe direct ou le nombre de vertex externes du graphe dual.\\

Exprimer chaque indice du graphe en termes d'un ensemble $\cI'$ d'indices indépendants est donc analogue au problème de \og{}momentum routing\fg{} dans un graphe de Feynman commutatif. La solution n'est pas canonique. Néanmoins un bon moyen de distribuer les moments consiste à choisir un arbre générateur enraciné $\cT_{\mu}$ dans le graphe dual, avec $V'-1$ lignes, et d'utiliser le complémentaire $\cL_{\mu}$ (les lignes de boucles) comme l'ensemble des différences indépendantes. L'indice $\mu$ représente l'attribution des échelles dans le graphe ; le choix de l'arbre est contraint par cette attribution. Dans la suite, nous allons montrer que la somme sur les indices de référence coûte $M^{4i}$ et que les sommes sur les différences d'indices dans $\cL_{\mu}$ coûtent $\cO(1)$ grâce à la borne \eqref{thsummax}. Nous devons donc optimiser l'arbre $\cT_{\mu}$ afin que les indices de référence appartiennent aux lignes d'échelles les plus basses possibles. C'est exactement le contraire de l'optimisation habituelle dans les théories commutatives.

\subsection{Optimisation de l'arbre}
\label{sec:optim-de-larbre}

Une attribution d'échelles $\mu = \{i_\delta\}$ définit un ordre parmi les lignes (duales) :
\begin{align}
\delta_1 \les \delta_2 \les \dotsb
\les \delta_I \qquad \text{si} \qquad 
i_{\delta_1} \les i_{\delta_2} \les\dotsb\les
i_{\delta_I}.
\label{order}
\end{align}
En cas d'égalité, nous faisons un choix arbitraire. Soient $\delta_1^{\mathcal{T}}$ la plus basse ligne dans l'ordre défini ci-dessus qui n'est pas un tadpole et $G^{i_{\delta^{\mathcal{T}}_1}}_{m_{\delta^{\mathcal{T}}_1};
  n_{\delta^{\mathcal{T}}_1}; k_{\delta^{\mathcal{T}}_1},
  l_{\delta^{\mathcal{T}}_1}}$ le propagateur correspondant. Cette ligne $\delta^{\mathcal{T}}_1$ joint deux vertex $v^\pm_1$ et forme le premier segment de l'arbre. Soient $\kL$ l'ensemble des lignes du graphe et $\kL_1= \kL\setminus (\delta_1 \cup\dotsb\cup\delta^{\mathcal{T}}_1)$ et $\mathcal{T}_1 =\delta^{\mathcal{T}}_1 \cup v^+_1 \cup v^-_1$. Nous identifions la plus basse ligne $\delta^{\mathcal{T}}_2$ de $\kL_1$ qui ne forme pas de boucle si elle est rajoutée à $\mathcal{T}_1$. Nous définissons $\kL_2= \kL \setminus
(\delta_1 \cup \dotsb \cup \delta^{\cT}_2)$ et
\begin{itemize}
\item $\mathcal{T}_2 = \mathcal{T}_1 \cup \delta^{\mathcal{T}}_2 \cup 
v^+_2$ si $\delta^{\mathcal{T}}_2$ relie un vertex $v^+_2$ à $\mathcal{T}_1$,
\item  $\mathcal{T}_2 = \mathcal{T}_1 \cup \delta^{\mathcal{T}}_2 \cup 
v^+_2 \cup v^-_2$ si $\delta^{\mathcal{T}}_2$ joint deux vertex $v^\pm_2 \notin \mathcal{T}_1$.
\end{itemize}
\medskip
\noindent
À la $n$\ieme{} étape, nous identifions la plus basse ligne $\delta^{\mathcal{T}}_n$ de $\kL_{n-1}$ qui ne forme pas de boucle si elle est rajoutée à $\mathcal{T}_{n-1}$. Nous définissons $\kL_n= \kL\setminus(\delta_1 \cup \dotsb \cup \delta^{\cT}_n)$ et
\begin{itemize}
\item $\mathcal{T}_n = \mathcal{T}_{n-1} \cup \delta^{\mathcal{T}}_n\cup 
v^+_n$ si $\delta^{\mathcal{T}}_n$ relie un vertex $v^+_n$ à $\mathcal{T}_{n-1}$,
\item  $\mathcal{T}_n = \mathcal{T}_{n-1} \cup \delta^{\mathcal{T}}_n \cup 
v^+_n \cup v^-_n$ si $\delta^{\mathcal{T}}_n$ joint deux vertex $v^\pm_n \notin \mathcal{T}_{n-1}$,
\item  $\mathcal{T}_n = \mathcal{T}_{n-1} \cup \delta^{\mathcal{T}}_n$ 
si $\delta^{\mathcal{T}}_n$ connecte deux sous ensembles disjoints de $\mathcal{T}_{n-1}$.
\end{itemize}
La $(V'-1)$\ieme{} étape fournit l'arbre optimisé $\mathcal{T}_\mu =
\mathcal{T}_{V'-1}$. En conséquence, toute ligne $\delta^{\mathcal{L}}_j \in \mathcal{L}_\mu$ qui joint deux vertex \emph{différents} $v^{\pm}_j$ a un indice d'échelle $i_{\delta_j^{\mathcal{L}}}$ \emph{supérieur ou égal} à tous les indices d'échelle des lignes d'arbre joignant $v^{\pm}_j$. 

\subsection{Attribution des indices}
\label{sec:attr-des-indic}

Nous choisissons un des $B\ges 1$ vertex externes du graphe dual comme racine $v_0$ de l'arbre optimisé $\cT_{\mu}$. Si le graphe est un graphe du vide \ie{} avec $B=0$, nous choisissons n'importe quel vertex. Nous renommons les vertex de l'arbre de telle sorte que tout vertex dans le sous-arbre au-dessus du vertex $v_{n}$ se nomme $v_{p}$ avec $p>n$.\\

\begin{figure}[!hbtp]
  \centering
  \includegraphics[scale=1]{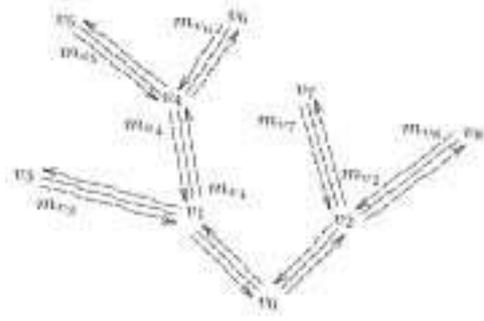}
  \caption{Indices de référence}
  \label{fig:main-index}
\end{figure}

L'ordre (\ref{order}) sur les lignes du graphe va nous permettre de choisir l'indice de référence $m$ à chaque vertex. Si $v$ est un vertex interne, nous appelons $\delta_v$ la plus basse ligne attachée à $v$. Par construction de l'arbre (voir section précédente), soit $\delta_v$ est un tadpole soit $\delta_{v}$ appartient à l'arbre. Nous choisissons l'indice \emph{sortant} de la ligne $\delta_{v}$ comme l'indice de référence $m_{v}$ du vertex $v$. Soit $\mathcal{G}_{\mathcal{M}}$ l'ensemble des lignes portant un indice de référence. Soit $\delta_{vv'}$ une ligne reliant les vertex $v$ et $v'$. Si $\delta_{vv'}=\delta_{v}=\delta_{v'}$ \ie{} $\delta_{vv'}$ est la plus basse ligne en $v$ et en $v'$, alors $\delta_{vv'}$ porte deux indices de références. Dans ce cas, elle apparaîtra deux fois dans $\mathcal{G}_{\mathcal{M}}$. Ainsi $\mathcal{G}_{\mathcal{M}}$ contient $V'-B$ éléments. Si $v$ est un vertex externe, nous choisissons un indice externe sortant comme indice de référence. La figure \ref{fig:main-index} montre une situation typique d'un arbre et de ses indices de référence.\\

\begin{figure}[!hbtp]
  \centering
  \includegraphics[scale=1]{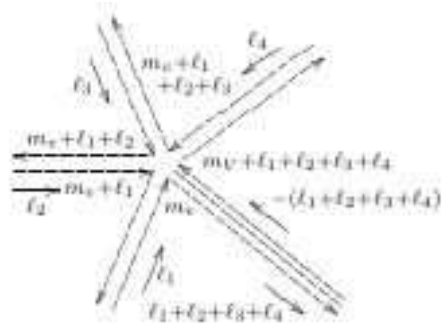}
  \caption{Attribution des indices}
  \label{fig:index-attribution-dual}
\end{figure}

\noindent
Tout indice du graphe s'écrit donc, de façon unique, en termes de
\begin{itemize}
\item $V'-B$ indices de références $m_v$,
\item $B$ indices de références aux vertex externes,
\item $L'$ moments anguaires internes $\ell_\delta,\,\delta\in\mathcal{L}_\mu$,
\item $N$ moments angulaires externes $\wp_\epsilon,\,\epsilon\in\cN$ où $\mathcal{N}$ est l'ensemble des lignes externes.
\end{itemize}
\medskip
Voici comment procéder. Nous commençons par les feuilles \ie{} les vertex (différents de la racine) qui ont une coordination $1$. Les feuilles de la figure \ref{fig:main-index} sont les vertex $v_3,v_5,v_6,v_7$ et $v_8$. Pour ces vertex, en commençant par l'indice de référence $m_{v}$ (à gauche de l'unique ligne de $\cT_{\mu}$ au vertex $v$ qui descend vers la racine, à moins qu'un tadpole en $v$ soit la ligne la plus basse) qui est égal à l'indice entrant de la ligne juste après $\delta_{v}$ dans le sens horaire, nous calculons tous les autres indices en tournant autour de $v$ dans le sens horaire et en ajoutant les moments angulaires associés aux lignes de boucles $\delta_1,\dotsc,\delta_k$ et aux lignes externes $\epsilon_1,\dotsc,\epsilon_{k'}$. Cela donne $m_v +
\ell_1$, $m_v + \ell_1 + \ell_2,\dotsc$ jusqu'à $m_v +
\ell_1 + \dotsb + \ell_{k+k'}$ qui se trouve à droite de $\delta_{v}$. Parmi les $\ell_j$ peuvent se trouver des moments externes. Par cyclicité du vertex, le moment angulaire associé à $\delta_{v}$ est $-(\ell_1 + \dotsb +\ell_{k+k'})$. La figure \ref{fig:index-attribution-dual} donne un exemple d'attribution des indices pour une feuille particulière. Après avoir procéder ainsi pour toutes les feuilles, nous continuons avec la prochaine couche de vertex ($v_2$ et $v_4$ sur la figure \ref{fig:main-index}) et ainsi de suite.

\noindent
Tout indice de sommation à un vertex $v$ est donc égal à $m_v$ plus une combinaison linéaire de moments angulaires $\ell_\delta,\,\delta\in\mathcal{L}_v \cup \mathcal{N}_v$ où $\cL_{v}$ (resp. $\cN_{v}$) est l'ensemble des lignes de boucles (resp. lignes externes) qui sont attachées au sous-arbre au-dessus de $v$ \ie{} les lignes dont au moins une extrémité $v'$ est telle que l'unique chemin dans l'arbre de $v'$ à $v_{0}$ passe par $v$.

L'ensemble $\mathcal{I}'$ des indices de sommation indépendants peut s'écrire comme l'union de deux sous-ensembles :
\begin{itemize}
\item l'ensemble $\mathcal{M}_\mu =\{m_v\}$ des indices de référence aux vertex internes qui contient $V'-B$ éléments,
\item l'ensemble $\mathcal{J}_\mu= \{\ell_\delta,\; \delta \in \mathcal{L}_\mu\}$ des moments angulaires qui contient $L'$ éléments.  
\end{itemize}
L'amplitude d'un graphe $G$ peut donc s'écrire :
\begin{align} 
  A_G &= \sum_{\mu} \sum_{\mathcal{M}_\mu, \mathcal{J}_\mu} 
\prod_{\delta \in G}
  G^{i_{\delta}}_{m_{\delta}(\mathcal{M}_\mu,\mathcal{J}_\mu),
n_{\delta}(\mathcal{M}_\mu,\mathcal{J}_\mu);
k_{\delta}(\mathcal{M}_\mu,\mathcal{J}_\mu),
l_{\delta}(\mathcal{M}_\mu,\mathcal{J}_\mu)}\; 
\chi(\mathcal{M}_\mu,\mathcal{J}_\mu)\;,
\label{AG}
\end{align}
où la somme sur $\mathcal{M}_\mu$ parcourt $\N^{|\cM_{\mu}|}$ et celle sur $\mathcal{J}_\mu$ parcourt $\Z^{|\cJ_{\mu}|}$. La fonction $\chi(\mathcal{M}_\mu,\mathcal{J}_\mu)$ est la fonction caractéristique imposant que tous les indices $\{m_{\delta}(\mathcal{M}_\mu,\mathcal{J}_\mu)$, $n_{\delta}(\mathcal{M}_\mu,\mathcal{J}_\mu)$ ; $k_{\delta}(\mathcal{M}_\mu,\mathcal{J}_\mu)$, $l_{\delta}(\mathcal{M}_\mu,\mathcal{J}_\mu)\}$ soient positifs. La dépendance de l'amplitude $A_G$ en les indices externes ($B$ indices de référence aux vertex externes et $N$ moments angulaires externes) n'est pas explicitement donnée.

\subsection{Comptage de puissance}

Nous allons maintenant montrer que toutes les sommes sur les différences d'indices peuvent être effectuées gratuitement grâce à \eqref{thsummax} \emph{en utilisant le propagateur $G^{i_\delta}$ pour effectuer la somme sur $\ell_\delta$}. Les sommes sur ces moments étant entrelacées, nous avons besoin de maximiser les autres propagateurs $G^{i_{\delta'}}$ sur $\ell_\delta$. Pour cela, nous devons choisir un ordre sur les lignes.

Nous introduisons un nouvel ordre $>$ sur l'ensemble de boucles $\mathcal{L}_\mu$. Soit $v_\delta$ le plus haut vertex (dans l'arbre) auquel $\delta \in\mathcal{L}_\mu$ est accrochée. Nous écrirons $\delta_1 > \delta_2$ si
\begin{itemize}
\item $v_{\delta_1} > v_{\delta_2}$ ou
\item en tournant autour de $v_{\delta_1} = v_{\delta_2}$ dans le sens horaire à partir de l'indice de référence, nous rencontrons $\delta_1$ puis $\delta_2$ ou la ligne d'arbre qui descend à la racine ou
\item en tournant autour de $v_{\delta_1} = v_{\delta_2}$ dans le sens horaire à partir de l'indice de référence, nous rencontrons la ligne d'arbre qui va à la racine puis $\delta_2$ et enfin $\delta_1$.
\end{itemize}
Nous orientons les lignes $\delta \in \mathcal{L}_\mu$ de telle sorte que les indices $m_{\delta}(\mathcal{M}_\mu,\mathcal{J}_\mu),
l_{\delta}(\mathcal{M}_\mu,\mathcal{J}_\mu)$ soient attachés à $v_\delta$. Leur différence est précisemment $\ell_{\delta}$ :
\begin{equation}   
l_{\delta}(\mathcal{M}_\mu,\mathcal{J}_\mu)
-m_{\delta}(\mathcal{M}_\mu,\mathcal{J}_\mu) = \ell_{\delta}
\end{equation}
Pour les tadpoles $\delta\in\mathcal{L}_\mu$, nous définissons les deux indices $m,l$ comme ceux de la première demi-ligne de $\delta$ dans l'ordre cyclique entre l'indice de référence et la ligne d'arbre qui va à la racine. Si cette ligne d'arbre est rencontrée avant les deux demi-lignes du tadpole, nous choisissons pour $m,l$ les indices de la seconde demi-ligne.

Soient $\mathcal{J}_\mu^{\delta +} =\{ \ell_{\delta'} \in 
\mathcal{J}_\mu, \delta' > \delta\}$ et $\mathcal{J}_\mu^{\delta -} 
=\{ \ell_{\delta'} \in \mathcal{J}_\mu, \delta >\delta'\}$. L'ordre $>$ implique que les indices $m_{\delta}(\mathcal{M}_{\mu},\mathcal{J}_{\mu}),l_{\delta}(\mathcal{M}_{\mu},\mathcal{J}_{\mu})$
en $v_\delta$ sont des fonctions $m_{\delta}(\mathcal{M}_{\mu},\mathcal{J}^{\delta +}_{\mu}),
m_{\delta}(\mathcal{M}_{\mu},\mathcal{J}^{\delta +}_{\mu})+\ell_\delta$ \emph{indépendantes} des indices dans $\mathcal{J}_\mu^{\delta -}$.  
Ainsi pour $\delta \in \mathcal{L}_\mu$ et avec les indices de $\mathcal{M}_{\mu}$ et $\mathcal{J}_{\mu}^{\delta^+}$ fixés,
\begin{align} 
 &\max_{\ell_{\delta'} \in \mathcal{J}_\mu^{\delta -}}
G^{i_{\delta}}_{m_{\delta}(\mathcal{M}_{\mu},  \mathcal{J}_{\mu}),
n_{\delta}(\mathcal{M}_{\mu}, \mathcal{J}_{\mu});
k_{\delta}(\mathcal{M}_{\mu}, \mathcal{J}_{\mu}),
l_{\delta}(\mathcal{M}_{\mu}, \mathcal{J}_{\mu})}
\nonumber\\ 
\les&\max_{n_\delta,k_\delta}G^{i_{\delta}}_{m_{\delta}(\mathcal{M}_{\mu}, 
\mathcal{J}^{\delta+}_{\mu}),n_\delta;k_\delta,
m_{\delta}(\mathcal{M}_{\mu},\mathcal{J}^{\delta +}_{\mu})+
\ell_\delta},
\end{align}
où, dans le membre de droite, la somme est sur tous les indices
$n_\delta, k_\delta$ dans $\N$. Il est instructif de regarder l'exemple de la figure \ref{fig:index-attribution-dual} où $\delta_1>\delta_2>\delta_3>\delta_4$. Les indices $m_2,l_2$ dépendent de l'indice de référence $m_v$ et des moments angulaires des lignes plus hautes (dans le sens $>$) : $\ell_{\delta_1} \in \mathcal{J}_\mu^{2+}$ mais $m_2,l_2$ ne dépendent pas de $\ell_{\delta_3},\ell_{\delta_4} \in
\mathcal{J}_\mu^{2-}$.\\

Pour une attribution d'échelles $\mu$ dans (\ref{AG}) et avec les indices dans $\mathcal{M}_{\mu}$ fixés, nous pouvons écrire les sommes télescopiques sur les moments de $\mathcal{J}_{\mu}$. Soient $\delta_{L'} > \delta_{L'-1} > \dotsb >\delta_2 >\delta_1$ les lignes de boucles. Les sommes sont effectuées des lignes les plus basses aux lignes les plus hautes :
\begin{align} 
&\sum_{\mathcal{J}_\mu} 
\prod_{\delta \in G}
  G^{i_{\delta}}_{m_{\delta}(\mathcal{M}_\mu,\mathcal{J}_\mu),
n_{\delta}(\mathcal{M}_\mu,\mathcal{J}_\mu);
k_{\delta}(\mathcal{M}_\mu,\mathcal{J}_\mu),
l_{\delta}(\mathcal{M}_\mu,\mathcal{J}_\mu)}\; 
\chi(\mathcal{M}_\mu,\mathcal{J}_\mu)
\nonumber
\\*
& \les\sum_{\mathcal{J}^{\delta_1 +}_\mu} \Bigg\{
\max_{\ell_{\delta_1}} \left(\prod_{\delta \in \mathcal{T}_\mu}
  G^{i_{\delta}}_{m_{\delta}(\mathcal{M}_\mu,\mathcal{J}_\mu),
n_{\delta}(\mathcal{M}_\mu,\mathcal{J}_\mu);
k_{\delta}(\mathcal{M}_\mu,\mathcal{J}_\mu),
l_{\delta}(\mathcal{M}_\mu,\mathcal{J}_\mu)} \right) 
\nonumber
\\*
& \times 
\max_{\ell_{\delta_1}} \left(\prod_{\delta \in \mathcal{L}_\mu\,,\; 
\delta > \delta_1 }
  G^{i_{\delta}}_{m_{\delta}(\mathcal{M}_\mu,\mathcal{J}_\mu),
n_{\delta}(\mathcal{M}_\mu,\mathcal{J}_\mu);
k_{\delta}(\mathcal{M}_\mu,\mathcal{J}_\mu),
l_{\delta}(\mathcal{M}_\mu,\mathcal{J}_\mu)}\right)
\nonumber
\\*
& \times 
\sum_{\ell_{\delta_1}} 
G^{i_{\delta_1}}_{m_{\delta_1}(\mathcal{M}_\mu,\mathcal{J}^{\delta_1+}_\mu),
n_{\delta_1}(\mathcal{M}_\mu,\mathcal{J}_\mu);
k_{\delta_1}(\mathcal{M}_\mu,\mathcal{J}_\mu),
m_{\delta_1}(\mathcal{M}_\mu,\mathcal{J}^{\delta_1+}_\mu)
+\ell_{\delta_1}}
\Bigg\}\chi(\mathcal{M}_\mu,\mathcal{J}_\mu) 
\nonumber
\\
&\les \sum_{\mathcal{J}^{\delta_2+}_\mu} \Bigg\{
\max_{\ell_{\delta_1},\ell_{\delta_2}} 
\left(\prod_{\delta \in \mathcal{T}_\mu}
  G^{i_{\delta}}_{m_{\delta}(\mathcal{M}_\mu,\mathcal{J}_\mu),
n_{\delta}(\mathcal{M}_\mu,\mathcal{J}_\mu);
k_{\delta}(\mathcal{M}_\mu,\mathcal{J}_\mu),
l_{\delta}(\mathcal{M}_\mu,\mathcal{J}_\mu)} \right) 
\nonumber
\\
& \times 
\max_{\ell_{\delta_1},\ell_{\delta_2}} 
\left(\prod_{\delta \in \mathcal{L}_\mu\,,\; 
\delta > \delta_2 }
  G^{i_{\delta}}_{m_{\delta}(\mathcal{M}_\mu,\mathcal{J}_\mu),
n_{\delta}(\mathcal{M}_\mu,\mathcal{J}_\mu);
k_{\delta}(\mathcal{M}_\mu,\mathcal{J}_\mu),
l_{\delta}(\mathcal{M}_\mu,\mathcal{J}_\mu)}\right)
\nonumber
\\
& \times 
\sum_{\ell_{\delta_2}} \left(\max_{\ell_{\delta_1}}
G^{i_{\delta_2}}_{m_{\delta}(\mathcal{M}_\mu,\mathcal{J}^{\delta_2+}_\mu),
n_{\delta_2}(\mathcal{M}_\mu,\mathcal{J}_\mu);
k_{\delta_2}(\mathcal{M}_\mu,\mathcal{J}_\mu),
m_{\delta_2}(\mathcal{M}_\mu,\mathcal{J}^{\delta_2+}_\mu)
+\ell_{\delta_2}}\right)
\nonumber
\\*
& \times 
\max_{\ell_{\delta_2}} \sum_{\ell_{\delta_1}} 
\left(\max_{n_{\delta_1},k_{\delta_1}}
G^{i_{\delta_1}}_{m_{\delta_1}(\mathcal{M}_\mu,\mathcal{J}^{\delta_1+}_\mu),
n_{\delta_1}; k_{\delta_1},
m_{\delta_1}(\mathcal{M}_\mu,\mathcal{J}^{\delta_1+}_\mu)+\ell_{\delta_1}}
\right)\Bigg\}\chi(\mathcal{M}_\mu,\mathcal{J}_\mu) \;,
\end{align}
et ainsi de suite jusqu'à
\begin{align}  
&\sum_{\mathcal{J}_{\mu}} \prod_{\delta \in G} 
G^{i_\delta}_{m_\delta(\mathcal{M}_{\mu},\mathcal{J}_{\mu}), 
n_\delta(\mathcal{M}_{\mu},\mathcal{J}_{\mu});
k_\delta(\mathcal{M}_{\mu},\mathcal{J}_{\mu}),l_\delta(\mathcal{M}_{\mu},\mathcal{J}_{\mu})} 
\chi(\mathcal{M}_{\mu}, \mathcal{J}_{\mu})
\nonumber
\\ &
\les \prod_{\delta \in \mathcal{T}_\mu} 
\left(\max_{\mathcal{J}_\mu}
G^{i_\delta}_{m_\delta(\mathcal{M}_{\mu},\mathcal{J}_{\mu}), 
n_\delta(\mathcal{M}_{\mu},\mathcal{J}_{\mu});
k_\delta(\mathcal{M}_{\mu},\mathcal{J}_{\mu}),
l_\delta(\mathcal{M}_{\mu},\mathcal{J}_{\mu})} \right)
\nonumber
\\
& \times \prod_{\delta \in \mathcal{L}_\mu}\left\{
\max_{\mathcal{J}_\mu^{\delta +}} 
\left(\sum_{\ell_\delta \ges - m_{\delta}(\mathcal{M}_{\mu},
\mathcal{J}_{\mu}^{\delta+})} \left(\max_{n_\delta,k_\delta} 
G^{i_{\delta}}_{
m_{\delta}(\mathcal{M}_{\mu},\mathcal{J}_{\mu}^{\delta+}), n_\delta ;
k_\delta , m_{\delta}(\mathcal{M}_{\mu},\mathcal{J}_{\mu}^{\delta+}) +
\ell_\delta }\right)\right)\right\}.
\label{telescsums}
\end{align}
Les contraintes de positivité de $\chi$ ont été utilisées pour déterminer l'intervalle de sommation correct des $\ell_{\delta}$.\\

Nous obtenons ainsi une borne sur la somme sur $\mathcal{J}_\mu$ de (\ref{AG}). Pour les lignes d'arbre $\delta \in \mathcal{T}_\mu$, dont tous les indices dépendent de $\mathcal{J}_\mu$, la borne \eqref{telescsums} donne grâce à \eqref{th1}
\begin{align}
\max_{\mathcal{J}_\mu} G^{i_\delta}_{m_{\delta}(\mathcal{M}_\mu,
\mathcal{J}_\mu),n_{\delta}(\mathcal{M}_\mu,\mathcal{J}_\mu);
k_{\delta}(\mathcal{M}_\mu,\mathcal{J}_\mu),
l_{\delta}(\mathcal{M}_\mu,\mathcal{J}_\mu)} 
\les K M^{-2i_\delta}, 
\qquad \delta \in \mathcal{T}_\mu\;,~~\delta \notin \mathcal{G}_\mu.
\label{DT-1}
\end{align}
Si l'un des indices de $\delta \in \mathcal{T}_\mu$ est un indice de référence au vertex $v$, nous avons
\begin{align}
\max_{\mathcal{J}_\mu} G^{i_\delta}_{m_v,
n_{\delta}(\mathcal{M}_\mu,\mathcal{J}_\mu);
k_{\delta}(\mathcal{M}_\mu,\mathcal{J}_\mu),
l_{\delta}(\mathcal{M}_\mu,\mathcal{J}_\mu)} 
\les K M^{-2i_\delta} e^{-c M^{-2i_\delta} \|m_{v}\|}, 
\qquad \delta \in \mathcal{T}_\mu \cap \mathcal{G}_\mu.
\label{DT-2}
\end{align}
Si deux indices de $\delta$ sont des indices de référence en $v,v'$, nous écrirons
\begin{align}
\max_{\mathcal{J}_\mu} G^{i_\delta}_{m_v,
n_{\delta}(\mathcal{M}_\mu,\mathcal{J}_\mu);
m_{v'},l_{\delta}(\mathcal{M}_\mu,\mathcal{J}_\mu)} 
\les K M^{-2i_\delta} e^{-c M^{-2i_\delta} (\|m_{v}\|+\|m_v{'}\|)}, 
\qquad \delta \in \mathcal{T}_\mu,~~
\delta \in \mathcal{G}_\mu\setminus \{\delta\}.
\label{DT-3}
\end{align}
Puis chaque propagateur correspondant à une ligne $\delta \in \mathcal{L}_{\mu}$ donne, par (\ref{telescsums}),
un facteur $K M^{-2i_\delta}$,
\begin{align}
\max_{\mathcal{J}_\mu^{\delta+}} \left( \sum_{\ell_\delta} \left(
    \max_{\mathcal{J}_\mu^{\delta -}} 
G^{i_{\delta}}_{
m_{\delta}(\mathcal{M}_{\mu},\mathcal{J}_{\mu}^{\delta+}), 
n_\delta(\mathcal{M}_{\mu},\mathcal{J}_{\mu}) ;
k_\delta(\mathcal{M}_{\mu},\mathcal{J}_{\mu}) , 
m_{\delta}(\mathcal{M}_{\mu},\mathcal{J}_{\mu}^{\delta+}) +
\ell_\delta }
\right)\right) 
&\les K'M^{-2i_\delta},
\nonumber
\\*
\delta\in\mathcal{L}_\mu,~~ &\delta\notin\mathcal{G}_\mu.
\label{DL-1}
\end{align}
Si $\delta\in\mathcal{L}_{\mu}$ est un tadpole en $v_{i}$ qui a l'indice d'échelle le plus bas parmi les lignes de $v_{i}$, nous obtenons grâce à \eqref{thsummax}
\begin{align}
\max_{\mathcal{J}_\mu^{\delta+}} \left( \sum_{\ell_\delta} \left(
    \max_{\mathcal{J}_\mu^{\delta -}} 
G^{i_{\delta}}_{
m_v, n_\delta(\mathcal{M}_{\mu},\mathcal{J}_{\mu}) ;
k_\delta(\mathcal{M}_{\mu},\mathcal{J}_{\mu}) , 
m_v + \ell_\delta }
\right)\right) 
&\les K'M^{-2i_\delta}\,e^{-c'M^{-2i_\delta}\|m_{v}\|},
\quad
\delta \in \mathcal{L}_\mu \cap \mathcal{G}_\mu.
\label{DL-2}
\end{align}
Finalement il y a aussi des indices externes $m_\epsilon,n_\epsilon$ (fixés). Chacun d'eux donne, grâce à \eqref{th1} et \eqref{thsummax} $e^{-cM^{-2i_\epsilon}\|m_\epsilon\|}$ et $e^{-cM^{-2i_\epsilon}\|n_\epsilon\|}$ respectivement.

Finalement la somme sur $\mathcal{J}_\mu$ dans (\ref{AG}) est bornée par
\begin{align}
 A_G &\les \sum_{\mu} \sum_{m_1,\dots,m_{V'-B} \in \N^2} 
\left(\prod_{\delta' \in \mathcal{G}_\mu}
e^{-c M^{-2i_{\delta'}} \|m_{v(\delta')}\|}\right)
\left(\prod_{\delta \in G}  K M^{-2i_{\delta}} \right) 
\nonumber
\\*
& \qquad \times 
\left(\prod_{\epsilon=1}^N e^{-cM^{-2i_\epsilon}\|m_\epsilon\|}\right)
\left(\prod_{\epsilon=1}^N e^{-cM^{-2i_\epsilon}\|n_\epsilon\|}\right),
\label{AG-1}
\end{align}
où $m_{v(\delta')}$ est l'indice de référence de la ligne $\delta' \in
\mathcal{G}_\mu$. Après avoir sommé sur $m_1,\dotsc,m_{V'-B}$, nous avons
\begin{align}
 A_G &\les \sum_{\mu} \frac{K^I}{c^{2(V'-B)}} 
\Big(M^{-2\sum_{\delta \in G} i_{\delta}} \Big)
\Big(M^{4\sum_{\delta' \in \mathcal{G}_\mu} i_{\delta'}}\Big)
\left(\prod_{\epsilon=1}^N e^{-cM^{-2i_\epsilon}\|m_\epsilon\|}\right)\!
\left(\prod_{\epsilon=1}^N e^{-cM^{-2i_\epsilon}\|n_\epsilon\|}\right).
\label{AG-2}
\end{align}

Comme expliqué dans la section \ref{sec:multiscalephi4}, nous définissons les composantes connexes : un sous-graphe est une composante connexe si toutes ses lignes internes sont au-dessus de toutes ses lignes externes (au sens de (\ref{order})). Le comptage de puissance se factorise alors dans les composantes connexes : 
\begin{align}
  A_G\les& K'^{V}\sum_{\mu}\prod_{i,k}M^{\omega(G^{i}_{k})}
  \left(\prod_{\epsilon=1}^N e^{-cM^{-2i_\epsilon}\|m_\epsilon\|}\right)\!
  \left(\prod_{\epsilon=1}^N e^{-cM^{-2i_\epsilon}\|n_\epsilon\|}\right)\\
  \text{avec }\omega(G^{i}_{k})=&4(V'_{i,k}-B_{i,k})-2I_{i,k}=4(F_{i,k}-B_{i,k})-2I_{i,k}\\
  =&(4-N_{i,k})-4(2g_{i,k}+B_{i,k}-1)\notag
\end{align}
et $N_{i,k}$, $V_{i,k}$, $I_{i,k}=2V_{i,k}-\frac{N_{i,k}}{2}$, $F_{i,k}$ et $B_{i,k}$ sont respectivement les nombres de pattes externes, de vertex, de propagateurs (internes), de faces et de faces brisées de la composante connexe $G^{i}_{k}$ ; $g_{i,k}= 1 - \frac{1}{2} (V_{i,k}-I_{i,k}+F_{i,k})$ est son genre. Nous avons alors
\begin{thm}
\label{pc-slice}
  La somme sur les attributions d'échelles $\mu$ dans (\ref{AG-2}) converge si $\forall i,k,\, \omega(G^{i}_{k}) <0$.
\end{thm}
Nous retrouvons donc le comptage de puissance obtenu dans \cite{GrWu03-1}. Nous constatons que les outils de l'analyse multi-échelles (décomposition du propagateur, optimisation de l'arbre, composantes connexes, arbre de Gallavotti) s'adaptent très bien à la base matricielle et aux modèles de matrices dynamiques.

\section{Étude de propagateurs}
\label{sec:boite-outils}

Nous donnons ici les résultats que nous avons obtenus dans \cite{toolbox05}. Dans cet article, nous avons calculé les noyaux en espace $x$ et dans la base matricielle d'opérateurs généralisant le noyau de Mehler \eqref{eq:Mehler}. Puis nous avons procédé à une étude fine des comportements de ces noyaux dans la base matricielle. Ce travail est notamment utile pour étudier le modèle de Gross-Neveu \ncf{} dans la base matricielle.

\subsection{Noyau bosonique}

Soient $x\wedge x'=x_0x'_1-x_1x'_0$ et $x\cdot x'=x_0x'_0+x_1x'_1$. Le lemme suivant généralise le noyau de Mehler \cite{simon79funct} :
\begin{lemma}
  \label{HinXspace}Soit l'opérateur $H$ :
  \begin{equation}
    H=\frac{1}{2}\Big(-\Delta+ \Omega^2x^2-2\imath B(x_0\partial_1-x_1\partial_0)\Big).
  \end{equation}
  Le noyau, en $x$, de l'opérateur $e^{-tH}$ est :
  \begin{equation}
    e^{-tH}(x,x')=\frac{\Omega}{2\pi\sinh\Omega t}e^{-A},\label{eq:propaxboson}
  \end{equation}
  \begin{equation}
    A=\frac{\Omega\cosh\Omega t}{2\sinh\Omega t}(x^2+x'^2)-
    \frac{\Omega\cosh Bt}{\sinh\Omega t}x\cdot x'-\imath
    \frac{\Omega\sinh Bt}{\sinh\Omega t}x\wedge x'.
  \end{equation}
\end{lemma}
\begin{proof}
  Nous pouvons soit vérifier directement que $P=e^{-tH}$ est solution de
  \begin{align}
    \frac{dP}{dt}+HP=&0\label{eq:diff}
  \end{align}
  soit postuler que la solution est de la forme $f(t)\,e^{-g(t)(x^2+x'^2)-h(t)x\cdot x'-i(t)x\wed x'}$, utiliser l'équation \eqref{eq:diff} pour dériver un système d'équations différentielles couplées pour les fonctions $f,g,h$ et $i$ et résoudre ce système. Voir \cite{toolbox05} pour une preuve directe.
\end{proof}
\begin{rem}
  Le noyau de Mehler correspond à $B=0$. De plus, la limite $\Omega=B\to 0$ donne le noyau de la chaleur habituel.
\end{rem}

Pour retrouver les résultats de la section \ref{sec:phi4-matrixbase}, nous commençons par effectuer les changements $\Omega\rightarrow 
\frac{2\Omega}{\theta}$ et $B\rightarrow\frac{2B}{\theta}$ dans $H$ (voir lemme \ref{HinXspace}). Le cas $B=0$ donne donc exactement le propagateur \eqref{eq:propaPhimatrix}. De plus $L_{2}=-\imath(x_0\partial_1-x_1\partial_0)$ commute avec le laplacien et avec $x^{2}$ si bien que pour calculer $e^{-tH}$ dans la base matricielle, il nous suffit de calculer $e^{-tL_{2}}$ et de faire le produit avec \eqref{eq:propaPhimatrix}. Le calcul de $e^{-tL_{2}}$ est simple car $L_{2}$ est diagonal dans la base matricielle :
\begin{align}
  H_{m,m+h;l+h,l}=&\frac{2}{\theta}
  (1+\Omega^2)(2m+h+1)\delta_{m,l}-\frac{4Bh}{\theta}~\delta_{m,l}\label{eq:HBini}
  \\
  &\hspace{-1cm}-\frac{2}{\theta}(1-\Omega^2)[\sqrt{(m+h+1)(m+1)}~\delta_{m+1,l}+
  \sqrt{(m+h)m}~\delta_{m-1,l}].\nonumber
  \label{HinMat}
\end{align}
Remarquons qu'à $\Omega=1$, $H$ est diagonal.
\begin{lemma}
  Soit $H$ donné par l'équation \eqref{HinMat}. Son inverse est :
  \begin{align}
    H^{-1}_{m,m+h;l+h,l}=&\frac{\theta}{8\Omega} \int_0^1 d\alpha\,  
\dfrac{(1-\alpha)^{\frac{\mu_0^2 \theta}{8 \Omega}+(\frac{D}{4}-1)}}{  
(1 + C\alpha )^{\frac{D}{2}}} (1-\alpha)^{-\frac{4B}{8\Omega}h}\prod_{s=1}^{\frac{D}{2}} 
G^{(\alpha)}_{m^s, m^s+h^s; l^s + h^s, l^s},\label{eq:propbosonmatrix}
\\
 G^{(\alpha)}_{m, m+h; l + h, l}
&= \left(\frac{\sqrt{1-\alpha}}{1+C \alpha} 
\right)^{m+l+h} \sum_{u=\max(0,-h)}^{\min(m,l)}
   {\cal A}(m,l,h,u)\ 
\left( \frac{C \alpha (1+\Omega)}{\sqrt{1-\alpha}\,(1-\Omega)} 
\right)^{m+l-2u},\notag
\end{align}
où ${\cal A}(m,l,h,u)=\sqrt{\binom{m}{m-u}
\binom{m+h}{m-u}\binom{l}{l-u}\binom{l+h}{l-u}}$ et $C$ est une fonction de $\Omega$ : $C(\Omega)=
\frac{(1-\Omega)^2}{4\Omega}$.
\end{lemma}

\subsection{Noyau fermionique}

La théorie fermionique libre à deux dimensions est définie par le lagrangien :
\begin{equation}
\label{eq:LGN}
{\cal L}=\psib(x)\lbt\ps+\mu\rbt\psi(x). 
\end{equation}
Le propagateur de la théorie $\lbt\ps + \mu\rbt^{-1}(x,y)$ peut être calculé par la méthode du noyau de la chaleur :
\begin{align}
  \lbt\ps
  +\mu\rbt^{-1}(x,y)&=\lbt-\ps+\mu\rbt\lbt\lbt\ps+\mu\rbt\lbt-\ps+\mu\rbt\rbt^{-1}(x,y)\nonumber\\
  &=\lbt-\ps+\mu\rbt\lbt p^{2}+\mu^{2}\rbt^{-1}(x,y)\\
  &=\lbt-\ps+\mu\rbt\int_{0}^{\infty}\frac{dt}{4\pi t}\,
  e^{-\frac{(x-y)^{2}}{4t}-\mu^{2}t}\\
  &=\int_{0}^{\infty}\frac{dt}{4\pi
    t}\,\lbt\frac{-\imath}{2t}(\xs-\ys)+\mu\rbt e^{-\frac{(x-y)^{2}}{4t}-\mu^{2}t}.
\end{align}
Sur l'espace de Moyal, nous souhaitons modifier l'action libre en ajoutant un terme de vulcanisation. Cette procédure évite le mélange UV/IR dangereux (voir section \ref{sec:renorm-et-vulc}) et permet la renormalisation. Ainsi l'action libre devient\footnote{Dans cette section, nous souhaitons uniquement étudier le propagateur du modèle de Gross-Neveu \ncf{}. Une définition précise du modèle complet est donnée au chapitre \ref{cha:GN}.}
\begin{equation}\label{sfree}
S_{\text{\emph{free}}} = \int d^2 x \psib^a(x)\lbt\ps +\mu+\Omega\xts\rbt\psi^a(x)
\end{equation}
où $\xt=2\Theta^{-1}x$, $\Theta=\lbt\begin{smallmatrix}\phantom{-}0&\theta\\-\theta&0\end{smallmatrix}\rbt$ et
$a$ est un indice de couleurs entre $1$ et $N$. Le propagateur $G$ étant diagonal dans cet indice, nous l'omettrons dans la suite. Pour calculer le propagateur, nous écrivons encore une fois :
\begin{eqnarray}
G&=& \lbt\ps+\mu+\Omega\xts\rbt^{-1}=\lbt-\ps+\mu-\Omega\xts\rbt  . Q^{-1},
\nonumber\\
Q&=& \lbt\ps+\mu+\Omega\xts\rbt\lbt-\ps+\mu-\Omega\xts\rbt
  \nonumber\\
  &=&\mathds{1}_{2}\otimes\lbt
  p^{2}+\mu^{2}+\frac{4\Omega^{2}}{\theta^{2}}x^{2}\rbt
  +\frac{4\imath\Omega}{\theta}\gamma^{0}\gamma^{1}\otimes{\text{Id}}
  +\frac{4\Omega}{\theta}\mathds{1}_{2}\otimes  L_{2},
\end{eqnarray}
où $L_{2}=x^{0}p_{1}-x^{1}p_{0}$. Pour inverser $Q$ nous utilisons la méthode de Schwinger :
\begin{lemma}
\label{FermioXspace} Nous avons :
\begin{eqnarray}
G(x,y) &=& -\frac{\Omega}{\theta\pi}\int_{0}^{\infty}\frac{dt}{\sinh(2\Ot
t)}\, e^{-\frac{\Ot}{2}\coth(2\Ot t)(x-y)^{2}+\imath\Ot
x\wedge y}
\\ 
&&    \lb\imath\Ot\coth(2\Ot t)(\xs-\ys)+\Omega(\xts-\yts)- \mu \rb
e^{-2\imath\Ot
t\gamma^{0}\gamma^{1}}e^{-t\mu^{2}}  \nonumber
\end{eqnarray}
Il sera également pratique d'exprimer $G$ en termes de commutateurs :
\begin{eqnarray}    
G(x,y)  &=&-\frac{\Omega}{\theta\pi}\int_{0}^{\infty}dt\,\lb \imath\Ot\coth(2\Ot
t)\lsb\xs, \Gamma^t  \rsb(x,y) \right.
\nonumber\\
&&
\left. +\Omega\lsb\xts, \Gamma^t \rsb(x,y)  -\mu \Gamma^t (x,y)  \rb
e^{-2\imath\Ot t\gamma^{0}\gamma^{1}}e^{-t\mu^{2}}, 
\label{xfullprop}
\end{eqnarray}
où
\begin{eqnarray}
\Gamma^t (x,y)  &=&
\frac{1}{\sinh(2\Ot t)}\,
e^{-\frac{\Ot}{2}\coth(2\Ot t)(x-y)^{2}+\imath\Ot x\wedge y}
\end{eqnarray}
avec $\Ot=\frac{2\Omega}{\theta}$ et $x\wedge y=x^{0}y^{1}-x^{1}y^{0}$.\\
\end{lemma}
La méthode employée pour démontrer ce résultat est identique à celle utilisée pour le lemme \ref{HinXspace}. La preuve du lemme \ref{FermioXspace} est donnée en annexe B de \cite{toolbox05}.\\

Nous allons maintenant donner l'expression du propagateur fermionique \eqref{xfullprop} dans la base matricielle. Soit $L_{2}=-\imath(x^{0}\partial_{1}-x^{1}\partial_{0})$. L'inverse de la forme quadratique
\begin{equation}
\Delta= Q- \frac{4 \imath \Omega} {\theta} \gamma^0\gamma^1= p^{2}+\mu^{2}+\frac{4\Omega^{2}}{\theta^2} x ^{2} +\frac{4B}{\theta}L_{2}
\end{equation}
est donnée par le résultat \eqref{eq:propbosonmatrix} de la section précédente :
\begin{align}
  \Gamma_{m, m+h; l + h, l} 
  &= \frac{\theta}{8\Omega} \int_0^1 d\alpha\,  
  \dfrac{(1-\alpha)^{\frac{\mu^2 \theta}{8 \Omega}-\frac{1}{2}}}{  
    (1 + C\alpha )} 
  \Gamma^{\alpha}_{m, m+h; l + h, l}\,\label{eq:propinit}
\\
  \Gamma^{(\alpha)}_{m, m+h; l + h, l}
  &= \left(\frac{\sqrt{1-\alpha}}{1+C \alpha} 
  \right)^{m+l+h}\left( 1-\alpha\right)^{-\frac{Bh}{2\Omega}} \label{eq:propinit-b}\\
  &
  \sum_{u=0}^{\min(m,l)} {\cal A}(m,l,h,u)\ 
  \left( \frac{C \alpha (1+\Omega)}{\sqrt{1-\alpha}\,(1-\Omega)} 
  \right)^{m+l-2u}.\notag
\end{align}
Le propagateur fermionique $G$ (\ref{xfullprop}) dans la base matricielle peut se déduire du noyau \eqref{eq:propinit}. Il suffit de prendre $B= \Omega$, d'ajouter le terme manquant en $\gamma^0 \gamma^1$ et de calculer l'action de $-\ps-\Omega\xts+\mu$ sur $\Gamma$. Nous devons donc évaluer $\lsb x^{\nu},\Gamma\rsb$ dans la base matricielle :
\begin{align}
  \lsb x^{0},\Gamma\rsb_{m,n;k,l}=&2\pi\theta\sqrt\frac{\theta}{8}\lb\sqrt{m+1}
  \Gamma_{m+1,n;k,l}-\sqrt{l}\Gamma_{m,n;k,l-1}+\sqrt{m}\Gamma_{m-1,n;k,l}
\right.\nonumber\\
&-\sqrt{l+1}\Gamma_{m,n;k,l+1}+\sqrt{n+1}\Gamma_{m,n+1;k,l}-\sqrt{k}
\Gamma_{m,n;k-1,l}\nonumber\\
&\left.+\sqrt{n}\Gamma_{m,n-1;k,l}-\sqrt{k+1}
  \Gamma_{m,n;k+1,l}\rb,\label{x0Gamma}\\
  \lsb
  x^{1},\Gamma\rsb_{m,n;k,l}=&2\imath\pi\theta\sqrt\frac{\theta}{8}\lb\sqrt{m+1}
  \Gamma_{m+1,n;k,l}-\sqrt{l}\Gamma_{m,n;k,l-1}-\sqrt{m}
  \Gamma_{m-1,n;k,l} \right.
\nonumber\\
&+\sqrt{l+1}\Gamma_{m,n;k,l+1}
-\sqrt{n+1}\Gamma_{m,n+1;k,l}+\sqrt{k}\Gamma_{m,n;k-1,l}
\nonumber\\
&\left.+\sqrt{n}\Gamma_{m,n-1;k,l}-\sqrt{k+1}\Gamma_{m,n;k+1,l}\rb.
  \label{x1Gamma}
\end{align}
Ceci nous permet de démontrer :
\begin{lemma}Soit $G_{m,n;k,l}$ le noyau, dans la base matricielle, de l'opérateur\\
$\lbt\ps+\Omega\xts+\mu\rbt^{-1}$. Nous avons :
\begin{align}
G_{m,n;k,l}=& 
-\frac{2\Omega}{\theta^{2}\pi^{2}} \int_{0}^{1} 
d\alpha\, G^{\alpha}_{m,n;k,l},\label{eq:propaFermiomatrix}
\\
G^{\alpha}_{m,n;k,l}=&\lbt\imath\Ot\frac{2-\alpha}{\alpha}\lsb\xs,
\Gamma^{\alpha}\rsb_{m,n;k,l}
+\Omega\lsb\slashed{\tilde{x}},\Gamma^{\alpha}\rsb_{m,n;k,l} - \mu\,\Gamma^{\alpha}_{m,n;k,l}\rbt
\nonumber\\
&\times\lbt\frac{2-\alpha}{2\sqrt{1-\alpha}}
\mathds{1}_{2}-\imath\frac{\alpha}{2\sqrt{1-\alpha}}\gamma^{0}\gamma^{1}
\rbt.\label{eq:matrixfullprop}
\end{align}
où $\Gamma^{\alpha}$ est donné par (\ref{eq:propinit-b}) et les commutateurs par les formules (\ref{x0Gamma}) et (\ref{x1Gamma}).
\end{lemma}
Les deux premiers termes de l'équation (\ref{eq:matrixfullprop}) contiennent des commutateurs et seront regroupés sous l'appellation $G^{\alpha, {\rm comm}}_{m,n;k,l}$. Le dernier terme sera $G^{\alpha, {\rm mass}}_{m,n;k,l}$ :
\begin{align}\label{commterm}
G^{\alpha, {\rm comm}}_{m,n;k,l}=& \lbt\imath\Ot\frac{2-\alpha}{\alpha}\lsb\xs,
\Gamma^{\alpha}\rsb_{m,n;k,l} +\Omega\lsb\slashed{\tilde{x}},\Gamma^{\alpha}\rsb_{m,n;k,l} \rbt  \nonumber\\
&\times\lbt\frac{2-\alpha}{2\sqrt{1-\alpha}}
\mathds{1}_{2}-\imath\frac{\alpha}{2\sqrt{1-\alpha}}\gamma^{0}\gamma^{1} \rbt,\\
\notag\\
G^{\alpha, {\rm mass}}_{m,n;k,l}=& - \mu\, \Gamma^{\alpha}_{m,n;k,l}
\times\lbt\frac{2-\alpha}{2\sqrt{1-\alpha}}
\mathds{1}_{2}-\imath\frac{\alpha}{2\sqrt{1-\alpha}}\gamma^{0}\gamma^{1} \rbt.\label{massterm}
\end{align}

\subsection{Bornes}
\label{sec:bornes}

Nous allons appliquer l'analyse multi-échelles pour étudier le comportement du propagateur \eqref{eq:matrixfullprop} et revisiter plus finement les bornes \eqref{th1} à \eqref{thsummax}. La décomposition en échelles est faite comme en section \ref{sec:bornes-sur-le} :
\begin{equation}
  \label{eq:slices}
  \int_{0}^1 d\alpha =\sum_{i=1}^\infty \int_{M^{-2i}}^{M^{-2(i-1)}} 
  d\alpha
\end{equation}
et conduit au propagateur suivant dans la tranche $i$ :
\begin{align}
  \Gamma^i_{m,m+h,l+h,l} 
  &=\frac{\theta}{8\Omega}  \int_{M^{-2i}}^{M^{-2(i-1)}} d\alpha\; 
  \dfrac{(1-\alpha)^{\frac{\mu_0^2 \theta}{8 \Omega}-\frac{1}{2}}}{  
    (1 + C\alpha )} 
  \Gamma^{(\alpha)}_{m, m+h; l + h, l}\;.
  \label{prop-slice-i}
\end{align}
\begin{eqnarray}
G_{m,n;k,l}&=& \sum_{i=1}^\infty G^i_{m,n;k,l} \ ; \ G^i_{m,n;k,l} = 
-\frac{2\Omega}{\theta^{2}\pi^{2}} \int_{M^{-2i}}^{M^{-2(i-1)}} 
d\alpha\, G^{\alpha}_{m,n;k,l}  
\label{eq:matrixfullpropsliced}
\end{eqnarray}
Nous séparons $G$ comme nous l'avons fait dans les équations (\ref{commterm}) et  (\ref{massterm}). Soient $h= n-m$ et $p=l-m$. Sans perte de généralité, nous supposerons $h \ges 0 $ et $p\ges 0$. Ainsi le plus petit des quatre entiers $m,n,k,l$ est $m$ et le plus grand est $k=m+h+p$. Nous pouvons alors énoncer le principal résultat de cette section :

\begin{thm}\label{maintheorem}
Sous les conditions $h =n-m\ges 0 $ et $p=l-m \ges 0$, il existe $K,c\in\R_{+}$ ($c$ dépend de $\Omega$) tels que le propagateur de Gross-Neveu dans une tranche $i$ obéisse à la borne
\begin{eqnarray}\label{mainbound1}  
\vert G^{i,{\rm comm}}_{m,n;k,l}\vert&\les&   
K M^{-i} \bigg( \chi(\alpha k>1)\frac{\exp \{- \frac{c p ^2  }{1+ kM^{-2i}}
- \frac{ c M^{-2i}}{1+k} (h - \frac{k}{1+C})^2 \}}{(1+\sqrt{ kM^{-2i}}) }  
\nonumber\\
&&+ \min(1,(\alpha k)^{p})e^{- c k M^{-2i} - c  p }\bigg).
\end{eqnarray}
Le terme de masse a une borne un peu différente :
\begin{align} \label{mainbound2}  
\vert  G^{i,{\rm mass}}_{m,n;k,l}\vert\les&   
K M^{-2i} \bigg( \chi(\alpha k>1) \frac{\exp \{- \frac{c p ^2  }{1+ kM^{-2i}}
- \frac{ c M^{-2i}}{1+k} (h - \frac{k}{1+C})^2 \}}{1+\sqrt{ kM^{-2i}}}\notag
\\
&+\min(1,(\alpha k)^{p}) e^{- c k M^{-2i} - c  p }\bigg).
\end{align}
\end{thm}
\begin{proof}
  Nous souhaitons donner les étapes principales de la preuve car nous pensons que cette étude pourra servir sur d'autres espaces que le plan de Moyal et renvoyons à \cite{toolbox05} pour les détails. L'analyse du propagateur \eqref{eq:propaFermiomatrix} révèle qu'il existe une région dans les indices $k,p$ et $h$ où le propagateur n'a pas le comportement d'échelles \eqref{th1}.\\

Nous écrivons le propagateur sous la forme :
\begin{equation}
  \Gamma=\int_{0}^{1}d\alpha\,\frac{(1-\alpha)^{-1/2}}{1+C\alpha}\Gamma^{\alpha}
\end{equation}
avec
\begin{align}
  \Gamma^{\alpha}=&\Big{(}\frac{\sqrt{1-\alpha}}{1+C\alpha}\Big{)}^{2m+p}
  \frac{1}{(1+C\alpha)^h}\sum_{u=o}^{m}
  \Big{(}\frac{\alpha\sqrt{C(1+C)}}
  {\sqrt{(1-\alpha)}}\Big{)}^{2m+p-2u}{\cal A}(m,m+p,h,u)\label{expressgamma}\\
  =&e^{(2m+p)\ln\frac{\sqrt{1-\alpha}}{1+C\alpha}-
    h\ln(1+C\alpha)}\sum_{0\les v=m-u\les m}e^{(2v+p)\ln
    \frac{\alpha\sqrt{C(1+C)}}{\sqrt{1-\alpha}}}{\cal A}(m,m+p,h,u).
\end{align}
Nous nous restreignons au régime $\alpha \ll C \ll 1$ \ie{} à la zone \og{}ultraviolette\fg{} et à $\Omega$ proche de $1$. Nous définissons les variables réduites $x = v/k$,  $y =h/k$, $z=p/k$. Celles-ci sont contenues dans le simplexe $0 \les x, y, z \les 1$,  $0 \les x+ y+ z \les 1$. En utilisant l'approximation de Stirling et en remplaçant la somme sur $v$ par une intégrale qui, à une constante multiplicative près, constitue une borne supérieure rigoureuse, nous avons
\begin{align}\label{integralinx}
  \Gamma^{\alpha}\les&\int_0^{1-y-z} dx  \frac{[(1-y)(1-z)(1-z-y)]^{1/4}}
  {[x(x+z)(1-x-z) (1-x-y-z)]^{1/2}} e^{k g(x,y,z)}\\
  \text{où }g=&(2-2y-z)\ln\frac{\sqrt{1-\alpha}}{1+C\alpha}
  +(2x+z)\ln\frac{\alpha\sqrt{C(1+C)}}{\sqrt{1-\alpha}}- y \ln(1+C\alpha)
  \nonumber\\
  &+ \frac{1-y}{2}\ln(1-y) +\frac{1-z}{2}\ln(1-z)  +\frac{1-y-z}{2}\ln(1-y-z)
  \nonumber\\
  &-x\ln x-(x+z)\ln(x+z)  -(1-x-z)\ln(1-x-z) 
  \nonumber\\
  &-(1-x-y-z)\ln(1-x-y-z).
\end{align}
Nous avons alors montré
\begin{lemma}\label{uniquemaximum}
  La fonction $g$ est concave dans tout le simplexe et son seul point critique est $x_0 = \frac{C\alpha}{1 + C\alpha}$, $y_0= \frac{1}{1 + C}$, $z=0$ où $g=0$.
\end{lemma} 
Puis le simplexe est divisé en deux régions. La première correspond à $\delta x=\vert x-x_0 \vert\ll\alpha $, $\delta y=\vert
  y-y_0\vert\ll\cO(1)$, $z\ll\alpha$. Dans la deuxième, le complémentaire de la première dans le simplexe, le propagateur retrouve une borne similaire au cas $\Phi^{4}$. Dans la première région, nous utilisons l'approximation hessienne et montrons
  \begin{equation}\label{auxbound3} 
    \Gamma^{\alpha}\les  K \frac{\exp \{- \frac{c}{1+\alpha k} p^2  
      - \frac{ c \alpha}{1+ k} (h - \frac{k}{1+C})^2  ) \}}{1 + \sqrt{\alpha k}}.
  \end{equation}
 En dehors, la concavité de $g$ nous permet de borner $g(x,y,z)$ par son approximation linéaire et de montrer
 \begin{equation}
  \Gamma^{\alpha}\les  K e^{- c \alpha k - c  p }.\label{eq:auxbound4}
 \end{equation}
 Il reste enfin à évaluer l'effet des commutateurs. Le commutateur $[\slashed x , \Gamma]$ contient des termes du type
\begin{align}
&\sqrt{m+1}\Gamma_{m+1,n;k,l}-\sqrt{l}\Gamma_{m,n;k,l-1}\nonumber\\
=&\lbt\sqrt{m+1}-   \sqrt{l}\rbt\Gamma_{m,n;k,l-1} 
+ \sqrt{m+1}\lbt\Gamma_{m+1,n;k,l}-\Gamma_{m,n;k,l-1}\rbt. \label{eq:typnum}
\end{align}
Le premier terme est le plus facile à borner. Il est non nul si $p=l-m-1 \ges 1$. Dans ce cas,
\begin{equation}
  \sqrt{l} -\sqrt{m+1}\les  \frac{2p}{1 + \sqrt{l}}.
\end{equation}
En utilisant, le lemme suivant
\begin{lemma}\label{lemma:dl}
Soit $(m,l,h)\in\N^{3}$ avec $p=l-m-1 \ges 1$. Nous avons :
\begin{itemize}
\item dans la région critique
\begin{equation}\label{inboundd}
      \Gamma_{m,m+h;k,m+p}\les K\alpha\,\Gamma_{m,m+h;m+p-1+h,m+p-1},
    \end{equation}
  \item en dehors de la région critique
    \begin{equation}\label{outboundd}
      \Gamma_{m,m+h;k,m+p}\les K\alpha \sqrt{kl}\,\Gamma_{m,m+h;m+p-1+h,m+p-1},
    \end{equation}
  \end{itemize}
\end{lemma}
nous avons :
\begin{itemize}
\item dans la région critique, $l =k-h \simeq \frac{C}{1+C} k$ et $\frac{2p}{1 + \sqrt{l}} \les\cO(1) \frac{p}{1 + \sqrt k}  $. Un facteur $\alpha$ supplémenaire vient de (\ref{inboundd}) si bien que nous avons une borne en $\cO(1) \frac{\alpha  p}{1 + \sqrt k}$. En utilisant une fraction de la décroissance $ e^{-c \alpha p^2/k}$ dans \eqref{auxbound3}, $\cO(1) \frac{\alpha  p}{1 + \sqrt k}\les\sqrt{\alpha}$.

\item En dehors de la région critique, nous avons un facteur $\alpha \sqrt{kl}\frac{2p}{1 + \sqrt{l}}$ qui vient de (\ref{outboundd}). Nous pouvons utiliser $p e^{-cp} \les  e^{-c'p}$ et $\sqrt{\alpha k} e^{-c\alpha k} \les  e^{-c'\alpha k}$ ce qui donne le facteur $\sqrt\alpha$ attendu pour passer\footnote{La région $\alpha k\les 1$ ne nécssite pas une analyse aussi fine et ne sera pas détaillée ici.} de \eqref{eq:auxbound4} à \eqref{mainbound1}.
\end{itemize}
Nous nous intéressons maintenant aux termes impliquant des différences de $\Gamma$. Nous utilisons les identités,
\begin{align}
  {\cA}(m,l,h,u)&=\frac{\sqrt{ml}}{u}{\cal A}(m-1,l-1,h+1,u-1),\text{ for
    $u\ges 1$},\label{eq:A-1}\\
  {\cA}(m,l,h,u)&=\frac{\sqrt{m(m+h)}}{m-u}{\cal A}(m-1,l,h,u),\label{eq:Am-1}\\
  {\cA}(m,l,h,u)&=\frac{\sqrt{m(m+h)l(l+h)}}{(m-u)(l-u)}
  {\cA}(m-1,l-1,h,u)\label{eq:Aml-1}
\end{align}
avec $h=n-(m+1)$ et $p=l-(m+1)$. Ainsi $\Gamma_{m+1,n;k,l}=\Gamma_{m+1,m+1+h;l+h,l}$ et $\Gamma_{m,n;k,l-1}=\Gamma_{m,m+h+1;l+h,l-1}$. Nous pouvons alors démontrer
\begin{equation}
  \sqrt{m+1}\lbt\Gamma_{m+1,n;k,l}-\Gamma_{m,n;k,l-1}\rbt
        \les K\sqrt{\alpha}\Gamma^{\alpha}_{m-1,m+h;l+h-1,l-2}  
\end{equation}
qui achève la preuve du théorème \ref{maintheorem}.
\end{proof}
\begin{rem}
  Nous pouvons répéter l'analyse ci-dessus et l'appliquer au propagateur de la théorie $\Phi^{4}$. Nous obtenons alors
  \begin{equation}\label{eq:boundphi4}  
    G^i_{m,n;k,l}\les K M^{-2i}\min\lbt 1,(\alpha k)^{p}\rbt e^{-c(M^{-2i}k+p)}
  \end{equation}
qui permet de retrouver les bornes \eqref{th1} à \eqref{thsummax}.
\end{rem}

\section{Propagateurs et renormalisabilité}
\label{sec:prop-et-renorm}

Dans cette section, nous revenons sur le comportement du propagateur d'une théorie matricielle nécessaire à l'obtention d'un comptage de puissance renormalisable. Nous donnons aussi notre avis sur l'étude des théories de champs non commutatives dans la base matricielle en comparaison de l'espace $x$ (ou $p$).\\

Dans le cadre de l'analyse multi-échelles et avec les conventions de la section précédente, les définitions des exposants $\delta_{0}$ \eqref{eq:delta0} et $\delta_{1}$ \eqref{eq:delta1} deviennent
\begin{align}
  \max_{k,p,h}G^{i}_{k-h-p,k-p;k,k-h}\les& KM^{-i\delta_{0}},\label{eq:delta0multiscale}\\
  \max_{k}\sum_{p}\max_{h}G^{i}_{k-h-p,k-p;k,k-h}\les& KM^{-i\delta_{1}}.\label{eq:delta1multiscale}
\end{align}
Nous souhaitons étudier le rôle de ces exposants sur la renormalisabilité d'une théorie. Nous supposons donc que nous avons affaire à un modèle matriciel dynamique avec pour contraintes la conservation du moment angulaire \eqref{eq:conservationindices} et les comportements d'échelle \eqref{eq:delta0multiscale} et \eqref{eq:delta1multiscale} du propagateur. Dans la section \ref{sec:analyse-multi-echell-matrix}, nous avons exposé une méthode permettant de retrouver relativement simplement le comptage de puissance de la théorie $\Phi^{4}$. Pour cela, nous avons utilisé trois bornes supplémentaires sur le propagateur.

Pour une ligne de boucle du graphe dual portant un indice de référence, nous avions
\begin{align}
  \sum_{p,k}\max_{h}G^{i}_{k-h-p,k-p;k,k-h}\les&KM^{(D-\delta_{1})i}\label{eq:boundLoop}
\end{align}
où $D$ est la dimension de l'espace. Pour le modèle $\Phi^4_{4}$, $D=4$. Pour toute ligne d'arbre portant un indice de référence, nous avons utilisé
\begin{align}
  \sum_{k}\max_{p,h}G^{i}_{k-h-p,k-p;k,k-h}\les&KM^{(D-\delta_{0})i}.\label{eq:boundtree1}
\end{align}
Enfin, pour les lignes d'arbre portant deux indices de référence, nous avons eu besoin de
\begin{align}
  \sum_{k,h}\max_{p}G^{i}_{k-h-p,k-p;k,k-h}\les&KM^{(2D-\delta_{0})i}.\label{eq:boundtree2}
\end{align}
Considérons une théorie matricielle d'interaction $\Tr\phi^{4}$ et dont le propagateur obéit à la conservation \eqref{eq:conservationindices} et aux bornes \eqref{eq:delta0multiscale} à \eqref{eq:boundtree2} (c'est le cas du modèle \eqref{eq:SPhi4matrix}). En répétant l'analyse de la section \ref{sec:analyse-multi-echell-matrix}, nous démontrons le comptage de puissance suivant
\begin{align}
  A_{G^{i}}\les&K^{V'}M^{-i(\delta_{1}L'+\delta_{0}(V'-1)-D(V'-B))}.\label{eq:powcountgen1}
\end{align}
La borne \eqref{eq:powcountgen1} est donnée dans le cas d'un graphe $G$ dont toutes les lignes sont d'échelle $i$ (comptage monotranche). Ce résultat est également valable pour toute composante connexe mais nous avons souhaité alléger les notations. En utilisant
\begin{itemize}
\item $L'=I-V'+1$ (pour un graphe connexe),
\item $V'=F=2-2g-V+I$ (caractéristique d'Euler),
\item $4V=2I+N$ (théorie $\Phi^{4}$),
\end{itemize}
la borne \eqref{eq:powcountgen1} devient
\begin{align}
    A_{G^{i}}\les&K^{V}M^{-i\omega},\label{eq:powcountgen2}\\
    \omega=&(\delta_{0}+\delta_{1}-D)V+\frac{D-\delta_{0}}{2}N-(D-\delta_{0}+\delta_{1})+2g(D-\delta_{0}+\delta_{1})+D(B-1).\notag
\end{align}
Nous constatons alors qu'une condition nécessaire à la renormalisabilité d'un tel modèle de matrices est
\begin{align}
  \delta_{0}+\delta_{1}\ges D.\label{eq:condnecessRenorm}
\end{align}
Dans \cite{GrWu03-1}, Grosse et Wulkenhaar avaient également remarqué que les indices $\delta_{0}$ et $\delta_{1}$ doivent être suffisament grands par rapport à la dimension de l'espace. Si $\delta_{0}+\delta_{1}>D$, la théorie est super-renormalisable. En cas d'égalité, elle est juste renormalisable. Dans \cite{GrWu04-3}, le comportement du propagateur matriciel de la théorie $\Phi^{4}_{4}$ à $\Omega=0$ a été estimé numériquement. Il a été trouvé $\delta_{0}=1$ et $\delta_{1}=0$. Nul doute qu'une étude fine, similaire à celle effectuée sur le propagateur du modèle de Gross-Neveu (voir section \ref{sec:boite-outils}), donnerait le même résultat. Rappelons que la théorie $\Phi^{4}$, en l'absence de vulcanisation, souffre du mélange UV/IR qui la rend non renormalisable. Dans la base matricielle, la solution à ce phénomène semble claire. Il faut trouver un propagateur tel que $\delta_{0}+\delta_{1}=D$. Toute la difficulté réside alors dans le choix du propagateur.\\

Considérons le propagateur \eqref{eq:propaFermiomatrix} du modèle de Gross-Neveu \ncf{}. Nous avons vu en section \ref{sec:bornes} qu'il existe deux régions de l'espace des indices du propagateur où celui-ci a des comportements très différents. Dans l'une d'elles, le propagateur se comporte comme celui de la théorie $\Phi^{4}$ et conduit donc au même comptage de puissance renormalisable. Dans la région critique, le propagateur est différent. Nous avons 
\begin{align}
  G^{i}\les&K\frac{M^{-i}}{1+\sqrt{ kM^{-2i}}}\,e^{- \frac{c p ^2  }{1+ kM^{-2i}}
    -\frac{ c M^{-2i}}{1+k} (h - \frac{k}{1+C})^2}.
\end{align}
La borne précédente, obtenue par une méthode du type point col, est très précise dans le sens où elle reproduit fidèlement le comportement du propagateur (nous pouvons également montrer une borne inférieure du même type). Ce comportement obéit aux équations \eqref{eq:delta0multiscale} et \eqref{eq:delta1multiscale} avec $\delta_{0}=\delta_{1}=1$. Ainsi le modèle est régulier au sens de la définition \ref{defn:modele-regulier}. Mais nous ne pouvons pas en conclure que la théorie a un comptage de puissance renormalisable car le propagateur ne reproduit pas la borne \eqref{eq:boundtree2}. Cette borne est utile pour les lignes d'arbre du graphe dual qui portent deux indices de référence. Nous constatons donc que le propagateur de Gross-Neveu ne permet pas de sommer deux indices de référence avec un seul propagateur. Ceci conduit également à du mélange UV/IR dans le sens suivant.

\begin{figure}[!htbp]
  \begin{center}
    \subfloat[À l'échelle $i$]{{\label{fig:sunseti}}\includegraphics[scale=.7]{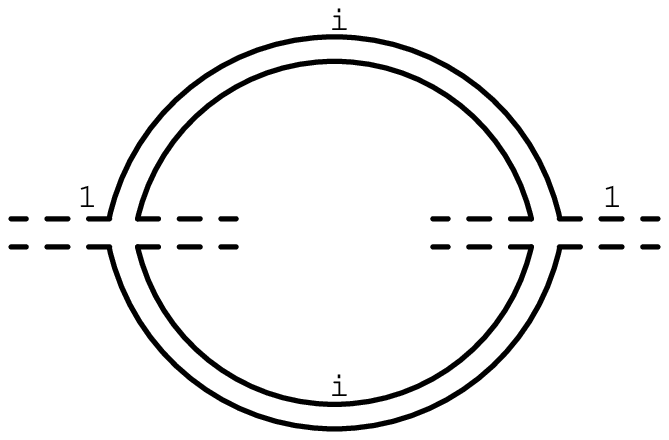}}\qquad
    \subfloat[À l'échelle $j$]{{\label{fig:sunsetj}}\includegraphics[scale=.7]{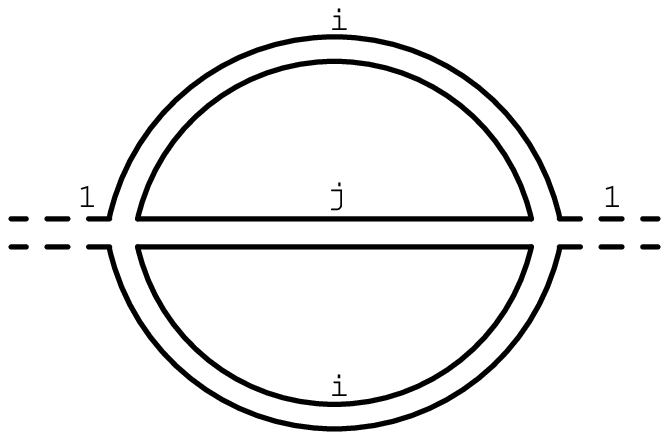}}
  \end{center}
  \caption{Coucher de soleil}
  \label{figsunset}
\end{figure}

Considérons le graphe de la figure \ref{fig:sunsetj} où les deux lignes externes portent un indice $i\gg 1$ et la ligne interne un indice $j<i$. Le propagateur \eqref{eq:propaFermiomatrix} a $\delta_{0}=\delta_{1}$ ce qui signifie que le modèle correspondant est quasi-local (si on fixe les indices d'un côté du propagateur, nous pouvons sommer sur les indices situés à l'autre extrémité sans perdre de bon facteur de comptage de puissance). Ainsi il ne reste plus qu'à sommer un indice par face interne. 

Sur le graphe de la figure \ref{fig:sunseti}, si les deux lignes qui se trouvent à l'intérieur sont de vraies lignes externes, le graphe possède deux faces brisées et nous n'avons aucun indice à sommer. Ainsi en utilisant deux fois la borne \eqref{eq:delta0multiscale}, nous obtenons $A_{G}\les M^{-2i}$. La somme sur $i$ est convergente et nous retrouvons le même comportement que la théorie $\Phi^{4}$ \ie{} les graphes avec plusieurs faces brisées ($B\ges 2$) sont convergents. Cependant si les deux lignes se trouvant à l'intérieur appartiennent à une ligne d'échelle $j<i$ (voir figure \ref{fig:sunsetj}), le résultat est différent. En effet, à l'échelle $i$, nous retrouvons le graphe de la figure \ref{fig:sunseti}. Si nous voulons maintenir le résultat précédent ($M^{-2i}$), il faut pouvoir sommer les deux indices des faces internes de la figure \ref{fig:sunsetj} avec le propagateur d'échelle $j$. Or ce n'est pas possible puisque celui-ci ne permet justement de n'en sommer qu'un seul. Ainsi une des deux faces doit être sommée avec un propagateur d'échelle $i$ :
\begin{align}
  \sum_{k,h}M^{-2i-j}\,e^{-M^{-2i}k}\frac{e^{-\frac{ c M^{-2j}}{1+k} (h - \frac{k}{1+C})^2}}{1+\sqrt{ kM^{-2j}}}\les KM^{j}.
\end{align}
La somme sur $i$ est ici logarithmiquement divergente. Le graphe de la figure \ref{fig:sunseti} est convergent si relié à des vraies pattes externes et divergent s'il est un sous-graphe d'un graphe d'échelle plus basse. Le comptage de puissance d'un graphe dépend donc des échelles inférieures à la plus basse des échelles du graphe. Nous étudierons ce phénomène en grand détail dans la section \ref{sec:multiscaleGN}. Nous l'avons aussi appelé mélange UV/IR car le comptage de puissance d'un graphe ne se factorise plus dans les composantes connexes individuelles. Il serait intéressant d'étudier cette autre forme de mélange dans la base matricielle. Voici en tous cas ce que nous pouvons déjà en dire.\\

La nouveauté vient du fait que le propagateur \eqref{eq:propaFermiomatrix} ne permet pas de sommer deux indices de références. Le problème se pose donc uniquement pour les propagateurs (duaux) reliant deux faces internes $a$ et $b$, et dont l'indice d'échelle est le plus bas de tous les propagateurs accrochés aux faces $a$ et $b$. Ainsi seules les composantes connexes avec plusieurs faces brisées sont concernées. Notons également que le problème ne se pose pas dans une composante monotranche.

Nous avons par ailleurs remarqué que la meilleure façon d'optimiser les sommes dans la base matricielle est de choisir un arbre dual \emph{minimisé} (voir section \ref{sec:analyse-multi-echell-matrix}) \ie{} avec les lignes les plus basses possible. Au contraire, en espace $x$ (ou $p$), l'arbre est \emph{maximisé} : ses lignes sont les plus hautes possible ou autrement dit il est sous arbre dans chaque composante connexe. Nous pouvons le comprendre encore autrement : quel que soit le vertex considéré, la plus haute ligne qui lui est accrochée est une ligne d'arbre
\begin{align}
  \forall v\in G,\,\exists l\in\cT\tq i_{l}=e_{v}\defi\max_{l'\in v}i_{l'}.\label{eq:Tmax}
\end{align}
Au contraire, un arbre minimisé signifie
\begin{align}
  \forall v\in G,\,\exists l\in\cT\tq i_{l}=i_{v}\defi\min_{l'\in v}i_{l'}.\label{eq:Tmin}
\end{align}
Ainsi, dans un graphe dual de la base matricielle, si l'arbre (minimisé) est sous-arbre dans une composante, celle-ci est monotranche. Le nouveau mélange UV/IR vient donc uniquement des composantes connexes dans lesquelles l'arbre n'est pas sous-arbre. J'avoue ne pas savoir encore exactement si c'est important et quel rôle peut jouer cette remarque dans la compréhension des modèles \ncf{}s dans la base matricielle.

De plus, remarquons que le graphe de la figure \ref{fig:sunseti} n'est pas renormalisable par un contreterme du lagrangien (voir chapitre \ref{cha:GN}). Sa divergence (logarithmique) ne peut donc pas être absorbée dans une redéfinition de la constante de couplage. Heureusement, il se trouve que la renormalisation du graphe à deux points de la figure \ref{fig:sunsetj} régularise non seulement la divergence quadratique de la fonction à deux points mais aussi la sous-divergence logarithmique de la fonction à quatre points. Dans la base matricielle, ceci est possible grâce au fait que la soustraction de Taylor permettant d'identifier la partie divergente du graphe \ref{fig:sunsetj} ne fait intervenir que les propagateurs de la face externe (voir \cite{GrWu04-3} pour des exemples).\\

Nous allons finir par quelques brèves remarques concernant la base matricielle. Commençons par deux inconvénients de la base matricielle par rapport à l'espace direct. Nous verrons aux chapitres \ref{cha:le-modele-phi4_4} et \ref{cha:GN} que la notion d'\emph{orientabilité} d'un graphe est très importante (voir section \ref{sec:graph-orientNCphi4}). Seuls les graphes non orientables souffrent de mélange UV/IR. Ainsi les théories du type $\bar{\phi}\star\phi\star\bar{\phi}\star\phi$ qui ne contiennent que des graphes orientables sont renormalisables sans vulcanisation (nous reviendrons sur ce point en section \ref{sec:renorm-et-vulc}). Nous n'avons pas encore identifié comment se traduit l'orientabilité d'un graphe dans la base matricielle.

Le second inconvénient est essentiellement technique. Nous verrons au chapitre \ref{cha:GN} que la renormalisation du modèle de Gross-Neveu \ncf{} nécessite d'utiliser la parité de certaines intégrales. De manière générale, il me semble moins simple de travailler avec des sommes discrètes qu'avec des intégrales. De plus, l'inversion de la forme quadratique \eqref{eq:formequadMatrixPhi4} \ie{} le calcul du propagateur \eqref{eq:propaPhimatrix} est très compliquée. Saura-t-on refaire cette inversion dans d'autres cas très différents ? Notons au passage que le calcul des propagateurs \eqref{eq:propaxboson} et \eqref{xfullprop} en espace $x$ n'est pas complètement trivial non plus mais néanmoins plus simple que dans la base matricielle. Le problème vient essentiellement du fait que nous sommes habitués à l'espace direct.\\

Outre ces inconvénients (mineurs), la base matricielle présente un certain nombre d'avantages \emph{à long terme}. Ici \emph{à long terme} signifie qu'il est, pour l'instant, plus simple de calculer en espace $x$ mais que la base matricielle pourrait être utile dans le but de mieux comprendre les théories de champs \ncv{}s (et commutatives). En effet, je pense que le principal atout de cette base est qu'elle ne fait pas intervenir l'espace sous-jacent de manière explicite. Autrement dit, nous pourrions prendre l'action \eqref{eq:SPhi4matrix} comme point
de départ sans savoir qu'elle correspond à l'action \eqref{actionphi4-chapmatrix}. Jusqu'à présent et dans la limite de mes connaissances, il me semble que seules les théories de champs sur des déformations ont été étudiées. Par déformation, j'entends que l'algèbre (\ncv{}) de fonctions ou de distributions est un espace vectoriel de fonctions usuelles muni d'un produit \ncf{}. C'est le cas de toutes les déformations isospectrales (voir \cite{Gayral2004cs,gayral05}). Ces déformations sont pratiques pour transposer ce que nous savons faire sur espace commutatif. Mais si nous voulons un jour unifier la mécanique quantique et la relativité générale, nous devons être capables d'écrire une théorie qui ne s'appuie pas sur un espace prédéfini. La base matricielle pourrait nous habituer à travailler sans espace et à comprendre comment se traduisent, dans ce cadre, les notions notamment nécessaires à la renormalisation telles que la localité. De plus, il est possible que l'existence d'une base matricielle ne soit pas restreinte au plan de Moyal. Il me semble qu'il suffit de pouvoir définir des opérateurs de création et d'annihilation et on peut ensuite construire une base matricielle à partir de l'idempotent $\exp -\bar{a}a$. Néanmoins je ne crois pas à l'existence d'une telle base pour des espaces plus généraux dans la mesure où sa construction fait intervenir explicitement le produit point à point et donc se sert du caractère déformé de l'espace.

Les théories de champs sur plan de Moyal souffrent de mélange UV/IR. À partir de la \og{}définition\fg{} même de l'algèbre de Moyal, $\lsb x^{\mu},x^{\nu}\rsb=\imath\Theta^{\mu\nu}$, il est clair qu'il est impossible de se restreindre à une zone de petites distances. Sur un espace commutatif, la région ultraviolette est clairemement identifée. Par exemple, si $\alpha$ est le paramètre de Schwinger, $\alpha$ proche de zéro correspond à cette région. La région infrarouge l'est également ($\alpha\to\infty$). Ces deux régions sont séparées et une masse non nulle arrête le flot dans l'infrarouge. Sur espace \ncf{}, nous avons vu que, même en présence du terme additionnel de vulcanisation, certains modèles (Gross-Neveu) présentent encore du mélange UV/IR. Celui-ci couple les différentes échelles du problème mais n'empêche pas la renormalisabilité de la théorie. De plus, le flot du groupe de renormalisation est arrêté dans la zone $\alpha\to\infty$ même à masse nulle. Ainsi les régions ultraviolette et infrarouge ne sont pas aussi clairement identifiables que sur espace commutatif. Dans la base matricielle, les indices des propagateurs sont dans $\N$. Il n'y a qu'une seule région à l'infini qui pourrait être à la fois l'ultraviolet et l'infrarouge.

Bien que par certains côtés, les calculs dans la base matricielle soient plus compliqués, elle permet de simplifier l'interaction. Nous verrons dans les deux prochains chapitres que les oscillations présentes dans l'interaction (équation \eqref{eq:int-Moyal-even}) contiennent beaucoup d'information concernant le comptage de puissance et les contretermes de la théorie. Au moins pour le modèle $\Phi^{4}$, l'interaction dans la base matricielle est très simple ($\Tr\phi^{4}$) et ne contient pratiquement plus aucune information. L'essentiel provient alors du propagateur. Dans ce formalisme, nous pouvons calculer le comptage de puissance facilement, notamment toute la dépendance topologique. Pour l'instant, dans l'espace $x$, seul un calcul exact (voir \cite{gurauhypersyman}) le permet. Notons aussi que la base matricielle a permis des calculs non perturbatifs (développement des fonctions de corrélations en puissance du genre du graphe) mais néanmoins restreints à des modèles possédant une certaine structure soluble (voir \cite{Grosse2006tc,Grosse2006qv,Grosse2005ig}).\\

Enfin, je pense qu'il faudrait mieux caractériser le mélange UV/IR dangereux dans la base matricielle. Nous avons vu qu'une condition nécessaire à son apparition est $\delta_{0}+\delta_{1}<D$. Cette condition n'est pas suffisante car une théorie dont la contrepartie commutative est non renormalisable, comme $\phi^{4}_{6}$ sur plan de Moyal, la remplirait également. Nous pourrions étudier la théorie \eqref{actionphi4-chapmatrix} à $\Omega=0$ mais avec l'interaction $\bar{\phi}\star\phi\star\bar{\phi}\star\phi$. Cette théorie est renormalisable (voir la section \ref{sec:renorm-et-vulc}) mais son propagateur est tel que $\delta_{0}+\delta_{1}<D$. Cette étude nous permettrait également de clarifier la notion d'orientabilité dans la base matricielle.


%% file: phi4-arXiv.tex
\chapter{Le modèle $\Phi^{4}_{4}$}
\label{cha:le-modele-phi4_4}
\epigraph{Ces idées qui survolent l'espace et qui tout à coup, se heurtent aux parois du crâne.}{Émile-Michel Cioran}

La théorie $\Phi^{4}_{4}$ (voir l'équation \eqref{action}) constitue un premier modèle simple à étudier en espace $x$. C'est le modèle dont Grosse et Wulkenhaar ont montré la renormalisabilité. C'est donc un bon moyen de développer des outils en espace $x$. Nous avons vu au chapitre précédent comment analyser cette théorie dans la base matricielle. Celle-ci a de nombreux avantages sur l'espace direct que j'ai résumés en section \ref{sec:prop-et-renorm}. Néanmoins je pense que l'espace $x$ est un intermédiaire de qualité. Mon opinion est que, tôt ou tard, il faudra plus ou moins s'abstraire de l'espace. Pour effectuer cette transition, l'étude des théories de champs \ncv{}s en espace $x$ peut s'avérer utile. En effet, l'espace direct nous permet de comparer facilement le comportement (entre autres du point de vue du groupe de renormalisation) d'une théorie \ncv{} avec son homologue commutatif. Puis nous pourrions traduire cette expérience dans la base matricielle ou dans un langage ne faisant pas intervenir explicitement l'espace sous-jacent.

Au-delà de la perturbation, seules les techniques constructives \cite{Riv1} permettent de définir une théorie des champs. La théorie constructive s'appuie sur l'espace $x$. Pour définir une théorie des champs non perturbativement et sans utiliser explicitement l'espace $x$, il faudrait commencer par développer des techniques constructives dans la base matricielle. En attendant, il y a un bon espoir de pouvoir \emph{construire} la théorie $\Phi^{4}_{4}$ \ncv{} au moins en espace $x$. En effet, bien que les théories de champs sur espace \ncf{} souffrent de divergences (ultraviolettes), il semble que les flots soient régularisés. C'est, en tous cas, ce que l'on constate pour $\Phi^{4}$. La fonction $\beta_{\lambda}$ a été calculée dans \cite{GrWu04-2} à l'ordre d'une boucle. Au contraire du modèle commutatif, asymptotiquement libre dans l'infrarouge, la théorie \ncv{} a un flot borné. Ceci devrait permettre de définire non perturbativement le modèle $\Phi^{4}$ \ncf{}.
\newpage
\section{La théorie $\Phi^{4}$}

\subsection{Le lagrangien}
\label{sec:lagrangienphi4}

Dans ce chapitre, nous étudions une théorie $\Phi^{4}$. Il s'agit d'une
théorie scalaire réelle avec interaction quartique. Elle est écrite sur
l'espace de Moyal quadri-dimensionnel $\R^{4}_{\Theta}$.
Sa fonctionnelle action, introduite dans \cite{GrWu04-3} est
\begin{equation}\label{action}
S[\phi] = \int d^4x \Big( -\frac{1}{2} \partial_\mu \phi
\star \partial^\mu \phi + \frac{\Omega^2}{2} (\tilde{x}_\mu \phi )
\star (\tilde{x}^\mu \phi ) + \frac{1}{2} m^2
\,\phi \star \phi
+ \frac{\lambda}{4} \phi \star \phi \star \phi \star
\phi\Big)(x)
\end{equation}
avec $\xt_\mu=2(\Theta^{-1}x)_{\mu}$. Nous nous placerons toujours dans le cas
euclidien. La métrique employée est donc
$g_{\mu\nu}=\delta_{\mu\nu}$.\\

Le propagateur $C$ de la théorie $\Phi^{4}$ non commutative est le noyau de
l'inverse de l'opérateur $-\Delta+\Omega^{2}\xt^{2}+m^{2}$. 
Dans notre cas, ce noyau est connu comme sous le nom de noyau de Mehler \cite{simon79funct,toolbox05}
\begin{equation}
  \label{eq:Mehler}
  C(x,y)=\frac{\Omega^{2}}{\theta^{2}\pi^{2}}\int_{0}^{\infty}\frac{dt}{\sinh^{2}(2\Ot
  t)}\,e^{-\frac{\Ot}{2}\coth( 2\Ot t)(x-y)^{2}-\frac{\Ot}{2}\tanh(2\Ot t)(x+y)^{2}-m^{2}t}.
\end{equation}
Le vertex de la théorie $\Phi^{4}$ non commutative est composé d'une fonction
delta et d'une oscillation\footnote{La différence de signe dans l'oscillation entre les équations \eqref{eq:interaction-phi4} et \eqref{eq:int-Moyal-even} est non pertinente comme le montre \eqref{eq:int-positive}.} (voir corollaire \ref{cor:int-Moyal}):
\begin{align}
  \int dx\, \phi^{\star
    4}(x)=&\int\prod_{i=1}^{4}dx_{i}\,\phi(x_{i})\,\delta(x_{1}-x_{2}+x_{3}-x_{4})e^{\imath\varphi},
  \label{eq:interaction-phi4}\\
  \varphi=&\sum_{i<j=1}^{4}(-1)^{i+j+1}x_{i}\wed x_{j}.\nonumber
\end{align}
Grâce à la fonction delta, l'oscillation peut être écrite de plusieurs
façons.
\begin{subequations}
  \begin{align}
    \delta(x_{1}-x_{2}+x_{3}-x_{4})e^{\imath\varphi}=&\delta(x_{1}-x_{2}+x_{3}-x_{4})e^{\imath
      x_{1}\wed x_{2}+\imath x_{3}\wed x_{4}}\\
    =&\delta(x_{1}-x_{2}+x_{3}-x_{4})e^{\imath
      x_{4}\wed x_{1}+\imath x_{2}\wed x_{3}}\\
    =&\delta(x_{1}-x_{2}+x_{3}-x_{4})\exp\imath(x_{1}-x_{2})\wed(x_{2}-x_{3}).\label{eq:oscill-trans}
  \end{align} 
\end{subequations}
L'interaction est réelle et positive\footnote{Une autre façon de le montrer est, à partir de \eqref{eq:Moyal-involution}, $\overline{\phi^{\star 4}}=\phi^{\star 4}$.} :
\begin{align}
  &\int\prod_{i=1}^{4}dx_{i}\phi(x_{i})\,\delta(x_{1}-x_{2}+x_{3}-x_{4})e^{\imath\varphi}\label{eq:int-positive}\\
  =&\int dk\lbt\int dxdy\,\phi(x)\phi(y)e^{\imath k(x-y)+\imath x\wed y}\rbt^{\!\!2}\in\R_{+}.\notag
\end{align}
Elle est également invariante par translation comme l'indique l'équation \eqref{eq:oscill-trans}.
Dans la suite de ce chapitre nous démontrons, en espace $x$, le théorème suivant
\begin{thm}[BPHZ]\label{thm:BPHZPhi4}
  La théorie quantique des champs définie par l'action \eqref{action} est
  renormalisable à tous les ordres de perturbation.
\end{thm}
Dans toute la suite de ce chapitre, nous utiliserons l'analyse multi-échelles (voir la section \ref{sec:multiscalephi4}).

\subsection{Orientation et variables d'un graphe}
\label{sec:graph-orientNCphi4}

La fonction delta \eqref{eq:interaction-phi4} de l'interaction nous
indique que le vertex est un parallèlogramme. Pour simplifier, nous le
représenterons soit sous forme d'un losange (Fig. \ref{fig:vertex}) soit comme
un carré.\\

\begin{floatingfigure}[p]{2cm}
  \centering
  \includegraphics[scale=1]{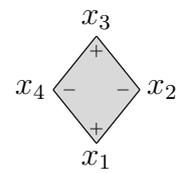}
  \caption[Le vertex de la théorie]{Un vertex}
  \label{fig:vertex}
\end{floatingfigure}
\noindent
Nous associons un signe, $+$ ou $-$, à chacune des quatre positions du
vertex. Ce signe alterne d'une position à l'autre et reflète les signes
intervenant dans l'argument de la fonction delta. Par exemple, la fonction
delta associée au vertex de la figure \ref{fig:vertex} doit être pensée comme
$\delta(x_{1}-x_{2}+x_{3}-x_{4})$ et non $\delta(-x_{1}+x_{2}-x_{3}+x_{4})$.
Le vertex étant invariant par permutation cyclique, nous pouvons choisir le
signe de l'une des quatre positions. Les signes des trois autres sont alors fixés. Nous dirons qu'une
ligne est \textbf{orientable} si elle joint un point $+$ à un point $-$. Dans
le cas contraire nous la qualifierons de \textbf{non orientable}. Par définition, un graphe est
orientable si toutes ses lignes le sont. Nous particulariserons les lignes
orientables en leur joignant une flèche allant du $-$ vers le $+$. Les
positions $-$ seront donc définies sortantes et les $+$ entrantes.\\

\begin{figure}[!htbp]
  \centering 
  \subfloat[Orientation d'un arbre]{{\label{fig:orientedtree}}\includegraphics[scale=1]{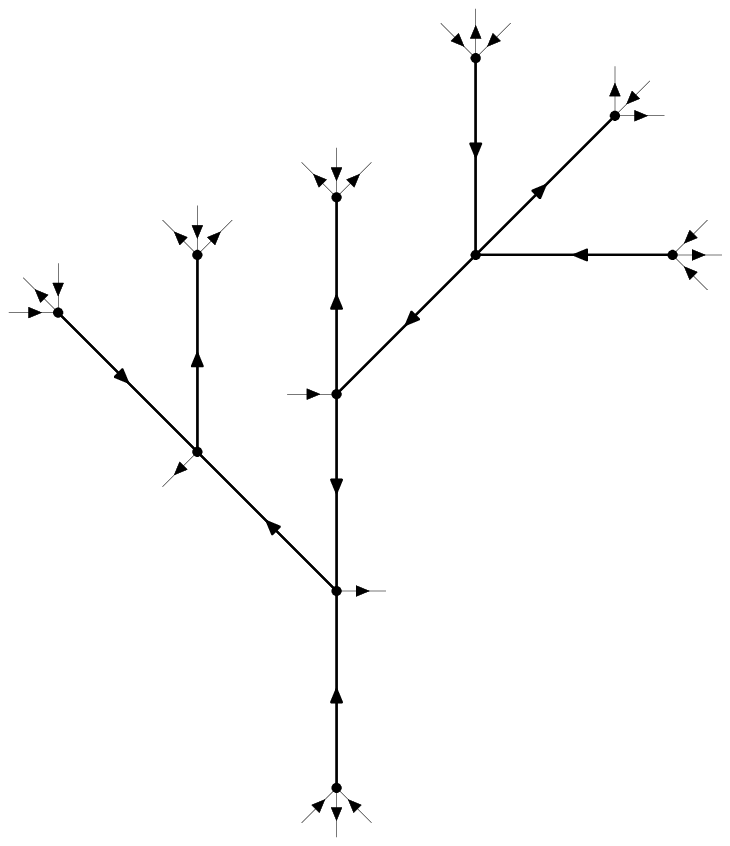}}\qquad
  \subfloat[Ordre total]{\label{fig:totalorder}\includegraphics[scale=.8]{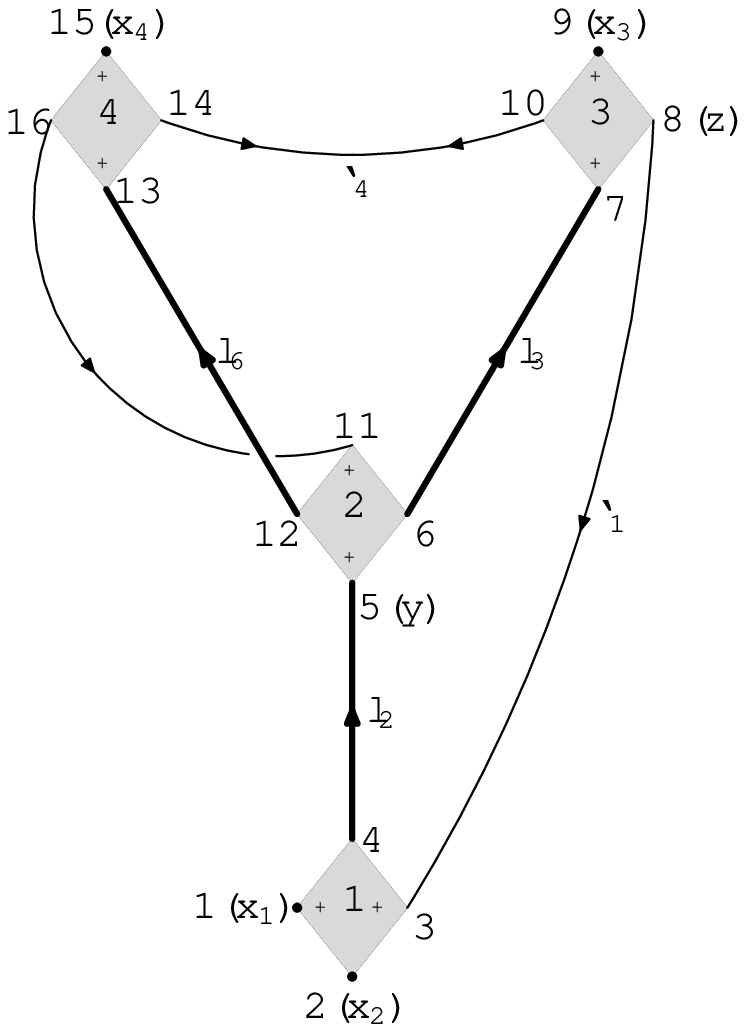}}
  \caption{Orientabilité et ordre}
  \label{fig:trees}
\end{figure}

Soit un graphe $G$. Nous choisissons un arbre optimal $\cT$ générateur
enraciné. L'orientation du graphe \ie{} l'attribution des signes à chaque vertex est
déterminée par l'orientation de l'arbre. Au vertex racine, nous choisissons
une position à laquelle nous attribuons un signe $+$, une position
entrante. Quand le graphe n'est
pas un graphe du vide, il est pratique de choisir comme racine un vertex
possédant une ou plusieurs pattes externes. Dans ce cas, nous choisissons une
position externe pour ce signe $+$. Quelle que soit cette position, une fois
son signe fixé, les signes des trois autres sommets de la racine sont
déterminés. Ainsi en imposant l'orientabilité des lignes de
l'arbre\footnote{C'est possible grâce à l'absence de lignes de boucles.}, nous
induisons une attribution des signes ou orientation des vertex et des lignes
de l'arbre. Chaque ligne possède une et une seule flèche. Ces flèches sont
alternativement entrantes et sortantes autour d'un vertex (figure
\ref{fig:orientedtree}). Remarquons qu'avec cette procédure, un arbre est toujours orientable (et orienté). Les lignes de boucles peuvent alors être orientables ou pas.
\begin{defn}[Ensembles de lignes]\label{defn:sets}
  Nous définissons\\

  \ensuremath{\begin{array}{lcl}
      \cT&=&\lb\text{lignes d'arbre}\rb,\\
      \cL&=&\lb\text{lignes de boucles}\rb=\cL_{0}\cup\cL_{+}\cup\cL_{-}
      \text{ avec}\\ 
      \cL_{0}&=&\lb\text{lignes de boucles } (+,-)\text{ ou }(-,+)\rb,\\
      \cL_{+}&=&\lb\text{lignes de boucles }(+,+)\rb,\\
      \cL_{-}&=&\lb\text{lignes de boucles }(-,-)\rb.
    \end{array}}
\end{defn}

Il est pratique de munir l'ensemble des variables de vertex d'un ordre total. Pour cela, nous commençons par la position racine et tournons autour de l'arbre dans le sens trigonométrique. Nous numérotons les positions dans l'ordre dans lequel elles sont rencontrées, voir la figure \ref{fig:totalorder}. Alors nous pouvons ordonner (partiellement) les lignes internes et les positions externes.
\begin{defn}[Relations d'ordre]\label{defn:relations}
  Soient $i<j$ et $p<q$ dans $\N$. Pour toutes lignes
  $l=(i,j),\, l'=(p,q)\in\cT\cup\cL$, pour toute position externe $x_{k}$, nous définissons\\
  
  \ensuremath{\begin{array}{ccccrl}
      l&\prec&l'&&\text{si}&i<j<p<q\\
      l&\prec&k&&&i<j<k\\
      l&\subset&l'&&&p<i<j<q\\
      k&\subset&l&&&i<k<j\text{: ``$l$ contracte au-dessus de $x_{k}$''}\\
      l&\ltimes&l'&&&i<p<j<q.
    \end{array}}\\
  
  \noindent
  Nous étendons ces définitions aux ensembles de lignes définis en \ref{defn:sets}. Par exemple, nous écrirons $\cL_{0}\ltimes\cL_{+}$
  au lieu de $\lb (\ell,\ell')\in\cL_{0}\times\cL_{+},\, \ell\ltimes
  \ell'\rb$. Nous définissons également l'ensemble suivant. Soient $S_{1}$ et
  $S_{2}$ deux ensembles de lignes,
  \begin{align}
    S_{1}\lrtimes S_{2}=&\lb (l,l')\in S_{1}\times S_{2},\, l\ltimes l'\text{
      ou }l \rtimes l'\rb.
  \end{align}
\end{defn}

Par exemple, sur la figure \ref{fig:totalorder}, $\ell_{1}\prec\ell_{4}$,
$l_{2}\subset\ell_{1}$, $l_{3}\succ x_{1}$. Remarquons aussi qu'avec de telles conventions de signes, une ligne orientable joint toujours une position paire ($-$) à une position impaire ($+$). Nous allons maintenant définir de nouvelles variables. Celles-ci seront relatives aux lignes du graphe alors que les variables utilisées jusqu'à maintenant sont des variables de vertex. Toute ligne orientable $l$ joint une position sortante $x_{l-}$ à une position entrante $x_{l+}$. Nous définissons $\mathbf{u_{l}}=x_{l+}-x_{l-}$ comme la différence entre les positions entrante et sortante. Pour les lignes non orientables, $u_{l}$ est aussi la différence entre ses deux extrémités mais le signe est arbitraire et donné dans la définition \ref{defn:longshort}. Les variables $u_{l}$ sont appelées variables \emph{courtes}. Les variables \emph{longues} sont définies comme la somme des deux extrémités des lignes. Nous les désignerons par $\mathbf{v_{l}}=x_{l+}+x_{l-}$ pour les lignes d'arbre et $\mathbf{w_{\ell}}=x_{\ell+}+x_{\ell-}$ pour les boucles.
\begin{defn}[Variables courtes et longues]\label{defn:longshort}
  Soient $i<j$. Pour toute ligne $l=(i,j)\in\cT\cup\cL$,
  \begin{align}
    u_{l}=&
    \begin{cases}
      (-1)^{i+1}s_{i}+(-1)^{j+1}s_{j}&\forall l\in\cT\cup\cL_{0},\\
      s_{i}-s_{j}&\forall l\in\cL_{+},\\
      s_{j}-s_{i}&\forall l\in\cL_{-}.
    \end{cases}\\
    v_{l}=&s_{i}+s_{j}\qquad\forall l\in\cT\\
    w_{l}=&s_{i}+s_{j}\qquad\forall l\in\cL.
  \end{align}
\end{defn}
Avec cette définition, le propagateur correspondant à une ligne $l$ s'écrit :
\begin{equation}
  \label{eq:Mehler2}
  C_{l}(u_{l},v_{l})=\frac{\Omega^{2}}{\theta^{2}\pi^{2}}\int_{0}^{\infty}\frac{dt_{l}}{\sinh^{2}(2\Ot
  t_{l})}\,e^{-\frac{\Ot}{2}\coth( 2\Ot t_{l})u_{l}^{2}-\frac{\Ot}{2}\tanh(2\Ot t_{l})v_{l}^{2}-m^{2}t}.
\end{equation}
Le signe cyclique aux vertex et l'ordre induit sur les positions par la rotation autour de l'arbre nous permet de donner un signe à chaque ligne:
\begin{defn}[Signe d'une ligne]\label{defn:signe} 
  Soient $i<j$. Pour toute ligne $l=(i,j)\in\cT\cup\cL$,\\
  
  \ensuremath{\begin{array}{cclccll}
      \veps(l)&=&+1&&\forall l\in&\cT\cup\cL_{0}&\text{ si $i$ est pair}\\
      &=&+1&&&\cL_{-}\\
      &=&-1&&&\cT\cup\cL_{0}&\text{ si $i$ est impair}\\
      &=&-1&&&\cL_{+}.
    \end{array}
  }
\end{defn}

\subsection{Résolution des fonctions delta}
\label{sec:resol-deltaNCPhi4}
Nous nous donnons ici une règle pour résoudre de façon optimale les fonctions delta de vertex. De plus cela nous permettra de factoriser la fonction delta globale (voir \eqref{eq:interaction-phi4}) pour chaque sous-graphe à quatre points.
Une telle procédure est appelée, en anglais, \og position routing\fg{}. C'est
l'équivalent en espace $x$ du \og momentum routing\fg{}. Il n'existe pas de
manière canonique d'effectuer une telle distribution. Cependant nous pouvons
rejeter l'arbitraire du procédé dans un choix d'arbre. Ce choix est cependant
contraint (mais pas fixé) par l'attribution des indices. Ainsi étant donné un
graphe $G$, nous pouvons choisir un arbre générateur enraciné (rooted spanning
tree). Une fois ce choix fait, il existe une procédure \emph{canonique} de
résolution des fonctions delta. Il est pratique d'introduire un système de
\textbf{branches}. À chaque ligne $l$ de l'arbre nous associons une branche
$\mathbf{b(l)}$ formée des vertex situés \emph{au-dessus} de $l$. Voici comment nous définissons \emph{au-dessus}. À chaque
vertex $\nu$, il existe une unique ligne d'arbre descendant vers la racine.
Notons-la $\mathbf{l_{\nu}}$. \textit{A contrario}, à chaque ligne d'arbre $l$
correspond un unique vertex $\nu$ tel que $l_{\nu}=l$. Nous définissons également
$\mathbf{\cP_{\nu}}$ comme l'unique ensemble de lignes de l'arbre joignant $\nu$ à
la racine. Ainsi la branche $b(l)$ est l'ensemble de vertex défini par
\begin{equation}
  \label{eq:branchset}
  b(l)=\lb\nu\in G\tqs l\in P_{\nu}\rb.
\end{equation}
Sur la figure \ref{fig:totalorder}, la branche $b(l_{2})=\lb 2,3,4\rb$.
Nous pouvons maintenant remplacer l'ensemble des fonctions delta de vertex par
un nouvel ensemble associé aux branches. Si le graphe $G$ a $n$ vertex,
l'arbre est composé de $n-1$ lignes. Il est donc constitué de $n-1$ branches.
À chaque vertex $\nu$, nous remplaçons la fonction
$\delta_{\nu}(\sum_{i=1}^{4}(-1)^{i+1}x_{\nu_{i}})$ par $\delta(\sum_{\nu'\in
  b(l_{\nu})}\sum_{i=1}^{4}(-1)^{i+1}x_{\nu'_{i}})$. \textit{Stricto sensu} il
n'existe pas de branche contenant la racine de l'arbre (ce serait l'arbre tout
entier) si bien qu'il faut rajouter à ces $n-1$ nouvelles fonctions delta, la
fonction de racine définie par $\delta_{G}(\sum_{\nu'\in G}\sum_{i=1}^{4}(-1)^{i+1}x_{\nu'_{i}})$. Nous avons ainsi redéfini $n$ fonctions delta. Ce changement dans la distribution des positions est clairement triangulaire. C'est la
structure en arbre qui l'assure.\\

Précisons maintenant les arguments de ces nouvelles fonctions delta en termes
des variables courtes et longues. Dans ce but, il est commode de définir l'ensemble
$\mathbf{\kb(l)}$ de lignes bouclant à l'intérieur d'une branche $b(l)$ donnée :
\begin{equation}
  \label{eq:branch-lines}
  \kb(l)=\lb l'=(x_{\nu},x_{\nu'})\in G\tqs\nu,\nu'\in b(l)\rb.
\end{equation}
Il existe également des lignes $l=(x_{\nu},x_{\nu'})$ avec $\nu\in b(l)$ et
$\nu'\notin b(l)$. De même $b(l)$ peut contenir des positions externes. Nous
noterons $\mathbf{\cX(l)}$ l'ensemble des positions externes de la branche
$b(l)$ et des extrémités des lignes bouclant à l'extérieur de cette branche.
La définition des variables courtes et longues entraîne donc, pour $\nu$ fixé,
\begin{equation}
  \label{eq:deltabranch}
  \sum_{\nu'\in
    b(l_{\nu})}\sum_{i=1}^{4}(-1)^{i+1}x_{\nu'_{i}}=\sum_{l\in(\cT\cup\cL_{0})\cap\kb(l_{\nu})}u_{l}
  +\sum_{\ell\in\cL_{+}\cap\kb(l_{\nu})}w_{\ell}-\sum_{\ell\in\cL_{-}\cap\kb(l_{\nu})}w_{\ell}+\sum_{e\in\cX(l_{\nu})}\eta(e)x_{e}
\end{equation}
où $\eta(e)=1$ si la position $e$ est entrante et $-1$ sinon. À titre
d'exemple, la fonction delta associée à la branche $b(l_{2})$ de la figure
\ref{fig:totalorder} est
\begin{equation}
  \label{eq:deltabranch2}
  \delta(y-z+x_{3}+x_{4}+u_{l_{3}}+u_{\ell_{5}}+u_{l_{6}}-w_{\ell_{4}}).
\end{equation}
De même la fonction delta de la branche complète est
\begin{equation}
  \label{eq:deltaroot-ex}
  \delta(x_{1}-x_{2}+x_{3}+x_{4}+u_{\ell_{1}}+u_{l_{2}}+u_{l_{3}}+u_{\ell_{5}}+u_{l_{6}}-w_{\ell_{4}}).
\end{equation}
Notons finalement le cas particulier de la fonction delta de racine :
\begin{equation}
  \label{eq:deltaroot}
  \delta_{G}\Big(\sum_{l\in\cT\cup\cL_{0}}u_{l}
  +\sum_{\ell\in\cL_{+}}w_{\ell}-\sum_{\ell\in\cL_{-}}w_{\ell}+\sum_{e\in\cE(G)}\eta(e)x_{e}\Big)
\end{equation}
où $\mathbf{\cE(G)}$ est l'ensemble des points externes de $G$. Remarquons que
si le graphe est \textit{orientable} ($\cL_{+}=\cL_{-}=\emptyset$) alors la fonction
delta de racine \eqref{eq:deltaroot} ne contient que les points externes et la somme de toutes les
variables $u_{l}$ du graphe.\\

Nous allons maintenant utiliser les $n-1$ fonctions delta de branches pour résoudre les longues variables $v_{l},\,l\in\cT$ de l'arbre. C'est le choix optimal. Les intégrations sur les longues variables $v_{l}$ ou $w_{\ell}$ coûtent $M^{2i_{l}}$. De plus, l'arbre étant choisi optimal, les $v_{l}$ sont les variables les plus longues. D'après \eqref{eq:deltabranch}, nous avons
\begin{equation}
  \delta_{b(l)}\Big(\sum_{l'\in(\cT\cup\cL_{0})\cap\kb(l)}u_{l'}
  +\sum_{\ell\in\cL_{+}\cap\kb(l)}w_{\ell}-\sum_{\ell\in\cL_{-}\cap\kb(l)}w_{\ell}+\sum_{e\in\cX(l)}\eta(e)x_{e}\Big).\label{eq:deltabranchfinale}
\end{equation}
Il existe $e_{l}\in\cX(l)$ tel que $x_{e_{l}}=\frac
12(\eta(e_{l})u_{l}+v_{l})$ (voir définition \ref{defn:longshort}). Ce point externe est une extrémité de la ligne $l$. Ainsi $\delta_{b(l)}$ donne
\begin{align}
  v_{l}=&-\eta(e_{l})u_{l}-2\eta(e_{l})\Big(\sum_{l'\in(\cT\cup\cL_{0})\cap\kb(l)}u_{l'}
  +\sum_{\ell\in\cL_{+}\cap\kb(l)}w_{\ell}-\sum_{\ell\in\cL_{-}\cap\kb(l)}w_{\ell}\label{eq:deltavl}\\
  &+\sum_{e\in\cX(l)\setminus\{e_{l}\}}\eta(e)x_{e}\Big).\notag
\end{align}
Nous avons alors utilisé $n-1$ fonctions delta (une par ligne d'arbre). La dernière est conservée (la fonction delta de racine). Elle est l'équivalent de la conservation globale des moments dans les théories des champs habituelles.

\section{Le facteur de rosette}\label{sec:rosfactNCPhi4}

Cette section est une préparation au comptage de puissance. Elle nous sera également utile dans le chapitre \ref{cha:GN} sur le modèle de Gross-Neveu \ncf{}. Dans la section précédente, nous avons constaté que les oscillations, venant de l'interaction, s'expriment en fonction des variables de vertex. Au contraire, les propagateurs utilisent plus naturellement les variables de lignes $u$ et $v\,(w)$. Evidemment ces deux ensembles de variables sont équivalents. Cependant il n'est pas très commode d'utiliser deux jeux de variables. Nous allons donc réexprimer les oscillations en termes des variables courtes et longues.

Strictement parlant, nous n'aurons pas besoin dans ce chapitre de l'expression exacte de l'oscillation totale\footnote{Cependant toute l'information nous sera utile pour le modèle de Gross-Neveu.}. En réalité, il nous faut seulement quelques informations concernant les graphes non planaires ($g\ges 1$) ou avec plusieurs faces brisées ($B\ges 2$). Il se trouve qu'un travail similaire a déjà été effectué par Filk \cite{Filk1996dm}. Dans cet article, Filk travaillait en espace des moments avec le propagateur habituel \ie{} l'inverse du laplacien. Ainsi nous pouvons retrouver ses résultats en mettant toutes les variables $u$ à zéro dans les expresions qui suivront. Ceci correspond à la conservation des moments. Remarquons aussi qu'en espace $p$ la fonction delta de vertex est $\delta(p_{1}+p_{2}+p_{3}+p_{4})$. Or nous avons vu dans la section \ref{sec:graph-orientNCphi4} que c'est l'alternance de signes dans l'argument de la fonction delta de vertex qui est responsable de la notion d'orientation d'un graphe \ncf{}. C'est pourquoi dans l'article de Filk il n'est pas fait mention de lignes orientables ou non orientables. L'espace des moments n'est pas adapté pour faire une telle distinction.\\

Dans la suite nous nommerons {\bfseries facteur de rosette} l'ensemble des
oscillations de vertex ajoutées à la fonction delta de racine. De plus nous
désignerons par $l$ une ligne d'arbre et par $\ell$ une ligne de
boucle\footnote{Si une ligne appartient à un ensemble contenant des lignes
  d'arbre et de boucles, nous la noterons $l$.}. La première étape vers une récriture complète des oscillations de vertex consiste en une \og{}réduction de l'arbre\fg{}. Il s'agit d'exprimer les variables de l'arbre en fonction des $u$ et $v$. Soit un graphe $G$ d'ordre $n$. Il contient $2(n-1)$ positions dans l'arbre. Les $2(n+1)$ positions de boucles et variables externes restantes sont désignées par $s_{j}$. En utilisant l'invariance cyclique des vertex et les fonctions delta, nous obtenons (voir \cite{xphi4-05} pour une preuve):

\begin{lemma}[Réduction de l'arbre]\label{lemma:Filk1}
  Le facteur de rosette après le premier mouvement de Filk est \cite{Filk1996dm,xphi4-05}:
  \begin{align}
    &\delta(s_1-s_2+\dots-s_{2n+2}+\sum_{l\in \cT}u_l)\exp{\imath\varphi}\\
    \nonumber\\
    \text{où }\varphi=&\sum_{i<j=0}^{2n+2}(-1)^{i+j+1}s_i\wed s_j+\frac
    12\sum_{l\in\cT}
    \veps(l)v_l\wed u_l- \sum_{\cT\prec\cT}u_l\wed u_{l'}\nonumber\\
    &+\sum_{\lb l\in\cT,\, i\prec l\rb} u_l\wed (-1)^{i+1} s_i+\sum_{\lb
      l\in\cT,\, i\succ l\rb}(-1)^{i+1}s_i\wed u_l\nonumber
  \end{align}
  et $\veps(l)$ obéit à la définition \ref{defn:signe}.
\end{lemma}
L'étape suivante consiste à exprimer toutes les variables de boucles avec les $u$ et $w$. Dans \cite{xphi4-05}, nous l'avons fait pour les graphes \emph{planaires réguliers} ($g=0$ et $B=1$). Dans la suite nous aurons besoin du cas général\footnote{En fait nous n'aurons besoin que du cas orientable. Cependant il reste à prouver que les graphes non orientables du modèle de Gross-Neveu sont convergents. Pour cela l'expression des oscillations pour un graphe complètement général sera utile.}. Nous désignerons les (vraies) variables externes par $s_{j_{k}},\,k\in\lnat 1,N\rnat\defi \lsb
1,N\rsb\cap\Z$ et écrirons $\comp\cL_{0}\defi\cL_{+}\cup\cL_{-}$.
\begin{lemma}\label{exactoscill}
  Le facteur de rosette d'un {\bfseries graphe général} est :
  \begin{align}
    &\delta\big(\sum_{k=1}^{N}(-1)^{j_{k}+1}s_{j_{k}}+\sum_{l\in\cT\cup\cL_{0}}u_l+\sum_{\ell\in\cL_{+}}w_{\ell}-\sum_{\ell\in\cL_{-}}w_{\ell}\big)%
    \,\exp\imath\varphi\\
    \nonumber\\
    \text{avec }\varphi=&\ \varphi_{E}+\varphi_{X}+\varphi_{U}+\varphi_{W},\nonumber\\
    \varphi_{E}=&\ \sum_{k<l=1}^{N}(-1)^{j_{k}+j_{l}+1}s_{j_{k}}\wed s_{j_{l}},\nonumber\\
    \nonumber\\
    \varphi_{X}=&\ \sum_{k=1}^{N}\sum_{\substack{((\cT\cup\cL_{0})\prec
        j_{k})\\\cup(\comp\cL_{0}\supset j_{k})}}(-1)^{j_{k}+1}s_{j_{k}}\wed u_{l}
    +\sum_{(\cT\cup\cL_{0})\succ
      j_{k}}u_{l}\wed (-1)^{j_{k}+1}s_{j_{k}},\nonumber\\
    \nonumber\\
    \varphi_{U}=&\ \frac 12\sum_{\cT}\veps(l)v_{l}\wed u_{l}+\frac 12\sum_{\cL}\veps(\ell)w_{\ell}\wed u_{\ell}\nonumber\\
    &+\frac 12\sum_{\cL_{0}\ltimes\cL_{0}}\veps(\ell)w_{\ell}\wed
    u_{\ell'}+\veps(\ell')w_{\ell'}\wed u_{\ell}+\frac
    12\sum_{\cL_{0}\ltimes\comp\cL_0}\veps(\ell)w_{\ell}\wed
    u_{\ell'}-\veps(\ell')w_{\ell'}\wed u_{\ell}\nonumber\\
    &+\frac 12\sum_{\cL_{0}\rtimes\comp\cL_0}-\veps(\ell)w_{\ell}\wed
    u_{\ell'}+\veps(\ell')w_{\ell'}\wed u_{\ell}\nonumber\\
    &+\frac
    12\sum_{\substack{(\cL_{+}\lrtimes\cL_{-})\\\cup(\cL_{+}\ltimes\cL_{+})\cup(\cL_{-}\ltimes\cL_{-})}}u_{\ell}
    \wed\veps(\ell')w_{\ell'}+u_{\ell'}\wed\veps(\ell)w_{\ell}\nonumber\\
    &+\sum_{\substack{((\cT\cup\cL_{0})\subset\cL_{0})\\\cup((\cT\cup\cL_{0})\succ\comp\cL_0)}}\veps(\ell')
    w_{\ell'}\wed
    u_{l}+\sum_{\substack{(\comp\cL_{0}\subset\comp\cL_{0})\\
        \cup((\cT\cup\cL_{0})\prec\comp\cL_0)}}u_{l}\wed\veps(\ell')w_{\ell'}
    \nonumber\\
    &+\sum_{\substack{(\cT\cup\cL_{0})\prec(\cT\cup\cL_{0})}}u_{l'}\wed
    u_{l}+\sum_{\substack{(\cT\cup\cL_{0})\subset\comp\cL_0}}u_{l}\wed
    u_{\ell'}\nonumber\\
    &+\frac
    12\sum_{\substack{(\cL_{0}\ltimes\cL_{0})\\\cup(\cL_{+}\ltimes\cL_{+})\cup(\cL_{-}\ltimes\cL_{-})}}
    u_{\ell'}\wed u_{\ell}+\frac
    12\sum_{\substack{(\cL_{0}\lrtimes\comp\cL_0)\\\cup(\cL_{+}\rtimes\cL_{-})\cup(\cL_{-}\rtimes\cL_{+})}}u_{\ell}\wed
    u_{\ell'},\nonumber\\
    \nonumber\\
    \varphi_{W}=&\ \sum_{\substack{(\comp\cL_{0}\prec
        j_{k})\\\cup(\cL_{0}\supset j_{k})}}\veps(\ell)w_{\ell} \wed (-1)^{j_{k}+1}s_{j_{k}}
    +\sum_{\substack{\comp\cL_{0}\succ
        j_{k}}}(-1)^{j_{k}+1}s_{j_{k}}\wed \veps(\ell)w_{\ell}\nonumber\\
    &+\frac
    12\sum_{\substack{(\cL_{0}\ltimes\cL_{0})\\\cup(\comp\cL_{0}\ltimes\comp\cL_0)
        \cup(\cL_{0}\lrtimes\comp\cL_0)}}\veps
    (\ell')w_{\ell'}\wed\veps(\ell)w_{\ell}+\sum_{\substack{(\cL_{0}\supset\comp\cL_0)\\\cup(\comp\cL_{0}
        \prec\comp\cL_0)}}\veps
    (\ell')w_{\ell'}\wed\veps(\ell)w_{\ell},\nonumber
  \end{align}
  où $l$($\ell$) appartient toujours à l'ensemble de gauche.
\end{lemma}

\begin{proof}
  Comme expliqué dans la section \ref{sec:resol-deltaNCPhi4}, la fonction $\delta$
  de racine est donnée par
  \begin{equation}
    \delta\big(\sum_{k=1}^{N}(-1)^{j_{k}+1}s_{j_{k}}+\sum_{l\in\cT\cup\cL_{0}}u_l+\sum_{\ell\in\cL_{+}}w_{\ell}-\sum_{\ell\in\cL_{-}}w_{\ell}\big).
  \end{equation}
  On exprime maintenant les variables des champs de boucles en termes des
  variables $u$ et $w$. Ainsi le terme quadratique dans les variables externes
  est
  \begin{equation}
    \sum_{k<l=1}^{N}(-1)^{j_{k}+j_{l}+1}s_{j_{k}}\wed s_{j_{l}}\, .
  \end{equation}
  
  Soit une variable externe $s_{j_{k}}$. Les termes linéaires dans cette
  variable sont :
  \begin{align}
    \varphi_{j_{k}}=&\sum_{i<j_{k}}(-1)^{i+1}s_i\wed (-1)^{j_{k}}s_{j_{k}}+\sum_{i>j_{k}}(-1)^{j_{k}}s_{j_{k}}\wed (-1)^{i+1}s_i\nonumber\\
    &+\sum_{\cT\succ j_{k}}(-1)^{j_{k}}s_{j_{k}}\wed u_l +\sum_{\cT\prec
      j_{k}} u_l \wed (-1)^{j_{k}}s_{j_{k}}
  \end{align}
  o\`u les $s_{i}$ sont toutes des variables de boucles. Soit une ligne de
  boucle $\ell=(i,j)\prec j_{k}$.\\
  Sa contribution à $\varphi_{j_{k}}$ est :
  \begin{align}
    \label{eq:termlin1}
    &\lsb (-1)^{i+1}s_{i}+(-1)^{j+1}s_{j}\rsb\wed (-1)^{j_{k}}s_{j_{k}}.
  \end{align}
  Le résultat en termes des variables $u_{\ell}$ et $w_{\ell}$ dépend de
  l'orientabilité de la ligne de boucle. À partir des définitions
  \ref{defn:longshort} et \ref{defn:signe}, on a
  \begin{align}
    &\lsb (-1)^{i+1}s_{i}+(-1)^{j+1}s_{j}\rsb\wed (-1)^{j_{k}}s_{j_{k}}\\
    =&\ u_{\ell}\wed (-1)^{j_{k}}s_{j_{k}}&&\text{si }\ell\in\cL_{0}\nonumber\\
    =&\ -\veps(l)w_{\ell}\wed (-1)^{j_{k}}s_{j_{k}}&&\text{si
    }\ell\in\cL_{+}\cup\cL_{-}.\nonumber
  \end{align}
  De la même façon si une ligne boucle au-dessus de la variable externe
  $s_{j_{k}}$, sa contribution à $\varphi_{j_{k}}$ est :
  \begin{align}
    \label{eq:termlin2}
    &\lsb (-1)^{i+1}s_{i}+(-1)^{j}s_{j}\rsb\wed (-1)^{j_{k}}s_{j_{k}}\\
    =&\ -\veps(l)w_{\ell}\wed (-1)^{j_{k}}s_{j_{k}}&&\text{si
    }\ell\in\cL_{0}\nonumber\\
    =&\ u_{\ell}\wed (-1)^{j_{k}}s_{j_{k}}&&\text{si
    }\ell\in\cL_{+}\cup\cL_{-}.\nonumber
  \end{align}
  Finalement le terme linéaire en $s_{j_{k}}$ est
  \begin{align}
    \varphi_{j_{k}}=&\sum_{\substack{((\cT\cup\cL_{0})\prec
        j_{k})\\\cup(\comp\cL_{0}\supset j_{k})}}u_{l}\wed
    (-1)^{j_{k}}s_{j_{k}}+\sum_{(\cT\cup\cL_{0})\succ
      j_{k}}(-1)^{j_{k}}s_{j_{k}}\wed
    u_{l}\\
    &+\sum_{\substack{(\comp\cL_{0}\prec j_{k})\\\cup(\cL_{0}\supset
        j_{k})}}(-1)^{j_{k}}s_{j_{k}}\wed \veps(\ell)w_{\ell}
    +\sum_{\substack{\comp\cL_{0}\succ j_{k}}}\veps(\ell)w_{\ell}\wed
    (-1)^{j_{k}}s_{j_{k}}.\nonumber
  \end{align}
  
  \bigskip Considérons une ligne de boucle $\ell=(p,q)$. Sa contribution au
  facteur de rosette se décompose en un terme \og pur boucle\fg{} et un terme
  \og arbre-boucle\fg. Nous détaillerons le premier. Le second est obtenu par la même méthode. Le terme pur boucle est :
  \begin{align}
    \varphi_{bb}=&\sum_{i<p}(-1)^{i+1}s_i\wed (-1)^p
    s_p+\sum_{\substack{p<i\\i\neq q}}(-1)^ps_p\wed
    (-1)^{i+1}s_i+(-1)^{p+q+1}s_{p}\wed s_{q}\nonumber\\
    &+\sum_{\substack{i<q\\i\neq p}}(-1)^{i+1}s_i\wed (-1)^q
    s_q+\sum_{q<i}(-1)^qs_q\wed
    (-1)^{i+1}s_i\nonumber\\
    =&\sum_{i<p}(-1)^{i+1}s_i\wed [(-1)^{p}s_p+(-1)^qs_q]+\sum_{q<i}[(-1)^ps_p+(-1)^{q}s_q]\wed (-1)^{i+1}s_i\nonumber\\
    &+\sum_{p<i<q}(-1)^{i+1}s_{i}\wed [(-1)^{p+1}s_p+(-1)^q
    s_q]+(-1)^{p+q+1}s_p\wed s_q \ .
  \end{align}
  Six possibilités s'offrent alors à une autre ligne de boucle $\ell'=(i,j)$.
  Elle peut suivre $\ell$ ou la précéder, la contenir ou être contenue par
  elle, la croiser par la gauche ou par la droite. De plus les lignes $\ell$
  et $\ell'$ peuvent être orientables ou pas. Nous n'allons pas exhiber toutes ces différentes contributions mais nous allons donner la
  méthode employée pour les obtenir grâce à deux exemples. Le lecteur
  remarquera que la procédure est complètement
  semblable à celle employée pour le terme $\varphi_{j_{k}}$.\\
  \\
  \noindent
  Soit $(\ell,\ell')\in\cL_{0}^{2}$ tel que $\ell'\ltimes\ell$. La ligne
  $\ell'$ croise alors $\ell$ par la gauche comme définie en
  \ref{defn:relations}. Le terme correspondant est :
  \begin{align}
    &\ (-1)^{i+1}s_i\wed [(-1)^{p}s_p+(-1)^qs_q]+(-1)^{j+1}s_{j}\wed
    [(-1)^{p+1}s_p+(-1)^qs_q]\nonumber\\
    =&\ (-1)^{i+1}s_i\wed (-u_{\ell})+(-1)^{j+1}s_{j}\wed
    (-\veps(\ell)w_{\ell})\nonumber\\
    =&\ \frac 12\lbt u_{\ell}\wed u_{\ell'}+\veps(\ell')w_{\ell'}\wed
    u_{\ell}+\veps(\ell)w_{\ell}\wed u_{\ell'}+\veps(\ell)w_{\ell}\wed
    \veps(\ell')w_{\ell'}\rbt.
  \end{align}
  De même si $\ell\in\cL_{0}$, $\ell'\in\cL_{+}$ telles que
  $\ell\subset\ell'$, on a :
  \begin{align}
    &\ (-1)^{i+1}s_i\wed [(-1)^{p}s_p+(-1)^qs_q]+[(-1)^ps_p+(-1)^{q}s_q]\wed
    (-1)^{j+1}s_j\nonumber\\
    =&\ (-1)^{i+1}s_i\wed(-u_{\ell})+(-u_{\ell})\wed
    (-1)^{j+1}s_j=u_{\ell}\wed u_{\ell'}
  \end{align}
  On procède de la même manière pour les autres contributions et on obtient le
  facteur \og pur boucle\fg{} suivant :
  \begin{align}
    \varphi_{bb}=&\ \frac 12\sum_{\cL}\veps(\ell)w_{\ell}\wed u_{\ell}\\
    &+\sum_{\substack{(\cL_{0}\subset\cL_{0})\\\cup(\cL_{0}\succ\comp\cL_0)}}\veps(\ell')w_{\ell'}\wed
    u_{\ell}+\sum_{\substack{(\cL_{0}\prec\comp\cL_0)\cup(\comp\cL_{0}\subset\comp\cL_{0})}}
    u_{\ell}\wed\veps(\ell')w_{\ell'}\nonumber\\
    &+\frac 12\sum_{\cL_{0}\ltimes\cL_{0}}\veps(\ell)w_{\ell}\wed
    u_{\ell'}+\veps(\ell')w_{\ell'}\wed u_{\ell}+\frac
    12\sum_{\cL_{0}\ltimes\comp\cL_0}\veps(\ell)w_{\ell}\wed
    u_{\ell'}-\veps(\ell')w_{\ell'}\wed u_{\ell}\nonumber\\
    &+\frac 12\sum_{\cL_{0}\rtimes\comp\cL_0}-\veps(\ell)w_{\ell}\wed
    u_{\ell'}+\veps(\ell')w_{\ell'}\wed u_{\ell}\nonumber\\
    &+\frac
    12\sum_{\substack{(\cL_{+}\lrtimes\cL_{-})\\\cup(\cL_{+}\ltimes\cL_{+})\cup(\cL_{-}\ltimes\cL_{-})}}u_{\ell}
    \wed\veps(\ell')w_{\ell'}+u_{\ell'}\wed\veps(\ell)w_{\ell}\nonumber\\
    &+\frac
    12\sum_{\substack{(\cL_{0}\ltimes\cL_{0})\cup(\comp\cL_{0}\ltimes\comp\cL_0)\\
        \cup(\cL_{0}\lrtimes\comp\cL_0)}}\veps
    (\ell')w_{\ell'}\wed\veps(\ell)w_{\ell}+\sum_{\substack{(\cL_{0}\supset\comp\cL_0)\\\cup(\comp\cL_{0}
        \prec\comp\cL_0)}}\veps
    (\ell')w_{\ell'}\wed\veps(\ell)w_{\ell}\nonumber\\
    &+\sum_{\substack{\cL_{0}\prec\cL_{0}}}u_{\ell'}\wed
    u_{\ell}+\sum_{\substack{\cL_{0}\subset\comp\cL_0}}u_{\ell}\wed
    u_{\ell'}\nonumber\\
    &+\frac
    12\sum_{\substack{(\cL_{0}\ltimes\cL_{0})\\\cup(\cL_{+}\ltimes\cL_{+})\cup(\cL_{-}\ltimes\cL_{-})}}
    u_{\ell'}\wed u_{\ell}+\frac
    12\sum_{\substack{(\cL_{0}\lrtimes\comp\cL_0)\\\cup(\cL_{+}\rtimes\cL_{-})\cup(\cL_{-}\rtimes\cL_{+})}}u_{\ell}\wed
    u_{\ell'}\nonumber
  \end{align}
  
  Enfin il reste le terme \og arbre-boucle\fg{} :
  \begin{align}
    \varphi_{ab}=&\sum_{\lb l'\in\cT,\, l'\prec
      p\rb}u_{l'}\wed(-1)^{p}s_{p}+\sum_{\lb l'\in\cT,\, l'\succ
      p\rb}(-1)^{p}s_p\wed
    u_{l'}\\
    &+\sum_{\lb l'\in\cT,\, l'\prec q\rb}u_{l'}\wed(-1)^{q}s_{q}+\sum_{\lb
      l'\in\cT,\, l'\succ
      q\rb}(-1)^{q}s_q\wed u_{l'}\nonumber\\
    =&\sum_{\lb l'\in\cT,\, l'\prec p\rb}u_{l'}\wed[(-1)^{p}s_p+(-1)^q
    s_q]+\sum_{\lb l'\in\cT,\, l'\succ q\rb}\lsb(-1)^{p}s_p+(-1)^q s_q\rsb\wed u_{l'}\nonumber\\
    &+\sum_{\lb l'\in\cT,\, p\prec l'\prec q\rb}u_{l'}\wed\lsb(-1)^{p+1}s_p+(-1)^{q}s_q\rsb\nonumber\\
    =&\ \sum_{\substack{\cL_{0}\succ\cT}}u_{\ell}\wed
    u_{l'}+\sum_{\substack{(\cL_{0}\prec\cT)\\\cup(\comp\cL_{0}\supset\cT)}}u_{l'}\wed
    u_{\ell}\nonumber\\
    &+\sum_{\substack{(\cL_{0}\supset\cT)\\\cup(\comp\cL_{0}\prec\cT)}}\veps(\ell)w_{\ell}\wed
    u_{l'}+\sum_{\comp\cL_{0}\succ\cT}u_{l'}\wed\veps(\ell)w_{\ell}\nonumber.
  \end{align}
\end{proof}

\begin{cor}\label{sec:oscillOrient}
  Le facteur de rosette d'un {\bfseries graphe orientable} est
  \begin{align}
    &\delta\big(\sum_{k=1}^{N}(-1)^{j_{k}+1}s_{j_{k}}+\sum_{l\in\cT\cup\cL}u_l\big)%
    \,\exp\imath\varphi\\
    \nonumber\\
    \text{avec }\varphi=&\ \varphi_{E}+\varphi_{X}+\varphi_{U}+\varphi_{W},\nonumber\\
    \varphi_{E}=&\ \sum_{k<l=1}^{N}(-1)^{j_{k}+j_{l}+1}s_{j_{k}}\wed s_{j_{l}},\nonumber\\
    \nonumber\\
    \varphi_{X}=&\ \sum_{k=1}^{N}\sum_{\substack{(\cT\cup\cL)\prec
        j_{k}}}(-1)^{j_{k}+1}s_{j_{k}}\wed u_{l}+\sum_{(\cT\cup\cL)\succ
      j_{k}}u_{l} \wed (-1)^{j_{k}+1}s_{j_{k}},\nonumber\\
    \nonumber\\
    \varphi_{U}=&\ \frac 12\sum_{\cT}\veps(l)v_{l}\wed u_{l}+\frac 12\sum_{\cL}\veps(\ell)w_{\ell}\wed u_{\ell}\nonumber\\
    &+\frac 12\sum_{\cL\ltimes\cL}\veps(\ell)w_{\ell}\wed
    u_{\ell'}+\veps(\ell')w_{\ell'}\wed
    u_{\ell}+\sum_{\substack{(\cT\cup\cL)\subset\cL}}\veps(\ell')w_{\ell'}\wed
    u_{l}\nonumber\\
    &+\sum_{\substack{(\cT\cup\cL)\prec(\cT\cup\cL)}}u_{l'}\wed u_{l}+\frac
    12\sum_{\substack{\cL\ltimes\cL}}u_{\ell'}\wed u_{\ell},\nonumber\\
    \nonumber\\
    \varphi_{W}=&\ \sum_{\substack{\cL\supset
        j_{k}}}\veps(\ell)w_{\ell} \wed (-1)^{j_{k}+1}s_{j_{k}}+\frac
    12\sum_{\substack{\cL\ltimes\cL}}\veps(\ell')w_{\ell'}\wed\veps(\ell)w_{\ell}.\nonumber
  \end{align}
\end{cor}
\begin{proof}
  Il suffit de faire $\cL_{+}=\cL_{-}=\emptyset$ dans l'expression générale du
  lemme \ref{exactoscill}.
\end{proof}
\newpage
\begin{cor}\label{sec:oscillRG}
  Soit un {\bfseries graphe régulier planaire} ($g=0$ et $B=1$). Son facteur
  de rosette est \cite{xphi4-05}
  \begin{align}
    &\delta\big(\sum_{k=1}^{N}(-1)^{k+1}x_{k}+\sum_{l\in\cT\cup\cL}u_l\big)%
    \,\exp\imath\varphi\label{eq:rosette-planreg}\\
    \nonumber\\
    \text{avec }\varphi=&\ \varphi_{E}+\varphi_{X}+\varphi_{U},\nonumber\\
    \varphi_{E}=&\ \sum_{i<j=1}^{N}(-1)^{i+j+1}x_{i}\wed
    x_{j},\nonumber\\
    \nonumber\\
    \varphi_{X}=&\ \sum_{k=1}^{N}\sum_{\substack{(\cT\cup\cL)\prec
        k}}(-1)^{k+1}x_{k}\wed u_{l}+\sum_{(\cT\cup\cL)\succ k}u_{l} \wed (-1)^{k+1}x_{k},\nonumber\\
    \nonumber\\
    \varphi_{U}=&\ \frac 12\sum_{\cT}\veps(l)v_{l}\wed u_{l}+\frac
    12\sum_{\cL}\veps(\ell)w_{\ell}\wed u_{\ell}\nonumber\\
    &+\sum_{(\cT\cup\cL)\subset\cL}\veps(\ell')w_{\ell'}\wed
    u_{l}+\sum_{\substack{(\cT\cup\cL)\prec(\cT\cup\cL)}}u_{l'}\wed
    u_{l}.\nonumber
  \end{align}
\end{cor}

\begin{proof}
  Le graphe n'ayant qu'une seule face brisée, il y a toujours un nombre pair
  de champs entre deux variables externes. Dans ce cas, $j_{k}$ et $k$ ont
  même parité. Ainsi, en effectuant le changement de variables $s_{j_{k}}\to
  x_{k}$, le terme quadratique dans les variables externes s'écrit :
  \begin{equation}
    \sum_{i<j=1}^{N}(-1)^{i+j+1}x_i\wed x_j\ .
  \end{equation}
  De plus les contraintes $g=0$ et $B=1$ impliquent que le graphe est
  orientable ($\cL=\cL_{0}$). En effet, considérons une ligne de boucle $\ell$
  (ce sont les seules à pouvoir être non orientables) reliant $s_{i}$ à
  $s_{i+2p}$. Ces deux positions ont alors même parité. Entre les deux bouts
  de la ligne $\ell$ se trouve un nombre impair de positions. Ainsi soit
  $\ell$ boucle au-dessus d'une variable externe et $B\ges 2$, soit une autre
  ligne de boucle
  la croise et $g\ges 1$.\\
  Finalement en ôtant du résultat du lemme \ref{exactoscill} les termes
  concernant les croisements, les lignes bouclant au-dessus de variables
  externes et les lignes non orientables, on obtient
  (\ref{eq:rosette-planreg}).
\end{proof}

L'information principale à retenir de cette section est, pour les graphes orientables,
\begin{itemize}
\item si le graphe est non planaire, le facteur de rosette contient des oscillations du type $w\wed w$,
\item si le graphe a plus d'une face brisée (il existe des lignes de boucles qui contractent au-dessus de points externes), le facteur de rosette contient des oscillations du type $x\wed w$.
\end{itemize}

\section{Comptage de puissance}
\label{sec:comptage-de-puisancephi4}

La première étape d'une analyse multi-échelles consiste à \og{}découper\fg{} le propagateur en tranches puis obtenir une borne supérieure dans chaque tranche :
\begin{align}
  C_{l}=&\sum_{i=0}^{\infty}C^{i}_{l},\,C^{i}_{l}=
  \begin{cases}
    {\displaystyle\int_{M^{-2i}}^{M^{-2(i-1)}}dt\,C_{l}(t;\phantom{u})}&\text{si }i\ges 1\vspace{.3cm}\\
     {\displaystyle\int_{1}^{\infty}dt\,C_{l}(t;\phantom{u})}&\text{si }i=0.\\
  \end{cases}
\end{align}
\begin{lemma}
  Pour tout $i\in\N$, il existe $K,\,k\in\R_{+}$ tels que
  \begin{align}
    \label{eq:propaboundNCphi4}
    \labs C_{l}^{i}(u_{l},v_{l})\rabs\les&K M^{2(i+1)}e^{-kM^{i+1}\labs u_{l}\rabs-kM^{-i-1}\labs v_{l}\rabs}.
  \end{align}
  Cette borne est aussi valable si $m=0$.
\end{lemma}
La preuve du lemme précédent est similaire à celle du lemme \ref{lem:bornepropaphi4comm}. Notons cependant une différence essentielle. La tranche $i=0$ représente la zone infrarouge de la théorie. Dans une théorie massive commutative, le propagateur, dans cette tranche, obéit à la même borne que dans les autres tranches. On dit que la masse arrête le flot dans l'infrarouge. Toute la région infrarouge peut être traitée d'un seul coup, nul besoin de découper cette zone comme nous l'avons fait pour l'ultraviolet. Si on étudie une théorie de masse nulle, on est obligé de diviser cette dernière tranche pour étudier le comportement infrarouge avec soin. Si on ne le fait pas, on obtient des amplitudes infinies dans cette tranche. Autrement dit les propagateurs ne sont plus suffisants pour intégrer sur les points internes du graphe. Dans la théorie $\Phi^{4}$ \ncv{} tout comme dans le modèle de Gross-Neveu, le propagateur à masse nulle obéit à la même borne dans toute les tranches y compris l'infrarouge. Ce comportement est dû au terme supplémentaire en $\xt^{2}$ dans le propagateur. Dans ces théories, la facteur de divergence globale du propagateur $t^{-D/2}$ est remplacé par $\sinh^{-D/2}t$ qui se comporte comme $e^{-Dt/2}$ à $t$ grand. Ceci remplace la masse. Le potentiel harmonique $\xt^{2}$ est bien, en ce sens, un potentiel confinant qui élimine les divergences infrarouges usuelles.\\

La suite de cette section est dédiée à la preuve du lemme suivant.
\begin{lemma}[Comptage de puissance]\label{lem:compt-puissNCPhi4}
  Soit $G$ un graphe orientable connexe. Quel que soit $\Omega\in\left] 0,1\rsb$, il existe $K\in\R$ tel que son
  amplitude amputée $A_{G}^{\mu}$ intégrée contre des fonctions test est bornée par
  \begin{align}
    \labs A_{G}^{\mu}\rabs\les&K^{n}\prod_{i,k}M^{-\omega(G^{i}_{k})}\label{eq:compt-boundPhi4}\\
    \text{avec } \omega(G^{i}_{k})=&
    \begin{cases}
      N&\text{si $G^i_{k}$ est non orientable,}\\
      &\text{ou si $G^i_{k}$ est orientable, $g=0$ et $B\ges 2$,}\\
      N+4&\text{si $G^i_{k}$ est orientable, $g\ges 1$,}\\
      N-4&\text{si $G^i_{k}$ est orientable, $g=0$ et $B=1$.}
    \end{cases}
  \end{align}
\end{lemma}
\begin{proof}
  L'amplitude amputée d'un graphe $G$ connexe avec l'attribution d'échelles $\mu$, intégrée contre des fonctions test s'écrit
   \begin{align}
    A^{\mu}_{G}=\int&\prod_{i=1}^{N}dx_{i}\,f_{i}(x_{i})\delta_{G}\prod_{l\in\cT}du_{l}dv_{l}\,\delta_{b(l)}C^{i_{l}}_{l}(u_{l},v_{l})
    \prod_{\ell\in\cL}du_{\ell}dw_{\ell}\,C^{i_{\ell}}_{\ell}(u_{\ell},w_{\ell})e^{i\varphi}
  \end{align}
  où $\varphi$ est l'oscillation totale des vertex et où nous avons utilisé les notations de la section \ref{sec:resol-deltaNCPhi4} pour les fonctions delta. Nous allons tout d'abord montrer comment obtenir la borne en $N-4$ qui est la plus simple. Pour celle-ci, nous n'avons pas besoin des oscillations. Ainsi en prenant la valeur absolue de l'amplitude et en utilisant la borne \eqref{eq:propaboundNCphi4},
  \begin{align}
    \labs A^{\mu}_{G}\rabs\les K^{n}\prod_{l\in G}M^{2(i_{l}+1)}\int&\prod_{i=1}^{N}dx_{i}\,f_{i}(x_{i})\delta_{G}\prod_{l\in\cT}du_{l}dv_{l}\,\delta_{b(l)}
    e^{-kM^{i_{l}+1}\labs u_{l}\rabs-kM^{-i_{l}-1}\labs v_{l}\rabs}\\
    &\prod_{\ell\in\cL}du_{\ell}dw_{\ell}\,e^{-kM^{i_{\ell}+1}\labs u_{\ell}\rabs-kM^{-i_{\ell}-1}\labs w_{\ell}\rabs}.\notag
  \end{align}
  Les fonctions delta de branches $\delta_{b(l)},\,l\in\cT$ nous permettent d'intégrer sur les longues variables de l'arbre. Les décroissances exponentielles obtenues (en remplaçant $v_{l}$ par sa valeur dans les propagateurs) sont bornées par $1$. La dernière fonction delta de racine (de l'arbre de Gallavotti) est utilisée pour intégrer sur une patte externe. Les autres sont intégrées avec des fonctions test. Nous avons alors
  \begin{align}
    \labs A^{\mu}_{G}\rabs\les K^{n}\prod_{l\in G}M^{2(i_{l}+1)}\int&\prod_{l\in\cT}du_{l}\,e^{-kM^{i_{l}+1}\labs u_{l}\rabs}
    \prod_{\ell\in\cL}du_{\ell}dw_{\ell}\,e^{-kM^{i_{\ell}+1}\labs u_{\ell}\rabs-kM^{-i_{\ell}-1}\labs w_{\ell}\rabs}.
  \end{align}
  Remarquons que nous écrivons $K$ pour toute constante inessentielle. Ainsi $K$ prendra différentes valeurs au fur et à mesure de la preuve. Pour toute ligne $l\in G$, l'intégration sur la variable $u_{l}$ donne $\cO(M^{-4(i_{l}+1)})$ et l'intégration sur $v_{l}$ (ou $w_{l}$) donne $\cO(M^{4(i_{l}+1)})$. Pour toute ligne de boucle $\ell\in\cL$, le produit de ces deux intégrations est d'ordre $\cO(1)$. Ainsi l'amplitude de $G$ est bornée par
  \begin{align}
    \labs A^{\mu}_{G}\rabs\les K^{n}\prod_{l\in G}M^{2(i_{l}+1)}\prod_{l\in\cT}M^{-4(i_{l}+1)}.\label{PCNCPhi4-avtdistrib}
  \end{align}
  De façon complètement standard \cite{Riv1}, nous distribuons le comptage de puissance parmi les composantes connexes.
  \begin{align}
    \prod_{l\in G}M^{2(i_{l}+1)}=&\prod_{l\in G}\prod_{i=0}^{i_{l}}M^{2}=\prod_{l\in
      G}\prod_{\substack{(i,k)\in\N^{2}/\\l\in
        G^{i}_{k}}}M^{2}=\prod_{(i,k)\in\N^{2}}\,\prod_{l\in G^{i}_{k}}M^{2},\\
    \prod_{l\in\cT}M^{-4(i_{l}+1)}=&\prod_{l\in\cT}\prod_{\substack{(i,k)\in\N^{2}/\\l\in
        G^{i}_{k}}}M^{-4}=\prod_{(i,k)\in\N^{2}}\,\prod_{l\in\cT^{i}_{k}}M^{-4}.
  \end{align}
Ainsi, en utilisant $4n=2I+N$, (\ref{PCNCPhi4-avtdistrib}) devient
\begin{align}
  \labs A_{G}^{\mu}\rabs\les&K^{n(G)}\prod_{(i,k)\in\N^{2}}M^{-\frac
    12\omega(G^{i}_{k})},\label{eq:PC-planregPhi4}\\
  \text{avec }\omega(G^{i}_{k})=&N(G^{i}_{k})-4\label{eq:degre-conv-compconnNCPhi4}
\end{align}
ce qui prouve la partie du lemme \ref{lem:compt-puissNCPhi4} conernant les graphes planaires réguliers.\\

Intéressons-nous maintenant aux graphes non orientables. Quelle que soit leur topologie, nous souhaitons prouver qu'ils convergent au moins comme $M^{-Ni}$. Pour ces graphes-ci non plus nous n'utiliserons pas les oscillations de vertex et pouvons prendre la valeur absolue de l'amplitude. Nous avons vu dans la section \ref{sec:resol-deltaNCPhi4} que la fonction delta de racine d'un graphe non orientable contient, en plus des points externes et des variables $u$ du graphe, les variables $w$ des lignes (de boucle) non orientables (voir \eqref{eq:deltaroot}). Ainsi au lieu d'utiliser cette dernière fonction delta pour intégrer un point externe, nous pouvons résoudre une variable $w$ d'une ligne non orientable. Par rapport au comptage de puissance \eqref{eq:degre-conv-compconnNCPhi4}, nous gagnons alors $M^{-4i}$ si l'échelle minimum de $G$ est $i$. Ceci prouve que les graphes non orientables sont convergents. Néanmoins nous avons besoin de plus que cela. Nous souhaiterions obtenir que toutes les composantes connexes (\ie{} les sous-graphes à renormaliser) non orientables sont convergentes. Pour cela nous devons optimiser l'emploi des fonctions delta. En effet, il n'est pas toujours optimal d'utiliser toutes les fonctions delta de branches pour résoudre les longues variables de l'arbre et la dernière fonction pour une ligne non orientable.

Nous allons parcourir l'arbre de Gallavotti-Nicol\`o branche par branche en partant des feuilles et en descendant vers sa racine. Nous choisissons donc arbitrairement un ordre sur les branches de l'arbre puis un ordre total sur les noeuds de l'arbre, compatible avec l'ordre sur les branches. La figure \ref{fig:GNtree} donne un exemple d'un tel ordre. Soit $G^{i_{1,1}}_{k_{1,1}}$ une composante connexe non orientable telle que pour tout $G^{j}_{k'}\subset G^{i_{1,1}}_{k_{1,1}}$, $G^{j}_{k'}$ est orientable. Ainsi $G^{i_{1,1}}_{k_{1,1}}$ est la plus haute composante connexe non orientable de sa branche\footnote{Par branche, nous entendons ici l'unique chemin dans l'arbre de Gallavotti reliant $G^{i_{1,1}}_{k_{1,1}}$ à $G$}. Supposons que $G^{i_{1,1}}_{k_{1,1}}$ soit la première composante connexe non orientable rencontrée. Nous notons $\scP_{1}$ sa branche. Nous utilisons alors la fonction $\delta_{G}$ pour intégrer sur une variable $w_{\ell},\,\ell\in\cL^{i_{1,1}}_{k_{1,1},\pm}$. Cette fonction étant précédemment utilisée pour intégrer sur un point externe, la gain par rapport à \eqref{eq:PC-planregPhi4} est $M^{-4i_{1,1}}$. Ce facteur fait passer le degré de convergence $\omega$ de toutes les composantes connexes (non orientables) entre $G^{i_{1,1}}_{k_{1,1}}$ et $G$ de $N-4$ à $N$. Puis nous parcourons la deuxième branche $\scP_{2}$. Elle rencontre $\scP_{1}$ en un noeud $G^{i_{2,3}}_{k_{2,3}}$. Soit $G^{i_{2,2}}_{k_{2,2}}$ l'unique descendant de $G^{i_{2,3}}_{k_{2,3}}$ dans $\scP_{2}$. S'il existe une composante non orientable $G^{i_{2,1}}_{k_{2,1}}$ dans $\scP_{2}\setminus\scP_{1}$, nous utilisons la fonction delta correspondant à l'unique ligne d'arbre reliant $G^{i_{2,2}}_{k_{2,2}}$ à $G^{i_{2,3}}_{k_{2,3}}$ pour intégrer sur une ligne non orientable de $G^{i_{2,1}}_{k_{2,1}}$. Le gain $M^{-4(i_{2,1}-i_{2,3})}$ rend convergentes toutes les composantes connexes entre $G^{i_{2,1}}_{k_{2,1}}$ et $G^{i_{2,3}}_{k_{2,3}}$. Nous procédons de même pour toutes les branches de l'arbre. Le gain obtenu dans la branche $\scP_{j+1}$ est $M^{-4(i_{j+1,1}-i_{j+1,3})}$. Il rend convergentes toutes les composantes connexes non orientables dans $\scP_{j+1}\setminus\scP_{j}$. À titre d'exemple, supposons que le noeud $1$ de la figure \ref{fig:GNtree} soit non orientable. En utilisant $\delta_{5}$ pour intégrer sur une ligne non orientable de $1$, nous augmentons le degré de convergence $\omega$ de $4$ pour les composantes $1,2,3,4$ et $5$. Supposons ensuite que $8$ soit également non orientable. Cette fois-ci nous utilisons l'unique ligne d'arbre dans $3$ ``reliant'' $8$ à $3$. Les noeuds $7$ et $8$ deviennent alors convergents. Nous avons ainsi prouvé que le degré de convergence de toute composante connexe non orientable est $N$.
\begin{figure}[!htbp]
  \centering 
  \includegraphics[scale=1]{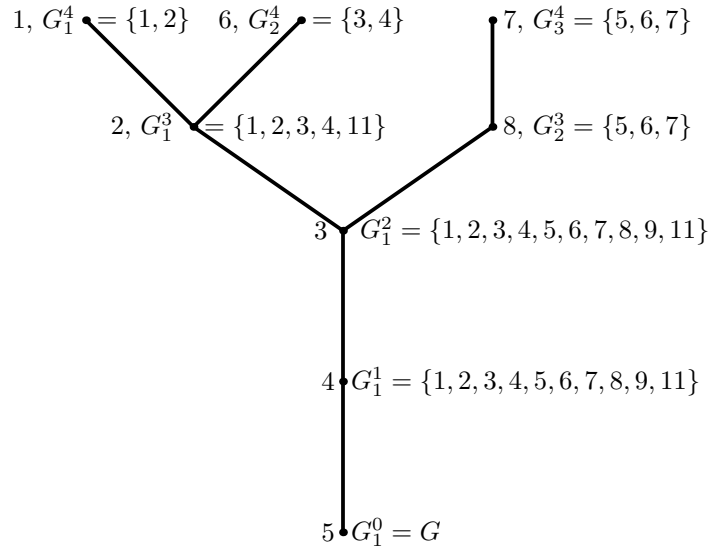}
  \caption{Parcours dans l'arbre de Gallavotti-Nicol\`o}
  \label{fig:GNtree}
\end{figure}

Il reste à démontrer le lemme \ref{lem:compt-puissNCPhi4} pour les graphes non planaires ou avec plusieurs faces brisées. Considérons un graphe $G$ non planaire orientable. Soit $\G{i}{k}\subset G$ une composante connexe non planaire. Le lemme \ref{sec:oscillOrient} nous apprend qu'il existe une oscillation $\sum_{\cL^{i}_{k}\ltimes\cL^{i}_{k}}\veps(\ell')w_{\ell'}\wed\veps(\ell)w_{\ell}$. Soit $\ell$ une ligne parmi celles qui se croisent. L'oscillation $\veps(\ell)w_{\ell}\wed\Big(\sum_{\cL^i_{k}\ltimes\ell}\veps(\ell')w_{\ell'}-\sum_{\cL^i_{k}\rtimes\ell}\veps(\ell')w_{\ell'}\Big)\defi \veps(\ell)w_{\ell}\wed\cW_{\ell}$ nous permet d'obtenir une fonction du type $\exp-M^{i_{\ell}}\labs\cW_{\ell}\rabs$ en intégrant sur $w_{\ell}$. Ainsi l'intégration sur l'un des $w_{\ell'}$ avec $\ell'$ qui croise $\ell$ donnera $M^{-4i_{\ell}}$ au lieu de $M^{4i_{\ell'}}$. Le gain est donc $M^{-4(i_{\ell'}+i_{\ell'})}\les M^{-8i}$. Ce facteur fait passer le degré de convergence de toutes les composantes connexes entre $\G{i}{k}$ et $G$ de $N-4$ à $N+4$. En procédant ainsi pour toutes les branches de l'arbre de Gallavotti, nous prouvons que toutes les composantes connexes non planaires sont convergentes comme $M^{-N-4}$.\\

Considérons enfin $\ccl{G}{i}{k}$ une composante connexe planaire orientable avec $B\ges 2$. Il existe donc, d'après le lemme \ref{sec:oscillOrient}, une ligne de boucle $\ell$ qui contracte au-dessus de points externes à $\G{i}{k}$. Ceci donne lieu à une oscillation $\veps(\ell)w_{\ell}\wed\sum_{k\subset\ell}(-1)^{k+1}x_{k}$. Ces points sont soit de vrais points externes soit des extrémités de lignes d'échelles strictement inférieures à $i$. De façon identique au cas non planaire, cette oscillation fournit une décroissance implémentant $\labs\sum_{k\subset\ell}(-1)^{k+1}x_{k}\rabs\les M^{-i_{\ell}}$. S'il existe parmi ces points un point externe à $G$, l'intégration sur celui-ci donne $M^{-4i_{\ell}}$ au lieu de $\cO(1)$. Cela rend convergentes toutes les composantes connexes entre $\G{i}{k}$ et $G$. S'il n'existe pas de points externes parmi ces points, l'orientabilité et la planarité du graphe nous assure que $\sum_{k\subset\ell}(-1)^{k+1}x_{k}$ est une somme de variables $u$. Soit $u_{\ell_{0}}$ la variable de plus basse échelle. L'intégration sur $u_{\ell_{0}}$ fournit $M^{-4(i_{\ell}-i_{\ell_{0}})}$. Le degré de convergence de toutes les composantes (avec plus de deux faces brisées) entre $\G{i}{k}$ et $G^{i_{\ell_{0}}}_{k'}$ passe de $N-4$ à $N$ ce qui achève la démonstration du lemme \ref{lem:compt-puissNCPhi4}.
\end{proof}

\section{Renormalisation}
\label{sec:renormalisationNCPhi4}

Dans cette section , nous ne considèrerons que des sous-graphes divergents \ie{} des sous-graphes planaires avec deux ou quatre pattes externes et une seule face brisée ($g=0,N=2\text{ ou }4,B=1$).
\subsection{La fonction à quatre points}\label{ren4pt}

Considérons un sous-graphe à quatre points nécessitant d'être renormalisé. Il est donc un noeud de l'arbre de Gallavotti-Nicol\`o :
il existe $(i,k)\in\N^{2}$ tel que $N(G^{i}_{k})=4$. Les quatre points externes du graphe amputé sont désignés par $x_{1},x_{2},x_{3}$ et $x_{4}$. Nous définissons également $Q$, $R$ et $S$ trois matrices antisymétriques de tailles respectives $4\times l(G^{i}_{k})$,
$l(G^{i}_{k})\times l(G^{i}_{k})$ et $[n(G^{i}_{k})-1]\times l(G^{i}_{k})$ où $n(G)-1$ est le nombre de boucles d'un graphe à quatre points et $n$ vertex. L'amplitude associée à la composante connexe $G^{i}_{k}$ est
\begin{align}
A(G^{i}_{k})(x_{1},x_{2},x_{3},x_{4})=&\int
\prod_{l\in\cT^{i}_{k}}  du_{l}\, C_{l}(x, u, w) 
\prod_{\ell\in\cL^{i}_{k}}du_{\ell}dw_{\ell}\,C_{\ell}(u_\ell, w_\ell) \label{eq:4pt-ini}\\
&\hspace{-3cm}\delta\big(x_{1}-x_{2}+x_{3}-x_{4}+\sum_{l\in G^{i}_{k}} u_l\big)
e^{\imath \lbt \sum_{p<q}(-1)^{p+q+1}x_{p}\wed x_{q}+XQU+URU+USW \rbt }.\notag
\end{align}
La forme exacte de l'oscillation $\sum_{p<q} (-1)^{p+q+1}x_{p}\wed
x_{q}$  provient du corollaire \ref{sec:oscillRG}. À partir de ce même corollaire et de \eqref{exactvalue} ci-dessous, nous pourrions obtenir les expressions exactes des matrices $Q$, $R$ et $S$ mais ce n'est pas nécessaire. Le point important est l'absence d'oscillation du type $X\wed W$ (car $B=1$) et $W\wed W$ (car $g=0$). $C_{l}$ est le propagateur de la ligne $l$. Pour les lignes de boucles, $C_{\ell}$ s'exprime en fonction de $u_\ell$ et $w_\ell$ par la formule \eqref{eq:Mehler2}. Par contre, pour les lignes de l'arbre $\ccl{\cT}{i}{k}$, $v_{l}$ a été remplacé par des variables $u$, par des points externes et de longues variables de boucles grâce au système de fonctions delta de branches, voir \eqref{eq:deltabranchfinale} et \eqref{eq:deltavl}. Plus précisement, soit $l\in\cT^i_k$ avec $i(l)\ges i$, nous pouvons écrire
\begin{equation}\label{exactvalue}
v_l =  X_l + W_l + U_l 
\end{equation} 
où
\begin{equation}\label{xvalue}
X_l = -2\eta(e_{l})\sum_{e\in\cE(l)}\eta(e)x_{e}
\end{equation}
et $\cE(l)$ est un ensemble indiçant les vrais points externes (d'échelles $<i$) de la branche $b(l)$,
\begin{equation}\label{wvalue}
W_l = -\eta(e_{l})\sum_{\ell\in\cX(l)\setminus\cE(l)}\eta(e_{\ell})w_{\ell}
\end{equation}
avec $\cX(l)\setminus\cE(l)$ l'ensemble des lignes apparaissant comme des pattes externes pour $b(l)$ (d'échelles $j\ges i$) et
\begin{equation}\label{uvalue}
U_l = -\eta(e_{l})u_{l}-2\eta(e_{l})\sum_{l'\in(\cT\cup\cL_{0})\cap\kb(l)}u_{l'}-\eta(e_{l})\sum_{\ell\in\cX(l)\setminus\cE(l)}u_{\ell}
\end{equation}
est une combinaison linéaire de variables courtes $u_{l'}$. Le propagateur pour une ligne $l\in\cT^{i}_{k}$ devient
\begin{align}
C^{i_{l}}_{l}(u_{l},X_{l},U_{l},W_{l})=\frac{\Omega^{2}}{\theta^{2}\pi^{2}}\int_{M^{-2i_{l}}}^{M^{-2(i_{l}-1)}}&\frac{dt_{l}}{\sinh^{2}(2\Ot
  t_{l})}\,e^{-\frac{\Ot}{2}\coth( 2\Ot t_{l})u_{l}^{2}}\label{propatree}\\
&e^{-\frac{\Ot}{2}\tanh(2\Ot t_{l})(U_{l}+W_{l}+X_{l})^{2}-m^{2}t}.\notag
\end{align}
Soit $e= \max_{1\les p\les 4} e_p$ le plus haut indice externe du sous-graphe $G^{i}_{k}$. Celui-ci étant un noeud de l'arbre de Gallavotti-Nicol\`o, nous avons $e<i$. Nous évaluons $A(G^{i}_{k})$ contre des champs externes $\phi^{\les e}(x_p)$ :
\begin{align}
A(G^{i}_{k})=\int&\prod_{p=1}^{4}dx_{p}\,\phi^{\les e}(x_{p})\,
A(G^{i}_{k})(x_{1},x_{2},x_{3},x_{4})\\
=\int&\prod_{p=1}^{4}dx_{p }\,\phi^{\les e}(x_{p})e^{\imath\varphi_{E}}
\prod_{l\in\cT^{i}_{k}}du_{l}\,C_{l}(u_{l}, sX_l, U_l, W_l)\label{eq:ampli-Taylor}\\
&\prod_{\ell \in\cL^{i}_{k}}du_{\ell}dw_{\ell}\,C_{\ell}(u_\ell, w_\ell)\ 
\delta\big(\Delta+s\sum_{l\in G^{i}_{k}}u_{l}\big)
e^{\imath sXQU+\imath URU+\imath USW}\,\Big|_{s=1}.\notag
\end{align}
avec $\Delta = x_{1}-x_{2}+x_{3}-x_{4}$ et $\varphi_{E}=\sum_{p<q=1}^{4}(-1)^{p+q+1}x_{p}\wed x_{q}$. 

L'équation \eqref{eq:ampli-Taylor} est construite de telle sorte qu'à $s=0$ toute la dépendance en les variables externes $x_{i}$ factorise en dehors des intégrales sur les variables $u,w$ et soit de la forme du vertex initial $\phi\star\phi\star\phi\star\phi$ ( voir (\ref{action})). Donc nous procédons à un développement de Taylor au premier ordre en $s$ et prouvons que les termes de reste sont inessentiels. Soit $\mathfrak{U}=\sum_{l\in G^{i}_{k}}u_{l}$ et
\begin{align}
{\mathfrak R}(s)=&-\sum_{l\in\cT^{i}_{k} }{\textstyle\frac{\Ot}{2}}\tanh(2\Ot t_{l}) \Big\{
s^2 X_l^2 + 2sX_l\big[ W_l+U_l\big] \Big\} 
\notag\\
\defi& - s^2  {\mathcal A} X.X  -  2s {\mathcal A}X.(W + U ).
\end{align} 
où ${\cal A}_l = \frac{\Ot}{2}\tanh(2\Ot t_{l})$ et $X. Y$ signifie $\sum_{l\in\cT^{i}_{k} } X_lY_l $. Ainsi
\begin{align}
A(G^{i}_{k})=&\int\prod_{p=1}^{4}dx_{p}\,\phi^{\les e}(x_{p})\, e^{\imath\varphi_{E}}
\prod_{l\in\cT^{i}_{k} }du_{l}\,C_{l}(u_l, U_l, W_l) \notag\\
&\Big[ \prod_{\ell \in\cL^{i}_{k}}du_{\ell}dw_{\ell}\,C_{\ell}(u_\ell, w_\ell) \Big]\,e^{\imath URU+\imath USW}\\
&\hspace{-2cm}
\Big\{ \delta(\Delta)+ \int_{0}^{1}ds\big[ \mathfrak{U}.\nabla\delta(\Delta+s\mathfrak{U})
+\delta(\Delta+s\mathfrak{U})\lbt\imath XQU  + {\mathfrak R}' (s)\rbt\big] 
e^{\imath sXQU + {\mathfrak R}(s)}  \Big\}.\notag
\end{align}
où $C_{l}(u_l, U_l, W_l) $ est donné par (\ref{propatree}) mais pris en $X_l=0$.\\

Le terme d'ordre $0$, $\tau A$, est de la forme du vertex initial \eqref{eq:interaction-phi4} multiplié par un nombre réel indépendant des variables externes $x_{i}$. Il est asymptotiquement indépendant de l'indice d'échelle $i$ si bien que la somme sur $i$ à $e$ fixé est logarithmiquement divergente. C'est bien ce à quoi nous nous attendions pour la fonction à quatre points. Il reste alors à vérifier que le terme de reste $(1-\tau)A$ converge quand $i-e \to \infty$. Celui-ci contient trois types de termes :\\
\begin{itemize}
\item Le terme $\mathfrak{U}.\nabla\delta(\Delta+t\mathfrak{U})$. En intégrant par parties sur une variable externe, le gradient $\nabla$ agit sur un champ externe $\phi^{\les e}$ et donne au plus $M^{e}$. $\mathfrak{U}$ apporte au moins $M^{-i}$.
\item  Le terme $XQU$. $X$ donne au plus $M^e$ et $U$ au moins $M^{-i}$.
\item Le terme ${\mathfrak R}'(s)$ se décompose en trois termes $\cA X.X$, $\cA X.U$ et $\cA X.W$. $\cA_{l}$ donne au moins $M^{-2 i_{l}}$, $X$ donne au plus $M^{e}$, $U$ apporte au moins $M^{-i}$ et $X_{l}W_{l}$ donne au plus $M^{e+i_{l}}$. Seul le dernier point est un peu subtil. Si $l\in\cT^i_k$, remarquons que $\cT^i_k$ étant sous-arbre dans toutes les sous-composantes connexes de $G^i_k$, toutes les variables $w_{\ell'},\,\ell'\in b(l)$ qui apparaissent dans $W_{l}$ ont des échelles inférieures ou égales à $i_{l}$. Dans le cas contraire, ces lignes auraient été choisies pour $\cT^{i}_{k}$ à la place de $l$.
\end{itemize}

\medskip
\noindent
En conclusion, comme $i_{l}\ges i$, le reste de Taylor $(1-\tau)A$ améliore le comptage de puissance de la composante connexe $G_{k}^{i}$ d'au moins $M^{-(i-e)}$ ce qui rend $(1-\tau)A(G^{i}_{k})$ convergent.

\subsection{La fonction à deux points}\label{Ren2pt}

Nous considérons les noeuds de l'arbre de Gallavotti tels que $N(G^{i}_{k})=2$. Les deux points externes sont notés $x$ et $y$. En utilisant la fonction delta globale qui est ici $\delta\big(x-y + {\mathfrak U}\big)$, nous remarquons que l'oscillation externe $e^{\imath x\wed y}$ peut être absorbée dans une redéfinition du terme $e^{\imath sXQU}$ ce que nous faisons à partir de maintenant. L'amplitude est
\begin{align}
  A(G^{i}_{k}) =&\int dxdy\,\phi^{\les e}(x)\phi^{\les e}(y) \delta\big(x-y+ {\mathfrak U}\big)\label{2point1}\\
  &\prod_{\ell\in\cL^{i}_{k}}du_{\ell}dw_{\ell}C_{\ell}(u_\ell, w_\ell)
  \prod_{l\in\cT^{i}_{k}}du_{l}\,C_{l}(u_l, X_l, U_l, W_l) \ e^{\imath XQU+\imath URU+\imath USW}.\notag
\end{align}
Nous écrivons tout d'abord l'identité
\begin{align}
\phi^{\les e}(x)\phi^{\les e}(y) =&
\frac 12 \Big( [\phi^{\les e}(x)]^2 + [\phi^{\les e}(y) ]^2 - 
[\phi^{\les e}(y) - \phi^{\les e}(x)]^2 \Big),
\label{eq:2pt-sym}
\end{align}
la développons comme
\begin{align}\label{symdev}
\phi^{\les e}(x)\phi^{\les e}(y)=& 
\frac 12 \Big\{ [\phi^{\les e}(x)]^2 + [\phi^{\les e}(y) ]^2 
- \Big[   (y-x)^\mu\nabla_\mu \phi^{\les e}(x)\\
&+ \int_{0}^{1}ds (1-s)  (y-x)^\mu (y-x)^\nu \nabla_\mu \nabla_\nu
\phi^{\les e}(x + s(y-x))  \Big]^2 \Big\}
\nonumber
\end{align}
et la substituons dans (\ref{2point1}). Le premier terme $A_0$ est une combinaison symétrique où les deux champs externes sont au même point. Considérons le cas avec deux champs en $x$ \ie{} le terme avec $[\phi^{\les e}(x)]^2 $. Nous intégrons sur $y$ en utilisant la fonction delta. Nous effectuons ensuite un développement de Taylor de la fonction suivante à l'ordre $3$ en $s$
\begin{equation}
  \label{eq:f}
  f(s)=  e^{\imath sXQ U   + {\mathfrak R}(s)}
\end{equation} 
où ${\mathfrak R}(s)= - [ s^2\cA X . X + 2s\cA X . (W + U )]$. Nous obtenons
\begin{align}
A_0 =& \frac 12 \int dx\,[\phi^{\les e}(x)]^2\,e^{\imath  (URU+ USW)} \notag\\
&\prod_{\ell \in\cL^{j}_{k}}du_{\ell}dw_{\ell}C_{\ell}(u_\ell, w_\ell)
 \prod_{l\in\cT^{i}_{k}}du_{l}C_{l}(u_{l}, U_l, W_l) \notag\\
&\Big( f(0)+f'(0)+\frac 12f''(0)+\frac 12\int_{0}^{1}ds\,(1-s)^{2}f^{(3)}(s)\Big).\label{2point2}
\end{align}
Pour évaluer cette expression, nous la décomposons en trois termes. Soient $A_{0,0},A_{0,1},A_{0,2}$ les termes d'ordre zéro, un et deux du développement de Taylor et $A_{0,R}$ le terme de reste. Tout d'abord
\begin{align}
A_{0,0}=&\int dx\,[\phi^{\les e}(x)]^2\,e^{\imath (URU+ USW)}
\prod_{\ell \in\cL^{i}_{k}}du_{\ell}dw_{\ell}C_{\ell}(u_\ell, w_\ell)\prod_{l\in\cT^{i}_{k}}du_{l}C_{l}(u_{l}, U_l, W_l)
\end{align}
est quadratiquement divergent et contribue à la renormalisation de la masse. Ensuite
\begin{align}
A_{0,1}=\frac 12 \int& dx [\phi^{\les e}(x)]^2\,e^{\imath (URU+ USW)}
\prod_{\ell \in\cL^{i}_{k}}du_{\ell}dw_{\ell}C_{\ell}(u_\ell, w_\ell)\notag\\
&\prod_{l\in\cT^{i}_{k}}du_{l}C_{l}(u_{l}, U_l, W_l)\big(\imath XQU  +  {\mathfrak R}' (0)\big) 
\end{align}
vaut identiquement zéro. En effet, les intégrales sur les variables $u$ et $w$ sont impaires. $A_{0,2}$ est plus compliqué :
\begin{align}\label{eqtwo2}
A_{0,2}=& \frac 12 \int dx [\phi^{\les e}(x)]^2\,e^{\imath (URU+ USW)}
\prod_{\ell \in\cL^{i}_{k}}du_{\ell}dw_{\ell}C_{\ell}(u_\ell, w_\ell)\notag\\
&\prod_{l\in\cT^{i}_{k}}du_{l}C_{l}(u_{l}, U_l, W_l)\Big( -( XQU)^2- 4\imath XQU {\cal A}X . (W + U )\notag\\
&  -2 \cA X . X +4  [\cA X . (W + U )] ^2 \Big).
\end{align}
Les quatre termes en $(XQU)^2$, $XQU\cA X.W$, $\cA X . X$ et $[{\cal A}X . W ]^2$ sont logarithmiquement divergents et contribuent à la renormalisation de la fréquence harmonique $\Ot$ dans (\ref{action}). Les termes en $x^\mu x^\nu$ avec $\mu\ne\nu$ sont nuls par parité et les termes en $(x^\mu)^2$ ont le même coefficient. Les autres termes $XQU\cA X.U$,
 $({\cA}X . U)({\cA}X . W)$, $[{\cA}X . U ]^2 $ et $A_{0,R}$ sont inessentiels.

Pour les termes en $A_{0}(y)$ (contenant $\int dx [\phi^{\les e}(y)]^2$) nous effectuons une analyse similaire mais cette fois-ci la fonction delta sert à intégrer sur $x$ si bien que $Q$, $S$, $R$ et ${\mathfrak R}$ changent mais pas la conclusion.\\
 
Puis nous considérons le terme en $\big[ (y-x)^\mu\nabla_\mu \phi^{\les e}(x) \big]^2$ dans (\ref{symdev}) pour lequel nous ne développons la fonction $f$ qu'au premier ordre. L'intégration sur $y$ remplace $y-x$ par un facteur ${\mathfrak U}$ :
\begin{align}
A_{1}=\frac 12\int& dx\, \big[ {\mathfrak U}^\mu\nabla_\mu \phi^{\les e}(x)\big]^2 
\,e^{\imath (URU+ USW)}\prod_{\ell \in\cL^{i}_{k}}du_{\ell}dw_{\ell}C_{\ell}(u_\ell, w_\ell)\notag\\
&\prod_{l\in\cT^{i}_{k}}du_{l}C_{l}(u_{l}, U_l, W_l) \big( f(0)+\int_0^1 ds  f'(s)\big).
\end{align}
Le premier terme est
\begin{align}
A_{1,0}=\frac 12\int& dx\,\big[ {\mathfrak U}^\mu\nabla_\mu \phi^{\les e}(x) \big]^2 
\,e^{\imath (URU+ USW)}\prod_{\ell \in\cL^{i}_{k}}du_{\ell}dw_{\ell}C_{\ell}(u_\ell, w_\ell)\notag\\
&\prod_{l\in\cT^{i}_{k}}du_{l}C_{l}(u_{l}, U_l, W_l).
\end{align}
Les termes avec $\mu \ne \nu$ sont nuls par parité. Les autres constituent un contreterme proportionnel au laplacien. Le comptage de puissance de ce facteur $A_{1,0}$ est amélioré, par rapport à $A$, d'au moins $M^{-2(i-e)}$ ce qui le rend logarithmiquement divergent comme nous nous y attendons pour le contreterme de fonction d'onde. Le terme de reste $A_{1,R}^{x}$ a un facteur supplémentaire $M^{-(i-e)}$ qui vient de $\int_0^1 ds  f'(s)$. Il est donc convergent. Finalement les termes dans $A_{R}$ avec trois ou quatre gradients dans (\ref{symdev}) sont également convergents. En effet, les améliorations sont de plusieurs types :
\begin{itemize}
\item Il y a des termes en $\kU^3$ avec $\nabla^3 $. Les gradients agissent sur la variable externe $x$ et apportent $M^{3e}$ alors que $\mathfrak{U}^3$ donne au moins $M^{-3i}$.
\item Enfin les termes avec quatre gradients sont encore plus petits.
\end{itemize}
Ainsi le comptage de puissance de l'amplitude renormalisée $A_{R}$ est amélioré, par rapport à $A_{0}$, d'un facteur $M^{-3(i-e)}$ et devient convergent. Ceci achève la preuve du théorème \ref{thm:BPHZPhi4}.

\section{Un modèle LSZ modifié}
\label{sec:modele-lsz}

Dans \cite{Langmann2003if,Langmann2003cg}, E.~Langmann, R.~Szabo et K.~Zarembo ont introduit un modèle bosonique complexe dans un champ magnétique uniforme. Ce modèle est exactement soluble. Dans \cite{xphi4-05}, nous avons démontré la renormalisabilité d'un modèle LSZ modifié et quelque peu généralisé. Il consiste en une théorie bosonique scalaire complexe dans un champ magnétique uniforme plus un terme harmonique à la Grosse-Wulkenhaar. L'interaction quartique est du type Moyal. L'action est donnée par
\begin{align}
  S=&\int\frac 12\bar{\phi}\lbt -D^{\mu}D_{\mu}+\Ot^{2}x^{2}+\mu_{0}^{2}\rbt\phi
 +\lambda\,\bar{\phi}\star\phi\star\bar{\phi}\star\phi\label{eq:lszaction}
\end{align}
où $D_{\mu}=\partial_{\mu}-\imath B_{\mu\nu}x^{\nu}$ est la dérivée covariante. Le facteur $1/2$ est quelque peu inhabituel dans une théorie complexe mais nous permet d'utiliser directement le propagateur calculé dans \cite{toolbox05} avec $\Ot^{2}\rightarrow\omega^{2}=\Ot^{2}+B^{2}$. En développant la partie quadratique de l'action, nous obtenons une partie cinétique du type $\Phi^{4}$ plus un terme de moment angulaire :
\begin{align}
  \bar{\phi}D^{\mu}D_{\mu}\phi
  +\Ot^{2}x^{2}\bar{\phi}\phi=&\,\bar{\phi}\lbt\Delta -\omega^{2}x^{2}-2BL_{4}\rbt\phi
\end{align}
avec $L_{4}=x^{0}p_{1}-x^{1}p_{0}+x^{2}p_{3}-x^{3}p_{2}= x\wed\nabla$. Ici la matrice antisymétrique $B$ a été mise sous sa forme canonique\begin{equation}
  \label{eq:Bform}
  B=\begin{pmatrix}\begin{matrix}0&-1\\1&\phantom{-}0\end{matrix}&(0)\\
    (0)&\begin{matrix}0&-1\\1&\phantom{-}0\end{matrix}
    \end{pmatrix}.
\end{equation}
En espace $x$, l'interaction est la même que pour la théorie $\Phi^{4}$. Le fait d'utiliser des champs complexes sélectionne seulement les graphes orientables. À $\Ot=0$, le modèle est similaire à celui de Gross-Neveu qui est traité dans le chapitre \ref{cha:GN}. Pour $B=\theta^{-1}$, nous retrouvons le modèle intégrable initial \cite{Langmann2003if,Langmann2003cg}.\\

Le propagateur correspondant à l'action (\ref{eq:lszaction}) a été calculé dans \cite{toolbox05} à deux dimensions. La généralisation aux dimensions supérieures à deux est directe :
\begin{align}
  C(x,y)=&\int_{0}^{\infty}dt\,\frac{\omega^{2}}{(2\pi\sinh\omega
    t)^{2}}\ \exp-\frac\omega 2\lbt\frac{\cosh Bt}{\sinh\omega
    t}(x-y)^{2}\right.\label{eq:lszprop}\\
&\left.+\frac{\cosh\omega t-\cosh Bt}{\sinh\omega
    t}(x^{2}+y^{2})+\imath\frac{\sinh Bt}{\sinh\omega t}x\wed y\rbt.\nonumber
\end{align}
Remarquons que pour $\Ot\neq 0$, ce propagateur, dans une tranche, obéit à la même borne \eqref{eq:propaboundNCphi4} que le propagateur de $\Phi^{4}$ \eqref{eq:Mehler}. De plus, les phases supplémentaires $\exp\imath x\wed y$ sont de la forme $\exp\imath u_{l}\wed v_{l}$. Ces termes ne jouaient aucun rôle ni dans le comptage de puissance ni dans la renormalisation de $\Phi^{4}$. Nous pouvons donc conclure que le lemme \ref{lem:compt-puissNCPhi4} s'applique au modèle \eqref{eq:lszaction}. Nous avons également prouvé dans \cite{xphi4-05} que toutes les divergences de ce modèle sont de la forme du lagrangien initial. Il faut notamment renormaliser le moment angulaire $L_{4}$. Remarquons cependant que si nous avions considéré une théorie réelle avec dérivée covariante, qui correspondrait à un champ scalaire neutre dans un champ magnétique, le flot du moment angulaire aurait été nul. Ce terme n'aurait pas été renormalisé (la formule de symétrisation \eqref{eq:2pt-sym} aurait été applicable). Seul le potentiel harmonique aurait flotté. Il semble donc que la renormalisation ``distingue'' la vraie théorie dans laquelle c'est un champ \emph{chargé} qui doit être couplé au champ magnétique.


%% file: GN-arXiv.tex
\chapter{Le modèle de Gross-Neveu non~commutatif}
\label{cha:GN}
\epigraph{L'homme intelligent se mesure à ce qu'il ne sait pas comprendre.}{Édouard Herriot}

\section{Introduction}

En plus de la théorie $\Phi^{4}_{4}$, du modèle LSZ modifié \cite{xphi4-05} et de théories supersymétriques, nous connaissons aujourd'hui plusieurs théories des champs \ncv{}s renormalisables. Néanmoins elles sont soit super-renormalisables ($\Phi^{4}_{2}$
\cite{GrWu03-2}) soit étudiées en un point particulier de l'espace des paramètres où elles sont solubles ($\Phi^{3}_{2},\Phi^{3}_{4}, \Phi^{3}_{6}$ \cite{Grosse2005ig,Grosse2006qv,Grosse2006tc}, les modèles du type LSZ \cite{Langmann2003if,Langmann2003cg,Langmann2002ai}). Bien que seulement logarithmiquement divergent pour des raisons de parité, le modèle de Gross-Neveu \ncf{} est une théorie des champs juste renormalisable comme $\Phi^{4}_{4}$. Le fait qu'il puisse être interprété comme une théorie des champs fermionique non locale dans un champ magnétique constant est une de ses caractéristiques les plus intéressantes. Ainsi, en plus de renforcer la procédure de vulcanisation pour obtenir des théories \ncv{}s renormalisables, le modèle de Gross-Neveu pourrait s'avérer utile pour étudier l'effet Hall quantique. C'est aussi un bon premier candidat pour une étude constructive \cite{Riv1} des théories \ncv{}s, les modèles fermioniques étant généralement plus simples à construire que les théories bosoniques. Enfin sa version commutative étant asymptotiquement libre et présentant une génération spontanée de masse \cite{Mitter1974cy,Gross1974jv,KMR}, une étude de la \og{}physique\fg{} de ce modèle serait intéressante.\\

Dans ce chapitre, nous démontrons la renormalisabilité perturbative du modèle de Gross-Neveu \ncf{} à tous les ordres \cite{RenNCGN05}. Pour des raisons techniques uniquement, nous nous restreindrons au cas orientable. Remarquons dès à présent que les graphes non orientables sont convergents dans $\Phi^{4}$ et ne jouent donc aucun rôle dans la renormalisation. Le calcul de quelques graphes nous incite à penser que c'est aussi le cas dans le modèle de Gross-Neveu. Il reste cependant à le prouver à tous les ordres.

Le modèle présente du mélange UV/IR même après la vulcanisation. Celui-ci apparaît comme un couplage entre les échelles du graphe : certains graphes convergents de la fonction à quatre points deviennent (logarithmiquement) divergents quand ils sont insérés dans un graphe à deux points. Ces sous-graphes à quatre points ne sont pas renormalisables par un contreterme \og{}local\fg{}\footnote{Par \og{}local\fg{} nous entendons \og{}de la forme du vertex initial\fg{}.}. Cependant ces composantes \emph{critiques} sont régularisées par la renormalisation de la fonction à deux points correspondante. Ainsi malgré ce mélange, le modèle est renormalisable. Mais le modèle massif nécessite l'introduction d'un contreterme de la forme $\delta m\,\psib\imath\gamma^{0}\gamma^{1}\psi$. Le modèle à masse nulle est renormalisable sans ce contreterme.\\

Dans la section \ref{sec:model-notationsGN}, nous présentons le modèle et fixons les notations. Nous énonçons le théorème principal (BPHZ). La section \ref{sec:from-oscill-decr} est dédiée à la principale difficulté technique de la preuve. Dans la section \ref{sec:multiscaleGN}, nous donnons le comptage de puissance grâce à une analyse multi-échelles. Dans la section \ref{sec:renorm-GN}, nous démontrons que tous les sous-graphes divergents sont renormalisables par des contretermes de la forme du lagrangien initial. Enfin l'appendice \ref{sec:a-propos-GN} contient certains détails techniques et la preuve de l'invariance par translation des graphes du vide orientables.

\section{Modèle et notations}
\label{sec:model-notationsGN}

Le modèle de Gross-Neveu \ncf{} ($\GN$) est une théorie des champs fermionique en interaction quartique sur le plan de Moyal $\R^{2}_{\Theta}$. La matrice antisymétrique $\Theta$ est donnée par
\begin{align}
  \Theta=&
  \begin{pmatrix}
    0&-\theta\\\theta&0
  \end{pmatrix}.
\end{align}
L'action du modèle est
\begin{align}\label{eq:actfunctGN}
  S[\psib,\psi]=&\int
  dx\lbt\psib\lbt-\imath\slashed{\partial}+\Omega\xts+m+\imath\delta m\,\theta\gamma\Theta^{-1}\gamma\rbt\psi+V_{\text{o}}(\psib,\psi)
  +V_{\text{no}}(\psib,\psi)\rbt(x)
\end{align}
où $\xt=2\Theta^{-1}x$ et $V=V_{\text{o}}+V_{\text{no}}$ est la partie interaction donnée ci-après. Le terme en $\delta m$ sera traité perturbativement comme un contreterme. Il apparait à l'ordre de deux boucles (voir section \ref{critic-comp-2pts}). Nous utilisons la métrique euclidienne et la notation de Feynman $\slashed{a}=\gamma^{\mu}a_{\mu}$. Les matrices $\gamma^{0}$ et $\gamma^{1}$ constituent une représentation bidimensionnelle de l'algèbre de Clifford $\{\gamma^{\mu},\gamma^{\nu}\}=-2\delta^{\mu\nu}$. Remarquons que les $\gamma^{\mu}$ sont alors anti-hermitiens : $\gamma^{\mu\dagger}=-\gamma^{\mu}$.

\paragraph{Propagateur}
Le propagateur correspondant à l'action \eqref{eq:actfunctGN} est donné par le lemme suivant :
\begin{lemma}[Propagateur 1 \cite{toolbox05}]\label{xpropa1GN}
  Le propagateur du modèle de Gross-Neveu est
  \begin{align}
    C(x,y)=&\int d\mu_{C}(\psib,\psi)\,\psi(x)\psib(y)=\lbt-\imath\slashed{\partial}+\Omega\xts+m\rbt^{-1}(x,y)\\
    =&\ \int_{0}^{\infty}dt\, C(t;x,y),\notag\\
    C(t;x,y)=&\ -\frac{\Omega}{\theta\pi}\frac{e^{-tm^{2}}}{\sinh(2\Ot t)}\,
    e^{-\frac{\Ot}{2}\coth(2\Ot t)(x-y)^{2}+\imath\Omega x\wed y}\\
    &\times\lb\imath\Ot\coth(2\Ot t)(\xs-\ys)+\Omega(\xts-\yts)-m\rb
    e^{-2\imath\Omega t\gamma\Theta^{-1}\gamma}\notag
  \end{align}
  avec $\Ot=\frac{2\Omega}{\theta}$ et $x\wed y=2x\Theta^{-1}y$.\\
Nous avons aussi $e^{-2\imath\Omega t\gamma\Theta^{-1}\gamma}=\cosh(2\Ot t)\mathds{1}_{2}-\imath\frac{\theta}{2}\sinh(2\Ot
  t)\gamma\Theta^{-1}\gamma$.
\end{lemma}
Si nous voulons étudier un modèle à $N$ \emph{couleurs}, nous pouvons considérer ce propagateur diagonal dans les indices de couleur.

\paragraph{Interactions}
Concernant la partie interaction $V$, rappelons tout d'abord (voir le corollaire \ref{cor:int-Moyal}) que $\forall
f_{1},f_{2},f_{3},f_{4}\in\cA_{\Theta}$,
\begin{align}
  \int dx\,\lbt f_{1}\star f_{2}\star f_{3}\star
  f_{4}\rbt(x)=&\frac{1}{\pi^{2}\det\Theta}\int\prod_{j=1}^{4}dx_{j}f_{j}(x_{j})\,
  \delta(x_{1}-x_{2}+x_{3}-x_{4})e^{-\imath\varphi},\label{eq:interaction-GN}\\
  \varphi=&\sum_{i<j=1}^{4}(-1)^{i+j+1}x_{i}\wed x_{j}.
\end{align}
Ce produit est non local et seulement invariant par permutations cycliques. Ainsi, au contraire du modèle de Gross-Neveu commutatif pour lequel il n'existe qu'une seule interaction (locale) possible, le modèle $\GN$ a, au moins, six interactions différentes :
les interactions \emph{orientables} 
\begin{subequations}\label{eq:int-orient}
  \begin{align}
    V_{\text{o}}=\phantom{+}&\frac{\lambda_{1}}{4}\sum_{a,b}\int
    dx\,\lbt\psib_{a}\star\psi_{a}\star\psib_{b}\star\psi_{b}\rbt(x)\label{eq:int-o-1}\\
    +&\frac{\lambda_{2}}{4}\sum_{a,b}\int
    dx\,\lbt\psi_{a}\star\psib_{a}\star\psi_{b}\star\psib_{b}\rbt(x)\label{eq:int-o-2}\\
    +&\frac{\lambda_{3}}{4}\sum_{a,b}\int
    dx\,\lbt\psib_{a}\star\psi_{b}\star\psib_{a}\star\psi_{b}\rbt(x),&\label{eq:int-o-3}
  \end{align}
\end{subequations}
où les $\psi$ alternent avec les $\psib$ et les interactions \emph{non orientables} 
\begin{subequations}\label{eq:int-nonorient}
  \begin{align}
    V_{\text{no}}=\phantom{+}&\frac{\lambda_{4}}{4}\sum_{a,b}\int
    dx\,\lbt\psib_{a}\star\psib_{b}\star\psi_{a}\star\psi_{b}\rbt(x)\label{eq:int-no-1}&\\
    +&\frac{\lambda_{5}}{4}\sum_{a,b}\int
    dx\,\lbt\psib_{a}\star\psib_{b}\star\psi_{b}\star\psi_{a}\rbt(x)\label{eq:int-no-2}\\
    +&\frac{\lambda_{6}}{4}\sum_{a,b}\int
    dx\,\lbt\psib_{a}\star\psib_{a}\star\psi_{b}\star\psi_{b}\rbt(x).\label{eq:int-no-3}
  \end{align}
\end{subequations}
Toutes ces interactions ont le même noyau en espace $x$ par l'équation \eqref{eq:interaction-GN}. Les indices $a,b$ sont les indices de spin et prennent des valeurs dans $\{0,1\}$ (ou $\{\uparrow,\downarrow\}$). Ils peuvent aussi contenir des indices de couleur entre $1$ et $N$. Pour des raisons uniquement techniques, nous nous restreindrons aux interactions orientables. Une telle qualification devrait être claire grâce au chapitre \ref{cha:le-modele-phi4_4} et en particulier la section \ref{sec:graph-orientNCphi4}. En effet, à chaque vertex, les signes $+$ et $-$ (de la fonction delta) alternent. Il en est de même pour les champs $\psi$ et $\psib$. Or un $\psi$ ne peut contracter qu'à un $\psib$. Ainsi il est toujours possible de choisir une orientation telle que le graphe soit orientable. Ce n'est évidemment pas le cas pour les interactions non orientables. En fait, tous les graphes construits avec les interactions orientables sont orientables mais tous les graphes orientables ne sont pas faits d'interactions orientables. En effet, les interactions non orientables produisent non seulement tous les graphes non orientables mais aussi des graphes orientables. Ce chapitre est principalement dédié à la preuve du
\begin{thm}[BPHZ pour $\GN$]\label{thm:BPHZGN}
  La théorie quantique des champs définie par l'action \eqref{eq:actfunctGN} avec $V=V_{\text{o}}$ est renormalisable à tous les ordres de perturbation.
\end{thm}

\paragraph{Analyse multi-échelles} Encore une fois, nous utiliserons l'analyse multi-échelles \cite{Riv1}. Nous \og{}découpons\fg{} le propagateur :
\begin{align}
  C_{l}=&\sum_{i=0}^{\infty}C^{i}_{l},\,C^{i}_{l}=
  \begin{cases}
    {\displaystyle\int_{M^{-2i}}^{M^{-2(i-1)}}dt\,C_{l}(t;\phantom{u})}&\text{if }i\ges 1\vspace{.3cm}\\
     {\displaystyle\int_{1}^{\infty}dt\,C_{l}(t;\phantom{u})}&\text{if }i=0.\\
  \end{cases}
\end{align}
Le lemme suivant donne une borne sur le propagateur $C^i$ dans chaque tranche $i$.
\begin{lemma}
  Pour tout $i\in\N$, il existe $K,\,k\in\R_{+}$ tels que
  \begin{align}
    \label{eq:propaboundGN}
    \labs C^{i}(x,y)\rabs\les&K M^{i}e^{-kM^{i}\labs x-y\rabs}.
  \end{align}
  Cette borne est également valable si $m=0$ (et $\Omega\neq 0$).
\end{lemma}
Encore une fois, tout comme pour le modèle $\Phi^{4}$ (voir chapitre \ref{cha:un-theoreme-bphz}), nous ne démontrons pas directement le théorème \ref{thm:BPHZGN} mais la finitude ordre par ordre de la série effective.

\paragraph{Orientation et variables d'un graphe}
\label{sec:orient-et-variGN}

Nous utiliserons toutes les définitions de la section \ref{sec:graph-orientNCphi4}. Notons cependant les différences suivantes. Nous avons muni chaque ligne d'un graphe d'un signe (voir définition \ref{defn:signe}). Celui-ci est $+$ si, en tournant autour de l'arbre, la ligne va d'un $-$ à un $+$. Il est $-$ si c'est le contraire. Dans un modèle avec champs complexes, nous pouvons définir un autre signe \emph{à priori} indépendant du premier. Il vaudra $+$ si la ligne va d'un $\psi$ vers un $\psib$ et $-$ si c'est le contraire :

\begin{defn}[Signe d'une ligne $2$]\label{defn:signeGN} 
  Soient $i<j$. Pour toute ligne $l=(i,j)\in\cT\cup\cL$,
  \begin{align}
    \epsilon(l)=&
    \begin{cases}
      +1&\text{si }\psi(x_{i})\psib(x_{j})\\
      -1&\text{si }\psib(x_{i})\psi(x_{j}).
    \end{cases}\notag
  \end{align}
\end{defn}
Remarquons que si le graphe est orientable, pour toute ligne $l\in G$, $\veps(l)=\epsilon(l)$.

\begin{cor}[Propagateur 2]\label{xpropa2GN}
  Avec les définitions \ref{defn:longshort}, \ref{defn:signe} et \ref{defn:signeGN}, le
  propagateur correspondant à la ligne $l$ s'écrit
  \begin{align}
    C_{l}(u_{l},v_{l})=&\ \int_{0}^{\infty}dt_{l}\, C(t_{l};u_{l},v_{l})\\
    C(t_{l};u_{l},v_{l})=&\ 
    \frac{\Omega}{\theta\pi}\frac{e^{-t_{l}m^{2}}}{\sinh(2\Ot t_{l})}\,
    e^{-\frac{\Ot}{2}\coth(2\Ot
      t_{l})u_{l}^{2}-\imath\frac{\Omega}{2}\epsilon(l)\veps(l)u_{l}\wed
      v_{l}}
    \label{xfullpropGN}\\ 
    &\times\lb\imath\Ot\coth(2\Ot
    t_{l})\epsilon(l)\veps(l)\us_{l}+\Omega\epsilon(l)\veps(l)\uts_{l}+m\rb
    e^{-2\imath\Omega t_{l}\gamma\Theta^{-1}\gamma}\nonumber
  \end{align}
  avec $\Ot=\frac{2\Omega}{\theta}$ et où $v_{l}$ sera remplacé par $w_{\ell}$
  si le propagateur correspond à une ligne de boucle.
\end{cor}

\paragraph{Les fonctions delta}
\label{sec:les-fonctions-deltaGN}

Nous échangeons les fonctions delta de vertex initiales pour le système de branches défini en \ref{sec:resol-deltaNCPhi4}. Toutefois, au contraire de la théorie $\Phi^{4}$, nous ne résoudrons pas ces nouvelles fonctions delta. Pour le modèle de Gross-Neveu, nous devons utiliser les oscillations de vertex et de propagateur avec soin. Il se trouve qu'il est plus pratique d'exprimer les fonctions delta comme des intégrales oscillantes. Pour toute ligne $l$ d'un graphe orientable, nous écrirons
\begin{align}
  \label{eq:delta-int0}
  \delta_{b(l)}\Big(\sum_{l'\in\kb(l)}u_{l'}+\sum_{e\in\cX(l)}\eta(e)x_{e}\Big)=&\int\frac{d^2p_{l}}{(2\pi)^2}\,e^{\imath p_{l}\cdot(\sum_{l'\in\kb(l)}u_{l'}+\sum_{e\in\cX(l)}\eta(e)x_{e})}.
\end{align}
Bien sûr, résoudre les longues variables de l'arbre étant la façon opimale d'utiliser les fonctions delta, nous voudrions garder le fait que l'intégration sur les $v_{l}$ soit d'ordre $\cO(1)$. Après quelques manipulations sur les oscillations \eqref{eq:delta-int0} (voir la section \ref{sec:masselottes}), nous obtiendrons des décroissances pour les variables $v_{l}$ et $p_{l}$. Pour toute ligne d'arbre $l$, nous intègrerons sur $v_{l}$ et $p_{l}$, le résultat sera borné par $\cO(1)$.

\section{Des oscillations aux décroissances}
\label{sec:from-oscill-decr}

Cette section a pour but d'expliquer comment utiliser les oscillations de vertex et de propagateurs pour obtenir assez de décroissances pour intégrer sur toutes les variables internes du graphe. Après avoir exploité correctement ces oscillations, nous prendrons la valeur absolue de l'amplitude pour obtenir une borne supérieure qui constituera le comptage de puissance.

\subsection{Les masselottes}
\label{sec:masselottes}

Contrairement au cas $\Phi^{4}$ \cite{xphi4-05}, le propagateur $C^{i}$ du modèle de Gross-Neveu
(\ref{xfullpropGN}) ne contient pas de termes de la forme $\exp-M^{-2i}w^{2}$ que nous appellerons \textbf{masselottes}\footnote{MÉCAN. Petite masse métallique
  agissant par inertie, par gravité ou par force centrifuge, dans divers
  dispositifs. \emph{À l'extrémité de la lame est une masselotte dont le but
    est d'entretenir le plus longtemps possible les oscillations} (A. LECLERC,
  \emph{Télégr. et téléph.}, 1924, p.249), Trésor de la Langue Française
  informatisé, \href{http://www.lexilogos.com/}{http://www.lexilogos.com/}.}.
Ici la masselotte est remplacée par une oscillation du type $u\wed w$. Bien que les masselottes
ne soient pas présentes dès le départ \ie{} dans le propagateur, elles sont
créées au fur et à mesure des intégrations sur les variables internes $u$ par
un mécanisme du type
\begin{equation}
  \label{eq:masscreation}
  \int d^{2}u\, e^{-M^{2i}u^{2}+\imath u\wed w}=K M^{-2i}\, e^{-kM^{-2i}w^{2}}.
\end{equation}
Soit $G$ un graphe connexe. Son amplitude vaut
\begin{align}
  A_{G}=\int&\prod_{i=1}^{N}dx_{i}\,f_{i}(x_{i})\delta_{G}\prod_{l\in\cT}du_{l}dv_{l}\,\delta_{b(l)}C_{l}(u_{l},v_{l})
  \prod_{\ell\in\cL}du_{\ell}dw_{\ell}\,C_{\ell}(u_{\ell},w_{\ell})e^{i\varphi}.
\end{align}
Les points $x_{i},\,i\in\lnat 1,N\rnat$ sont les positions externes.
Concernant les fonctions delta, nous avons utilisé les notations de la section
\ref{sec:resol-deltaNCPhi4}. L'oscillation totale $\varphi$ des vertex est donnée
par le lemme \ref{exactoscill}. Il est pratique de séparer le propagateur en
deux parties. Nous définissons, pour toute ligne $l\in G$,
$\mathbf{\bar{C}}_{l}(u_{l})$ par
$C_{l}(u_{l},v_{l})=\bar{C}_{l}(u_{l})\,
e^{-\imath\frac{\Omega}{2}\epsilon(l)\veps(l)u_{l}\wed v_{l}}$. Il faudra
remplacer $v$ par $w$ pour les lignes de boucles. Cette séparation nous permet
de regrouper les oscillations des propagateurs avec celles des vertex.
L'oscillation totale $\varphi_{\Omega}$ se déduit simplement de $\varphi$ en
remplaçant les termes $\frac 12\veps(l)v_{l}\wed u_{l}$ par
$\frac 12\veps(l)(1+\epsilon(l)\Omega)v_{l}\wed u_{l}$. Là encore, nous
remplacerons $v$ par $w$ pour les termes relatifs aux boucles. L'amplitude
d'un graphe devient alors
\begin{align}
  A_{G}=\int&\prod_{i=1}^{N}dx_{i}\,f_{i}(x_{i})\delta_{G}\prod_{l\in\cT}du_{l}dv_{l}\,\delta_{b(l)}\bar{C}_{l}(u_{l})
  \prod_{\ell\in\cL}du_{\ell}dw_{\ell}\,\bar{C}_{\ell}(u_{\ell})e^{i\varphi_{\Omega}}.
\end{align}
Au contraire de la théorie $\Phi^4$, nous ne résoudrons pas les fonctions delta de branches. À la place, nous gardons $\delta_{G}$ mais exprimons les $n-1$ autres fonctions delta par des intégrales oscillantes :
\begin{align}
  \label{eq:delta-int}
  \delta_{b(l)}\Big(\sum_{l'\in\kb(l)}u_{l'}+\sum_{e\in\cX(l)}\eta(e)x_{e}\Big)=&\int\frac{d^2p_{l}}{(2\pi)^2}\,e^{\imath p_{l}\cdot(\sum_{l'\in\kb(l)}u_{l}+\sum_{e\in\cX(l)}\eta(e)x_{e})}.
\end{align}
Dans la section \ref{sec:resol-deltaNCPhi4}, nous avons vu qu'il existe $e_{l}\in\cX(l)$ tel que $x_{e_{l}}=\frac
12(\eta(e_{l})u_{l}+v_{l})$. Remarquons que $\eta(e_{l})=\veps(l)$. Ainsi
\begin{align}
  \sum_{l'\in\kb(l)}u_{l'}+\sum_{e\in\cX(l)}\eta(e)x_{e}=&\frac 12(u_{l}+\veps(l)v_{l})+\sum_{l'\in\kb(l)}u_{l'}+\sum_{e\in\cX(l)\setminus\{e_{l}\}}\eta(e)x_{e}.
\end{align}
Dans la suite nous utiliserons une notation suplémentaire. Pour toute ligne $l\in\cT$, nous définissons $\nu_{l}$ comme l'unique vertex tel que $l=l_{\nu}$ où $l_{\nu}$ est défini dans la section \ref{sec:resol-deltaNCPhi4}. $\nu_{l}$ est le vertex juste au-dessus de $l$ dans l'arbre. Nous écrirons $\varphi'_{\Omega}$ pour l'oscillation totale où nous ajoutons les nouvelles oscillations provenant des fonctions delta\footnote{Notons que ces oscillations sont invariantes sous $p_{l}\to -p_{l}$ pour toute ligne $l\in G$ indépendamment.}. L'amplitude du graphe s'écrit
\begin{align}
  A_{G}=\int&\prod_{i=1}^{N}dx_{i}\,f_{i}(x_{i})\delta_{G}\prod_{l\in\cT}du_{l}dv_{l}dp_{l}\,\bar{C}_{l}(u_{l})
  \prod_{\ell\in\cL}du_{\ell}dw_{\ell}\,\bar{C}_{\ell}(u_{\ell})e^{i\varphi'_{\Omega}}.\label{eq:amplitude-avt-massel}
\end{align}
Remarquons que nous avons omis les facteurs $2\pi$ tout comme nous
l'avons fait avec les facteurs de vertex
$\frac{-\lambda}{4\pi^{2}\det\Theta}$. Pour obtenir les masselottes,
nous pourrions tenter d'intégrer sur les variables $u_{l}$. Ce calcul
exact serait l'équivalent de l'équation \eqref{eq:masscreation}. Il
faudrait intégrer $2n-N/2$ gaussiennes couplées où $n$ est le nombre
de vertex du graphe. Nous obtiendrions des gaussiennes en des
variables $\cW_{\ell}$ qui seraient des combinaisons linéaires des
variables $w_{\ell'}$. Outre la difficulté inhérente au calcul
lui-même, il faudrait ensuite démontrer que les décroissances obtenues
sont indépendantes. Pour des graphes généraux, ceci est relativement
difficile. Ainsi, plutôt que d'effectuer un calcul exact, nous allons
contourner la difficulté en exploitant les oscillations de propagateur
et de vertex avant d'intégrer sur les variables $u_{l}$, $v_{l}$ et
$w_{\ell}$. La suite de cette section est dédiée à la preuve du
\begin{lemma}\label{lem:masselottes}
  Soit $G$ un graphe orientable à $n$ vertex et $\mu$ une attribution d'échelles. Pour tout $\Omega\in\lsb 0,1\right[$, il existe
  $K\in\R$ tel que l'amplitude \eqref{eq:amplitude-avt-massel}, amputée,
  intégrée contre des fonctions tests, avec l'attribution $\mu$ soit bornée uniformément en $n$ par
  \begin{align}
    \labs A^{\mu}_{G}\rabs\les&K^{n}\int
    dx_{1}\,g_{1}(x_{1}+\{a\})\delta_{G}\prod_{i=2}^{N}dx_{i}\,g_{i}(x_{i})\prod_{l\in G}da_{l}\,M^{2i_{l}}\Xi(a_{l})\\
    &\qquad\prod_{l\in\cT}du_{l}d\cV_{l}dp_{l}\,M^{i_{l}}e^{-M^{2i_{l}}(u_{l}-\veps(l)a_{l})^{2}}\prod_{\mu=0}^{1}\frac{1}{1+M^{-2i_{l}}\cV^{2}_{l,\mu}}
    \frac{1}{1+M^{2i_{l}}p^{2}_{l,\mu}}\notag\\
    &\qquad\prod_{\ell\in\cL}
    du_{\ell}d\cW_{\ell}M^{i_{\ell}}
    e^{-M^{2i_{\ell}}(u_{\ell}+\{a\})^{2}}
    \prod_{\mu=0}^{1}\frac{1}{1+M^{-2i_{\ell}}\cW^{2}_{\ell,\mu}}\notag\\
    \text{avec }\veps(l)\cV_{l}=&{\textstyle\frac
    12}(1+\epsilon(l)\Omega)\veps(l)v_{l}+\sum_{\ell'\supset l}\veps(\ell')w_{\ell'}-{\textstyle\frac 12}\pt_{l}-\sum_{l'\in\cP_{\kv_{l}}}\pt_{l'},\label{eq:Vl}\\
    \veps(\ell)\cW_{\ell}=&{\textstyle\frac
    12}(1+\epsilon(\ell)\Omega)\veps(\ell)w_{\ell}+\sum_{\ell'\supset\ell}\veps(\ell')w_{\ell'}+
    \sum_{\ell'\ltimes\ell}\veps(\ell')w_{\ell'}\label{eq:Wl}
  \end{align}
  et $\pt=\frac 12\Theta p$, $g_{i},\,i\in\lnat 1,N\rnat$ et $\Xi$ sont des fonctions test telles que $\|g_{i}\|\les\sup_{0\les p\les 2}\|f^{(p)}_{i}\|$.
\end{lemma}

\medskip Rappelons que nous nous restreignons aux graphes
\emph{orientables}. Nous introduisons une fonction de Schwartz $\xi\in\cS(\R^{2})$ qui va mimer la
décroissance du propagateur sur une échelle $M^{-i_{l}}$. Il est clair que
cette fonction, si elle remplaçait le propagateur, aurait le même effet \ie{}
créerait une masselotte en $w$\footnote{Nous donnons ici un exemple utilisant
  une \og borne\fg{} sur le propagateur mais nous faisons ensuite le calcul
  exact.} :
\begin{align}
  \int du_{l}\, M^{2i_{l}}\xi(u_{l}M^{i_{l}})e^{\imath u_{l}\wed
    w_{l}}=\hat{\xi}(M^{-i_{l}}\wt_{l})
\end{align}
où $\hat{\xi}$ est la transformée de Fourier de $\xi$. Or $\cS$ étant stable
par transformation de Fourier, $\hat{\xi}(M^{-i}\wt_{l})$ est bien une
fonction à décroissance rapide sur une échelle $M^{-i_{l}}$.

Rappelons que nous voulons obtenir une décroissance en $v_{l}$ sans
intégrer sur $u_{l}$. Nous utilisons
\begin{align}
  1=&\int d^{2}a_{l}\, \coth(2\Ot t_{l})\xi(a_l\coth^{1/2}(2\Ot t_{l})).\label{eq:un}
\end{align}
Le couplage entre ce $1$ et le reste du graphe se fait par l'intermédiaire
d'un changement de variables \emph{ad hoc}. Pour le déterminer, nous avons deux
impératifs. Nous devons d'une part obtenir des décroissances indépendantes et
d'autre part il faut que pour toute ligne $l$, la décroissance
correspondante soit d'échelle\footnote{Dans certains cas particuliers, une ligne de
  boucle aura une masselotte d'échelle supérieure à son propre indice. Ces cas
  forment une seule classe de graphes que nous caractériserons en détail plus
  loin (voir section \ref{subsec:brokenfaces}).} $M^{i_{l}}\simeq\coth^{1/2}(2\Ot t_{l})$.\\

Nous allons fabriquer les masselottes ligne par ligne. Soient $x_{1}$ la position racine et
$l$ une ligne d'arbre. Nous effectuons le changement de variables
\begin{align}
  \lb
  \begin{aligned}
    u_{l}\to&u_{l}-\veps(l)a_{l},\\
    x_{1}\to&x_{1}+\eta(1)\veps(l)a_{l}.
  \end{aligned}\right.\label{eq:chgtvarVl}
\end{align}
Il n'est pas difficile de vérifier que $\varphi'_{\Omega}\to\varphi'_{\Omega}+a_{l}\wed\cV_{l}+a_{l}\wed(U_{l}+A_{l}+X_{l})$
où $\cV_{l}$ est donné par (\ref{eq:Vl}) et $U_{l}$, $A_{l}$ et
$X_{l}$ sont respectivement des combinaisons linéaires de variables $u$, $a$ et
$x$ externes. Notons que ce changement de variables laisse la fonction delta globale inchangée.
En écrivant seulement les termes de l'amplitude $A_{G}$ dépendant de $a_{l}$, on a
\begin{align}
  A_{G,l}=\int&
  da_{l}\int_{M^{-2i_{l}}}^{M^{-2(i_{l}-1)}}dt_{l}\,\coth(2\Ot
  t_{l})\xi(a_{l}\coth^{1/2}(2\Ot t_{l}))\\
  &\lb\imath\Ot\coth(2\Ot
  t_{l})(\epsilon\veps)(l)(\us_{l}-\veps(l)\slashed{a}_{l})+\Omega(\epsilon\veps)(l)
  (\uts_{l}-\epsilon(l)\slashed{\tilde{a}}_{l})-m\rb\notag\\
  &e^{-\frac{\Ot}{2}\coth(2\Ot
    t_{l})(u_{l}-\veps(l)a_{l})^{2}}f_{1}(x_{1}+\eta(1)\veps(l)a_{l})\,
 	e^{\imath a_{l}\wed(\cV_{l}+U_{l}+A_{l}+X_{l})}\notag\\
  =\int&
  da_{l}dt_{l}\,\coth(2\Ot
  t_{l})\xi(a_{l}\coth^{1/2}(2\Ot t_{l}))\,
  e^{-\frac{\Ot}{2}\coth(2\Ot
    t_{l})(u_{l}-\veps(l)a_{l})^{2}}f_{1}(x_{1}+\eta(1)\veps(l)a_{l})\notag\\
  &\hspace{-.3cm}\lb\imath\Ot\coth(2\Ot
  t_{l})(\epsilon\veps)(l)(\us_{l}-\veps(l)\slashed{a}_{l})+\Omega(\epsilon\veps)(l)
  (\uts_{l}-\veps(l)\slashed{\tilde{a}}_{l})-m\rb e^{\imath a_{l}\wed(U_{l}+A_{l}+X_{l})}\nonumber\\
  &\prod_{\mu=0}^{1}\lbt\frac{\coth^{1/2}(2\Ot t_{l})+\frac{\partial}{\partial
      a_{l}^{\mu}}}{\coth^{1/2}(2\Ot
    t_{l})+\imath\cVt_{l,\mu}}\rbt^{\!\!\!2}  e^{\imath a_{l}\wed\cV_{l}}.\label{eq:int-part}
  \intertext{Nous intégrons par parties sur $a_{l}$. Les termes de bords sont nuls. Nous ne donnons ici que l'ordre de grandeur du résultat. Les détails du calcul sont reportés à l'annexe \ref{sec:integration-parts}.}
  A_{G,l}\simeq\int&da_{l}dt_{l}\,\coth(2\Ot t_{l})e^{\imath
    a_{l}\wed\cV_{l}}\prod_{\mu=0}^{1}\lbt\frac{1}{\coth^{1/2}(2\Ot
    t_{l})+\imath\cVt_{l,\mu}}\rbt^{\!\!\!2}\Xi(a_{l}\coth^{1/2}(2\Ot t_{l}))\notag\\
  &e^{-\frac{\Ot}{2}\coth(2\Ot
    t_{l})(u_{l}-\veps(l)a_{l})^{2}}e^{\imath a_{l}\wed(U_{l}+A_{l}+X_{l})} g_{1}(x_{1}+\eta(1)\veps(l)a_{l})\,\cO\big(\coth^{5/2}(2\Ot
  t)\big).\label{eq:intpart-result}
  \intertext{Nous obtenons ainsi la borne suivante}
  \labs A_{G,l}\rabs\les&
  KM^{-i_{l}}e^{-kM^{2i_{l}}(u_{l}-\veps(l)a_{l})^{2}}g_{1}(x_{1}+\eta(1)\veps(l)a_{l})
  \prod_{\mu}\frac{1}{1+M^{-2i_{l}}\cV^{2}_{l,\mu}}.\label{eq:borne-interm}
\end{align}

Nous expliquons maintenant comment obtenir les décroissances associées aux variables $p_{l}$. Nous commençons par effectuer le changement de variables $v_{l}\to\cV_{l}$ pour toute ligne d'arbre $l$. Le jacobien vaut $2^{-(n-1)}\prod_{l\in\cT}(1+\epsilon(l)\Omega)$. Il est non nul quelque soit $\Omega\in\lsb 0,1\right[$. L'oscillation totale devient
\begin{align}
  \label{eq:oscillapresVl}
  \varphi'_{\Omega}=&\varphi_{E}+\varphi_{X}+\varphi_{W}+\sum_{\cT}\veps(l)\cV_{l}\wed (u_{l}-\veps(l)a_{l})+\sum_{\cT}p_{l}\!\cdot\!(\sum_{l'\in\cL\cap\kb(l)}u_{l}+\sum_{e\in\cX(l)\setminus\{e_{l}\}}\eta(e)x_{e})\nonumber\\
    &+\sum_{\cT}(1+\epsilon(l)\Omega)^{-1}\veps(l)\cV_{l}\!\cdot\!p_{l}+WR_{1}P+PR_{2}P\notag\\
    &\hspace{-.4cm}+\frac 12\sum_{\cL}(1+\epsilon(\ell)\Omega)\veps(\ell)w_{\ell}\wed u_{\ell}+\frac 12\sum_{\cL\ltimes\cL}\veps(\ell)w_{\ell}\wed
    u_{\ell'}+\veps(\ell')w_{\ell'}\wed
    u_{\ell}+\sum_{\substack{\cL\subset\cL}}\veps(\ell')w_{\ell'}\wed
    u_{\ell}\nonumber\\
    &+\sum_{\substack{(\cT\cup\cL)\prec(\cT\cup\cL)}}u_{l'}\wed u_{l}+\frac
    12\sum_{\substack{\cL\ltimes\cL}}u_{\ell'}\wed u_{\ell}+AR_{3}A+AR_{4}U+AR_{5}X
\end{align}
où nous avons utilisé les notations du corollaire \ref{sec:oscillOrient} et les $R_{i},\,i\in\lnat 1,5\rnat$ sont des matrices anti-symétriques. En utilisant
\begin{align}
  e^{\imath(1+\epsilon(l)\Omega)^{-1}\veps(l)\cV_{l}\cdot p_{l}}=&\frac{M^{-i_{l}}+(1+\epsilon(l)\Omega)\veps(l)\frac{\partial}{\partial\cV_{l,\mu}}}{M^{-i_{l}}+\imath p_{l,\mu}}e^{\imath(1+\epsilon(l)\Omega)^{-1}\veps(l)\cV_{l}\cdot p_{l}}
\end{align}
et en intégrant par parties sur $\cV_{l}$, nous obtenons une décroissance en $p_{l}$ qui se comporte comme $\lbt 1+M^{2i_{l}}p_{l}^{2}\rbt^{-1}$. Nous finissons par les lignes de boucles. Nous voulons également obtenir des décroissances pour leurs longues variables $w$. Soit $\ell=(x_{\ell},x'_{\ell})\in\cL$ une ligne de boucle de $G$ avec $x_{\ell}\prec x'_{\ell}$. Nous faisons le changement\footnote{Ce changement de variables est quelque peu différent de celui que nous avons utilisé pour les lignes d'arbre (\ref{eq:chgtvarVl}). Cela permet une preuve simplifiée de l'indépendance des masselottes.}
\begin{align}
  \lb
  \begin{aligned}
    u_{\ell}\to&u_{\ell}-\veps(\ell)a_{\ell},\\
    w_{\ell}\to&w_{\ell}+a_{\ell},\\
    x_{1}\to&x_{1}+\eta(1)\veps(\ell)a_{\ell}.
  \end{aligned}\right.\label{eq:chgtvarWl}
\end{align}
Les changements concernant $u_{\ell}$ et $w_{\ell}$ correspondent à \og{}bouger\fg{} $x_{\ell}$. Il est facile de vérifier que \eqref{eq:chgtvarWl} implique $\varphi'_{\Omega}\to\varphi'_{\Omega}+a_{\ell}\wed\cW_{\ell}+a_{\ell}\wed(U_{\ell}+A_{\ell}+X_{\ell}+P_{\ell})$ où
$\cW_{\ell}$ est donné par \eqref{eq:Wl} et $U_{\ell}$, $A_{\ell}$, $X_{\ell}$ et $P_{\ell}$ sont respectivement des combinaisons linéaires de variables $u$, $a$, $x$ externes et $p$. Nous pouvons alors procéder au même type d'intégrations par parties que nous avons utilisées pour les $v_{l}$ et obtenir des bornes similaires à (\ref{eq:borne-interm}) ce qui prouve le lemme \ref{lem:masselottes}.

\paragraph{Indépendance des décroissances}
Rappelons que la procédure de fabrication des masselottes avait deux objectifs principaux. Tout d'abord nous voulions obtenir des décoissances d'indice $i_{l}$ pour toutes les variables $v_{l}\,(w_{l})$. C'est ce que nous avons obtenu avec le lemme \ref{lem:masselottes}. Ensuite nous devions obtenir des masselottes indépendantes. La procédure employée ci-dessus a été conçue pour rendre ce point transparent.\\

Dans la section \ref{sec:model-notationsGN}, la définition \ref{defn:relations} donne un moyen de munir les lignes du graphe d'un ordre partiel. Cet ordre était utile pour exprimer les oscillations de vertex en fonction des variables $u$, $v$ et $w$. Mais nous pouvons aussi définir un ordre total. Nous écrirons que $l<l'$ si la première extrémité (rencontrée en tournant autour de l'arbre dans le sens trigonométrique) de $l$ est rencontrée avant la première extrémité de $l'$. Ainsi pour toute ligne $l\in G$, $\cV_{l}\,(\cW_{l})$ ne dépend que des $v_{l'}$ et $w_{l'}$ avec $l'<l$. Soient $V\,(W)$ et $V'\,(W')$ les vecteurs contenant respectivement les variables $\veps(l)v_{l}\,(\veps(\ell)w_{\ell})$ et $\veps(l)\cV_{l}\,(\veps(\ell)\cW_{\ell})$. Soit $M^{-1}$ la matrice jacobienne du changement de variables $(\veps v\ \veps w)\to(\veps\cV\ \veps\cW)$: $(V'\ W')=M(V\ W)$. L'ordre défini ci-dessus permet de montrer que $M$ est triangulaire. Son déterminant vaut
\begin{align}
  \det M=2^{-(2n-N/2)}\prod_{l\in G}(1+\epsilon(l)\Omega).
\end{align}
Clairement $\forall\Omega\in\lsb 0,1\right [,\,\det M\neq 0$ et $M$ est inversible. Les décroissances en $\cV_{l}\,(\cW_{\ell})$ sont donc indépendantes.
\begin{rem}
  Avec les interactions non orientables (\ref{eq:int-nonorient}), nous n'avons pas (encore) été capables de trouver une procédure qui rende l'indépendance des masselottes transparente.
\end{rem}
\subsection{Non planarité}
\label{subsec:nonplanar}

Dans la section précédente, nous avons prouvé que les oscillations de vertex
et de propagateurs du modèle de Gross-Neveu permettent d'obtenir des
décroissances faibles similaires aux masselottes du propagateur de la théorie
$\Phi^{4}$ non commutative. Ici nous améliorons les décroissances dans le cas
d'un graphe non planaire. Pour cela, le lemme \ref{lem:masselottes} est
insuffisant. En effet, avant de prendre la valeur absolue de l'amplitude du
graphe, nous voulons encore exploiter les oscillations.\\

Soit $T^{-1}$ la matrice jacobienne du changement de variables $\veps w\to\veps\cW$: $W'=TW$. Définissons la matrice anti-symétrique $Q_{W}$ par $\varphi_{W}=WQ_{W}W$ où $\varphi_{W}$ est donné par le corollaire \ref{sec:oscillOrient}. Après le changement de variables $W\to W'=TW$, $\varphi_{W}=W'Q'_{W}W'$ avec $Q'_{W}=\trans{T}^{-1}Q_{W}T^{-1}$. $T$ étant inversible, le rang de $Q'_{W}$ est égal à celui de $Q_{W}$. Remarquons que $Q_{W}$ est la matrice d'intersection du graphe pour laquelle nous avons le résultat suivant $\rg
Q_{W}=2g$ \cite{gurauhypersyman,CheRoi}. Considérons un graphe non planaire. Le rang de $Q_{W}$ étant non nul, il existe une ligne de boucle $\ell$ telle que nous ayons une oscillation $\cW_{\ell}\wed\cW'_{\ell}$ avec $\cW'_{\ell}=\sum_{\ell'}Q'_{W,\ell\ell'}\cW_{\ell}+U+A+X+P$. Grâce au lemme \ref{lem:masselottes}, nous savons que $\cW_{\ell}$ décroît sur une \og{}longueur\fg{} $M^{i_{\ell}}$ avec la fonction $(1+M^{-2i_{\ell}}\cW_{\ell}^{2})^{-1}$. Par une intégration par parties semblable à \eqref{eq:int-part}, nous obtenons une décroissance en $\cW'_{\ell}$ sur une échelle $M^{-i_{\ell}}$. Cette décroissance sera utilisée pour intégrer sur un $\cW_{\ell'}$ contenu dans $\cW'_{\ell}$. Le résultat de cette intégration est d'ordre $M^{-2i_{\ell}}$ au lieu de $M^{2i_{\ell'}}$. Le gain est donc $M^{-2i_{\ell}-2i_{\ell'}}$.

\subsection{Faces brisées}
\label{subsec:brokenfaces}

Nous rappelons qu'une face brisée est une face à laquelle appartiennent des
points externes. Quand on ne considère pas de graphe du vide, il y a toujours
au moins une face brisée. Celle-ci est par définition la face externe. Dans
l'image de la rosette, les lignes qui bouclent au-dessus de points externes
forment des faces brisées supplémentaires. Celles-ci produisent des
oscillations du type $x\wed w$ (voir lemme \ref{sec:oscillOrient}). Dans le
cas planaire avec $B\ges 2$ faces brisées, nous allons nous servir de ces
oscillations pour obtenir de meilleures décroissances que celles du lemme
\ref{lem:masselottes}. Soit $Q_{X\!W}$ la matrice anti-symétrique représentant les oscillations entre les $x$ externes et les $w$. Après le changement de variables $W\to W'$, cette matrice devient
\begin{equation}
Q'_{X\!W}=Q_{X\!W}T^{-1}.
\end{equation}
Ainsi $\rg Q'_{X\!W}=\rg Q_{X\!W}$. Soit $I$ un ensemble d'entiers naturels consécutifs indiçant des
variables externes $x_{k},\, k\in I$. Celles-ci oscillent avec
les variables $w_{\ell},\,\ell\in B_{I}$ où $B_{I}$ est l'ensemble des lignes
bouclant au-dessus de ces variables. Vérifions maintenant que dans les
nouvelles variables, les $x_{k},\,k\in I$ oscillent seulement avec les
$\cW_{\ell},\,\ell\in B_{I}$. Pour cela, supposons que deux ensembles $X$
et $Y$ de variables externes oscillent avec deux ensembles $A$ et $B$
différents de lignes de boucles :
\begin{align}
  Q_{X\!W}&=
  \begin{pmatrix}
    A&0\\
    0&B
  \end{pmatrix},\quad T=
  \begin{pmatrix}
    C&0\\
    0&D
  \end{pmatrix}
\\
  Q'_{X\!W}&=Q_{X\!W}T^{-1}=
  \begin{pmatrix}
    AC^{-1}&0\\
    0&BD^{-1}
  \end{pmatrix}.
\end{align}
En effet, dans le cas planaire, les $\cW_{\ell}$ ne sont fonction que des
$w_{\ell'}$ avec $\ell'\supset\ell$. $T$ (et $T^{-1}$) est donc non
seulement triangulaire (inférieure) mais aussi diagonale par blocs. Ainsi
l'oscillation entre les variables externes $x_{k}$ et les $\cW_{\ell}$ est
\begin{equation}
  X_{I}Q_{X\!W}T^{-1}W'_{B_{I}}=\sum_{k\in
  I}\eta(k)x_{k}\wed\text{CL}(\cW_{\ell},\,\ell\in B_{I})
\end{equation}
où $\text{CL}$ signifie combinaison linéaire. Après les exemples des
masselottes et de la non planarité, il devrait être clair que cette
oscillation permet l'obtention d'une décroissance en les variables externes
d'échelle $M^{-\min_{\ell\in B_{I}}i_{\ell}}$. Si ces points sont de \og{}vrais\fg{} points externes (d'échelle $-1$, intégrés avec des fonctions test), nous utiliserons cette décroissance pour améliorer le comptage de puissance. Habituellement les points
externes sont intégrés contre des fonctions test (le résultat est d'ordre $1$) si bien que le gain est ici $M^{-2\min i_{\ell}}$.

\section{Comptage de puissance}
\label{sec:multiscaleGN}

Dans cette section, nous utilisons les décroissances obtenues dans les
sections précédentes en les adaptant au cas multi-tranches. Le lemme
\ref{lem:masselottes} stipule qu'il est possible d'obtenir $\labs\cL\rabs$
décroissances indépendantes équivalentes aux masselottes de la théorie
$\Phi^{4}$ plus $n-1$ masselottes pour les longues variables de l'arbre couplées à $n-1$ fortes décroissances (en les $p_{l}$).
La méthode employée pour obtenir le comptage de puissance de la théorie dépend de la
topologie du graphe considéré.\\

Nous ne considérons que des graphes avec au moins deux pattes externes. Les graphes du vide sont considérés en appendice \ref{sec:vacuum-graph}. Nous utilisons l'arbre de Gallavotti. Nous le parcourons des feuilles vers la
racine \ie{} de l'échelle de la coupure ultraviolette à l'échelle $0$. Soit une
composante connexe $G^{i}_{k}$ orientable. Pour toutes ses lignes, nous obtenons les masselottes par la méthode exposée dans la section \ref{sec:masselottes}. Si $G^{i}_{k}$ est planaire
régulier ($g=0$, $B=1$), nous utiliserons directement le lemme
\ref{lem:masselottes}. Si $G^{i}_{k}$ est non planaire ($g\ges 1$), nous
exploitons les oscillations du type $\cW\wed\cW$. Grâce à la procédure
expliquée dans la section \ref{subsec:nonplanar}, nous obtenons une
décroissance supplémentaire en un $\cW'_{\ell},\,\ell\in\cL^{i}_{k}$. Celle-ci
est au pire d'échelle $M^{-i}$. Nous procédons de la même façon pour toutes
les composantes connexes non planaires primitives \ie{} ne contenant pas de sous-composante non planaire. Les améliorations correspondantes sont indépendantes.\\

Si un noeud de l'arbre de Gallavotti est planaire mais possède plusieurs faces
brisées, nous regardons le nombre de pattes externes qu'il
a\footnote{\label{fn:N2B2}Il a été remarqué dans \cite{Chepelev2000hm} que
  les graphes orientables ne peuvent pas avoir $N=2$ et $B=2$. Un argument
  simple sur la rosette de Filk le montre également.}. Si $N(G^{i}_{k})\ges
6$, nous utilisons la borne du lemme \ref{lem:masselottes}. Lorsque
$N(G^{i}_{k})= 4$, le nombre de faces brisées est $1$ ou $2$.
Intéressons-nous au cas $B=2$. À l'échelle $i$, une ou plusieurs lignes
bouclent au-dessus de deux points externes notés $x'$ et $y'$. Contrairement à la théorie des champs commutative, le comptage de puissance de cette composante connexe dépend des échelles entre $i-1$ et $0$. Soit $\scP$ l'unique chemin dans l'arbre de Gallavotti-Nicol\`o reliant $G^{i}_{k}$ à $G$. S'il existe une échelle $i_{0}<i$ et une composante connexe $G^{i_{0}}_{k'}$ dans $\scP$ telle que $N(G^{i_{0}}_{k'})=2$ alors il existe des lignes d'échelles entre $i$ et $i_{0}$ joignant $x'$ à $y'$. Soient $I$ l'ensemble de ces lignes et $i_{m-1}$ l'échelle du premier noeud dans $\scP$ après $G^{i}_{k}$. Nous allons montrer que si $\card I=1$ alors $G^{i}_{k}$ est logarithmiquement divergente. Si $\card I\ges 2$ alors $G^{i}_{k}$ sera convergente comme $M^{-2(i-i_{m-1})}$. Enfin s'il n'existe pas une telle $G^{i_{0}}_{k'}$ alors $G^{i}_{k}$ sera aussi convergente comme $M^{-2(i-i_{m-1})}$.\\

\begin{figure}[htbp]
  \centering 
  \subfloat[Situation typique]{{\label{fig:newG-typ}}\includegraphics[scale=.8]{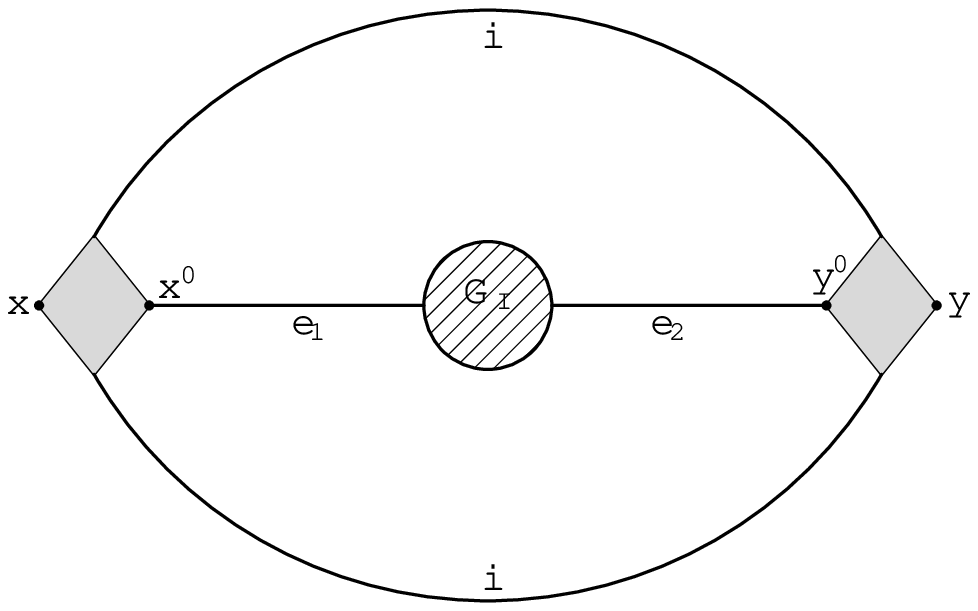}}\qquad
  \subfloat[L'insertion $I$]{\label{fig:insertion}\includegraphics[scale=.8]{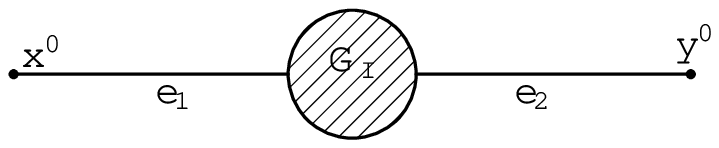}}
  \caption{Composante connexe (potentiellement) \emph{critique}}
  \label{fig:newG}
\end{figure}

Considérons la figure \ref{fig:newG} qui est plus simple que la situation générale mais présente toutes ses caractéristiques importantes. Nous définissons $I$ comme l'insertion composée des lignes $e_{1}$,
$e_{2}$ et du graphe $G_{I}$. Notons que $I$ peut être vide et $G_{I}$ non planaire. Les différentes échelles des lignes de $I$ sont $i_{0}<i_{1}<\dotsb<i_{m-1}(<i_{m}=i)$. La composante connexe d'échelle $i_{0}$ est notée $G^{i_{0}}_{k'}$. Nous écrivons également $\cL_{I}$ l'ensemble des lignes de boucles de l'insertion $I$.

\medskip
\noindent
Nous obtenons d'abord toutes les décroissances pour les variables $v_{l}$ et $p_{l}$ sauf pour la plus basse ligne d'arbre $t$ dans $I$. L'oscillation totale peut s'écrire
\begin{align}
  \varphi'_{\Omega}=&\varphi_{E}+\varphi_{X}+\sum_{\cT\setminus\{t\}}\veps(l)\cV_{l}\wed (u_{l}-\veps(l)a_{l})+\frac 12(1+\epsilon(t)\Omega)\veps(t)v_{t}\wed u_{t}\\
    &\hspace{-.5cm}+\sum_{\cT\setminus\{t\}}(1+\epsilon(l)\Omega)^{-1}\veps(l)\cV_{l}\!\cdot\!p_{l}+\sum_{\cL^i_{k}}\veps(\ell)\cW_{\ell}\wed
  (u_{\ell}-\veps(\ell)a_{\ell})+W'R_{1}P+PR_{2}P+PR_{3}U\notag\\
    &+\frac 12\sum_{\cL_{I}}(1+\epsilon(\ell)\Omega)\veps(\ell)w_{\ell}\wed u_{\ell}+\frac 12\sum_{\cL_{I}\ltimes\cL_{I}}\veps(\ell)w_{\ell}\wed u_{\ell'}+\veps(\ell')w_{\ell'}\wed u_{\ell}\nonumber\\
    &+\sum_{(\cL_{I}\cup \{t\})\subset\cL^i_{k}}\cW_{k}^{i}\wed u_{l}+\sum_{(\cL_{I}\cup \{t\})\subset\cL_{I}}\veps(\ell')w_{\ell'}\wed u_{\ell}
    +\sum_{k\subset\cL^i_{k}}\cW^{i}_{k}\wed\eta(k)x_{k}+\cW_{I}Q_{W\!X}X\notag\\
    &+\cW_{I}Q_{W}\cW_{I}+\sum_{\substack{(\cT\cup\cL)\prec(\cT\cup\cL)}}u_{l'}\wed u_{l}+\frac
    12\sum_{\substack{\cL_{I}\ltimes\cL_{I}}}u_{\ell'}\wed u_{\ell}+AR_{4}A+AR_{5}U+AR_{6}X\notag
\end{align}
où nous avons écrit $\cW_{I}$ ($\cW^{i}_{k}$) pour une combinaison linéaire de variables $\cW_{\ell},\,\ell\in\cL_{I}\,(\cL^{i}_{k})$. Choisissons une variable $\cW_{\ell},\,\ell\in\cL^{i}_{k}$. Nous utilisons l'oscillation $\cW_{\ell}\wed\big(\sum_{(\cL_{I}\cup \{t\})\subset\ell}u_{l}+\sum_{k\subset\ell}\eta(k)x_{k}\big)$ pour obtenir une décroissance $\ks$ impliquant $\big|\sum_{\cL_{I}\cup \{t\}}u_{l}+\sum_{k\subset\ell}\eta(k)x_{k}\big|\les M^{-i}$.

S'il y a des points externes survolés par la ligne $\ell$, il existe $k$ tel que $x_{k}\defi z\subset\ell$. Alors pour toute ligne de $\cL_{I}\cup\{t\}$, nous effectuons les changements de variables \eqref{eq:chgtvarVl} et \eqref{eq:chgtvarWl} mais où $z$ remplace $x_{1}$. Ces modifications laissent la fonction $\ks$ indépendante des variables $a_{l},\, l\in\cL_{I}\cup\{t\}$. Ceci permet donc d'obtenir une masselotte d'échelle $i_{l}$ pour toute ligne $l\in\cL_{I}\cup\{t\}$.

S'il n'y a pas de points externes à part $x$ et $y$ dans $G^{i_{0}}_{k'}$ (voir figure \ref{fig:newG}), la fonction $\ks$ ne dépend que de $\sum_{(\cL_{I}\cup \{t\})\subset\ell}u_{l}$ et  $G^{i_{0}}_{k'}$ est un graphe à deux points. Soit $\ell_{0}$ la plus basse ligne de $I$, $i_{\ell_{0}}=i_{0}$. Notons que c'est nécessairement une ligne de boucle. Pour toute ligne $\ell\in\cL_{I}\setminus\{\ell_{0}\}$, nous effectuons
\begin{align}
  \lb
  \begin{aligned}
    u_{\ell}\to& u_{\ell}-\veps(\ell)a_{\ell},\\
    w_{\ell}\to& w_{\ell}+a_{\ell},\\
    u_{\ell_{0}}\to& u_{\ell_{0}}+\veps(\ell)a_{\ell},\\
    w_{\ell_{0}}\to& w_{\ell}-\veps(\ell_{0})\veps(\ell)a_{\ell}.
  \end{aligned}\right.\label{eq:chgtvarWl-I}
\end{align}
Ces changements laissent $u_{\ell}+u_{\ell_{0}}$ (et $\ks$) fixe(s). Ainsi pour toute ligne
$\ell\in\cL_{I}\setminus\{\ell_{0}\}$, nous avons une décroissance en $\cW_{\ell}-\veps(\ell)\veps(\ell_{0})\cW_{\ell_{0}}$ d'indice $i_{\ell}$. Toutes ces fonctions sont indépendantes. Pour $\ell_{0}$, nous faisons
\begin{align}
    \lb
  \begin{aligned}
    u_{\ell_{0}}\to& u_{\ell_{0}}-\veps(\ell_{0})a_{\ell_{0}},\\
    w_{\ell_{0}}\to& w_{\ell_{0}}+a_{\ell_{0}},\\
    u_{t}\to&u_{t}+\veps(\ell_{0})a_{\ell_{0}}.
  \end{aligned}\right.\label{eq:chgtvarVl-I}
\end{align}
Nous obtenons une décroissance permettant d'intégrer sur $\cW_{\ell_{0}}$ au prix de $M^{i_{t}}\ges M^{i_{0}}$. Enfin, pour la ligne d'arbre $t$, nous utilisons le changement de variables \eqref{eq:chgtvarVl}. Celui-ci introduit $a_{t}$ dans $\ks$. La masselotte pour $\cV_{t}$ est donc d'échelle $M^{i}$. Heureusement la forte décroissance associée à $p_{t}$ est d'échelle $M^{-i}$. Nous retrouvons le fait que les longues variables de l'arbre ne coûtent rien.\\

\noindent
Qualifions de \textbf{critique} une composante connexe avec $N=4, g=0, B=2$ et dont l'insertion $I$ est réduite à une seule ligne.
 Nous pouvons maintenant prouver le lemme suivant
\begin{lemma}[Comptage de puissance]\label{lem:compt-puissGN}
  Soit $G$ un graphe connexe orientable. Quelque soit $\Omega\in\lsb 0,1\right[$, il existe $K\in\R$ tel que son
  amplitude amputée $A_{G}^{\mu}$ intégrée contre des fonctions test (voir
  (\ref{eq:amplitude-avt-massel})) soit bornée par
  \begin{align}
    \labs A_{G}^{\mu}\rabs\les&K^{n}\prod_{i,k}M^{-\frac 12\omega(G^{i}_{k})}\label{eq:compt-bound}\\
    \text{avec } \omega(G^{i}_{k})=&
    \begin{cases}
      N-4&\text{si ($N=2$ ou $N\ges 6$) et $g=0$,}\\
      &\text{si $N=4$, $g=0$ et $B=1$,}\\
      &\text{si $G^{i}_{k}$ est critique,}\\
      N&\text{si $N=4$, $g=0$, $B=2$ et $G^{i}_{k}$ non critique,}\\
      N+4&\text{si $g\ges 1$.}
    \end{cases}
  \end{align}
\end{lemma}
\begin{rem}
  Le comptage de puissance du lemme précédent n'est pas optimal mais il est
  suffisant pour prouver la renormalisabilité de la théorie. Après une étude
  du propagateur de la théorie dans la base matricielle \cite{toolbox05}, nous
  pourrions obtenir le comptage optimal notamment la dépendance en le
  genre. Concernant les faces brisées, la borne \eqref{eq:compt-bound} est
  presque optimale. Pour la fonction à quatre points, elle l'est. Mais pour les fonctions à six points ou plus, nous n'avons pas cherché à améliorer la borne simple donnée par le lemme \ref{lem:masselottes}. Néanmoins remarquons que pour ces fonctions, des situations semblables à celle rencontrée pour la fonction à quatre points peuvent se produire. Les points \og{}externes\fg{} dans les faces brisées supplémentaires peuvent n'être joints que par une seule ligne basse. Dans ce cas, la face brisée n'améliore pas le comptage de puissance même pour les fonctions à six points ou plus. C'est l'une des différences entre le modèle de Gross-Neveu et la théorie $\Phi^4$.
\end{rem}
\begin{proof}
  Le lemme \ref{lem:masselottes} nous permet de borner
  l'amplitude d'un graphe $G$ connexe orientable par
  \begin{align}
    \labs A^{\mu}_{G}\rabs\les&K^{n}\int
    dx_{1}\,g_{1}(x_{1}+\{a\})\delta_{G}\prod_{i=2}^{N}dx_{i}\,g_{i}(x_{i})\prod_{l\in G}da_{l}\,M^{2i_{l}}\Xi(a_{l})\label{eq:absbound-start}\\
    &\qquad\prod_{l\in\cT}du_{l}d\cV_{l}dp_{l}\,M^{i_{l}}e^{-M^{2i_{l}}(u_{l}-\veps(l)a_{l})^{2}}\prod_{\mu=0}^{1}\frac{1}{1+M^{-2i_{l}}\cV^{2}_{l,\mu}}
    \frac{1}{1+M^{2i_{l}}p^{2}_{l,\mu}}\notag\\
    &\qquad\prod_{\ell\in\cL}
    du_{\ell}d\cW_{\ell}M^{i_{\ell}}M^{i_{\ell}}
    e^{-M^{2i_{\ell}}(u_{\ell}+\{a\})^{2}}
    \prod_{\mu=0}^{1}\frac{1}{1+M^{-2i_{\ell}}\cW^{2}_{\ell,\mu}}\notag
  \end{align}
  où $K\in\R$ et $g_{i},\,i\in\lnat 1,N\rnat$ et $\Xi$ sont des fonctions de Schwartz. La fonction $\delta_{G}$ qui correspond à la racine
  de l'arbre, est donnée par (voir section \ref{sec:resol-deltaNCPhi4})
  \begin{equation}
    \delta_{G}\Big(\sum_{i\in\cE(G)}\eta(i)x_{i}+\sum_{l\in\cT\cup\cL}u_{l}\Big).
  \end{equation}
  Elle permet d'intégrer sur l'une des positions externes. Les autres sont
  intégrées avec les fonctions $g_{i}$. La borne \eqref{eq:absbound-start} sur la valeur absolue de
  l'amplitude devient donc
  \begin{align}
    \labs A^{\mu}_{G}\rabs\les&K^{n}\int
    \prod_{l\in G}da_{l}\,M^{2i_{l}}\Xi(a_{l})\prod_{\ell\in\cL}
    du_{\ell}d\cW_{\ell}M^{i_{\ell}}M^{i_{\ell}}
    e^{-M^{2i_{\ell}}(u_{\ell}+\{a\})^{2}}
    \prod_{\mu=0}^{1}\frac{1}{1+M^{-2i_{\ell}}\cW^{2}_{\ell,\mu}}\notag\\
    &\qquad\prod_{l\in\cT}du_{l}d\cV_{l}dp_{l}\,M^{i_{l}}e^{-M^{2i_{l}}(u_{l}-\veps(l)a_{l})^{2}}\prod_{\mu=0}^{1}\frac{1}{1+M^{-2i_{l}}\cV^{2}_{l,\mu}}
    \frac{1}{1+M^{2i_{l}}p^{2}_{l,\mu}}.\label{eq:absbound-2}
  \end{align}
  Les intégrations sur les variables $a_{\ell}$ coûtent $\cO(1)$. Pour toute
  ligne $l$ du graphe, l'intégration sur la variable $u_{l}$ correspondante
  rapporte $\cO(M^{-2i_{l}})$. L'intégration sur $v_{l}$ (resp. $w_{l}$) est d'ordre $\cO(M^{2i_{l}})$. Mais pour les lignes d'arbre, ce mauvais facteur est compensé par l'intégration sur $p_{l}$ qui donne $\cO(M^{-2i_{l}})$. Ainsi
  les boucles ne coûtent que $\cO(1)$ alors que les lignes d'arbre rapportent
  $\cO(M^{-2i_{l}})$. Nous avons donc la borne suivante
  \begin{align}
    \labs A_{G}\rabs\les& K^{n}\prod_{l\in
      G}M^{i_{l}}\prod_{l\in\cT}M^{-2i_{l}}\notag\\
    \les&K'^{n}\prod_{l\in
      G}M^{i_{l}+1}\prod_{l\in\cT}M^{-2(i_{l}+1)}.
  \end{align}
  Nous pouvons alors redistribuer le comptage de puissance dans les composantes
  connexes en suivant \cite{Riv1} :
  \begin{align}
    \prod_{l\in G}M^{i_{l}+1}=&\prod_{l\in G}\prod_{i=0}^{i_{l}}M=\prod_{l\in
      G}\prod_{\substack{(i,k)\in\N^{2}\tq\\l\in
        G^{i}_{k}}}M=\prod_{(i,k)\in\N^{2}}\,\prod_{l\in G^{i}_{k}}M,\\
    \prod_{l\in\cT}M^{-2(i_{l}+1)}=&\prod_{l\in\cT}\prod_{\substack{(i,k)\in\N^{2}\tq\\l\in
        G^{i}_{k}}}M^{-2}=\prod_{(i,k)\in\N^{2}}\,\prod_{l\in\cT^{i}_{k}}M^{-2}.
  \end{align}
Ainsi, en changeant $K'$ en $K$, l'amplitude d'un graphe $G$ connexe
orientable est bornée par
\begin{align}
  \labs A_{G}^{\mu}\rabs\les&K^{n(G)}\prod_{(i,k)\in\N^{2}}M^{-\frac
    12\omega(G^{i}_{k})},\\
  \text{où }\omega(G^{i}_{k})=&N(G^{i}_{k})-4\label{eq:degre-conv-compconn}
\end{align}
ce qui prouve la première partie du lemme \ref{lem:compt-puissGN}.\\

Si une composante connexe $G^{i}_{k}$ est non planaire, il existe $\ell,\ell'\in
G^{i}_{k}$ telles que l'intégration sur $\cW_{\ell}$ rapporte
$M^{-2i_{\ell'}}\les M^{-2i}$ au lieu de $M^{2i_{\ell}}$ (voir section \ref{subsec:nonplanar}). Le gain par rapport
à \eqref{eq:degre-conv-compconn} est d'au moins $M^{-4i}$. Le degré
superficiel de convergence devient alors $\omega(G^{i}_{k})=N(G^{i}_{k})+4$.\\

Enfin soit une composante connexe $G^{i}_{k}$ avec quatre pattes externes et deux
faces brisées. En utilisant les notations définies précedemment, si $G^{i_{0}}_{k'}$ a plus de deux points externes, nous utilisons la fonction $\ks$ pour intégrer sur l'un de ces points ce qui rapporte $M^{-2i}$ au lieu de $\cO(1)$. Soit $\mathbf{\scP}$ le chemin de l'arbre de Gallavotti-Nicolò qui relie $G^i_{k}$ à $G$. Le facteur $M^{-2i}$ améliore le degré superficiel de convergence de tous les noeuds de $\scP$ avec $N=4, B=2$. Il devient donc $\omega(G^{i}_{k})=N(G^{i}_{k})$. Si $G^{i_{0}}_{k'}$ est un graphe à deux points, nous utilisons $\ks$ pour intégrer sur la variable $u_{\ell_{0}}$ de la plus basse ligne $\ell_{0}$ de l'insertion $I$. Ceci rapporte $M^{-2i}$ au lieu de $M^{-2i_{0}}$. Le gain par rapport à \eqref{eq:degre-conv-compconn} est $M^{-2(i-i_{0})}$. Mais l'intégration sur $\cW_{\ell_{0}}$ coûte $M^{2i_{t}}$ au lieu de $M^{2i_{\ell_{0}}}$. Le gain total est donc seulement $M^{-2(i-i_{t})}$. Ce facteur supplémentaire permet d'améliorer le comptage de puissance de toutes les composantes connexes avec $N=4, B=2$ dans $\scP$ entre $G^{i}_{k}$ et l'échelle $i_{t}$. Leur comptage de puissance passe de $N-4$ à $N$. Mais notons qu'entre $i_{t}$ (l'échelle de la plus basse ligne d'arbre de $I$) et $i_{0}$, seules des lignes de boucles peuvent apparaître. Ainsi le nombre de points externes ne peut que décroitre \emph{strictement} dans $\scP$ de l'échelle $i_{t}$ à l'échelle $i_{0}$. $G^{i_{0}}_{k'}$ étant un graphe à deux points, il ne peut y avoir qu'une seule composante connexe divergente dans $\scP$ entre $i_{t}$ et $i_{0}$. C'est un graphe à quatre points avec $B=2$ à l'échelle $i_{1}$ (l'échelle la plus basse de $I$ supérieure à $i_{0}$). De plus, ça ne peut se produire que s'il n'y a qu'une seule ligne de boucle à l'échelle $i_{0}$. Cette composante est \emph{critique} (par définition) et nous ne pouvons pas améliorer son comptage de puissance qui reste $N-4$. Ceci achève la preuve du lemme \ref{lem:compt-puissGN}.
\end{proof}

\section{Renormalisation}
\label{sec:renorm-GN}

Grâce au comptage de puissance prouvé dans le lemme \ref{lem:compt-puissGN},
nous savons que seuls les sous-graphes planaires avec deux et quatre pattes
externes sont divergents. Plus précisément, les seuls graphes à deux points
divergents n'ont qu'une seule face brisée. D'autre part, les graphes à quatre
points divergents ont soit une seule face brisée soit sont \emph{critiques}
\ie{} ont $N=4,g=0,B=2$ et les deux points \og{}externes\fg{} qui
appartiennent à la seconde face brisée sont liés par une seule ligne
d'échelle plus basse. Nous allons prouver que les parties divergentes des
amplitudes correspondant à ces graphes sont de la forme du lagrangien
initial.

\subsection{La fonction à quatre points}
\label{subsec:4pt-fct}

\subsubsection{$B=1$}

Soit un sous-graphe planaire à quatre points et une face brisée nécessitant
d'être renormalisé. Il est donc un noeud de l'arbre de Gallavotti-Nicol\`o. Il
existe $(i,k)\in\N^{2}$ tel que
$N(G^{i}_{k})=4,g(G^{i}_{k})=0,B(G^{i}_{k})=1$. Les quatre points externes du
graphe amputé seront notés $x_{j},\,j\in\lnat 1,4\rnat$. L'amplitude associée
à la composante connexe $G^{i}_{k}$ est 
\begin{align}
  A^{\mu}_{G^{i}_{k}}(\{x_{j}\})=\int&\prod_{i=1}^{4}dx_{i}\,\psib_{e}(x_{1})\psi_{e}(x_{2})\psib_{e}(x_{3})
  \psi_{e}(x_{4})\delta_{G^{i}_{k}} e^{\imath\varphi'_{\Omega}}\\
  &\prod_{l\in\cT^{i}_{k}}du_{l}dv_{l}dp_{l}\,\bar{C}^{i_{l}}_{l}(u_{l})\prod_{\ell\in\cL^{i}_{k}}du_{\ell}dw_{\ell}\,
  \bar{C}^{i_{\ell}}_{\ell}(u_{\ell})\notag
\end{align}
où $e$ est le plus grand indice externe du sous-graphe $G^{i}_{k}$ et
$\psi_{e},\psib_{e}$ sont des champs d'indices inférieurs ou égaux à $e<i$.
Nous allons procéder à un développement de Taylor au premier ordre qui
permettra de découpler les variables externes $x_{j}$ des variables internes
$u$ et $p$ et d'identifier la partie divergente de l'amplitude. Nous
introduisons un paramètre $s$ en trois endroits différents. Tout d'abord, nous
développons la fonction $\delta_{G^{i}_{k}}$ comme
\begin{align}
  \delta_{G^{i}_{k}}\big(\Delta+s\kU\big)\Big|_{s=1}=&\delta(\Delta)+\int_{0}^{1}
  ds\,\kU\cdot\nabla\delta(\Delta+s\kU)\\
  \text{où }\Delta=&x_{1}-x_{2}+x_{3}-x_{4}\text{ et }\kU=\sum_{l\in G^{i}_{k}}u_{l}.\notag
\end{align}
Dans le cas d'un graphe orientable, les champs $\psib$ sont associés aux
positions impaires et les $\psi$ aux positions paires. De plus, si le graphe
est planaire régulier, le corollaire \ref{sec:oscillRG} nous permet de fixer
la forme exacte de la fonction delta de racine notamment l'alternance des
signes. Ce corollaire donne aussi l'oscillation externe $\varphi_{E}$. Le
reste de l'oscillation $\varphi'_{\Omega}$ est également développé. Dans notre
cas, l'oscillation totale est donnée par le corollaire \ref{sec:oscillRG} et
par les oscillations des fonctions delta de branches : 
\begin{align}
  \varphi'_{\Omega}(s=1)=\varphi_{E}+XQ_{X\!U}U+XQ_{X\!P}P+UQ_{U}U+PQ_{P}P+UQ_{U\!W}W+PQ_{PW}W.
\end{align}
Remarquons que $Q_{X\!W}=Q_{W}=0$ pour les graphes réguliers
planaires. Nous écrivons
\begin{align}
  &\exp\imath(XQ_{X\!U}U+XQ_{X\!P}P+UQ_{U}U+PQ_{P}P)\\
  =&1+\imath\int_{0}^{1}ds\,(XQ_{X\!U}U+XQ_{X\!P}P+UQ_{U}U+PQ_{P}P)e^{\imath s(XQ_{X\!U}U+XQ_{X\!P}P)+\imath UQ_{U}U+\imath PQ_{P}P}.\notag
\end{align}
Finalement nous développons aussi les propagateurs internes. Pour toute ligne
$l\in G^{i}_{k}$,
\begin{align}
  \bar{C}_{l}(u_{l},s=1)=&\frac{\Omega}{\theta\pi}\int_{0}^{\infty}\frac{dt_{l}\,e^{-t_{l}m^{2}}}{\sinh(2\Ot t_{l})}\,  
  e^{-\frac{\Ot}{2}\coth(2\Ot t_{l})u_{l}^{2}}\big(\imath\Ot\coth(2\Ot
  t_{l})\epsilon(l)\veps(l)\us_{l}\label{eq:taylor-propa4}\\
&\qquad+s\Omega\epsilon(l)\veps(l)\uts_{l}+sm\big)\big(\cosh(2\Ot t_{l})\mathds{1}_{2}-s\imath{\textstyle\frac{\theta}{2}}\sinh(2\Ot
  t_{l})\gamma\Theta^{-1}\gamma\big)\Big|_{s=1}\notag\\
  =&\frac{2\imath \Omega^{2}}{\theta^{2}\pi}\int_{0}^{\infty}\frac{dt_{l}\,e^{-t_{l}m^{2}}}{\tanh(2\Ot t_{l})}\,  
  e^{-\frac{\Ot}{2}\coth(2\Ot t_{l})u_{l}^{2}}\coth(2\Ot
  t_{l})\epsilon(l)\veps(l)\us_{l}\notag\\
  &+\frac{\Omega}{\theta\pi}\int_{0}^{1}ds\,\int_{0}^{\infty}\frac{dt_{l}\,
    e^{-t_{l}m^{2}}}{\sinh(2\Ot t_{l})}\,e^{-\frac{\Ot}{2}\coth(2\Ot
    t_{l})u_{l}^{2}}\notag\\
  &\qquad\times\Big\{(\Omega\epsilon(l)\veps(l)\uts_{l}+m)\big(\cosh(2\Ot
  t_{l})\mathds{1}_{2}-s\imath{\textstyle\frac{\theta}{2}}\sinh(2\Ot
  t_{l})\gamma\Theta^{-1}\gamma\big)\notag\\ 
  &\qquad-\imath{\textstyle\frac{\theta}{2}}\sinh(2\Ot t_{l})\big(\imath\Ot\coth(2\Ot
  t_{l})\epsilon(l)\veps(l)\us_{l}+s\Omega\epsilon(l)\veps(l)\uts_{l}+sm\big)\gamma\Theta^{-1}\gamma\Big\}.\notag
\end{align}
Le contreterme $\tau A$ associé à la composante connexe $G^{i}_{k}$ correspond
aux termes d'ordre $0$ des trois développements précédents :
\begin{align}
  \tau A^{\mu}_{G^{i}_{k}}=&\int\prod_{i=1}^{4}dx_{i}\,\psib_{e}(x_{1})\psi_{e}(x_{2})\psib_{e}(x_{3})
  \psi_{e}(x_{4})\delta(\Delta) e^{\imath\varphi_{E}}\\
  &\times\int\prod_{l\in\cT^{i}_{k}}du_{l}dv_{l}dp_{l}\,\bar{C}^{i_{l}}_{l}(u_{l},s=0)
  \prod_{\ell\in\cL^{i}_{k}}du_{\ell}dw_{\ell}\,
  \bar{C}^{i_{\ell}}_{\ell}(u_{\ell},s=0)e^{\imath\varphi'_{\Omega}(s=0)}\notag
  \intertext{où $\varphi_{E}=\sum_{i<j=1}^{4}(-1)^{i+j+1}x_{i}\wed
    x_{j}$. Ainsi le contreterme est de la forme}
  \tau
  A^{\mu}_{G^{i}_{k}}=&\int dx\,\lbt\psib_{e}\star\psi_{e}\star\psib_{e}\star\psi_{e}\rbt(x)\\
  &\times\int\prod_{l\in\cT^{i}_{k}}du_{l}dv_{l}dp_{l}\,\bar{C}^{i_{l}}_{l}(u_{l},s=0)
  \prod_{\ell\in\cL^{i}_{k}}du_{\ell}dw_{\ell}\,
  \bar{C}^{i_{\ell}}_{\ell}(u_{\ell},s=0)e^{\imath\varphi'_{\Omega}(s=0)}.\notag
\end{align}
Pour prouver que $\tau A$ est de la forme du vertex initial, il reste à
montrer que sa structure spinorielle est l'une de celles de l'équation
\eqref{eq:int-orient}. Outre les oscillations et les décroissances
exponentielles des propagateurs, le contreterme $\tau A$ contient
\begin{align}\label{eq:polyn}
  P=&\prod_{l\in G}\us_{l}=\prod_{l\in G}\lbt\gamma^{0}u_{l}^{0}+\gamma^{1}u_{l}^{1}\rbt=\prod_{i=1}^{2^{2n-N/2}}P_{i}.
\end{align}
Chacun des $2^{2n-N/2}$ termes $P_{i}$ de $P$ consiste à choisir pour chaque
ligne $l\in G$ soit $\gamma^{0}u^{0}_{l}$ soit $\gamma^{1}u_{l}^{1}$. Tout
$P_{i}$ possède $n_{i}^{0}$ $u^{0}$ et $n_{i}^{1}$ $u^{1}$. Remarquons que,
mis à part $P$, le contreterme $\tau A$ est invariant sous le changement
$\forall l\in G,\, u_{l}^{0}\to -u_{l}^{0}$ et $w_{l}^{1}\to -w_{l}^{1}$.
Ainsi seuls les $P_{i}$ avec $n_{i}^{0}$ pair sont non nuls. Avec le même type
d'argument, nous prouvons que $n_{i}^{1}$ doit également être pair. Chaque
terme de $\tau A$ contient donc un nombre pair de $\gamma^{0}$ et de
$\gamma^{1}$. Pour la fonction à quatre points, le développement de Taylor
\eqref{eq:taylor-propa4} est possible car le nombre de lignes internes est
pair (il vaut $2(n-1)$). Nous introduisons maintenant la notion de
\textbf{chaîne} et de \textbf{cycle}. 
\begin{defn}[Chaîne et cycle]\label{def:cycle-chaine}
  Nous dirons que deux champs sont dans la
  même chaîne
  \begin{itemize}
  \item s'ils appartiennent à un même produit scalaire à un vertex\footnote{Par
      exemple, les deux premiers champs de l'interction \eqref{eq:int-o-1}
      appartiennent au même produit scalaire.},
  \item s'ils sont reliés par un propagateur.
  \end{itemize}
Un cycle est une chaîne fermée.
\end{defn}
Ainsi les champs externes sont reliés par des chaînes. Les autres champs
appartiennent à des cycles. Les matrices $\gamma^{0}$ et $\gamma^{1}$ sont
donc réparties dans les chaînes et les cycles. Chaque cycle correspond à un
terme qui, à un signe près, vaut
$\Tr\lbt(\gamma^{0})^{p}(\gamma^{1})^{q}\rbt$. Ce terme est non nul seulement
si $p$ et $q$ sont pairs. Sachant que le nombre total de $\gamma^{0}$ est
pair, que le nombre total de $\gamma^{1}$ est pair et que chaque cycle
contient des nombres pairs de $\gamma^{0}$ et de $\gamma^{1}$, les différentes
chaînes du graphe se partagent un nombre pair de $\gamma^{0}$ et un nombre
pair de $\gamma^{1}$. La fonction à quatre points contient deux chaînes. Il y
a donc quatre possibilités pour la répartition des matrices gamma dans ces
deux chaînes. Chacune d'elles peut contenir un nombre pair ou impair de
$\gamma^{0}$ ou de $\gamma^{1}$.\\

En fonction du nombre et du type de vertex rencontrés le
long de ces chaînes, elles peuvent soit relier un $\psi$ à un $\psib$ soit
relier deux champs de même nature. Nous pouvons ainsi avoir affaire à douze
structures spinorielles différentes :
\begin{subequations}
  \begin{align}
    \psi\star\lsb\lbt\gamma^{0}\rbt^{2p}\lbt\gamma^{1}\rbt^{2q}\rsb\psib\star
    \psi\star\lsb\lbt\gamma^{0}\rbt^{2p'}\lbt\gamma^{1}\rbt^{2q'}\rsb\psib
    =&\pm\psi\star\mathds{1}\psib\star\psi\star\mathds{1}\psib,\\
    &\notag\\
    \psi\star\lsb\lbt\gamma^{0}\rbt^{2p+1}\lbt\gamma^{1}\rbt^{2q}\rsb\psib\star
    \psi\star\lsb\lbt\gamma^{0}\rbt^{2p'+1}\lbt\gamma^{1}\rbt^{2q'}\rsb\psib
    =&\pm\psi\star\gamma^{0}\psib\star\psi\star\gamma^{0}\psib,\\
    &\notag\\
    \psi\star\lsb\lbt\gamma^{0}\rbt^{2p}\lbt\gamma^{1}\rbt^{2q+1}\rsb\psib\star
    \psi\star\lsb\lbt\gamma^{0}\rbt^{2p'}\lbt\gamma^{1}\rbt^{2q'+1}\rsb\psib
    =&\pm\psi\star\gamma^{1}\psib\star\psi\star\gamma^{1}\psib,\\
    &\notag\\
    \psi\star\lsb\lbt\gamma^{0}\rbt^{2p+1}\lbt\gamma^{1}\rbt^{2q+1}\rsb\psib\star
    \psi\star\lsb\lbt\gamma^{0}\rbt^{2p'+1}\lbt\gamma^{1}\rbt^{2q'+1}\rsb\psib
    =&\pm\psi\star\gamma^{0}\gamma^{1}\psib\star\psi\star\gamma^{0}\gamma^{1}\psib.
  \end{align}
\end{subequations}
De la même façon, nous pouvons avoir
\begin{subequations}
  \begin{align}
    \pm\psib\star\mathds{1}\psi\star\psib\star\mathds{1}\psi,\\
    &\notag\\
    \pm\psib\star\gamma^{0}\psi\star\psib\star\gamma^{0}\psi,\\
    &\notag\\
    \pm\psib\star\gamma^{1}\psi\star\psib\star\gamma^{1}\psi,\\
    &\notag\\
    \pm\psib\star\gamma^{0}\gamma^{1}\psi\star\psib\star\gamma^{0}\gamma^{1}\psi,
  \end{align}
\end{subequations}
\begin{subequations}
  \begin{align}
    \pm\psib_{a}\star\psi_{c}\star\psib_{b}\star\psi_{d}\mathds{1}_{ab}\mathds{1}_{cd},\\
    &\notag\\
    \pm\psib_{a}\star\psi_{c}\star\psib_{b}\star\psi_{d}\gamma^{0}_{ab}\gamma^{0}_{cd},\\
    &\notag\\
    \pm\psib_{a}\star\psi_{c}\star\psib_{b}\star\psi_{d}\gamma^{1}_{ab}\gamma^{1}_{cd},\\
    &\notag\\
    \pm\psib_{a}\star\psi_{c}\star\psib_{b}\star\psi_{d}\lbt\gamma^{0}\gamma^{1}
    \rbt_{ab}\lbt\gamma^{0}\gamma^{1}\rbt_{cd}.
  \end{align}
\end{subequations}
Pour prouver que la divergence de la fonction à quatre points est bien de
l'une des formes des vertex originaux \eqref{eq:int-orient}, il est pratique de les récrire d'une manière différente.

\paragraph{Identités de Fierz non commutatives}

Une base de $M_{D}(\C)$ est fournie par la représentation de l'algèbre de
Clifford $\lb\gamma^{\mu},\gamma^{\nu}\rb=-D\delta^{\mu\nu}$ de dimension $D$. En
dimension $2$,
$\cB=\lb\Gamma^{0}=\mathds{1},\Gamma^{1}=\gamma^{0},\Gamma^{2}=\gamma^{1},\Gamma^{3}=\gamma^{0}\gamma^{1}\rb$
est une base de $M_{2}(\C)$. Ainsi soit $M\in M_{2}(\C)$,
\begin{align}\label{eq:cliff-decomp}
  M=&-\frac 12\sum_{A,B=0}^{3}\eta_{AB}\Tr(M\Gamma^{A})\Gamma^{B},\\
  \text{avec }\eta=&\diag(-1,1,1,1).\notag
\end{align}
Nous allons utiliser une telle décomposition pour récrire les interactions du modèle sous une autre forme. Par exemple, considérons
l'interaction \eqref{eq:int-o-2}. Si nous définissons
$M_{ab}=\psib_{b}\star\psi_{a}$ et utilisons \eqref{eq:cliff-decomp}, on a
\begin{align}
  \psib_{b}\star\psi_{a}=&-\frac 12\sum_{A,B}\eta_{AB}\psib_{b'}\star\psi_{a'}\Gamma^{A}_{b'a'}\Gamma^{B}_{ab}.
\end{align}
Ceci nous permet d'écrire
\begin{align}
  \int\psi_{a}\star\psib_{a}\star\psi_{b}\star\psib_{b}=\int\psib_{b}\star\psi_{a}\star\psib_{a}\star\psi_{b}
  =-\frac
  12\sum_{A,B}\eta_{AB}\int\psib\star\Gamma^{A}\psi\star\psib\star\Gamma^{B}\psi.
\end{align}
De la même façon, pour l'interaction \eqref{eq:int-o-3}, nous utilisons la
décomposition 
\begin{align}
 M_{ba}=&\psib_{a}\star\psi_{b}=-\frac 12\sum_{A,B}\eta_{AB}\psib_{a'}\star\psi_{b'}\Gamma^{A}_{a'b'}\Gamma^{B}_{ba}
\end{align}
et écrivons
\begin{align}
  \sum_{a,b}\int
  \psib_{a}\star\psi_{b}\star\psib_{a}\star\psi_{b}
  =&-\frac
  12\sum_{A,B}\eta_{AB}\int\psib\star\Gamma^{A}\psi\star\psib\star\!\!\!\phantom{\Gamma}^{t\!}\Gamma^{B}\psi\notag\\
  =&-\frac
  12\sum_{A,B}\int g^{3}_{AB}\psib\star\Gamma^{A}\psi\star\psib\star\Gamma^{B}\psi
\end{align}
avec $g^{3}_{AB}=\diag(-1,1,1,-1)$. Nous procédons de la même façon pour les trois
autres interactions. Les six interactions possibles sont résumées dans le
tableau \ref{tab:interactions}. Finalement, les trois interactions orientables
\eqref{eq:int-orient} s'écrivent comme combinaisons linéaires de
\begin{subequations}\label{eq:courants-or}
  \begin{align}
    &\int\psib\star\bbbone\psi\star\psib\star\bbbone\psi,\label{eq:courants-or-1}\\
    &\int\psib\star\gamma^{\mu}\psi\star\psib\star\gamma_{\mu}\psi\text{ et}\label{eq:courants-or-2}\\
    &\int\psib\star\gamma^{0}\gamma^{1}\psi\star\psib\star\gamma^{0}\gamma^{1}\psi\label{eq:courants-or-3}
  \end{align}
\end{subequations}
alors que les interactions non orientables \eqref{eq:int-nonorient} se
récrivent en fonction de
\begin{subequations}\label{eq:courants-no}
  \begin{align}
    &\int\psi\star\bbbone\psib\star\psib\star\bbbone\psi,\label{eq:courants-no-1}\\
    &\int\psi\star\gamma^{\mu}\psib\star\psib\star\gamma_{\mu}\psi\text{ et}\label{eq:courants-no-2}\\
    &\int\psi\star\gamma^{0}\gamma^{1}\psib\star\psib\star\gamma^{0}\gamma^{1}\psi.\label{eq:courants-no-3}
  \end{align}
\end{subequations}
Dans les équations \eqref{eq:courants-or-2} et \eqref{eq:courants-no-2}, la
somme sur $\mu$ est implicite.

\begin{landscape}
  \begin{table}\label{tab:interactions}
    \centering
    \begin{equation*}
      \begin{array}{|l|l|}
        \hline
        \multicolumn{2}{|c|}{\rule[-3pt]{0pt}{20pt}\text{\Large \bfseries Interactions du modèle de
            Gross-Neveu non commutatif}}\\
        \hline
        \hline
        \multicolumn{1}{|c|}{\rule[-3pt]{0pt}{15pt}\text{\large
            Orientables}}&\multicolumn{1}{|c|}{\text{\large Non orientables}}\\
        \hline
        \rule[5pt]{0pt}{15pt}{\displaystyle\bullet\phantom{=}\sum_{a,b}\int
        dx\,\lbt\psib_{a}\star\psi_{a}\star\psib_{b}\star\psi_{b}\rbt(x)\phantom{=\Tr\lbt\psib\cdot\psi\rbt^{2}}}&
        {\displaystyle\bullet\phantom{=}\sum_{a,b}\int
        dx\,\lbt\psib_{a}\star\psib_{b}\star\psi_{b}\star\psi_{a}\rbt(x)\phantom{=\Tr\lbt\psi\cdot\psib\rbt\lbt
        \psib\cdot\psi\rbt}}\\
        \rule[0pt]{0pt}{15pt}{\displaystyle\phantom{\bullet}=-\frac{1}{2}
        \sum_{A,B}\int g^{1}_{AB}\psib\star\Gamma^{A}\psi\star\psib\star\Gamma^{B}\psi}&
        {\displaystyle\phantom{\bullet}=-\frac{1}{2}
        \sum_{A,B}\int g^{1}_{AB}\psi\star\Gamma^{A}\psib\star\psib\star\Gamma^{B}\psi}\\
        &\\
        &\\
        {\displaystyle\rule[0pt]{0pt}{15pt}\bullet\phantom{=}\sum_{a,b}\int
        dx\,\lbt\psi_{a}\star\psib_{a}\star\psi_{b}\star\psib_{b}\rbt(x)\phantom{=\Tr\lbt\psi\cdot\psib\rbt^{2}}}&
        {\displaystyle\bullet\phantom{=}\sum_{a,b}\int
        dx\,\lbt\psib_{a}\star\psib_{a}\star\psi_{b}\star\psi_{b}\rbt(x)\phantom{=\Tr\lbt\psib\cdot\psib\rbt\lbt\psi\cdot\psi\rbt}}\\
        {\displaystyle\rule[0pt]{0pt}{15pt}\phantom{\bullet}=-\frac{1}{2}
        \sum_{A,B}\int g^{2}_{AB}\psib\star\Gamma^{A}\psi\star\psib\star\Gamma^{B}\psi}&
        {\displaystyle\phantom{\bullet}=-\frac{1}{2}
        \sum_{A,B}\int g^{2}_{AB}\psi\star\Gamma^{A}\psib\star\psib\star\Gamma^{B}\psi}\\
        &\\
        &\\
        {\displaystyle\rule[0pt]{0pt}{15pt}\bullet\phantom{=}\sum_{a,b}\int
        dx\,\lbt\psib_{a}\star\psi_{b}\star\psib_{a}\star\psi_{b}\rbt(x)}&
        {\displaystyle\bullet\phantom{=}\sum_{a,b}\int
        dx\,\lbt\psib_{a}\star\psib_{b}\star\psi_{a}\star\psi_{b}\rbt(x)}\\
        {\displaystyle\rule[-10pt]{0pt}{25pt}\phantom{\bullet}=-\frac{1}{2}
        \sum_{A,B}\int g^{3}_{AB}\psib\star\Gamma^{A}\psi\star\psib\star\Gamma^{B}\psi}&
        {\displaystyle\phantom{\bullet}=-\frac{1}{2}
        \sum_{A,B}\int
        g^{3}_{AB}\psi\star\Gamma^{A}\psib\star\psib\star\Gamma^{B}\psi}\\
        \hline
        \multicolumn{2}{|c|}{{\displaystyle\rule[2pt]{0pt}{15pt}g^{1}=\diag(-2,0,0,0),\ g^{2}=\eta=\diag(-1,1,1,1),
          \ g^{3}=\diag(-1,1,1,-1)}}\\
        \multicolumn{2}{|c|}{{\displaystyle\rule[-8pt]{0pt}{23pt}\forall A\in\lnat
            1,4\rnat,\,\Gamma^{A}\in\cB=\lb\Gamma^{0}=\mathds{1},\Gamma^{1}=\gamma^{0},\Gamma^{2}=\gamma^{1},
            \Gamma^{3}=\gamma^{0}\gamma^{1}\rb}}\\
          \hline
        \end{array}
      \end{equation*}
      \caption{Les interactions et leurs différentes écritures}
    \end{table}
  \end{landscape}
Nous montrons que quelque soit
$\Gamma^{C}\in\cB,\,\psi\star\Gamma^{C}\psib\star\psi\star\Gamma^{C}\psib$,
$\psib\star\Gamma^{C}\psi\star\psib\star\Gamma^{C}\psi$ et
$\psib_{a}\star\psi_{c}\star\psib_{b}\star\psi_{d}\Gamma^{C}_{ab}\Gamma^{C}_{cd}$
peuvent s'exprimer en fonction des interactions orientables du tableau
\ref{tab:interactions} grâce à une symétrie du modèle. 
\begin{align}
  \int\psi\star\Gamma^{C}\psib\star\psi\star\Gamma^{C}\psib=&\int\psib_{d}\star\psi_{a}\star\psib_{b}\star\psi_{c}
  \Gamma^{C}_{ab}\Gamma^{C}_{cd}\\
  =&-\frac
  12\sum_{A,B}\eta_{AB}\psib\star\Gamma^{A}\psi\star\psib\star\!\!\!\!\!\phantom{O}^{t\!}\Gamma^{C}
  \Gamma^{B}\!\!\!\!\phantom{O}^{t\!}\Gamma^{C}\psi\notag\\
  \phantom{O}^{t\!}\Gamma^{C}\Gamma^{B}\!\!\!\!\phantom{O}^{t\!}\Gamma^{C}=&
  \begin{cases}
    \Gamma^{B}&\text{si $\Gamma^{C}=\mathds{1}$}\\
    g_{BB'}\Gamma^{B'}&\text{avec $g=\diag(-1,1,1-1)$ si
      $\Gamma^{C}=\gamma^{0}\gamma^{1}$}\\
    g_{BB'}\Gamma^{B'}&\text{avec $g=\diag(-1,-1,1,1)$ si
      $\Gamma^{C}=\gamma^{0}$}\\
    g_{BB'}\Gamma^{B'}&\text{avec $g=\diag(-1,1,-1,1)$ si
      $\Gamma^{C}=\gamma^{1}$}
  \end{cases}
\end{align}
On a donc
\begin{align}
  \int\psi\star\Gamma^{C}\psib\star\psi\star\Gamma^{C}\psib=&-\frac
    12\sum_{A,B} g_{AB}\psib\star\Gamma^{A}\psi\star\psib\star\Gamma^{B}\psi\label{eq:inter1}\\
    \text{avec }g=&
  \begin{cases}
    \diag(-1,1,1,1)&\text{si }\Gamma^{C}=\mathds{1}\\
    \diag(1,1,1,-1)&\text{si }\Gamma^{C}=\gamma^{0}\gamma^{1}\\
    \diag(1,-1,1,1)&\text{si }\Gamma^{C}=\gamma^{0}\\
    \diag(1,1,-1,1)&\text{si }\Gamma^{C}=\gamma^{1}.
  \end{cases}\label{eq:or2-contreterme}
\end{align}
Si $\Gamma^{C}=\mathds{1}$ ou $\gamma^{0}\gamma^{1}$, l'interaction
\eqref{eq:inter1} s'écrit en fonction des interactions \eqref{eq:courants-or}.
Par contre, si $\Gamma^{C}=\gamma^{0}$ ou $\gamma^{1}$ indépendamment, c'est
impossible. Heureusement il existe une symétrie impliquant que les
contretermes associés à l'interaction \eqref{eq:inter1} pour
$\Gamma^{C}=\gamma^{0}$ et $\gamma^{1}$ sont égaux. En effet, chaque terme
$P_{i}$ du polynôme $P$ \eqref{eq:polyn} consiste à choisir, pour chaque ligne
du graphe, soit $\gamma^{0}$ soit $\gamma^{1}$. À chacun de ces termes est
canoniquement associé un terme $\bar{P}_{i}=P_{j},\,j\neq i$ pour lequel on a
fait exactement le choix inverse de $P_{i}$ pour chaque ligne. Ainsi pour
obtenir $\bar{P}_{i}$, nous considérons $P_{i}$ et changeons les $\gamma^{0}$
en $\gamma^{1}$, les $u_{l}^{0}$ en $u_{l}^{1}$ et vice-versa. Tout
contreterme associé à un $P_{i}$ est constitué d'un ensemble de matrices gamma
et des intégrales sur les variables $u_{l}$, $p_{l}$, $v_{l}$ et $w_{l}$. La rotation
\begin{align}
 \forall l\in G,\,u_{l}^{0}\to& u_{l}^{1}\\
 u_{l}^{1}\to& -u_{l}^{0}\notag\\
 w_{l}^{0}\,(v_{l}^0)\to& w_{l}^{1}\,(v_{l}^1)\notag\\
 w_{l}^{1}\,(v_{l}^1)\to& -w_{l}^{0}\,(-v_{l}^0)\notag
\end{align}
montre que les intégrales de $\bar{P}_{i}$ sont égales à celles de $P_{i}$ (le
nombre total de $u_{l}^{1}$ est pair). Examinons alors les produits de
matrices gamma. Soient $N\in\N$ et $\forall j\in\lnat 0,2N+1\rnat,\,n_{j}\in\N$. 
\begin{align}
  P_{\gamma}=&\prod_{i=0}^{N}\lbt\gamma^{0}\rbt^{n_{2i}}\lbt\gamma^{1}\rbt^{n_{2i+1}}\notag\\
  =&\prod_{i=0}^{N}(-1)^{\lsb\frac{n_{2i}}{2}\rsb+\lsb\frac{n_{2i+1}}{2}\rsb}\lbt\gamma^{0}
  \rbt^{\frac{1-(-1)^{n_{2i}}}{2}}\lbt\gamma^{1}\rbt^{\frac{1-(-1)^{n_{2i+1}}}{2}}.\label{eq:pgamma}
\end{align}
Chaque produit de $\gamma^{0}$ (resp. $\gamma^{1}$) a été réduit par la
relation $\lbt\gamma^{0}\rbt^{2}=\lbt\gamma^{1}\rbt^{2}=-\mathds{1}$. Le
produit $P_{\gamma}$ est donc égal, à un signe près, à un produit
\emph{alternant} $P_{\gamma}^{\text{a}}$ de $\gamma^{0}$ et de $\gamma^{1}$. De la même façon,
\begin{align}
  \bar{P}_{\gamma}=&\prod_{i=0}^{N}\lbt\gamma^{1}\rbt^{n_{2i}}\lbt\gamma^{0}\rbt^{n_{2i+1}}\notag\\
  =&\prod_{i=0}^{N}(-1)^{\lsb\frac{n_{2i}}{2}\rsb+\lsb\frac{n_{2i+1}}{2}\rsb}\lbt\gamma^{1}
  \rbt^{\frac{1-(-1)^{n_{2i}}}{2}}\lbt\gamma^{0}\rbt^{\frac{1-(-1)^{n_{2i+1}}}{2}}.
\end{align}
Remarquons que les signes en facteur de $P^{\text{a}}_{\gamma}$ et $\bar{P}^{\text{a}}_{\gamma}$ sont les mêmes.
Soient $n_{0}^{\text{a}}$ et $n_{1}^{\text{a}}$ le nombre total de
$\gamma^{0}$ (resp. $\gamma^{1}$) dans $P^{\text{a}}_{\gamma}$. Ce produit
$P^{\text{a}}_{\gamma}$ peut être
\begin{enumerate}
\item $\gamma^{0}\gamma^{1}\dotsm\gamma^{0}\gamma^{1}$,
  $n_{0}^{\text{a}}=n_{1}^{\text{a}}$.
  \begin{align}
    P^{\text{a}}_{\gamma}=&
    \begin{cases}
      (-1)^{p}\mathds{1}&\text{si }n_{0}^{\text{a}}=n_{1}^{\text{a}}=2p\\
      (-1)^{p}\gamma^{0}\gamma^{1}&\text{si }n_{0}^{\text{a}}=n_{1}^{\text{a}}=2p+1
    \end{cases}\label{eq:cycle1}
  \end{align}
\item $\gamma^{1}\gamma^{0}\dotsm\gamma^{1}\gamma^{0}$,
  $n_{0}^{\text{a}}=n_{1}^{\text{a}}$.
  \begin{align}
    P^{\text{a}}_{\gamma}=&
    \begin{cases}
      (-1)^{p}\mathds{1}&\text{si }n_{0}^{\text{a}}=n_{1}^{\text{a}}=2p\\
      (-1)^{p}\gamma^{1}\gamma^{0}&\text{si }n_{0}^{\text{a}}=n_{1}^{\text{a}}=2p+1
    \end{cases}\label{eq:cycle2}
  \end{align}
\item $\gamma^{0}\gamma^{1}\dotsm\gamma^{0}\gamma^{1}\gamma^{0}$,
  $n_{0}^{\text{a}}=n_{1}^{\text{a}}+1$.
  \begin{align}
    P^{\text{a}}_{\gamma}=&
    \begin{cases}
      (-1)^{p}\gamma^{0}&\text{si }n_{1}^{\text{a}}=2p\\
      (-1)^{p}\gamma^{1}&\text{si }n_{1}^{\text{a}}=2p+1
    \end{cases}\label{eq:chaine1}
  \end{align}
\item $\gamma^{1}\gamma^{0}\dotsm\gamma^{1}\gamma^{0}\gamma^{1}$,
  $n_{1}^{\text{a}}=n_{0}^{\text{a}}+1$.
  \begin{align}
    P^{\text{a}}_{\gamma}=&
    \begin{cases}
      (-1)^{p}\gamma^{1}&\text{si }n_{0}^{\text{a}}=2p\\
      (-1)^{p}\gamma^{0}&\text{si }n_{0}^{\text{a}}=2p+1
    \end{cases}\label{eq:chaine2}
  \end{align}
\end{enumerate}
Appliquons ces résultats aux cycles et chaînes d'un graphe. Tout d'abord,
remarquons que les nombres respectifs de $\gamma^{0}$ et de $\gamma^{1}$ dans
le polynôme alterné ont même parité que les nombres totaux dans $P_{\gamma}$.
Chaque cycle comporte un nombre pair de $\gamma^{0}$ et de $\gamma^{1}$ et
correspond donc à des situations du type \eqref{eq:cycle1} ou
\eqref{eq:cycle2} qui sont exactement symétriques sous l'échange
$\gamma^{0}\leftrightarrow\gamma^{1}$. Quand les deux chaînes du graphe
possèdent un nombre impair de $\gamma^{0}$ et pair de $\gamma^{1}$, nous
sommes dans la situation $3$ ou $4$. Ces situations sont symétriques sous l'échange
$\gamma^{0}\leftrightarrow\gamma^{1}$. Le signe relatif entre les polynômes
$P^{\text{a}}_{\gamma}$ et $\bar{P}^{\text{a}}_{\gamma}$ est $+$ et (surtout)
ne dépend que des parités des nombres totaux de $\gamma^{0}$ et
$\gamma^{1}$. Ce signe ne dépend pas de la configuration des produits de
matrices \ie{} qu'il ne dépend pas des $n_{j}$ dans \eqref{eq:pgamma}.\\ 

Ainsi le contreterme $\psi\Gamma^{C}\psib\psi\Gamma^{C}\psib$ ne peut être que
de la forme $\psi\mathds{1}\psib\psi\mathds{1}\psib$,
$\psi\gamma^{0}\gamma^{1}\psib\psi\gamma^{0}\gamma^{1}\psib$ ou
$\psi\gamma^{\mu}\psib\psi\gamma_{\mu}\psib$. Il en est de même pour les deux
autres contretermes $\psib\Gamma^{C}\psi\psib\Gamma^{C}\psi$ et
$\psib_{a}\psi_{c}\psib_{b}\psi_{d}\Gamma^{C}_{ab}\Gamma^{C}_{cd}$.
La somme des deux dernières interactions dans \eqref{eq:or2-contreterme} est
bien une combinaison linéaire des interactions initiales. Nous vérifierions de
la même façon pour
$\psib_{a}\psi_{c}\psib_{b}\psi_{d}\Gamma^{C}_{ab}\Gamma^{C}_{cd}$. Ceci
prouve que $\tau A$ est bien de la forme des vertex initiaux.\\

Comme nous nous y attendons pour la fonction à quatre points, $\tau A$ est
logarithmiquement divergent. Pour le vérifier, il suffit de réitérer la procédure utilisée en section \ref{sec:masselottes} avec les changements de variables (\ref{eq:chgtvarVl}) et \eqref{eq:chgtvarWl} mais sans
modifier $x_{1}$ (les variables externes sont découplées des variables
internes dans le contreterme). Le reste $(1-\tau)A$ est composé de quatre
types de termes différents. Chacun d'eux améliore le comptage de puissance et
rend $(1-\tau)A$ convergent quand $i-e\to\infty$ :
\medskip
\begin{itemize}
\item $\kU\cdot\nabla\delta(\Delta+s\kU)$. Par intégration par parties sur une
  variable externe, le gradient agit sur un champ externe et donne au plus
  $M^{e}$. $\kU$ apporte au moins $M^{-i}$.
\item $XQ_{X\!U}U$, $XQ_{X\!P}P$. $X$ apporte $M^{e}$ et $U$ (resp. $P$) $M^{-i}$.
\item $UQ_{U}U$, $PQ_{P}P$ fournissent au moins $M^{-2i}$.
\item Le développement \eqref{eq:taylor-propa4} des propagateurs apporte $M^{-i}$.
\end{itemize}
\medskip
En conclusion, ces termes de reste améliorent le comptage de puissance d'un
facteur au moins égal à $M^{-(i-e)}$ ce qui rend $(1-\tau)A$ convergent et non
pertinent pour la renormalisation.

\subsubsection{$B=2$, critique}

Le comptage de puissance démontré en \eqref{eq:compt-bound} laisse supposer
que les composantes connexes critiques divergent logaritmiquement. Des calculs
exacts de graphes simples et le comportement de la théorie dans la base
matricielle montrent que c'est bien le cas. Mais la partie divergente de
l'amplitude de ces graphes n'est pas de la forme du lagrangien initial.
Malgré cette divergence, nous ne renormalisons pas ces graphes. En fait, nous
montrerons dans la section \ref{subsec:2pt-fct} que la renormalisation de la
fonction à deux points correspondante est suffisante pour rendre le graphe
complet convergent, y compris la sous-divergence critique à quatre points.
Soient $i$ l'échelle de la composante critique et $j<i$ l'échelle de la
sous-fonction à deux points correspondante. Les termes de reste de la
renormalisation de cette fonction à deux points fourniront
$M^{-(i-e)}$ $(<M^{-(j-e)})$.

\subsection{La fonction à deux points}
\label{subsec:2pt-fct}

Soit un sous-graphe planaire à deux points nécessitant d'être renormalisé. Il
existe $(i,k)\in\N^{2}$ tel que $N(G^{i}_{k})=2,g(G^{i}_{k})=0$. Les deux points externes du
graphe amputé seront notés $x,y$. L'amplitude associée
à la composante connexe $G^{i}_{k}$ est
\begin{align}
  A^{\mu}_{G^{i}_{k}}(x,y)=&\int dxdy\,\psib_{e}(x)\psi_{e}(y)\delta_{G^{i}_{k}} e^{\imath\varphi'_{\Omega}}
  \prod_{l\in\cT^{i}_{k}}du_{l}dv_{l}dp_{l}\,\bar{C}^{i_{l}}_{l}(u_{l})\prod_{\ell\in\cL^{i}_{k}}du_{\ell}dw_{\ell}\,
  \bar{C}^{i_{\ell}}_{\ell}(u_{\ell}).\notag
\end{align}
Nous allons procéder à un développement de Taylor au second ordre. Tout
d'abord, nous développons la fonction $\delta_{G^{i}_{k}}$ comme
\begin{align}
  \delta_{G^{i}_{k}}\big(x-y+s\kU\big)\Big|_{s=1}=&\delta(x-y)+\kU\cdot\nabla\delta(x-y)+\int_{0}^{1}
  ds\,(1-s)(\kU\cdot\nabla)^{2}\delta(\Delta+s\kU)
\end{align}
où nous avons utilisé les mêmes notations que dans la section
précédente. L'oscillation entre $x$ et $y$ est $\exp\imath x\wed y$. Grâce à
la fonction $\delta_{G^{i}_{k}}$, nous absorbons cette oscillation dans
une redéfinition de la matrice $Q_{X\!U}$. Puis nous développons l'oscillation :
\begin{align}
  &\exp\imath(XQ_{X\!U}U+XQ_{X\!P}P+UQ_{U}U+PQ_{P}P)=1+\imath(XQ_{X\!U}U+XQ_{X\!P}P)\\
  &-\int_{0}^{1}ds\,\Big((1-s)(XQ_{X\!U}U+XQ_{X\!P}P)^{2}-\imath(UQ_{U}U+PQ_{P}P)\Big)\notag\\
  &\hspace{1cm}\times e^{\imath s(XQ_{X\!U}U+XQ_{X\!P}P+UQ_{U}U+PQ_{P}P)}.\notag
\end{align}
Finalement nous développons aussi les propagateurs internes. Pour toute ligne
$l\in G^{i}_{k}$,
\begin{align}
  \bar{C}_{l}(u_{l},s=1)=&\frac{\Omega}{\theta\pi}\int_{0}^{\infty}\frac{dt_{l}\,e^{-t_{l}m^{2}}}{\sinh(2\Ot t_{l})}\,  
  e^{-\frac{\Ot}{2}\coth(2\Ot t_{l})u_{l}^{2}}\big(\imath\Ot\coth(2\Ot
  t_{l})\epsilon(l)\veps(l)\us_{l}\label{eq:taylor-propa2}\\
&\qquad+s\Omega\epsilon(l)\veps(l)\uts_{l}+m\big)\big(\cosh(2\Ot t_{l})\mathds{1}_{2}-s{\textstyle\frac{\theta}{2}}\imath\sinh(2\Ot
  t_{l})\gamma\Theta^{-1}\gamma\big)\Big|_{s=1}\notag\\
  =&\frac{\Omega}{\theta\pi}\int_{0}^{\infty}\frac{dt_{l}\,e^{-t_{l}m^{2}}}{\tanh(2\Ot t_{l})}\,  
  e^{-\frac{\Ot}{2}\coth(2\Ot t_{l})u_{l}^{2}}\lbt\imath\Ot\coth(2\Ot
  t_{l})\epsilon(l)\veps(l)\us_{l}+m\rbt\notag\\
  &+\frac{\Omega}{\theta\pi}\int_{0}^{1}ds\,\int_{0}^{\infty}\frac{dt_{l}\,
    e^{-t_{l}m^{2}}}{\sinh(2\Ot t_{l})}\,e^{-\frac{\Ot}{2}\coth(2\Ot
    t_{l})u_{l}^{2}}\notag\\
  &\qquad\times\Big\{\Omega\epsilon(l)\veps(l)\uts_{l}\big(\cosh(2\Ot
  t_{l})\mathds{1}_{2}-s\imath{\textstyle\frac{\theta}{2}}\sinh(2\Ot
  t_{l})\gamma\Theta^{-1}\gamma\big)\notag\\ 
  &\qquad-\imath{\textstyle\frac{\theta}{2}}\sinh(2\Ot t_{l})\big(\imath\Ot\coth(2\Ot
  t_{l})\epsilon(l)\veps(l)\us_{l}+s\Omega\epsilon(l)\veps(l)\uts_{l}+m\big)\gamma\Theta^{-1}\gamma\Big\}.\notag
\end{align}
Le lecteur attentif aura remarqué que le développement
\eqref{eq:taylor-propa2} du propagateur est différent de celui pour la
fonction à quatre points \eqref{eq:taylor-propa4}. En effet, nous autorisons
ici le terme de masse à faire partie du développement à l'ordre $0$. La raison
est que le nombre de lignes internes de la fonction à deux points est impair
(il vaut $2n-1$). Pour le contreterme de masse, si tous les propagateurs
contribuent par un terme en $u$, le contreterme est nul. En réalité, le
comptage de puissance est obtenu quand un propagateur utilise le terme de
masse et tous les autres le terme $\us$. Ceci implique que la divergence de la
masse n'est en fait que logarithmique. Pour les contretermes de fonction
d'onde et contribuant à $\Omega\xts$, chaque propagateur contribue par
l'intermédiaire du terme dominant en $\us$. En effet, le contreterme $\tau A$
associé à la composante connexe $G^{i}_{k}$ correspond ainsi aux termes
d'ordre $0$ et $1$ des trois développements précédents :
\begin{align}
  \tau A^{\mu}_{G^{i}_{k}}=\phantom{\int}&\negthickspace\!\!\tau A_{m}+\tau A_{\ps}+\tau A_{\xts},\\
  \tau A_{m}=\int&dxdy\,\psib_{e}(x)\psi_{e}(y)\delta(x-y)
  \int\prod_{l\in\cT^{i}_{k}}du_{l}dv_{l}dp_{l}\,\bar{C}^{i_{l}}_{l}(u_{l},s=0)\label{eq:massterm}\\
  &\prod_{\ell\in\cL^{i}_{k}}du_{\ell}dw_{\ell}\,
  \bar{C}^{i_{\ell}}_{\ell}(u_{\ell},s=0)e^{\imath\varphi'_{\Omega}(s=0)},\notag\\
  \tau A_{\ps}=\int&dxdy\,\psib_{e}(x)\psi_{e}(y)\kU\cdot\nabla\delta(x-y)
  \prod_{l\in\cT^{i}_{k}}du_{l}dv_{l}dp_{l}\,\bar{C}^{i_{l}}_{l}(u_{l},s=0)\label{eq:waveterm}\\
  &\prod_{\ell\in\cL^{i}_{k}}du_{\ell}dw_{\ell}\,
  \bar{C}^{i_{\ell}}_{\ell}(u_{\ell},s=0)e^{\imath\varphi'_{\Omega}(s=0)},\notag\\
  \tau A_{\xts}=\imath\int&dxdy\,\psib_{e}(x)\psi_{e}(y)\delta(x-y)(XQ_{X\!U}U+XQ_{X\!P}P)\label{eq:Omegaterm}\\
  &\prod_{l\in\cT^{i}_{k}}du_{l}dv_{l}dp_{l}\,\bar{C}^{i_{l}}_{l}(u_{l},s=0)
  \prod_{\ell\in\cL^{i}_{k}}du_{\ell}dw_{\ell}\,
  \bar{C}^{i_{\ell}}_{\ell}(u_{\ell},s=0)e^{\imath\varphi'_{\Omega}(s=0)}.\notag
\end{align}
Le contreterme $\tau A_{m}$ renormalise la masse. Sa divergence est
logarithmique pour les raisons de parité évoquées plus haut. $\tau A_{\ps}$
constitue le contreterme de fonction d'onde.
\begin{align}
  \tau A_{\ps}=-\int&dx\,\psib_{e}(x)\nabla^{\mu}\psi_{e}(x)\kU^{\mu}
  \prod_{l\in\cT^{i}_{k}}du_{l}dv_{l}dp_{l}\,\bar{C}^{i_{l}}_{l}(u_{l},s=0)\\
  &\prod_{\ell\in\cL^{i}_{k}}du_{\ell}dw_{\ell}\,
  \bar{C}^{i_{\ell}}_{\ell}(u_{\ell},s=0)e^{\imath\varphi'_{\Omega}(s=0)}\notag
\end{align}
Comme pour la fonction à quatre points, ce terme contient le polynôme
\eqref{eq:polyn} ici de degré impair. Les matrices gamma de chacun de ses monômes
sont réparties en une chaîne et en cycles (voir définition
\ref{def:cycle-chaine}). Le nombre de matrices $\gamma^{0}$ et $\gamma^{1}$
dans chaque cycle est pair. Ainsi le nombre de matrices gamma dans la chaîne
liant les points externes est impair. Le terme en
$\psib_{e}\kU^{0}\partial_{0}\psi_{e}$ est non nul si le nombre de
$\gamma^{0}$ est impair de telle sorte que le nombre total de $u^{0}$ soit
pair. Le nombre de $\gamma^{1}$ est alors pair. Le contreterme correspondant
est donc de la forme $\psib_{e}\gamma^{0}\partial_{0}\psi_{e}$. Nous lui
associons le terme en $\psib_{e}\kU^{1}\partial_{1}\psi_{e}$ où nous
choisissons le monôme inverse ($\forall l\in G,\,
\gamma^{0}u_{l}^{0}\leftrightarrow\gamma^{1}u_{l}^{1}$). Par une rotation des
coordonnées, nous montrons que le contreterme est bien de la forme
$\psib_{e}\slashed{\nabla}\psi_{e}$. Sa divergence est logarithmique.\\

Le contreterme $\tau A_{\xts}$, également de divergence logarithmique,
renormalise le \og{}champ magnétique\fg{} $\Omega\xts$. Les différents termes contribuant à
cette renormalisation sont de la forme
$\int\psib_{e}\psi_{e}(x^{0}u^{1}-x^{1}u^{0})\dotsm$. Là encore, en associant
deux monômes correspondant à des choix opposés et par une rotation, nous
montrons que le contreterme est bien de la forme $\psib_{e}\xts\psi_{e}$. Remarquons que les termes $\int\psib_{e}\psi_{e}(x^{0}p_{0}+x^{1}p_{1})\dotsm$ sont nuls par parité sur $p$ (attention ici $p_{\mu}$ est le \og{}moment\fg{} associé à une ligne d'arbre et non une dérivée). Il
est facile de vérifier, à partir de \eqref{eq:waveterm} et
\eqref{eq:Omegaterm}, que les contretermes $\tau A_{\ps}$ et $\tau A_{\xts}$
sont anti-hermitiens. Ils sont donc bien de la forme $\psib\ps\psi$ et $\psib\xts\psi$.\\

Les termes de reste, regroupés dans $(1-\tau)A$, sont convergents. En effet,
\begin{itemize}
\item $(\kU\cdot\nabla)^{2}\delta$ fournit $M^{-2i}$ grâce à $\kU^{2}$ et
  $M^{2e}$ par intégration par partie sur un point externe,
\item $(XQ_{X\!U}U+XQ_{X\!P}P)^{2}$ donne $M^{-2(i-e)}$,
\item $UQ_{U}U+PQ_{P}P$ apporte $M^{-2i}$,
\item le développement des propagateurs fournit au moins $M^{-i}$.
\end{itemize}

\paragraph{Composantes critiques}
\label{critic-comp-2pts}

Considérons un graphe à deux points orientable d'échelle $j$ contenant un sous-graphe critique d'échelle $i>j$ (voir la définition section \ref{sec:multiscaleGN}). Cette composante à deux points est donc constituée d'un sous-graphe à quatre points d'échelle $i$ avec $g=0, B=2$ et d'une unique ligne (de boucle) d'échelle $j$. Nous renormalisons l'amplitude à deux points comme nous l'avons fait ci-dessus. Nous allons montrer que les termes de reste sont d'ordre $M^{-(i-e)}$ (et non $M^{-(j-e)}$) ce qui implique la convergence de l'amplitude renormalisée incluant sa sous-divergence à quatre points.\\

Nous procédons comme expliqué dans la section \ref{sec:multiscaleGN}. Jusqu'à l'échelle $i$, nous obtenons toutes les masselottes nécessaires pour les variables $v$ et $w$ et les décroissances associées pour les $p$. Nous avons ainsi une oscillation $\cW_{\ell}\wed u^{j}$ où $i_{\ell}=i$ et $u^{j}$ est la variable $u$ de l'unique ligne d'échelle $j$. Nous l'utilisons pour obtenir une fonction $\ks$ impliquant $\labs u^{j}\rabs\les M^{-i}$. Il reste à fabriquer la masselotte pour la variable $w^{j}$. Son $u^{j}$ associé étant maintenant d'ordre $M^{-i}$, nous ne pouvons pas obtenir une masselotte d'échelle $M^j$. Nous ne pouvons atteindre que $M^{i}$. Le gain obtenu avec la variable $u^{j}$ est perdu par la masselotte en $w^{j}$ et nous remarquons une fois encore que les composantes critiques sont divergentes. Cependant, toutes les variables $u$ sont maintenant bornées par $M^{-i}$ ce qui implique que les termes de reste, excepté le développement des propagateurs, donnent $M^{-2(i-e)}=M^{-2(i-j)}M^{-2(j-e)}$. Tous les restes des développements de propagateurs sauf celui correspondant à la ligne d'indice $j$ donnent au moins $M^{-i}$. Il y a un terme dans le reste du propagateur le plus bas ($\imath m{\textstyle\frac{\theta}{2}}\sinh(2\Ot t_{\ell})\gamma\Theta^{-1}\gamma$) qui ne rapporte que $M^{-2j}$. Ce n'est pas suffisant pour régulariser la sous-divergence à quatre points. La solution consiste à inclure ce terme dans le contreterme. Ainsi, seulement pour le propagateur le plus bas, nous utilisons un développement différent de \eqref{eq:taylor-propa2} :
\begin{align}
  \bar{C}_{l}(u_{l},s=1)=&\frac{\Omega}{\theta\pi}\int_{0}^{\infty}\frac{dt_{l}\,e^{-t_{l}m^{2}}}{\sinh(2\Ot t_{l})}\,  
  e^{-\frac{\Ot}{2}\coth(2\Ot t_{l})u_{l}^{2}}\big(\imath\Ot\coth(2\Ot
  t_{l})\epsilon(l)\veps(l)\us_{l}\label{eq:taylor-propa-crit}\\
&\qquad+s\Omega\epsilon(l)\veps(l)\uts_{l}+m\big)\big(\cosh(2\Ot t_{l})\mathds{1}_{2}-{\textstyle\frac{\theta}{2}}\imath\sinh(2\Ot
  t_{l})\gamma\Theta^{-1}\gamma\big)\Big|_{s=1}\notag\\
  =&\frac{\Omega}{\theta\pi}\int_{0}^{\infty}\frac{dt_{l}\,e^{-t_{l}m^{2}}}{\tanh(2\Ot t_{l})}\,  
  e^{-\frac{\Ot}{2}\coth(2\Ot t_{l})u_{l}^{2}}\lbt\imath\Ot\coth(2\Ot
  t_{l})\epsilon(l)\veps(l)\us_{l}+m\rbt\notag\\
  &\qquad\times\big(\cosh(2\Ot t_{l})\mathds{1}_{2}-{\textstyle\frac{\theta}{2}}\imath\sinh(2\Ot
  t_{l})\gamma\Theta^{-1}\gamma\big)\notag\\
  &+\frac{\Omega}{\theta\pi}\int_{0}^{1}ds\,\int_{0}^{\infty}\frac{dt_{l}\,
    e^{-t_{l}m^{2}}}{\sinh(2\Ot t_{l})}\,e^{-\frac{\Ot}{2}\coth(2\Ot
    t_{l})u_{l}^{2}}\notag\\
  &\qquad\times\Omega\epsilon(l)\veps(l)\uts_{l}\big(\cosh(2\Ot
  t_{l})\mathds{1}_{2}-\imath{\textstyle\frac{\theta}{2}}\sinh(2\Ot
  t_{l})\gamma\Theta^{-1}\gamma\big).\notag 
\end{align}
Ceci rend convergents les sous-graphes à deux et quatre points. Le prix à payer est un contreterme de la forme $\imath\delta m\,\theta\gamma\Theta^{-1}\gamma$. La preuve de ce dernier point est donné en appendice \ref{sec:modif-count-two}. Remarquons enfin que si $m=0$, $\tau A_{m}\equiv 0$ et aucun contreterme du type $\psib\gamma^{0}\gamma^{1}\psi$ n'est nécessaire.

\begin{rem}
  Il serait intéressant d'étudier et de comparer les flots de Gross-Neveu, $\Phi^4$, LSZ modifié et non modifié. Ces deux derniers ont-ils des flots ayant les mêmes caractéristiques. Et par rapport à $\Phi^4$ et Gross-Neveu ?
\end{rem}

\subsection{Renormalisabilité et vulcanisation}
\label{sec:renorm-et-vulc}

Dans ce chapitre, nous avons prouvé que le modèle de Gross-Neveu \ncf{}, défini par l'action (\ref{eq:actfunctGN}) restreinte aux interactions orientables, est renormalisable à tous les ordres de la théorie des perturbations. Nous avons d'abord calculé une borne sur l'amplitude amputée des graphes (voir lemme \ref{lem:compt-puissGN}). Ce comptage de puissance est celui d'une théorie renormalisable. Remarquons que cette borne est tout aussi valable à $\Omega =0$. Puis nous avons montré que les contretermes nécessaires sont de la forme du lagrangien initial. Cela signifie que le modèle de Gross-Neveu \ncf{} avec interactions orientables est renormalisable même sans la procédure de vulcanisation.\\

Définissons le mélange ultraviolet/infrarouge de la façon suivante. Nous dirons qu'un graphe à $N$ points présente du mélange si son amplitude amputée intégrée contre des fonctions test (ou contre $N$ propagateurs externes) est finie (ou convergente au sens de la somme sur les attributions d'échelles) mais si, inséré dans un autre graphe, son amplitude diverge. Cette définition est bien compatible avec l'appellation de mélange \emph{ultraviolet/infrarouge} dans la mesure où, dans le cadre de l'analyse multi-échelles, les seuls sous-graphes rencontrés sont des composantes connexes. Ainsi une composante connexe d'échelle $i$ présente du mélange UV/IR si sa divergence (ultraviolette) dépend des échelles inférieures à $i$ (infrarouges). Nous pouvons alors avoir affaire à deux types de mélange. Le premier est celui que nous avons rencontré dans le modèle de Gross-Neveu. Il concerne les composantes critiques. Ces graphes planaires à quatre points et deux faces brisées ne sont pas renormalisables par un contreterme du lagrangien initial. Néanmoins ils apparaissent uniquement comme sous-graphe dans la fonction à deux points dont la renormalisation régularise la sous-divergence à quatre points. Nous pouvons qualifier ce mélange de renormalisable. Il existe un second type de mélange qui, lui, est non renormalisable. C'est celui qui a empêché la renormalisation des théories des champs \ncv{} jusqu'à l'article \cite{GrWu04-3}. Les graphes présentant ce mélange non seulement sont non renormalisables par un contreterme du lagrangien initial mais encore ne peuvent pas être régularisés par la renormalisation du graphe dans lequel ils sont insérés. L'exemple typique est celui du \og{}tadpole\fg{} non planaire de $\Phi^{4}$ qui diverge dans les fonctions à $6$ points ou plus.\\

Le modèle de Gross-Neveu \ncf{} orientable est renormalisable et ne présente donc pas de mélange UV/IR dangereux \ie{} non renormalisable, même en l'absence de vulcanisation ($\Omega =0$). Remarquons d'ailleurs que le comptage de puissance (le lemme \ref{lem:compt-puissGN}) est aussi valable pour la théorie complète (avec $V=V_{\text{o}}+V_{\text{no}}$) mais restreinte aux graphes orientables. Cela suggère que le modèle complet pourrait être renormalisable à $\Omega =0$ si restreint aux graphes orientables. Bien sûr, la \og{}localité\fg{} des contretermes devrait être vérifiée. Ceci est immédiat si on considère la théorie $\Phi^{4}$ naïve
\begin{align}
  S[\phi]=&\int d^{4}x\,\big(\frac 12\phi(-\Delta+m^{2})\phi+\frac{\lambda}{4}\phi\star\phi\star\phi\star\phi\big)(x)
\end{align}
restreinte aux graphes orientables. En effet nous pourrions exploiter les oscillations de vertex exactement de la même façon que pour le modèle de Gross-Neveu montrant ainsi que les seuls graphes divergents sont planaires réguliers ou critiques. Puis nous démontrerions que les contretermes sont de la forme du lagrangien initial. Le travail de Filk \cite{Filk1996dm} nous assure que les amplitudes des graphes réguliers planaires sont les mêmes que dans la théorie commutative.
\begin{rem}
Filk a démontré l'équivalence de la théorie commutative et du secteur planaire de la théorie \ncv{} en espace des moments. L'absence de contretermes en $\xt^{2}$ (ou $d^{2}/dp^{2}$) provient de la conservation des moments le long des lignes du graphe. En effet, nous avons vu, en espace $x$, que les contretermes en $\nabla$ proviennent du développement de la fonction delta globale autour d'un parallèlogramme parfait (voir section \ref{subsec:2pt-fct}). Or la conservation des moments le long des propagateurs, en espace $p$, nous assure que la fonction delta globale est une conservation des moments exactes. Nous savons également que la fonction delta en moments devient l'oscillation en $x$ et vice-versa. Ainsi le contreterme en $\xt^{2}$ doit provenir des oscillations en $x$. C'est bien le cas, il provient du développement des oscillations du type $U\wed X$. Or le corollaire \ref{sec:oscillRG} nous fournit l'oscillation de vertex des graphes réguliers planaires. Celle-ci contient des termes $u\wed x$. Si on veut démontrer la renormalisabilité de $\Phi^{4}$ à $\Omega =0$ en espace $x$, ces oscillations ne devraient pas être présentes car elles donneraient naissance à des contretermes en $\xt^{2}$. En fait, nous pouvons vérifier que $\varphi_{X}=0$. En effet, l'oscillation totale a été écrite à partir du choix arbitraire d'un sens de rotation autour de l'arbre. En écrivant simplement que l'oscillation $\varphi$ est la demi somme des oscillations écrites en tournant dans le sens horaire et le sens trigonomètrique, nous montrons qu'il n'y a pas d'oscillation du type $U\wed X$\footnote{C'est valable plus généralement pour tous les graphes orientables.}. 
\end{rem}
Le fait qu'une théorie $\phi^{4}$ \ncv{} orientable soit renormalisable sans vulcanisation avait été conjecturé dans \cite{Chepelev2000hm}. L'orientabilité semble donc être une solution au mélange UV/IR non renormalisable et donc une alternative à la vulcanisation. Néanmoins, en l'absence d'arguments généraux en faveur des interactions orientables, nous devons également considérer les interactions non orientables pour lesquelles la seule solution connue pour les rendre renormalisables est la vulcanisation.


%% file: conclusion-arXiv.tex
\renewcommand{\theequation}{\arabic{equation}}
\chapter*{Conclusion et perspectives}
\addcontentsline{toc}{chapter}{\protect\numberline{}Conclusion et perspectives}
\chaptermark{Conclusion et perspectives} 
\epigraph{Wir müssen wissen. Wir werden wissen.}{Épitaphe de David Hilbert}

Cette thèse a pour cadre la renormalisation des théories de champs \ncv{}s. Plus précisément, nous avons étudié la renormalisabilité de plusieurs modèles définis sur un espace de Moyal. Jusqu'à récemment, les théories \ncv{}s étaient définies à partir de leurs homologues commutatifs en remplaçant le produit point par point par le produit de Moyal. Mais cette méthode ne fonctionne pas à cause d'un nouveau type de divergences, le mélange UV/IR qui rend ces modèles non renormalisables. En 2005, Harald Grosse et Raimar Wulkkenhaar ont publié la preuve, à tous les ordres de perturbation, de la renormalisabilité d'un modèle du type $\phi^{4}$ sur l'espace de Moyal quadri-dimensionnel \cite{GrWu04-3}. L'action de ce modèle est :
\begin{equation}
S[\phi] = \int d^4x \Big( -\frac{1}{2} \partial_\mu \phi
\star \partial^\mu \phi + \frac{\Omega^2}{2} (\tilde{x}_\mu \phi )
\star (\tilde{x}^\mu \phi ) + \frac{1}{2} m^2
\,\phi \star \phi
+ \frac{\lambda}{4} \phi \star \phi \star \phi \star
\phi\Big)(x)\notag
\end{equation}
avec $\xt_\mu=2(\Theta^{-1}x)_{\mu}$. La nouveauté réside dans le terme supplémentaire de potentiel harmonique. À ce jour, nous ne savons pas encore si la vulcanisation s'applique à d'autres espaces que le plan de Moyal. De plus, nous ne savons pas non plus interpréter convenablement ce terme supplémentaire. Afin de répondre à ces interrogations, nous devons d'abord étudier la procédure de vulcanisation sur plusieurs modèles afin d'en dégager les caractéristiques principales. Cette thèse s'inscrit dans cette démarche.\\

Les propriétés fines du groupe de renormalisation sont bien capturées par l'analyse multi-échelles (section \ref{sec:multiscalephi4}). Cet outil mathématique puissant permet de clarifier la structure de la série perturbative. À chaque ligne du graphe, nous attribuons une échelle. Plus celle-ci est grande, plus la longueur de la ligne est courte. Autrement dit, plus l'indice de la ligne est haut, plus le propagateur correspondant se trouve dans l'ultraviolet. Le premier message de l'analyse multi-échelles est que seuls divergent les sous-graphes dont toutes les lignes internes sont plus hautes que toutes les pattes externes. Ces graphes-ci sont renormalisables car ils \og{}ressemblent\fg{} à des contretermes locaux : ils sont quasi-locaux.

Ces composantes connexes sont naturellement organisées en forêt, au sens de Zimmermann ce qui d'ailleurs, règle le problème des divergences enchevêtrées. Pour rendre les graphes finis, il suffit de renormaliser les sous-graphes quasi-locaux. Dans ce cas, l'objet obtenu est une multi-série effective. Celle-ci développe n'importe quelle fonction de la théorie en puissances d'une infinité de constantes de couplage effectives, une par échelle. Cette série effective est manifestement le bon objet à considérer. Du point de vue mathématique, la série renormalisée (à la BPHZ) ne converge pas en général. Pour exprimer toute fonction comme une série entière en une unique constante renormalisée, nous sommes obligés de renormaliser les sous-graphes, convergents, qui ne sont pas quasi-locaux. Ces contributions créent les renormalons contribuant au comportement en $K^{n}n!$ de la série des perturbations ; le rayon de convergence de la série est alors nul. Dans les bons cas (fermions), la série est Borel sommable. Dans les mauvais ($\phi^{4}$), elle ne l'est pas (les renormalons forment une singularité sur l'axe réel positif du plan de Borel : $K>0$). Si le flot est borné, la série effective a un multi-rayon de convergence fini dans le cas fermionique et est Borel sommable dans le cas bosonique (à cause des singularités instantoniques). Du point de vue physique, la série effective suit les idées de Wilson sur le groupe de renormalisation. La constante effective $\lambda_{i}$ est la constante de couplage de la théorie effective à l'échelle $i$ \ie{} la théorie obtenue en moyennant sur les fluctuations d'échelles supérieures ou égales à $i+1$. Les constantes effectives forment un flot discret équivalent au flot continu habituel (voir \cite{DR} pour la construction d'un flot continu non perturbatif).\\

Dans cette thèse, nous avons utilisé l'analyse multi-échelles pour étudier la renormalisabilité de modèles \ncf{}s et en particulier, le phénomène de mélange UV/IR. Nous nous sommes principalement intéressés aux modèles $\Phi^{4}_{4}$ et Gross-Neveu, en espace $x$ et dans la base matricielle. La complexité technique de la preuve originale de la renormalisabilité de $\Phi^{4}_{4}$ dans la base matricielle nous a incité à redémontrer ce résultat en espace $x$. Bien sûr, il s'agissait seulement de s'habituer à la théorie des champs \ncv{} pour pouvoir ensuite étudier d'autres modèles, ce que j'ai fait avec le modèle de Gross-Neveu \ncf{}. Le travail sur $\Phi^{4}$ \cite{xphi4-05} nous a donné l'occasion de généraliser les résultats de Filk au cas de propagateurs qui ne conservent pas l'impulsion. Ceci nous permet d'exprimer les oscillations de vertex en termes des variables de lignes, adaptées à une méthode graphique. Ce résultat est utile car l'essentiel de la difficulté technique réside dans l'exploitation des oscillations. L'étude de ce modèle nous a aussi permis de définir la notion d'orientabilité d'un graphe \ncf{}. Tous les graphes non orientables sont convergents.

Le modèle de Gross-Neveu quant à lui réserve quelques surprises. Tout d'abord, la non localité du produit de Moyal implique l'existence de six interactions différentes. En nous restreignant aux trois interactions orientables, nous avons montré qu'elles sont stables sous l'action du groupe de renormalisation. Le propagateur, à $\Omega\neq 0$, est équivalent à une dérivée covariante en champ magnétique uniforme. Il est alors tentant d'interpréter le terme supplémentaire comme tel. Dans ce cas, comment ne pas penser à la théorie des cordes ? En effet, la théorie effective de champs qui résulte de cordres ouvertes sur une D-brane, en présence d'un champ de fond $B_{\mu\nu}$, est \ncv{}. Sa non commutativité est celle du plan de Moyal. La dérivée du modèle de Gross-Neveu est-elle covariante par rapport à ce champ $B$ ? 

La preuve de la renormalisabilité de ce modèle a mis en lumière la propriété d'orientabilité. Les interactions orientables ne conduisent qu'à des graphes orientables. Pour ceux-ci, il n'y a pas besoin de vulcanisation. Les oscillations de vertex suffisent. Ce résultat n'est pas trop étonnant. En effet, le mélange UV/IR est uniquement dû aux graphes non orientables (voir section \ref{sec:renorm-et-vulc}). C'est le cas du tadpole non planaire. Les graphes qui présentent du mélange sont dangereux non pas parce qu'ils sont finis à moments non exceptionnels et infinis dans une boucle mais surtout parce qu'ils ne sont pas renormalisables. Dans le cadre de l'analyse multi-échelles, le problème est clairement identifié. Seules les composantes connexes divergent. Pour une théorie commutative, cela implique que les graphes divergents sont quasi-locaux dans le sens où leur partie divergente est proche d'un contreterme local. Dans un modèle \ncf{},  une composante connexe n'est pas forcemment quasi-locale : la distance typique entre les points externes du graphe n'est pas nécessairement petite par rapport aux propagateurs externes. C'est la non localité du vertex qui en est responsable. En fait, nous avons montré que toutes les composantes connexes orientables sont quasi-locales. Pour les graphes non orientables tels que le tadpole non planaire, ce n'est pas le cas. Voilà schématiquement pourquoi un modèle avec interaction orientable est renormalisable sans vulcanisation. De plus, il semble que dans le cadre de la géométrie \ncv{}, seules les interactions orientables ont un sens.

Le modèle de Gross-Neveu a également révélé la présence d'un mélange UV/IR inoffensif \ie{} renormalisable. Il apparaît seulement dans les graphes à quatre points et deux faces brisées. Ceux-ci sont finis si les points externes sont intégrés contre des fonctions test ou des propagateurs externes mais divergent logarithmiquement quand ils sont insérés dans un plus grand graphe. Néanmoins ces graphes sont régularisés par la renormalisation du graphe dans lequel ils sont insérés. La présence de mélange n'est pas surprenante sur un espace dont les coordonnées obéissent à $\lsb x^{\mu},x^{\nu}\rsb=\imath\Theta^{\mu\nu}$. Ce qui est étonnant, c'est plutôt l'absence totale de mélange dans le modèle $\Phi^{4}_{4}$.

Notons aussi que la renormalisation des sous-graphes qui présentent du mélange oblige, pour le modèle massif, à introduire un contreterme proportionnel à $\psib\gamma^{5}\psi$ dont il serait bon d'expliquer l'apparition.\\

L'étude des modèles $\Phi^{4}_{4}$ et Gross-Neveu en espace $x$ nous a permis de mieux caractériser le mélange UV/IR. Le principal atout de cet espace est que nous y sommes habitués. Des techniques y sont développées pour lesquelles nous avons un certain savoir-faire. C'est d'ailleurs pourquoi l'étude constructive du modèle de Gross-Neveu \ncf{} sera faite en $x$. Les autres avantages de cet espace sont les suivants. Il permet de faire des comparaisons avec les théories commutatives. Il peut servir de première étape vers une formulation (et une quantification) des théories \ncv{}s indépendante de l'espace sous-jacent. Enfin, la seule restriction sur $\Omega$ dont j'ai eu besoin pour l'étude du modèle de Gross-Neveu est $\Omega< 1$ alors que les études menées dans la base matricielle nécessitent $2/3\les\Omega\les 1$. Cette restriction est quelque peu gênante si nous voulons construire la limite commutative.\\

La deuxième partie\footnote{En fait, dans l'ordre chronologique, il s'agit de la première partie.} de ma thèse a été consacrée à l'étude de $\Phi^{4}_{4}$ ainsi que du propagateur du modèle de Gross-Neveu, dans la base matricielle. Nous y avons adapté les méthodes de l'analyse multi-échelles. Il se trouve qu'il n'y a que peu de changements par rapport à l'espace $x$ dès lors qu'on considère le graphe dual (voir section \ref{sec:topologie-des-graphes}). Cette notion apparaît de façon centrale pour plusieurs raisons. Le graphe dual est l'objet naturel pour classer les variables indépendantes, c'est lui qui définit la quadrangulation de la variété sur laquelle le graphe (direct) est dessiné. Enfin le graphe dual est également important pour obtenir la représention paramétrique de $\Phi^{4}_{4}$ \cite{gurauhypersyman}.

L'étude du propagateur de Gross-Neveu a permis de développer une méthode de type point-col qui fournit une borne précise. Celle-ci remplace les quatre bornes nécesssaires au calcul du comptage de puissance. Il serait intéressant d'appliquer cette méthode au propagateur de $\Phi^{4}$ à $\Omega=0$. En effet, un des problèmes que j'aimerais aborder dans la suite est celui de la caractérisation du mélange UV/IR dans la base matricielle. Dans \cite{GrWu03-1}, Grosse et Wulkenhaar ont démontré un comptage de puissance général pour un modèle de matrices d'interaction $\Tr\phi^{4}$ et dont le propagateur obéit aux bornes \eqref{eq:delta0multiscale} et \eqref{eq:delta1multiscale} qui définissent les exposants $\delta_{0},\delta_{1}$. On constate qu'à $\Omega=0$, $\delta_{0}=1$ et $\delta_{1}=0$. Or une condition nécessaire à la renormalisabilité du modèle est $\delta_{0}+\delta_{1}\ges D=4$. On est alors tenté de tenir le raisonnement suivant. À $\Omega=0$, le modèle n'est pas renormalisable à cause du mélange UV/IR. Le fait que $\delta_{1}<\delta_{0}$ est équivalent au fait que le propagateur ne soit pas quasi-local. On peut donc relier la non localité du propagateur au mélange UV/IR. Cette image développée en section \ref{sec:comptage-matrix-general} est séduisante mais certainement incomplète. En effet, si l'on considère un modèle d'interaction $\Tr(\bar{\phi}\phi)^{2}$ avec $p^2$ pour propagateur, on a aussi $\delta_{0}=1$ et $\delta_{1}=0$ mais le modèle est renormalisable (c'est l'équivalent de Gross-Neveu débarassé de ses difficultés algébriques liées aux spineurs). Bien sûr, ici une partie de l'information se trouve dans l'interaction complexe. Ainsi une étude comparative de ce modèle et de $\Tr\phi^{4}$ à $\Omega=0$ devrait permettre de mieux cerner le mélange UV/IR et l'orientabilité dans la base matricielle. Entre autres, cette étude permettra de distinguer les mélanges renormalisable et non renormalisable.

D'ailleurs, le modèle de Gross-Neveu, renormalisable, présente du mélange UV/IR. Bien qu'inoffensif (du point de vue de la renormalisation), ce mélange couple les échelles de la théorie et déstabilise notre image du groupe de renormalisation allant de l'ultraviolet à l'infrarouge. La base matricielle pourrait bien être mieux adaptée aux théories \ncv{}s que l'espace $x$. En effet, les indices de matrices appartiennent à $\N$ et non à $\Z$. Pour ainsi dire, il n'y a plus qu'une seule direction à l'infini. Au lieu d'intégrer les contributions de la théorie de l'infiniment petit à l'infiniment grand, ne vaudrait-il pas mieux, dans un espace \ncf{}, considérer qu'on intègre \og{}de l'infiniment loin de nous à nous\fg{} ?

D'autre part, dans le but de quantifier la gravitation, il faudrait être capable d'écrire une théorie sans référence à l'espace sous-jacent. À plus court terme, nous aimerions étudier une théorie des champs sur un espace \ncf{} qui ne soit pas une déformation. Pour cela, nous devons nous passer de l'espace $x$. La base matricielle fait un pas dans cette direction.\\

Enfin, en plus des quelques thèmes de recherche que j'ai mentionnés jusque là, je voudrais donner quatre grandes directions que j'aimerais suivre. Tout d'abord, il me semble important de tester la généralité de la vulcanisation. Pour cela, il faut étudier la renormalisabilité de modèles simples sur d'autres espaces que le plan de Moyal. On pourrait commencer par le tore \ncf{}. De plus, il serait intéressant de refaire l'analyse très générale de \cite{Gayral2004cs} avec le bon propagateur de $\Phi^{4}$ et également dans le cadre d'un modèle complexe avec dérivée covariante afin de bien différencier les mélanges dangereux et inoffensif sur différents espaces. En parallèle, pour essayer d'interpréter le terme de vulcanisation, nous devons nous intéresser à des modèles plus \og{}réalistes\fg{}. Bien sûr, je pense à Yang-Mills. La tâche est très ardue, ce problème a déjà été abordé par plusieurs groupes.

Les théories de champs \ncv{}s ont été introduites dans l'espoir d'éviter les divergences ultraviolettes. Il est clair que ce n'est pas le cas. Cependant le terme de potentiel harmonique de $\Phi^{4}$ agit comme un régulateur sur le flot de la constante de couplage \cite{GrWu04-2}. Cette propriété de régulation du flot est-elle générique ? La première chose à faire est de calculer la fonction beta du modèle de Gross-Neveu.

Le fait que le flot du modèle $\phi^{4}_{4}$ commutatif soit non borné est le principal obstacle à une définition constructive du modèle. Sur le plan de Moyal, le flot est borné ce qui devrait permettre de construire $\Phi^{4}_{4}$. Avant cela, nous allons étudier le modèle de Gross-Neveu \ncf{} sous l'angle constructif. Les modèles fermioniques étant plus simples à construire que les théories bosoniques, cette étude permettra de se familiariser avec les méthodes constructives \ncv{}s.

Enfin, certaines théories des champs \ncv{}s apparaissent comme limites de théories des cordes. Jusqu'à présent, ces théories ne sont pas renormalisables. Existe-t-il des systèmes de cordes qui donnent lieu à des théories de champs vulcanisées ?


%% file: GNapp-arXiv.tex
\numberwithin{equation}{section}
\chapter{À propos du modèle de Gross-Neveu}
\label{sec:a-propos-GN}

\section{Intégration par parties}
\label{sec:integration-parts}
Nous reproduisons ici les détails du calcul montrant que la procédure formée du changement de variables \eqref{eq:chgtvarVl} et de l'intégration par parties \eqref{eq:int-part} permet d'obtenir une fonction décroissante de l'échelle voulue.
\begin{align}
  A_{G,l}=\int&
  da_{l}dt_{l}\,\coth(2\Ot
  t_{l})\xi(a_{l}\coth^{1/2}(2\Ot t_{l}))\,
  e^{-\frac{\Ot}{2}\coth(2\Ot
    t_{l})(u_{l}-\veps(l)a_{l})^{2}}f_{1}(x_{1}+\eta(1)\veps(l)a_{l})\notag\\
  &\hspace{-.3cm}\lb\imath\Ot\coth(2\Ot
  t_{l})(\epsilon\veps)(l)(\us_{l}-\veps(l)\slashed{a}_{l})+\Omega(\epsilon\veps)(l)
  (\uts_{l}-\veps(l)\slashed{\tilde{a}}_{l})-m\rb e^{\imath a_{l}\wed(U_{l}+A_{l}+X_{l})}\nonumber\\
  &\prod_{\mu=0}^{1}\lbt\frac{\coth^{1/2}(2\Ot t_{l})+\frac{\partial}{\partial
      a_{l}^{\mu}}}{\coth^{1/2}(2\Ot
    t_{l})+\imath\cVt_{l,\mu}}\rbt^{\!\!\!2} e^{\imath a_{l}\wed\cV_{l}}.\tag*{\eqref{eq:int-part}}
  \intertext{Soit $\kc_{l}\defi\coth(2\Ot t_{l})$.}
  A_{G,l}=\int&da_{l}\,\kc_{l}\, e^{\imath
    a_{l}\wed\cV_{l}}\prod_{\mu=0}^{1}\lbt\frac{1}{\sqrt{\kc_{l}}+\imath\cVt_{l,\mu}}\rbt^{2}\lbt\sqrt{\kc_{l}}
  -\frac{\partial}{\partial
    a_{l}^{\mu}}\rbt^{2}\xi(a_{l}\sqrt{\kc_{l}})f_{1}(x_{1}+\eta(1)\veps(l)a_{l})\notag\\
  &\hspace{-1.2cm}\lb\imath\Ot\kc_{l}(\epsilon\veps)(l)(\us_{l}-\veps(l)\slashed{a}_{l})+\Omega(\epsilon\veps)(l)
  (\uts_{l}-\veps(l)\slashed{\tilde{a}}_{l})-m\rb e^{-\frac{\Ot}{2}
    \kc_{l}(u_{l}-\veps(l)a_{l})^{2}+\imath a_{l}\wed(U_{l}+A_{l}+X_{l})}.\notag
\end{align}
Nous utiliserons les notations suivantes :
\begin{align}
  \lb l\rb=&\imath\Ot\kc_{l}(\epsilon\veps)(l)(\us_{l}-\veps(l)\slashed{a}_{l})
  +\Omega(\epsilon\veps)(l)(\uts_{l}-\veps(l)\slashed{\tilde{a}}_{l})-m,\\
  \lb l\rb'\!=&-\epsilon(l)\lbt\imath\Ot\kc_{l}\gamma^{\mu}+\Ot(-1)^{\mu+1}\gamma^{\mu+1}\rbt.
\end{align}
Calculons la première dérivée :
\begin{align}
  &\frac{\partial}{\partial
    a_{l}^{\mu}}\,e^{-\frac{\Ot}{2}
    \kc_{l}(u_{l}-\veps(l)a_{l})^{2}+\imath a_{l}\wed(U_{l}+A_{l}+X_{l})}\xi(a_{l}\sqrt{\kc_{l}})f_{1}(x_{1}+\eta(1)\veps(l)a_{l})\lb l\rb\\
  =&e^{-\frac{\Ot}{2}\kc_{l}(u_{l}-\veps(l)a_{l})^{2}+\imath a_{l}\wed(U_{l}+A_{l}+X_{l})} \Big\{\{l\}\big[\Ot\kc_{l}\veps(l)(u_{l}-\veps(l)a_{l})^{\mu}\xi f_{1}
  +\imath(\Ut_{l}+\At_{l}+\Xt_{l})_{\mu}\xi f_{1}\notag\\
  &+\sqrt{\kc_{l}}\xi'f_{1}+\eta(1)\veps(l)\xi f'_{1}\big]+\lb l'\rb\xi f_{1}\Big\}\notag\\
  \defi&e^{-\frac{\Ot}{2}\kc_{l}(u_{l}-\veps(l)a_{l})^{2}+\imath a_{l}\wed(U_{l}+A_{l}+X_{l})} \Big\{\{l\}B_{0}\Xi_{0}F_{0}+\{l'\}\Xi_{1}F_{1}\Big\}+\text{ termes sous-dominants}
\end{align}
où $B_{0}$ ne dépend pas de $a_{l}^{\mu+1}$ et, en utilisant une partie de la décroissance exponentielle en $u_{l}-\veps(l)a_{l}$, $B_{0}=\cO(\sqrt{\kc_{l}})$. $\Xi_{i},\,i=0,1$ (resp. $F_{i}$) est une combinaison linéaire de $\xi$ (resp. $f_{1}$) et de ses dérivées. Puis calculons la dérivée seconde :
\begin{align}
    &\frac{\partial^{2}}{\partial
    (a_{l}^{\mu})^{2}}\,e^{-\frac{\Ot}{2}
    \kc_{l}(u_{l}-\veps(l)a_{l})^{2}+\imath a_{l}\wed(U_{l}+A_{l}+X_{l})}\xi(a_{l}\sqrt{\kc_{l}})f_{1}(x_{1}+\eta(1)\veps(l)a_{l})\lb l\rb\\
  =&e^{-\frac{\Ot}{2}\kc_{l}(u_{l}-\veps(l)a_{l})^{2}+\imath a_{l}\wed(U_{l}+A_{l}+X_{l})} \Big\{\{l\}\big[\big(\Ot\kc_{l}\veps(l)(u_{l}-\veps(l)a_{l})^{\mu}\xi f_{1}
  +\imath(\Ut_{l}+\At_{l}+\Xt_{l})_{\mu}\xi f_{1}\notag\\
  &+\sqrt{\kc_{l}}\xi'f_{1}+\eta(1)\veps(l)\xi f'_{1}\big)\times\big(\Ot\kc_{l}\veps(l)(u_{l}-\veps(l)a_{l})^{\mu}
  +\imath(\Ut_{l}+\At_{l}+\Xt_{l})_{\mu}\big)\notag\\
  &-\Ot\kc_{l}\veps(l)\xi f_{1}+\Ot\kc_{l}^{3/2}\veps(l)(u_{l}-\veps(l)a_{l})^{\mu}\xi'f_{1}+\Ot\eta(1)\kc_{l}(u_{l}-\veps(l)a_{l})^{\mu}\xi f'_{1}\notag\\
  &+\imath\sqrt{\kc_{l}}(\Ut_{l}+\At_{l}+\Xt_{l})_{\mu}\xi' f_{1}+\imath\eta(1)\veps(l)(\Ut_{l}+\At_{l}+\Xt_{l})_{\mu}\xi f'_{1}\notag\\
  &+\kc_{l}\xi''f_{1}+2\sqrt\kc_{l}\eta(1)\veps(l)\xi'f'_{1}+\xi f''_{1}\big]\notag\\
  &+2\{l'\}\big[\Ot\kc_{l}\veps(l)(u_{l}-\veps(l)a_{l})^{\mu}\xi f_{1}
  +\imath(\Ut_{l}+\At_{l}+\Xt_{l})_{\mu}\xi f_{1}+\sqrt\kc_{l}\xi'f_{1}+\eta(1)\veps(l)\xi f'_{1}\big]\Big\}\notag\\
  \defi&e^{-\frac{\Ot}{2}\kc_{l}(u_{l}-\veps(l)a_{l})^{2}+\imath a_{l}\wed(U_{l}+A_{l}+X_{l})} \Big\{\{l\}\big(B_{1}\Xi_{2}^{2}F_{2}^{2}+B_{2}\Xi_{3}F_{3}\big)+2\{l'\}B_{3}\Xi_{4}F_{4}\Big\}\\
  &+\text{ termes sous-dominants}\notag
\end{align}
où $B_{1},B_{2}=\cO(\kc_{l})$, $B_{3}=\cO(\sqrt{\kc_{l}})$ et les $\Xi_{i}$ et $f_{i}$ sont définis comme précédemment. Encore une fois, les $B_{i}$ ne dépendent pas de $a_{l}^{\mu+1}$. Nous avons
\begin{align}
    &\prod_{\mu=0}^{1}\lbt\sqrt{\kc_{l}}-\frac{\partial}{\partial a_{l}^{\mu}}\rbt^{2}\,\xi f_{1}\{l\}e^{-\frac{\Ot}{2}\kc_{l}(u_{l}-\veps(l)a_{l})^{2}+\imath a_{l}\wed(U_{l}+A_{l}+X_{l})}\label{eq:order}\\
    =&\lbt\kc_{l}-2\sqrt\kc_{l}\frac{\partial}{\partial a_{l}^{1}}+\frac{\partial^{2}}{\partial (a_{l}^{1})^{2}}\rbt\Big(\kc_{l}\{l\}\xi f_{1}-2\sqrt\kc_{l}\big(\{l\}B_{0}\Xi_{0}F_{0}+\{l'\}\Xi_{1}F_{1}\big)\notag\\
    &+\{l\}\big(B_{1}\Xi_{2}^{2}F_{2}^{2}+B_{2}\Xi_{3}F_{3}\big)+2\{l'\}B_{3}\Xi_{4}F_{4}\Big)e^{-\frac{\Ot}{2}\kc_{l}(u_{l}-\veps(l)a_{l})^{2}+\imath a_{l}\wed(U_{l}+A_{l}+X_{l})}\notag\\
    &+\text{ termes sous-dominants.}\notag
\end{align}
Rappelons que pour tout $i=0,\dotsc,3$, $B_{i}$ ne dépend pas de $a_{l}^{1}$. Le lecteur courageux peut alors vérifier que l'équation \eqref{eq:order} donne des termes d'ordre $\cO(\kc_{l}^{5/2})$ ce qui implique \eqref{eq:intpart-result}.

\section{Les graphes du vide}
\label{sec:vacuum-graph}

\begin{figure}[htbp]
  \centering 
  \includegraphics[scale=1]{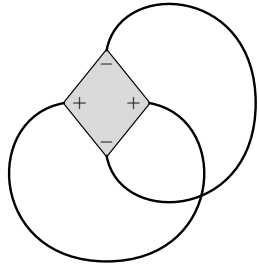}
  \caption[Un graphe du vide]{Exemple de graphe du vide non orientable}
  \label{fig:no-vacuum-ex}
\end{figure}

Dans cette annexe, nous calculons le comptage de puissance des graphes du vide
du modèle de Gross-Neveu orientable. Rappelons que dans les théories de champs
commutatives, les graphes du vide sont infinis dans une tranche \ie{} même en
présence de coupures ultraviolette et infrarouge. Ceci est dû à l'invariance par
translation de la théorie. En effet, considérons un graphe du vide $G$ de la
théorie $\phi^{4}$ commutative. Son amplitude s'écrit
\begin{align}
  A_{G}=&\ \lambda^{n}\int\prod_{v=1}^{n}dx_{v}\, \prod_{l\in G}C_{l}
\end{align}
où $n$ est le nombre de vertex de $G$. La covariance $C$ de la mesure
gaussienne associée à la théorie libre est l'inverse du Laplacien. Le noyau de
cet opérateur, le propagateur, est diagonal en espace des moments,
conserve l'impulsion et est donc invariant par translation en espace
$x$. Ainsi en choisissant un vertex $v_{0}$ au hasard, nous pouvons effectuer
le changement de variables de Jacobien $1$, $\forall v\neq v_{0},
x_{v}=y_{v}+x_{v_{0}}$. L'amplitude devient alors
\begin{align}
  A_{G}=&\ \lambda^{n}\int dx_{v_{0}} A(x_{v_{0}}),\\
  A(x_{v_{0}})=&\int\prod_{v\neq v_{0}}dy_{v}\, \prod_{l\in G}C_{l}\nonumber.
\end{align}
\`A cause de l'invariance par translation, $A(x_{v_{0}})$ est en fait
indépendant de $x_{v_{0}}$. Ainsi l'intégrale sur $x_{v_{0}}$ est infinie.
Au contraire, dans le cas de la théorie $\Phi^{4}$ non commutative, les graphes du vide
sont finis dans une tranche. Cependant la somme sur l'attribution des échelles
(l'équivalent de la limite coupure $\to\infty$ dans le formalisme
multi-échelles) est divergente comme $M^{8i}$. L'interaction quartique du
type Moyal est invariante par translation. En effet, elle s'écrit
\begin{align}
  &\delta(x_{1}-x_{2}+x_{3}-x_{4})\exp\imath\sum_{i<j=1}^{4}(-1)^{i+j+1}x_{i}\wed x_{j}\\
  =&\delta(x_{1}-x_{2}+x_{3}-x_{4})\exp\imath\lbt
  x_{1}\wed(x_{2}-x_{3})+x_{2}\wed x_{3}\rbt\nonumber\\
  =&\delta(x_{1}-x_{2}+x_{3}-x_{4})\exp\imath(x_{1}-x_{2})\wed(x_{2}-x_{3})\nonumber
\end{align}
Cette régularisation est donc uniquement due à la brisure de l'invariance
par translation par le terme harmonique $\xt^{2}$ du propagateur. Le
propagateur du modèle de Gross-Neveu, bien que brisant l'invariance par
translation, permet en fait d'obtenir une amplitude invariante par
translation. On est tenté d'avancer l'hypothèse suivante : $\ps
+\xts$ est une dérivée covariante correspondant à une théorie dans un champ
magnétique de fond constant. La physique est donc invariante par translation
mais pour l'écrire il faut fixer un potentiel vecteur qui brise cette
invariance. Concrètement nous vérifions l'invariance par translation de
l'amplitude d'un graphe quelconque de Gross-Neveu en effectuant le changement
de variables $\forall i,\, x_{i}\mapsto x_{i}+a$ et en vérifiant que le
résultat est indépendant de $a$.
\begin{align}
  A_{G}=&\lambda^{n}\int\prod_{l\in G}du_{l}dv_{l}\,C_{l}(u_{l},v_{l})\
  e^{\imath\varphi}\label{eq:AGN}\\
  =&\lambda^{n}\int\prod_{l\in G}du_{l}dv_{l}\,C_{l}(u_{l},v_{l}+2a)\
  e^{\imath\varphi}\nonumber
\end{align}
Dans l'équation (\ref{eq:AGN}), par souci de simplification, nous avons écrit $v_{l}$ quelle que soit la
ligne. Nous avons déjà remarqué que les oscillations de vertex sont
invariantes par translation. C'est pourquoi sous le changement de variables,
$\varphi$ reste indépendante de $a$. Si on considère une interaction du type
$\psib\psi\psib\psi$, les oscillations de propagateurs sont toujours
$\exp-\frac{\imath\Omega}{2}u_{l}\wed v_{l}$. Ainsi le changement
$v_{l}\mapsto v_{l}+2a_{l}$ implique la dépendance en $a$ de l'amplitude
\begin{equation}
  \label{eq:adepGN}
  \exp\imath\Omega a\wed\sum_{l\in G}u_{l}=1
\end{equation}
qui vaut $1$ car la somme de tous les $u$ est nulle par la fonction
delta globale\footnote{Pour une interaction de ce type, tous les graphes sont orientables.} \eqref{eq:deltaroot}. Ceci ne prouve pas que les graphes du vide sont infinis dans une tranche. En effet, une symétrie pourrait les rendre nuls. Néanmoins il est facile de constater sur l'exemple le plus simple qu'il existe au moins un graphe du vide infini. Pour une interaction non orientable, l'invariance par translation est de nouveau brisée. Le lecteur peut le vérifier sur l'exemple de la figure \ref{fig:no-vacuum-ex}.

\section{Contretermes (non) modifiés de la fonction à deux points}
\label{sec:modif-count-two}

Considérons une composante connexe à deux points $G^{j}_{k'}$ avec une sous-composante critique $G^{i}_{k}$. Nous allons prouver que si l'on met le terme $\gamma\Theta^{-1}\gamma$ de la plus basse ligne $\ell_{0}$ de $G^{j}_{k'}$ dans le contreterme, la partie divergente de la fonction à deux points reste de la forme du lagrangien initial \eqref{eq:actfunctGN}.

Pour simplifier nous utiliserons une notation allégée : $\exp-2\imath\Omega t_{\ell_{0}}\gamma\Theta^{-1}\gamma=$\\$\cosh(2\Ot t_{\ell_{0}})\bbbone_{2}-\imath\sinh(2\Ot t_{\ell_{0}})\gamma^{0}\gamma^{1}$. Comme expliqué dans la section \ref{subsec:4pt-fct}, les propagateurs d'un graphe à deux points se répartissent entre une chaîne et des cycles. Pour tout graphe $G$, écrivons $\bC$ l'ensemble des cycles et $\bCh$ l'ensemble des chaînes. Nous définissons également $T^{\mu}$ comme le nombre de variables $u^{\mu}$ provenant des développements de Taylor de la fonction delta et des oscillations\footnote{Par exemple, pour le terme de masse, le développement de Taylor ne donne pas de $u$ et $T^0=T^1=0$. Le contreterme de fonction d'onde donne $u^0\partial_{0}+u^1\partial_{1}$. Le premier terme aura $T^0=1$ et $T^1=0$, le second le contraire.}. Chaque cycle ou chaîne contient un produit de propagateurs. Soit $c\in\bC\,(\bCh)$,
\begin{align}
  P_{c}=&
  \begin{cases}
    \prod_{l\in c}(\imath\Ot\coth(2\Ot t_{l})\us_{l}+m)&\text{si }\ell_{0}\notin c\\
    (\imath\Ot\coth(2\Ot t_{\ell_{0}})\us_{\ell_{0}}+m)e^{-2\imath\Ot t_{\ell_{0}}\gamma^{0}\gamma^{1}}\prod_{l\in c\setminus\{\ell_{0}\}}(\imath\Ot\coth(2\Ot t_{l})\us_{l}+m)&\text{si }\ell_{0}\in c.
  \end{cases}
\end{align}
$P_{c}$ est la somme de différents termes : $P_{c}=\sum_{i=1}^{n}P_{c}^{i}$ où $n=3^{|c|}$ si $\ell_{0}\notin\bC$ et $n=2.3^{|c|+1}$ si $\ell_{0}\in\bC$ ($|c|=\card c$). Soit $|\gamma^{\mu}|_{c}^{i}$ le nombre total de $\gamma^{\mu}$ dans un terme $i$ fixé de $c\in\bC\,(\bCh)$. De la même façon, nous définissons $|u^{\mu}|_{c}^{i}$. Soit $i_{c}\in\lnat 1,n\rnat$ pour tout $c\in\bC\cup\bCh$. Le fait que les matrices gamma soient sans trace et les propriétés de parité des intégrales sur les variables $u$  impliquent deux contraintes :
\begin{align}
  \forall c\in\bC,\,\forall i\in\lnat 1,2^{|c|}\rnat,\,\forall\mu\in\{0,1\},\,|\gamma^{\mu}|^{i}_{c} \text{ est}&\text{ pair},\label{eq:cycle-constr}\\
  \forall\mu\in\{0,1\},\,\sum_{c\in\bC\cup\bCh}|u^{\mu}|^{i_{c}}_{c}+T^{\mu}\text{ est}&\text{ pair.}\label{eq:chain-constr}
\end{align}
À partir de maintenant, nous fixons une suite $(i_{c})_{c\in\bC\cup\bCh}$ à valeurs dans $\N$. Rappelons que pour les graphes de la fonction à deux points, $|\bCh|=1$ et que le nombre total de lignes internes est impair : $\sum_{c\in\bC\cup\bCh}|c|$ est impair. Pour $\ell_{0}$, nous choisirons toujours le terme $\gamma^{0}\gamma^{1}$ sinon l'analyse est la même que dans la section \ref{subsec:2pt-fct}. Dans la suite, nous appellerons \og{}contreterme de masse\fg{} l'expression (\ref{eq:massterm}) avec le développement (\ref{eq:taylor-propa-crit}), \og{}contreterme $\ps$ (ou $\xts$)\fg{} l'équation (\ref{eq:waveterm}) (ou (\ref{eq:Omegaterm})) encore une fois avec le développement (\ref{eq:taylor-propa-crit}).
\begin{enumerate}
\item Soit $c_{1}\in\bCh$. Si $|c_{1}|$ (le nombre de lignes dans la chaîne) est pair
  \begin{enumerate}
  \item et $\ell_{0}\in c_{1}$, $\sum_{c\in\bC}|c|$ est impair. L'équation (\ref{eq:cycle-constr}) implique $\forall\mu,\,\sum_{c\in\bC}|u^{\mu}|^{i_{c}}_{c}$ pair. Le nombre total de lignes dans les cycles étant impair, nous avons choisi la masse pour au moins une ligne dans $\bC$.
    \begin{itemize}
    \item Pour le contreterme de masse, $T^0=T^1=0$. L'équation (\ref{eq:chain-constr}) implique $|u^{\mu}|^{i_{c_{1}}}_{c_{1}}$ pair. Ainsi $|\gamma^{\mu}|^{i_{c_{1}}}_{c_{1}}$ impair pour $\mu=0$ et $1$. Le contreterme est proportionnel à $\gamma^0\gamma^1$.
    \item Pour les contretermes $\ps$ ou $\xts$, soit $\mu\in\Z_{2},\,T^\mu=1$ et $T^{\mu+1}=0$. $|\gamma^{\mu}|^{i_{c_{1}}}_{c_{1}}$ est pair et $|\gamma^{\mu+1}|^{i_{c_{1}}}_{c_{1}}$ est impair. Le nombre de lignes dans $c_{1}$ étant pair, au moins une ligne de $c_{1}$ \og{}a choisi\fg{} la masse. Ce terme est donc d'ordre $M^{-i}$. De tels termes donneraient $\pts$ ou $\xs$.
    \end{itemize}
  \item Soit $\ell_{0}\notin c_{1}$. L'équation (\ref{eq:cycle-constr}) implique $\forall\mu,\,\sum_{c\in\bC}|u^{\mu}|^{i_{c}}_{c}$ impair. Nous avons choisi la masse au moins une fois.
    \begin{itemize}
    \item Contreterme de masse : $|u^{\mu}|^{i_{c_{1}}}_{c_{1}}$ est impair. Ce contreterme est proportionnel à $\gamma^{0}\gamma^{1}$.
    \item Contreterme $\ps$ ($\xts$) : $|\gamma^{\mu}|^{i_{c_{1}}}_{c_{1}}$ est pair et $|\gamma^{\mu+1}|^{i_{c_{1}}}_{c_{1}}$ est impair. Ce terme donne $\pts$ ou $\xs$ mais est convergent comme $M^{-i}$ car $|c_{1}|$ est pair et au moins une ligne de $c_{1}$ porte un terme de masse. 
    \end{itemize}
  \end{enumerate}
\item Si $|c_{1}|$ est impair
  \begin{enumerate}
  \item Soit $\ell_{0}\in c_{1}$. $\sum_{c\in\bC}|u^{\mu}|_{c}^{i_{c}}$ est pair. 
    \begin{itemize}
    \item Contreterme de masse : les $|\gamma^{\mu}|^{i_{c_{1}}}_{c_{1}}$ sont impairs. Cela donne $\psib\gamma^{0}\gamma^{1}\psi$.
    \item Contreterme $\ps$ ($\xts$) : $|\gamma^{\mu}|^{i_{c_{1}}}_{c_{1}}$ est pair et $|\gamma^{\mu+1}|^{i_{c_{1}}}_{c_{1}}$ est impair. Ce terme donne $\pts$ ou $\xs$ mais est convergent comme $M^{-(i-j)}$. Le nombre de lignes dans $c_{1}$ étant impair, soit toutes les lignes de $c_{1}$ portent le terme en $u$ ou au moins deux d'entre elles ont la masse. 
    \end{itemize}
  \item Soit $\ell_{0}\notin c_{1}$. Les $\sum_{c\in\bC}|\gamma^{\mu}|^{i_{c}}_{c}$ sont impairs. Soit toutes les lignes de $\bC$ portent le terme en $u$ (le nombre total de lignes dans $\bC$ est pair) ou au moins deux d'entre elles ont la masse. Les termes correspondants sont d'ordre $M^{-(i-j)}$.
    \begin{itemize}
    \item Contreterme de masse : les $|\gamma^{\mu}|^{i_{c_{1}}}_{c_{1}}$ sont impairs. Nous obtenons $\psib\gamma^{0}\gamma^{1}\psi$.
    \item Contreterme $\ps$ ($\xts$) : $|\gamma^{\mu}|^{i_{c_{1}}}_{c_{1}}$ est pair et $|\gamma^{\mu+1}|^{i_{c_{1}}}_{c_{1}}$ est impair. Ce terme donne $\pts$ ou $\xs$.
    \end{itemize}
  \end{enumerate}
\end{enumerate}
En conclusion, le terme de masse ne donne que $\psib\gamma^{0}\gamma^{1}\psi$. Les contretermes $\ps$ et $\xts$ donnent $\psib\pts\psi$ et $\psib\xs\psi$ qui ne sont pas dans le lagrangian initial, mais ces termes sont convergents et peuvent donc être laissés dans l'amplitude renormalisée. Nous pouvons définir les nouveaux contretermes par
\begin{align}
  \tau'A_{m}=&\frac 12\Tr(\tau A_{m}),\\
  \tau'A_{\delta m}=&-\frac 12\gamma^{0}\gamma^{1}\Tr(\gamma^{0}\gamma^{1}\tau A_{m}),\\
  \tau'A_{\ps}=&-\frac{\ps}{2p^{2}}\Tr(\ps\tau A_{\ps}),\\
  \tau'A_{\xts}=&-\frac{\xts}{2\xt^{2}}\Tr(\xts\tau A_{\xts}).
\end{align}
Remarquons que si $m=0$, $\tau A_{m}\equiv 0$. Cela signifie que si la masse nue est nulle, la masse renormalisée l'est aussi et aucun contreterme du type $\psib\gamma^{0}\gamma^{1}\psi$ n'apparaît.

\section{Les tadpoles}
\label{sec:tadpolesGN}

Nous nous proposons ici de calculer les tadpoles du modèle de
Gross-Neveu. Nous considèrerons les six interactions possibles.

\subsection{Interactions orientables}
\label{subec:altint}

Avec ces interactions, seuls les tadpoles planaires sont possibles.

\subsubsection{$\psib_{a}\star\psi_{a}\star\psib_{b}\star\psi_{b}$}

\begin{figure}[!hbt]
  \centering
  \subfloat[Régulier]{\label{fig:tad-plan-reg11}\includegraphics[scale=1]{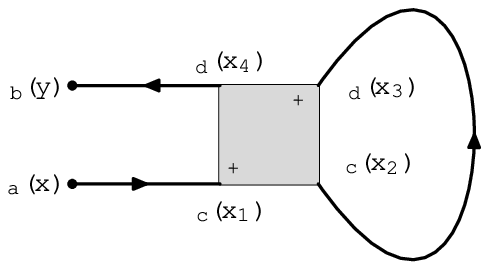}}\qquad
  \subfloat[Singulier]{\label{fig:tad-plan-sing11}\includegraphics[scale=1]{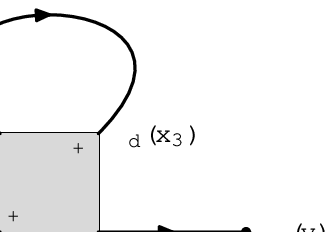}}
  \caption{Tadpoles planaires}
  \label{fig:tad-plan}
\end{figure}

L'amplitude amputée du graphe de la figure \ref{fig:tad-plan-reg11} est
\begin{align}
  A_{\ref{fig:tad-plan-reg11}}=\int&
  dx_{2}dx_{3}\,\delta(x_{1}-x_{2}+x_{3}-x_{4})e^{\imath x_{4}\wed
    x_{1}+\imath x_{2}\wed x_{3}}C_{cd}(x_{2},x_{3}).
\end{align}

En utilisant l'expression du propagateur donnée par le lemme \ref{xpropa1GN}, l'amplitude s'écrit
\begin{align}
  A_{\ref{fig:tad-plan-reg11}}=&-\frac{\Omega}{\theta\pi}\int dt dx_{2}
  \frac{e^{-tm^{2}}}{\sinh(2\Ot t)} e^{\imath x_{4}\wed
    x_{1}+\imath(1+\Omega)x_{2}\wed(x_{4}-x_{1})}e^{-\frac{\Ot}{2}\coth(2\Ot t)(x_{1}-x_{4})^{2}}\notag\\
  &\lsb\lbt\imath\Ot\coth(2\Ot
  t)(\xs_{1}-\xs_{4})+\Omega(\xts_{1}-\xts_{4})-m\rbt e^{-2\imath\Ot
    t\gamma^{0}\gamma^{1}}\rsb_{cd}\notag\\
  =&\frac{\theta\Omega m}{4\pi(1+\Omega)^{2}}\int\frac{e^{-tm^{2}}}{\sinh(2\Ot
    t)}\lsb e^{-2\imath\Ot
    t\gamma^{0}\gamma^{1}}\rsb_{cd}\delta(x_{1}-x_{4})\label{eq:ampli-amput-planartadreg11}
\end{align}
où nous avons utilisé, au sens des distributions, $\delta(Ax)=|\det
A|^{-1}\delta(x)$ pour toute matrice $A$ inversible.\\

L'amplitude amputée du graphe de la figure \ref{fig:tad-plan-sing11} est
\begin{align}
  A_{\ref{fig:tad-plan-sing11}}=&\int
  dx_{3}dx_{4}\,\delta(x_{1}-x_{2}+x_{3}-x_{4})e^{\imath x_{1}\wed
    x_{2}+\imath x_{3}\wed x_{4}}C_{dd}(x_{4},x_{3})\notag\\
  =&-\frac{\Omega}{\theta\pi}\int dtdx_{3} \frac{e^{-tm^{2}}}{\sinh(2\Ot t)}
  e^{\imath x_{1}\wed x_{2}+\imath(1-\Omega)x_{3}\wed(x_{1}-x_{2})}e^{-\frac{\Ot}{2}\coth(2\Ot t)(x_{1}-x_{2})^{2}}\notag\\
  &\lsb\lbt\imath\Ot\coth(2\Ot
  t)(\xs_{1}-\xs_{2})+\Omega(\xts_{1}-\xts_{2})-m\rbt
  e^{-2\imath\Ot
    t\gamma^{0}\gamma^{1}}\rsb_{dd}\notag\\
  =&\frac{\theta\Omega m}{4\pi(1-\Omega)^{2}}\int\frac{e^{-tm^{2}}}{\sinh(2\Ot
    t)}\lsb e^{-2\imath\Ot
    t\gamma^{0}\gamma^{1}}\rsb_{dd}\delta(x_{1}-x_{2}).\notag
\end{align}
Ici nous pouvons explicitement effectuer la somme sur l'indice $d$.
\begin{align}
  \sum_{d}A_{\ref{fig:tad-plan-sing11}}=&\frac{\theta\Omega m}{4\pi(1-\Omega)^{2}}\int\frac{e^{-tm^{2}}}{\tanh(2\Ot
  t)}\,\delta(x_{1}-x_{2}).\label{eq:ampli-amput-planartadsing11}
\end{align}
Remarquons que la limite $\Omega\to 1$ est singulière.

\subsubsection{$\psi_{a}\star\psib_{a}\star\psi_{b}\star\psib_{b}$}

\begin{figure}[!hbt]
  \centering
  \subfloat[Régulier]{\label{fig:tad-plan-reg12}\includegraphics[scale=1]{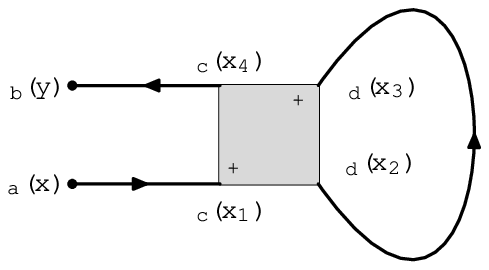}}\qquad
  \subfloat[Singulier]{\label{fig:tad-plan-sing12}\includegraphics[scale=1]{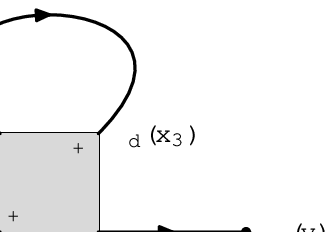}}
  \caption{Tadpoles planaires}
  \label{fig:tad-plan2}
\end{figure}

L'amplitude amputée du graphe de la figure \ref{fig:tad-plan-reg12} est
\begin{align}
  A_{\ref{fig:tad-plan-reg12}}=&\int
  dx_{2}dx_{3}\,\delta(x_{1}-x_{2}+x_{3}-x_{4})e^{\imath x_{4}\wed
    x_{1}+\imath x_{2}\wed x_{3}}C_{dd}(x_{2},x_{3})\\
  =&-\frac{\Omega}{\theta\pi}\int dt dx_{2}
  \frac{e^{-tm^{2}}}{\sinh(2\Ot t)} e^{\imath x_{4}\wed
    x_{1}+\imath(1+\Omega)x_{2}\wed(x_{4}-x_{1})}e^{-\frac{\Ot}{2}\coth(2\Ot t)(x_{1}-x_{4})^{2}}\notag\\
  &\lsb\lbt\imath\Ot\coth(2\Ot
  t)(\xs_{1}-\xs_{4})+\Omega(\xts_{1}-\xts_{4})-m\rbt e^{-2\imath\Ot
    t\gamma^{0}\gamma^{1}}\rsb_{dd}\notag\\
  =&\frac{\theta\Omega m}{4\pi(1+\Omega)^{2}}\int\frac{e^{-tm^{2}}}{\sinh(2\Ot
    t)}\lsb e^{-2\imath\Ot
    t\gamma^{0}\gamma^{1}}\rsb_{dd}\delta(x_{1}-x_{4}),\notag\\
  \sum_{d}A_{\ref{fig:tad-plan-reg12}}=&\frac{\theta\Omega m}{4\pi(1+\Omega)^{2}}\int\frac{e^{-tm^{2}}}{\tanh(2\Ot
    t)}\,\delta(x_{1}-x_{4}).\label{eq:ampli-amput-planartadreg12}
\end{align}

L'amplitude amputée du graphe de la figure \ref{fig:tad-plan-sing12} est
\begin{align}
  A_{\ref{fig:tad-plan-sing12}}=&\int
  dx_{3}dx_{4}\,\delta(x_{1}-x_{2}+x_{3}-x_{4})e^{\imath x_{1}\wed
    x_{2}+\imath x_{3}\wed x_{4}}C_{cd}(x_{4},x_{3})\notag\\
  =&-\frac{\Omega}{\theta\pi}\int dtdx_{3} \frac{e^{-tm^{2}}}{\sinh(2\Ot t)}
  e^{\imath x_{1}\wed x_{2}+\imath(1-\Omega)x_{3}\wed(x_{1}-x_{2})}e^{-\frac{\Ot}{2}\coth(2\Ot t)(x_{1}-x_{2})^{2}}\notag\\
  &\lsb\lbt\imath\Ot\coth(2\Ot
  t)(\xs_{1}-\xs_{2})+\Omega(\xts_{1}-\xts_{2})-m\rbt
  e^{-2\imath\Ot
    t\gamma^{0}\gamma^{1}}\rsb_{cd}\notag\\
  =&\frac{\theta\Omega m}{4\pi(1-\Omega)^{2}}\int\frac{e^{-tm^{2}}}{\sinh(2\Ot
    t)}\lsb e^{-2\imath\Ot
    t\gamma^{0}\gamma^{1}}\rsb_{cd}\delta(x_{1}-x_{2}).
\end{align}

\subsubsection{$\psib_{a}\star\psi_{b}\star\psib_{a}\star\psi_{b}$}

\begin{figure}[!hbt]
  \centering
  \subfloat[Régulier]{\label{fig:tad-plan-reg13}\includegraphics[scale=1]{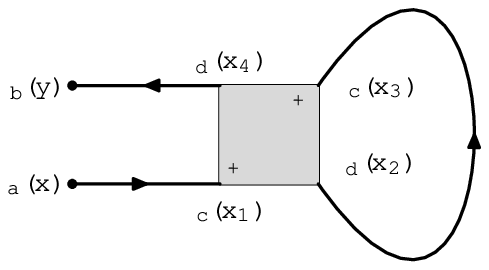}}\qquad
  \subfloat[Singulier]{\label{fig:tad-plan-sing13}\includegraphics[scale=1]{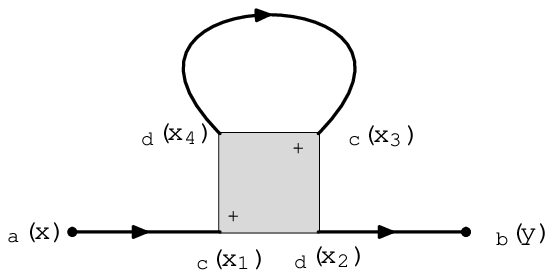}}
  \caption{Tadpoles planaires}
  \label{fig:tad-plan3}
\end{figure}

Le calcul étant très proche du cas de la figure \ref{fig:tad-plan}, nous
donnons directement le résultat. L'amplitude amputée du graphe de la figure \ref{fig:tad-plan-reg13} est
\begin{align}
  A_{\ref{fig:tad-plan-reg12}}=&\frac{\theta\Omega m}{4\pi(1+\Omega)^{2}}\int\frac{e^{-tm^{2}}}{\sinh(2\Ot
    t)}\lsb e^{-2\imath\Ot
    t\gamma^{0}\gamma^{1}}\rsb_{dc}\delta(x_{1}-x_{4}).\label{eq:ampli-amput-planartadreg13}
\end{align}

L'amplitude amputée du graphe de la figure \ref{fig:tad-plan-sing13} est
\begin{align}
  A_{\ref{fig:tad-plan-sing13}}=&\frac{\theta\Omega m}{4\pi(1-\Omega)^{2}}\int\frac{e^{-tm^{2}}}{\sinh(2\Ot
    t)}\lsb e^{-2\imath\Ot
    t\gamma^{0}\gamma^{1}}\rsb_{dc}\delta(x_{1}-x_{2}).
\end{align}

\subsection{Interactions non orientables}
\label{subec:couplint}

\subsubsection{$\psib_{a}\star\psib_{b}\star\psi_{a}\star\psi_{b}$}

\begin{figure}[!hbt]
  \centering
  \subfloat[Régulier]{\label{fig:tad-plan-reg21}\includegraphics[scale=1]{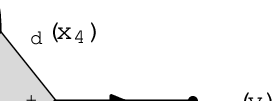}}\qquad
  \subfloat[Singulier]{\label{fig:tad-plan-sing21}\includegraphics[scale=1]{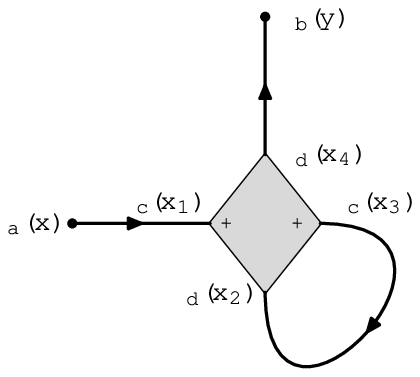}}
  \caption{Tadpoles planaires}
  \label{fig:tad-plan22}
\end{figure}

L'amplitude amputée du graphe de la figure \ref{fig:tad-plan-reg21} est
\begin{align}
  A_{\ref{fig:tad-plan-reg21}}=&\int
  dx_{1}dx_{4}\,\delta(x_{1}-x_{2}+x_{3}-x_{4})e^{\imath x_{4}\wed
    x_{1}+\imath x_{2}\wed x_{3}}C_{dc}(x_{4},x_{1})\\
  =&-\frac{\Omega}{\theta\pi}\int dt dx_{1}
  \frac{e^{-tm^{2}}}{\sinh(2\Ot t)} e^{\imath x_{2}\wed
    x_{3}+\imath(1+\Omega)(x_{3}-x_{2})\wed x_{1}}e^{-\frac{\Ot}{2}\coth(2\Ot t)(x_{2}-x_{3})^{2}}\notag\\
  &\lsb\lbt\imath\Ot\coth(2\Ot
  t)(\xs_{3}-\xs_{2})+\Omega(\xts_{3}-\xts_{2})-m\rbt e^{-2\imath\Ot
    t\gamma^{0}\gamma^{1}}\rsb_{dc}\notag\\
  =&\frac{\theta\Omega m}{4\pi(1+\Omega)^{2}}\int\frac{e^{-tm^{2}}}{\sinh(2\Ot
    t)}\lsb e^{-2\imath\Ot
    t\gamma^{0}\gamma^{1}}\rsb_{dc}\delta(x_{2}-x_{3}).\label{eq:ampli-amput-planartadreg21}
\end{align}

L'amplitude amputée du graphe de la figure \ref{fig:tad-plan-sing21} est
\begin{align}
  A_{\ref{fig:tad-plan-sing21}}=&\int
  dx_{2}dx_{3}\,\delta(x_{1}-x_{2}+x_{3}-x_{4})e^{\imath x_{4}\wed
    x_{1}+\imath x_{2}\wed x_{3}}C_{cd}(x_{3},x_{2})\notag\\
  =&-\frac{\Omega}{\theta\pi}\int dtdx_{2} \frac{e^{-tm^{2}}}{\sinh(2\Ot t)}
  e^{\imath x_{4}\wed x_{1}+\imath(1-\Omega)x_{2}\wed(x_{4}-x_{1})}e^{-\frac{\Ot}{2}\coth(2\Ot t)(x_{1}-x_{4})^{2}}\notag\\
  &\lsb\lbt\imath\Ot\coth(2\Ot
  t)(\xs_{4}-\xs_{1})+\Omega(\xts_{4}-\xts_{1})-m\rbt
  e^{-2\imath\Ot
    t\gamma^{0}\gamma^{1}}\rsb_{cd}\notag\\
  =&\frac{\theta\Omega m}{4\pi(1-\Omega)^{2}}\int\frac{e^{-tm^{2}}}{\sinh(2\Ot
    t)}\lsb e^{-2\imath\Ot
    t\gamma^{0}\gamma^{1}}\rsb_{cd}\delta(x_{1}-x_{4}).
\end{align}\\

\begin{figure}[!hbt]
  \centering
  \subfloat[Régulier]{\label{fig:tad-nonplan-reg21}\includegraphics[scale=1]{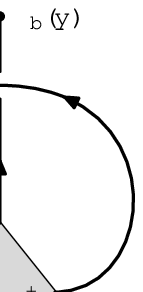}}\qquad
  \subfloat[Régulier]{\label{fig:tad-nonplan-sing21}\includegraphics[scale=1]{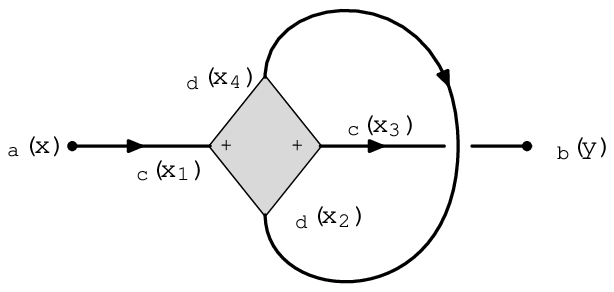}}
  \caption{Tadpoles non planaires}
  \label{fig:tad-nonplan}
\end{figure}

L'amplitude amputée du graphe de la figure \ref{fig:tad-nonplan-reg21} est
\begin{align}
  A_{\ref{fig:tad-nonplan-reg21}}=&\int
  dx_{1}dx_{3}\,\delta(x_{1}-x_{2}+x_{3}-x_{4})e^{\imath x_{1}\wed
    x_{2}+\imath x_{3}\wed x_{4}}C_{cc}(x_{3},x_{1})\\
  =&-\frac{\Omega}{\theta\pi}\int dt dx_{1} \frac{e^{-tm^{2}}}{\sinh(2\Ot t)}
  e^{\imath x_{1}\wed x_{2}+\imath (x_{2}-x_{1})\wed
    x_{4}+\imath\Omega(x_{2}+x_{4})\wed x_{1}}
  e^{-\frac{\Ot}{2}\coth(2\Ot t)(2x_{1}-x_{2}-x_{4})^{2}}\notag\\
  &\lsb\lbt\imath\Ot\coth(2\Ot
  t)(\xs_{2}+\xs_{4}-2\xs_{1})+\Omega(\xts_{2}+\xts_{4}-2\xts_{1})-m\rbt
  e^{-2\imath\Ot
    t\gamma^{0}\gamma^{1}}\rsb_{cc}\notag\\
  =&\frac{\Omega}{4\theta\pi}\int dt dX_{1} \frac{e^{-tm^{2}}}{\sinh(2\Ot t)}
  e^{\frac{\imath}{2}X_{1}\wed((1-\Omega)x_{2}-(1+\Omega)x_{4})}e^{-\frac{\Ot}{2}\coth(2\Ot
    t)X_{1}^{2}}\notag\\
  &\lsb\lbt\imath\Ot\coth(2\Ot
  t)\slashed{X}_{1}+\Omega\slashed{\tilde{X}}_{1}+m\rbt e^{-2\imath\Ot
    t\gamma^{0}\gamma^{1}}\rsb_{cc}\notag\\
  =&\frac{m}{4}\int dt\frac{e^{-tm^{2}}}{\cosh(2\Ot t)} e^{-\frac{\tanh(2\Ot
    t)}{8\Ot}((1-\Omega)x_{2}-(1+\Omega)x_{4})^{2}}\lsb e^{-2\imath\Ot
    t\gamma^{0}\gamma^{1}}\rsb_{cc},\notag\\
  \sum_{c}A_{\ref{fig:tad-nonplan-reg21}}=&\frac{m}{4}\int dt\, e^{-tm^{2}} e^{-\frac{\tanh(2\Ot
    t)}{8\Ot}((1-\Omega)x_{2}-(1+\Omega)x_{4})^{2}}.\label{eq:ampli-amput-planartadreg28}
\end{align}

L'amplitude amputée du graphe de la figure \ref{fig:tad-nonplan-sing21} est
\begin{align}
  A_{\ref{fig:tad-nonplan-sing21}}=&\int
  dx_{2}dx_{4}\,\delta(x_{1}-x_{2}+x_{3}-x_{4})e^{\imath x_{1}\wed
    x_{2}+\imath x_{3}\wed x_{4}}C_{dd}(x_{4},x_{2})\\
  =&-\frac{\Omega}{\theta\pi}\int dt dx_{2} \frac{e^{-tm^{2}}}{\sinh(2\Ot t)}
  e^{\imath x_{1}\wed x_{2}+\imath x_{3}\wed
    (x_{1}-x_{2})+\imath\Omega(x_{1}+x_{3})\wed x_{2}}
  e^{-\frac{\Ot}{2}\coth(2\Ot t)(2x_{2}-x_{1}-x_{3})^{2}}\notag\\
  &\lsb\lbt\imath\Ot\coth(2\Ot
  t)(\xs_{1}+\xs_{3}-2\xs_{2})+\Omega(\xts_{1}+\xts_{3}-2\xts_{2})-m\rbt
  e^{-2\imath\Ot
    t\gamma^{0}\gamma^{1}}\rsb_{dd}\notag\\
  =&\frac{\Omega}{4\theta\pi}\int dt dX_{2} \frac{e^{-tm^{2}}}{\sinh(2\Ot t)}
  e^{\frac{\imath}{2}X_{2}\wed((1-\Omega)x_{3}-(1+\Omega)x_{1})}e^{-\frac{\Ot}{2}\coth(2\Ot
    t)X_{2}^{2}}\notag\\
  &\lsb\lbt\imath\Ot\coth(2\Ot
  t)\slashed{X}_{2}+\Omega\slashed{\tilde{X}}_{2}+m\rbt e^{-2\imath\Ot
    t\gamma^{0}\gamma^{1}}\rsb_{dd}\notag\\
  =&\frac{m}{4}\int dt\frac{e^{-tm^{2}}}{\cosh(2\Ot t)} e^{-\frac{\tanh(2\Ot
    t)}{8\Ot}((1-\Omega)x_{3}-(1+\Omega)x_{1})^{2}}\lsb e^{-2\imath\Ot
    t\gamma^{0}\gamma^{1}}\rsb_{dd},\notag\\
  \sum_{d}A_{\ref{fig:tad-nonplan-sing21}}=&\frac{m}{4}\int dt\, e^{-tm^{2}} e^{-\frac{\tanh(2\Ot
    t)}{8\Ot}((1-\Omega)x_{3}-(1+\Omega)x_{1})^{2}}.\label{eq:ampli-amput-nonplanartadsing21}
\end{align}

\subsubsection{$\psib_{a}\star\psib_{b}\star\psi_{b}\star\psi_{a}$}

\begin{figure}[!hbt]
  \centering
  \subfloat[Régulier]{\label{fig:tad-plan-reg22}\includegraphics[scale=1]{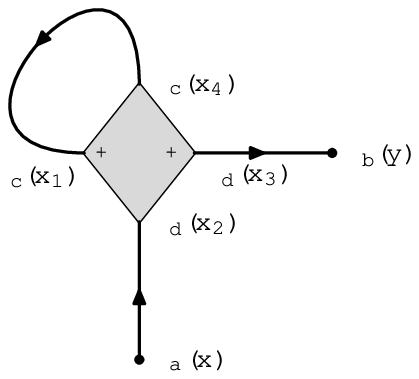}}\qquad
  \subfloat[Singulier]{\label{fig:tad-plan-sing22}\includegraphics[scale=1]{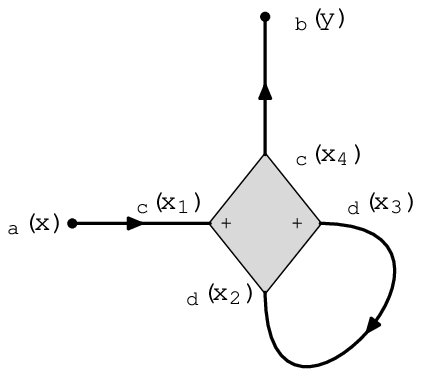}}
  \caption{Tadpoles planaires}
  \label{fig:tad-plan4}
\end{figure}

L'amplitude amputée du graphe de la figure \ref{fig:tad-plan-reg22} est
\begin{align}
  \sum_{c}A_{\ref{fig:tad-plan-reg22}}=&\frac{\theta\Omega m}{4\pi(1+\Omega)^{2}}\int\frac{e^{-tm^{2}}}{\tanh(2\Ot
    t)}\delta(x_{2}-x_{3}).\label{eq:ampli-amput-planartadreg22}
\end{align}

L'amplitude amputée du graphe de la figure \ref{fig:tad-plan-sing22} est
\begin{align}
  \sum_{d}A_{\ref{fig:tad-plan-sing22}}=&\frac{\theta\Omega m}{4\pi(1-\Omega)^{2}}\int\frac{e^{-tm^{2}}}{\tanh(2\Ot
    t)}\delta(x_{1}-x_{4}).
\end{align}\\

\begin{figure}[!hbt]
  \centering
  \subfloat[Régulier]{\label{fig:tad-nonplan-reg22}\includegraphics[scale=1]{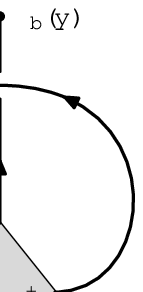}}\qquad
  \subfloat[Régulier]{\label{fig:tad-nonplan-sing22}\includegraphics[scale=1]{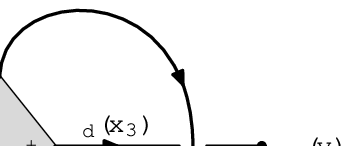}}
  \caption{Tadpoles non planaires}
  \label{fig:tad-nonplan2}
\end{figure}

L'amplitude amputée du graphe de la figure \ref{fig:tad-nonplan-reg22} est
\begin{align}
  A_{\ref{fig:tad-nonplan-reg22}}=&\frac{m}{4}\int dt\frac{e^{-tm^{2}}}{\cosh(2\Ot t)} e^{-\frac{\tanh(2\Ot
    t)}{8\Ot}((1-\Omega)x_{2}-(1+\Omega)x_{4})^{2}}\lsb e^{-2\imath\Ot
    t\gamma^{0}\gamma^{1}}\rsb_{dc}.
\end{align}

L'amplitude amputée du graphe de la figure \ref{fig:tad-nonplan-sing22} est
\begin{align}
  A_{\ref{fig:tad-nonplan-sing22}}=&\frac{m}{4}\int dt\frac{e^{-tm^{2}}}{\cosh(2\Ot t)} e^{-\frac{\tanh(2\Ot
    t)}{8\Ot}((1-\Omega)x_{3}-(1+\Omega)x_{1})^{2}}\lsb e^{-2\imath\Ot
    t\gamma^{0}\gamma^{1}}\rsb_{cd}.
\end{align}

\subsubsection{$\psib_{a}\star\psib_{a}\star\psi_{b}\star\psi_{b}$}

\begin{figure}[!hbt]
  \centering
  \subfloat[Régulier]{\label{fig:tad-plan-reg23}\includegraphics[scale=1]{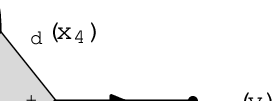}}\qquad
  \subfloat[Singulier]{\label{fig:tad-plan-sing23}\includegraphics[scale=1]{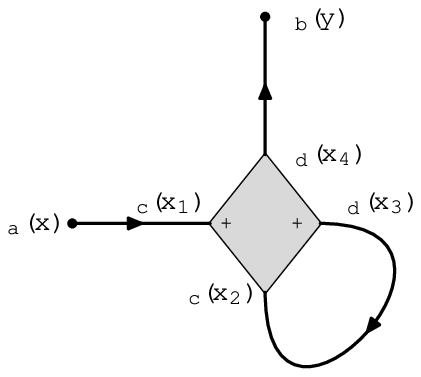}}
  \caption{Tadpoles planaires}
  \label{fig:tad-plan5}
\end{figure}

L'amplitude amputée du graphe de la figure \ref{fig:tad-plan-reg23} est
\begin{align}
  A_{\ref{fig:tad-plan-reg23}}=&\frac{\theta\Omega m}{4\pi(1+\Omega)^{2}}\int\frac{e^{-tm^{2}}}{\sinh(2\Ot
    t)}\lsb e^{-2\imath\Ot
    t\gamma^{0}\gamma^{1}}\rsb_{dc}\delta(x_{2}-x_{3}).\label{eq:ampli-amput-planartadreg23}
\end{align}

L'amplitude amputée du graphe de la figure \ref{fig:tad-plan-sing23} est
\begin{align}
  A_{\ref{fig:tad-plan-sing23}}=&\frac{\theta\Omega m}{4\pi(1-\Omega)^{2}}\int\frac{e^{-tm^{2}}}{\sinh(2\Ot
    t)}\lsb e^{-2\imath\Ot
    t\gamma^{0}\gamma^{1}}\rsb_{dc}\delta(x_{1}-x_{4}).
\end{align}\\

\begin{figure}[!hbt]
  \centering
  \subfloat[Régulier]{\label{fig:tad-nonplan-reg23}\includegraphics[scale=1]{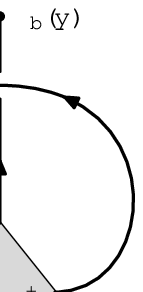}}\qquad
  \subfloat[Régulier]{\label{fig:tad-nonplan-sing23}\includegraphics[scale=1]{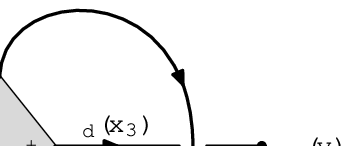}}
  \caption{Tadpoles non planaires}
  \label{fig:tad-nonplan3}
\end{figure}

L'amplitude amputée du graphe de la figure \ref{fig:tad-nonplan-reg23} est
\begin{align}
  A_{\ref{fig:tad-nonplan-reg23}}=&\frac{m}{4}\int dt\frac{e^{-tm^{2}}}{\cosh(2\Ot t)} e^{-\frac{\tanh(2\Ot
    t)}{8\Ot}((1-\Omega)x_{2}-(1+\Omega)x_{4})^{2}}\lsb e^{-2\imath\Ot
    t\gamma^{0}\gamma^{1}}\rsb_{dc}.
\end{align}

L'amplitude amputée du graphe de la figure \ref{fig:tad-nonplan-sing23} est
\begin{align}
  A_{\ref{fig:tad-nonplan-sing23}}=&\frac{m}{4}\int dt\frac{e^{-tm^{2}}}{\cosh(2\Ot t)} e^{-\frac{\tanh(2\Ot
      t)}{8\Ot}((1-\Omega)x_{3}-(1+\Omega)x_{1})^{2}}\lsb e^{-2\imath\Ot
    t\gamma^{0}\gamma^{1}}\rsb_{dc}.
\end{align}
